# Verifying Cryptographic Security Implementations in C

## Using Automated Model Extraction

Mihhail Aizatulin
Department of Computing
The Open University

A thesis submitted for the degree of

*Doctor of Philosophy*

January 2015

*To my family*

# Acknowledgements

I would like to thank my supervisors Andy Gordon, Jan Jürjens, and Bashar Nuseibeh for their guidance and support. This project was made possible by the Microsoft PhD Scholarship program. I am particularly grateful to Microsoft Research for providing me with an opportunity to be part of the research community in Cambridge—the weekly discussion organized by the Programming Principles and Tools group provided a lot of education and inspiration.

I would like to thank my family for putting up with my reclusive behaviour when staying with them in Estonia, and I promise to be much more sociable next time.

# Declaration

Unless explicitly stated otherwise the ideas and the work described in this dissertation are mine. The work on the symbolic execution (chapter 3) and the ProVerif verification (chapter 5) was presented at the ACM Conference on Computer and Communication Security (CCS'11) [Aizatulin et al., 2011b] and the work on the CryptoVerif verification (chapter 4) was presented at (CCS'12) [Aizatulin et al., 2012]. The theoretical framework developed in this dissertation is a significant improvement of the one used in those publications.

# Abstract

This thesis presents an automated method for verifying security properties of protocol implementations written in the C language. We assume that each successful run of a protocol follows the same path through the C code, justified by the fact that typical security protocols have linear structure. We then perform symbolic execution of that path to extract a model expressed in a process calculus similar to the one used by the CryptoVerif tool. The symbolic execution uses a novel algorithm that allows symbolic variables to represent bitstrings of potentially unknown length to model incoming protocol messages.

The extracted models do not use pointer-addressed memory, but they may still contain low-level details concerning message formats. In the next step we replace the message formatting expressions by abstract tupling and projection operators. The properties of these operators, such as the projection operation being the inverse of the tupling operation, are typically only satisfied with respect to inputs of correct types. Therefore we typecheck the model to ensure that all type-safety constraints are satisfied. The resulting model can then be verified with CryptoVerif to obtain a computational security result directly, or with ProVerif, to obtain a computational security result by invoking a computational soundness theorem.

In order to formalise the security properties of C programs and to prove the correctness of our approach we describe an embedding of C programs into the process calculus, such that C protocol participants can be executed as part of a larger system, described by the process calculus, that represents the environment and the attacker. We develop a security-preserving simulation relation that is preserved by embedding, and show that each step of our model transformation simulates the previous step, thus proving the overall soundness of the approach. Currently we only consider trace properties.

Our method achieves high automation and does not require user input beyond what is necessary to specify the properties of the cryptographic primitives and the desired security goals. We evaluated the method on several protocol implementations, totalling over 3000 lines of code. The biggest case study was a 1000-line implementation that was independently written without verification in mind. We found several flaws that were acknowledged and fixed by the authors, and were able to verify the fixed code without any further modifications to it.

# Contents

# Chapter 1

# Introduction

Cryptographic protocols are used to securely transmit data in computer networks. Typical goals are to establish a mutually known secret and ensure confidentiality and integrity of the data. Protocols like TLS [Dierks and Rescorla, 2008] or SSH [Ylonen and Lonvick, 2006] are widely used today and we rely on their correctness, both in terms of the logical structure and the implementation. Violations of both have occurred before. A famous example of a protocol with flawed structure is the Needham-Schroeder protocol, a serious flaw in which was discovered by Lowe almost two decades after its inception [Lowe, 1995]. Perhaps even more relevant to practice are errors in the implementations of protocols, such as the flaw in the OpenSSL implementation of TLS [CVE, a] that allowed a carefully forged certificate to bypass verification, or the more recent "Heartbleed" vulnerability [CVE, c] that allows the attacker to learn the secret key with minimal effort. The goal of this work is to verify absence of both types of flaws starting from the source code of a protocol implementation.

The approach that we propose draws on previous experience in security verification by extracting a high-level model from C code and then reusing existing tools ProVerif [Blanchet, 2009] or CryptoVerif [Blanchet, 2008] to verify that model. Consider, for example, the small piece of C code in figure 1.1 that generates a random value, tags it, and sends it encrypted with a one-time pad. Our method automatically extracts a much simpler model, shown under the code, that makes no use of pointers and memory. The model is written in slightly simplified CryptoVerif notation—it waits for an input from the attacker, so that the attacker can choose when each participant runs, chooses a random value *nonce* and outputs $XOR(0x01|nonce, pad)$, where $|$ denotes bitstring concatenation. The variable *pad* is free in the process—it is inserted by the user-provided model of the function `otp`, and is meant to be bound in the user-provided environment as shown below. The model is then further simplified to hide the bitstring concatenation operation—the expression $0x01|nonce$ is replaced by $tag(nonce)$, while making sure that the function *tag* satisfies the properties required to typecheck the process—we automatically infer and prove prove the type $tag\colon \texttt{fixed}_{20} \to \texttt{fixed}_{21}$. At this point the secrecy of *nonce*

```
unsigned char * payload = malloc(PAYLOAD_LEN);   // PAYLOAD_LEN = 20
if(payload == NULL) exit(1);
RAND_bytes(payload, PAYLOAD_LEN);
size_t msg_len = PAYLOAD_LEN + 1;
unsigned char * msg = malloc(msg_len);
if(msg == NULL) exit(1);
*msg = 0x01;                                     // add the tag
memcpy(msg + 1, payload, PAYLOAD_LEN);           // add the payload
unsigned char * pad = otp(msg_len);
xor(msg, pad, msg_len);                          // apply one-time pad
send(msg, msg_len);
```

$Q_1 = \mathbf{in}();\ \mathbf{new}\ nonce\colon \texttt{fixed}_{20};\ \mathbf{out}(XOR(0x01|nonce, pad))$

$tag\colon \texttt{fixed}_{20} \to \texttt{fixed}_{21}$
$Q_2 = \mathbf{in}();\ \mathbf{new}\ nonce\colon \texttt{fixed}_{20};\ \mathbf{out}(XOR(tag(nonce), pad))$

$XOR\colon \texttt{fixed}_{21} \times \texttt{fixed}_{21} \to \texttt{fixed}_{21}$
$C_E = !^N(\ \mathbf{in}();\ \mathbf{new}\ pad\colon \texttt{fixed}_{21};\ \mathbf{out}();\ []\ )$

Figure 1.1: Example C fragment, the extracted models, and the user-provided verification environment (the hole [] is filled with $Q_2$ for verification).

can be easily verified by CryptoVerif with respect to a user-provided environment, shown at the bottom of figure, that encodes our assumptions about the origin of *pad* and the way we invoke the program—we generate $N$ versions of *pad*, but only run the program once with each of them.

The model extraction is done by symbolic execution [King, 1976] of the C program, using symbolic expressions to over-approximate the sets of values that may be stored in memory during concrete execution. The main difference from existing symbolic execution algorithms, such as Cadar et al. [2008] or Godefroid et al. [2008], is that our symbolic variables represent bitstrings of potentially unknown length, whereas in previous algorithms a single variable corresponds to a single byte. The reason for this difference is that our symbolic expressions are used to capture the formats of the messages generated by the program, where often a field of the message has variable length, and the length is stored separately in another field. To represent these formats we allow expressions of the form $\mathrm{len}(e)$ denoting the length of an expression $e$. Our symbolic execution therefore needs to reason about lengths. For instance, if a memory location contains a concatenation $a|b$, and the program checks that $x = \mathrm{len}(a)$ for some $x$ then we know that extracting $\mathrm{len}(b)$ bytes at position $x$, written as $(a|b)\{x, \mathrm{len}(b)\}$, results in $b$.

To formally define and prove security for C programs we allow them to be executed in an environment defined by a high-level process calculus. For instance, if $Q$ is the C program in figure 1.1, $\tilde{Q}$ the extracted model, and $C_E$ the user-provided environment then our soundness results state that $C_E\{Q\}$ satisfies all the security properties of $C_E\{\tilde{Q}\}$. In our work we only consider *trace properties* of protocols, that is, properties that either hold or do not hold for any particular execution, in contrast to *observational equivalence properties* that hold with respect to the set of all executions. An example of a trace property is *"Every event $e_1$ is preceded by an event $e_2$"*. *Authentication* is usually formulated as a trace property of the form *"If A has accepted the message from B then either B has sent the message before or B has been corrupted by the attacker"*. *Weak secrecy* is a trace property saying that the attacker cannot obtain the full value of the secret—formally, there is no trace in which the secret is sent on the network in plaintext (in section 2.1 we show how to set up an environment for verification of weak secrecy of our example in figure 1.1). Unfortunately, weak secrecy does not prevent the attacker from having partial knowledge of the secret. Instead CryptoVerif uses *strong secrecy* based on observational equivalence, formulated as *"The attacker cannot reliably distinguish between a process that reveals the secret and a process that reveals a randomly chosen value.* Strong secrecy is an observational equivalence property, but not a trace property—it is impossible to tell whether it is satisfied by looking at a single trace only. Our method supports authentication and weak secrecy properties, but does not yet support strong secrecy. We believe that the latter can be covered without significant changes to the method, and discuss it as future work in section 7.2.

We assume that the protocol code follows a single execution path, without loops, and with any deviation immediately leading to termination. We choose to accept this limitation because many interesting protocols have this structure. For instance, the classic NSL protocol prescribes a fixed sequence of messages and the conditions to check after receiving each message. If a condition check fails, the participant simply stops execution of the protocol. As we show in one of our verification examples, such a protocol can be implemented by a C program that satisfies the assumptions of our method. Our method allows us to prove absence of security flaws in such programs. In particular, we show that there are no bugs related to parsing and formatting of messages, and no memory safety violations or integer overflows that lead to a vulnerability.

Some security bugs are closely related to non-trivial branching. For instance, Albrecht et al. [2009] show how to recover the plaintext of a secret in some configurations of the SSH protocol. The attack works by injecting the encrypted secret in place of the encrypted length field and then seeing how many blocks an SSH implementation will accept before triggering an error, thus finding out the decrypted value of the length field. This kind of attack cannot be expressed in our model since it relies on looping behaviour of both the implementation and the attacker. This is a general problem. In fact, the encryption scheme in question had been proved secure by Bellare et al. [2004], but using a model that was too abstract. A more realistic security model, proposed by Paterson and Watson [2010], is required to capture this kind of attacks. All of this suggests that even defining security models for complex branching behaviour is difficult, let alone extracting these models from code, which further justifies our restriction to linear protocols. We believe that our method can be generalised to more complex control flow, and discuss this as future work in section 7.1. Apart from the single path limitation we do not require use of any specific programming style.

We do not verify the code that implements the cryptographic primitives, such as encryption or hashing. Instead we assume that the functions implemented by the code possess the desired properties, say, that the encryption is indistinguishable with respect to a chosen ciphertext attack (IND-CCA). Such assumptions are necessary because no encryption scheme has been proven to be IND-CCA secure so far (however, Almeida et al. [2009] show how to verify that a C implementation of a primitive is equivalent to a reference implementation). Our verification method expects the user to provide models for cryptographic functions. These models are themselves written as C functions that instruct the symbolic execution about how to interpret a call to a primitive. This process is explained in more detail in section 3.1. In our example in figure 1.1 the user provides models for functions RAND_bytes, otp, and xor.

Our approach aims to achieve high automation: we do not expect the user to provide any annotations beyond what is necessary to specify the properties of the cryptographic operations and the desired security goals. In particular, we do not require an explicit specification of the protocol. For many industrial protocols no formal specifications are provided by the designer—these protocols are usually described by an RFC document, which is not meant to be machine-readable. Thus the implementation code is the first place where the protocol is defined precisely enough to be analysed, which justifies the use of a model extraction approach. Our experimental results are encouraging—we were able to analyse an externally written 1000-line implementation of a protocol without modifying it at all.

An important alternative approach is to develop a verified reference implementation of the protocol. This has been done for the TLS protocol—Bhargavan et al. [2013, 2014] describe a verified RFC-compliant implementation in F#. We hope that the effort required to produce such implementations will eventually become small enough so that every new protocol can be equipped with a formally verified reference implementation from the very start. In addition, if the efficiency of such implementations approaches the efficiency of their C counterparts, no further verification will be required, and our present work will become obsolete.

**Original Contributions and Thesis** Our two main contributions are:

- A theoretical framework, based on process calculus embedding, that allows to formalise cryptographic security of C programs.
- A model extraction algorithm, based on symbolic execution, that allows to automatically prove cryptographic security of C programs.

We use our theoretical framework to prove security of the model extraction algorithm. Theorems 3.2, 4.3 and 5.2 establish the correctness of our approach. In a nutshell, their significance is as follows: given C implementations $Q_1, \ldots, Q_n$ that satisfy our single-path assumption, and a CryptoVerif context $C_E$ that describes an execution environment, if symbolic execution of $Q_1, \ldots, Q_n$ yields models $\tilde{Q}_1, \ldots, \tilde{Q}_n$, the process $C_E\{\tilde{Q}_1, \ldots, \tilde{Q}_n\}$ is successfully translated to a CryptoVerif process $Q_{CV}$, and CryptoVerif successfully verifies $Q_{CV}$ against a trace property $\rho$ then $Q_1, \ldots, Q_n$ form a secure protocol implementation with respect to the environment $C_E$ and property $\rho$. More precisely, for any attacker process $Q_A$ the parallel composition $Q_A|C_E\{Q_1, \dots, Q_n\}$ has negligible probability of producing an execution trace that violates $\rho$. The same applies to verification with ProVerif, with additional computational soundness conditions. Notice how the process $C_E\{Q_1, \dots, Q_n\}$ mixes high-level model code with low-level implementation code—this is an important feature of our approach that enables a precise and convenient definition of security for C programs.

We did not aim for theoretical results regarding completeness; instead we evaluated the verification approach on a range of protocol implementations, including recent code for smart electricity meters by Rial and Danezis [2010]. We were able to find bugs in preexisting implementations or to verify them without having to modify the code, covering more than 3000 lines of C code in total. Chapter 6 provides details. Chapter 8 informally summarizes the types of security flaws and programs that our method applies to.

Overall, our work supports the following *thesis*:

> *Process calculus embedding combined with model extraction is an appropriate tool for analysing security of C programs.*

The work on the symbolic execution and the ProVerif verification was presented at the ACM Conference on Computer and Communication Security (CCS'11) [Aizatulin et al., 2011b] and the work on the CryptoVerif verification was presented at (CCS'12) [Aizatulin et al., 2012]. An invited paper at the workshop on Formal Aspects of Security and Trust (FAST 2011) contains a high-level overview of symbolic execution on an extended example [Aizatulin et al., 2011a]. Our implementation can be found at `http://github.com/tari3x/csec-modex`.

**Computational versus Symbolic Security Models** Cryptographic security models can be divided into two groups. *Symbolic* models [Dolev and Yao, 1983] define cryptographic operations by rewriting rules over a set of abstract terms. A typical rewriting rule would be $\mathit{decrypt}(\mathit{encrypt}(m, k), k) = m$, saying that the only way to extract information from an encryption is by having a valid key. In other words, symbolic models assume unbreakable cryptography. *Computational* models try to be closer to reality by viewing messages as bitstrings, and encryption primitives and the attacker as polynomial-time algorithms operating on those bitstrings [Blum and Micali, 1984; Goldwasser and Micali, 1984; Yao, 1982]. Security is then formulated in terms of probabilities of certain computation outcomes.

ProVerif [Blanchet, 2009] allows to prove security of protocols in the symbolic model. It works by converting the process that represents the protocol into a set of rewriting rules, adding to this set the rewriting rules for the cryptographic operations, and then runs a resolution algorithm on the resulting set. Proving a security property then amounts to making sure that a fact of the form $\mathit{attacker}(\mathit{secrect})$ is not derivable or that a fact of the form $\mathit{receive}(x)$ is only derivable from $\mathit{send}(x)$. CryptoVerif [Blanchet, 2008] works directly in the computational model. It starts from the process that represents the protocol and constructs a sequence of transformations such that each process in the sequence cannot be distinguished from the previous process in polynomial time with non-negligible probability, and the last process in the sequence is obviously secure.

We formalise security of C programs in the computational model, but we support verification of extracted models both with ProVerif and CryptoVerif. We use a *computational soundness result* of Backes et al. [2009] to transfer the ProVerif verification result back into the computational setting. In general, computational soundness principles have significant limitations. They only apply to a limited set of cryptographic primitives and often make assumptions that are not enforced in practice. For instance, Backes et al. [2009] require that all nonces are tagged and that every parsing operation checks its argument for full compliance with the message format, even when only extracting a single field of the message. To take another example, obtaining computational soundness for XOR in general is not possible [Unruh, 2010], although it is possible for a limited class of protocols [Küsters and Truderung, 2011]. So far our example in figure 1.1 cannot be analysed with ProVerif. The computational soundness result of Backes et al. [2009] covers public encryption and signatures, and only applies to one of six examples that we analysed.

CryptoVerif allows to obtain a verification result for a much wider range of crypto-algorithms: the protocols that we verified make use of XOR, Diffie-Hellman commitments, as well as operations including public-key encryption, MACs, and authenticated encryption. On the other hand, ProVerif is both easier to use and can deal with more sophisticated protocols than CryptoVerif. Even without a computational soundness result ProVerif shows that the protocol cannot be broken by a *symbolic* attacker who is only allowed to apply the rewriting rules of the symbolic model. This can be used as an initial sanity check when developing a new protocol or when dealing with a protocol that is too complex to be analysed with CryptoVerif or to support a computational soundness result. Another advantage of ProVerif is that it is often able to construct an explicit attack trace against the protocol, which CryptoVerif does not do.

**Description of the Method and Structure of the Dissertation** Our verification method proceeds in several steps (cf. Figure 1.2). The method takes as input the C source code of the protocol participants as well as a CryptoVerif or ProVerif *template file*, which contains the cryptographic assumptions about the primitives used by the implementation, the environment process which spawns the participants and generates shared cryptographic material, and a query for the property that the implementation is supposed to satisfy. The template file omits the actual model of the protocol participants. That model is extracted from C code by symbolic execution and rewriting. For each cryptographic function `f` called by the implementation the input also includes the C code of a function `f_proxy` that describes the correspondence of the C arguments of `f` to the formal arguments of the implemented primitive in the CryptoVerif or ProVerif model. The template file and the proxy functions form the trusted base of the verification.

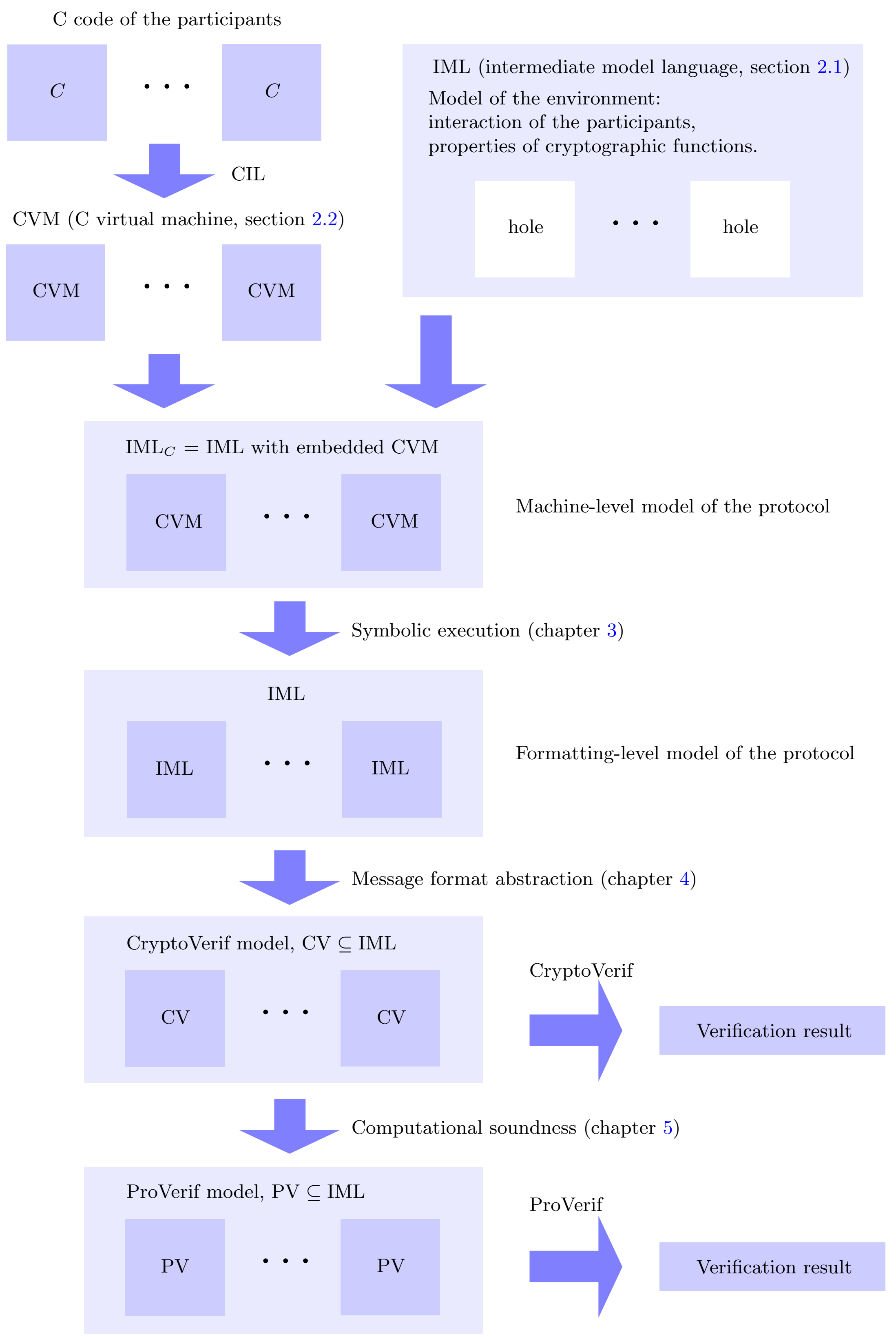

Figure 1.2: An outline of the method. Our simulation results allow us to perform the transformation steps for each participant in isolation.

In the first step, we compile the program down to a simple stack-based instruction language (CVM), using CIL [Necula et al., 2002] to parse and simplify the C input (section 3.1). In the next step we symbolically execute CVM programs to eliminate memory accesses and destructive updates, thus obtaining a model in an intermediate model language (IML)—a version of the applied pi calculus extended with bitstring manipulation primitives. For each allocated memory area the symbolic execution stores an expression describing how the contents of the memory area have been computed. For instance, a certain memory area might be associated with an expression $hmac(01|x, k)$, where $x$ is known to originate from the network, $k$ is known to be an environment variable, and $|$ denotes concatenation. The symbolic execution does not enter the functions that implement the cryptographic primitives, it uses the models provided by the proxy functions instead. We prove that the model simulates the original C program in any environment, implying that any trace of the C program in an environment is also a trace of the model in the same environment. The languages CVM and IML together with the simulation relation are introduced in chapter 2 and the symbolic execution is described in chapter 3.

The next step abstracts away the details of message formatting that are present in IML by replacing the bitstring manipulation expressions by application of new *formatting functions* for encoding and parsing bitstrings, so that, say, $\mathtt{0x01}|\,\mathrm{len}(x_1)|x_1|x_2$ becomes $conc(x_1, x_2)$ and $x\{5, x\{1, 4\}\}$ becomes $parse(x)$, where $b|b'$ denotes the concatenation of $b$ and $b'$ and $b\{p, l\}$ is the substring of $b$ starting at position $p$ of length $l$. A detail that we omit here is the encoding of the integer expression $\mathrm{len}(x)$ as a bitstring and the conversion of the bitstring $x\{1, 4\}$ to an integer. Our method is aware of integer encodings used in C programs and is sound with respect to integer overflows. By constructing suitable queries to an automatic prover Yices by Dutertre and Moura [2006], we check certain properties of the formatting functions that help CryptoVerif or ProVerif verify the protocol. These properties include parsing equations such as $parse(conc(x_1, x_2)) = x_1$, injectivity, or disjointness of ranges. These properties will often be satisfied only for inputs of appropriate types. For instance, $x_1$ in $conc(x_1, x_2)$ must be short enough for its length to fit in the length field. We therefore typecheck the model to make sure that the formatting functions will only be supplied with inputs of acceptable types. The formatting abstraction step is described in chapter 4.

After formatting abstraction we obtain a process in the CryptoVerif calculus. This can either be verified directly with CryptoVerif as described in chapter 4 or with ProVerif with application of a computational soundness result. The soundness result imposes certain conditions on the formatting functions that cannot be satisfied by a typical implementation. For instance, it requires that the concatenation operations should be defined for all bitstrings, which is impossible if they use a fixed-size length field. To overcome this we consider a relaxed set of soundness conditions that only hold for inputs of correct types. We then use our typechecking algorithm to make sure that inputs to all formatting functions will always be of correct types. This can be used to show that our process $Q$ satisfying the relaxed soundness conditions can be generalised to a process $\tilde{Q}$ that satisfies full soundness conditions, and such that $\tilde{Q}$ simulates $Q$. Verification with ProVerif is described in chapter 5.

**Challenges** It is important to understand why existing software verification tools cannot be readily applied to cryptographic security verification. Tools like Frama-C [Frama-C] or SLAM [Ball et al., 2004] allow to annotate the code with `assume` and `assert` statements and then aim to statically prove that the assertions hold in each run of the program that satisfies the assumptions.When trying to apply this approach to security properties one faces the following difficulties:

- The assertions are only allowed to refer to the internal state of the program in question. However, security properties are typically defined by reference to a state of some external system. For instance, secrecy refers to the knowledge of an unknown attacker and authentication is expressed through beliefs of the communication partner.
- Cryptographic verification has a different notion of what it means for an assertion to hold. In typical software verification an assertion needs to be valid in all possible program runs. Compared to that, cryptography is interested in events occurring with non-negligible probability, where this probability is influenced by the abilities of the attacker.

  An important step in many cryptographic proofs is elimination of collisions between independently chosen random values: if two "large" values $x$ and $y$ are independently chosen then the traces in which $x$ equals $y$ have negligible probability, therefore we can assume $x \neq y$. General-purpose verification tools are not designed to perform such a reasoning step.

An interesting theoretical challenge arises from the fact that cryptographic security definitions are given with respect to a varying security parameter, but a C implementation is written for a single value of the security parameter. In particular, it can only address a memory of a fixed size, a problem that is also discussed by Küsters et al. [2012]. In our work we deal with this problem by defining the semantics of IML with respect to a fixed security parameter only and then showing how to generalize the extracted IML model to arbitrary security parameters such that all the relevant properties of the model are preserved. All our security results are therefore a combination of two statements: 1) the insecurity of the C program is bounded by the insecurity of the model with respect to our chosen security parameter, and 2) the insecurity of the model is negligible with respect to an arbitrary security parameter. This is somewhat unsatisfactory because a negligible function can still be arbitrarily large in any given point, but this issue is always present when interpreting asymptotic results. One could read the two statements above as 1) the C program is a correct implementation of the model with respect to the chosen security parameter, and 2) the model is asymptotically secure with respect to an arbitrary security parameter.

The formal description of CryptoVerif uses an asymptotic model but the tool itself goes a step beyond that and provides concrete security bounds expressed in terms of probabilities of breaking the security primitives for each security parameter (as far as we know, this work has not been formalised yet). Our soundness theorems show that these bounds apply directly to the original C programs. The computational soundness result that we use with ProVerif only provides us with an asymptotic statement.

**Limitations** Given the complexity of the C language, analysis tools rely on front ends that translate the language into a much simpler representation. Examples are LLVM [Lattner, 2002] used by KLEE [Cadar et al., 2008] and CIL [Necula et al., 2002] used by Frama-C [Frama-C]. We wrote a plugin for CIL that instruments a C program to output its own CVM representation when run. As there is no formal proof of correctness for CIL, our formal results apply to CVM programs, and the link to C is made only informally, as discussed in section 3.1.

We only analyse a single execution path of the protocol. The CVM is produced by a single run $R$ of the program $Q$ and thus represents a pruned program $\tilde{Q}$ such that $\tilde{Q}$ produces the same run. In the pruned program every statement of the form $\mathsf{if}\,(\mathsf{c})\{\mathsf{A}\}\mathsf{else}\{\mathsf{B}\}$ is replaced by $\mathsf{if}\,(\mathsf{c})\{\mathsf{A}\}\mathsf{else}\{\,\mathsf{exit}\,(1)\}$ or $\mathsf{if}\,(\mathsf{c})\{\,\mathsf{exit}\,(1)\}\,\mathsf{else}\,\{\mathsf{B}\}$, depending on which branch was taken in $R$, and loops are unrolled the number of times they were executed in $R$. All our soundness results apply to the program $\tilde{Q}$. In the case of crypto-protocols we believe $\tilde{Q}$ to be a close approximation of $Q$, as the structure the protocols (such as those in the extensive SPORE repository [Project EVA, 2007]) is often linear. The code of libraries such as PolarSSL contains a lot of if-statements, but many of those have conditions that become constant for particular settings of configuration variables, or deal with cryptographic security checks and abort the execution immediately if those checks fail. In section 7.1 we discuss ideas for generalising our approach to multiple paths. The new approach will be able to provide a measure of difference between $Q$ and $\tilde{Q}$. For now we rely on a (straightforward) manual check that $Q = \tilde{Q}$: check that $Q$ contains no loops and all the conditional statements in $Q$ contain an exit statement in one of the branches without any attacker-observable actions preceding that statement. Verifying $\tilde{Q}$ increases confidence in the correctness of $Q$ even if $Q \neq \tilde{Q}$, because $\tilde{Q}$ is a program that can successfully execute a session of the protocol in place of $Q$.

Our work is currently limited to trace properties of protocols, such as authentication or weak secrecy. We leave treatment of observational equivalence properties, such as strong secrecy, to future work, as discussed in section 7.2.

## 1.1 Running Example: RPC-enc

Our example protocol, due to Fournet et al. [2011], is an encryption-based variant of the RPC protocol, already considered by [Bhargavan et al., 2010; Dupressoir et al., 2014]. In section 3.6 we shall demonstrate how a fragment of the IML model is extracted from the C code by symbolic execution, and in section 4.2 we shall use RPC-enc as a running example to describe the sequence of transformations from IML to CryptoVerif. Appendix C shows the full CryptoVerif model of

$$
\begin{array}{lll}
A & : & \mathsf{event}\ \mathit{client_begin}(A, B, \mathit{request}) \\
A \rightarrow B & : & A, \{\mathit{request}, k_S\}_{k_{AB}} \\
B & : & \mathsf{event}\ \mathit{server_reply}(A, B, \mathit{request}, \mathit{response}) \\
B \rightarrow A & : & \{\mathit{response}\}_{k_S} \\
A & : & \mathsf{event}\ \mathit{client_accept}(A, B, \mathit{request}, \mathit{response})
\end{array}
$$

Figure 1.3: Authenticated RPC: RPC-enc

the protocol.

In the following, $\{m\}_k$ stands for the encryption, using an authenticated encryption mechanism, of plaintext $m$ under key $k$ while the comma represents an injective pairing operation. The protocol narration in Figure 1.3 describes the process of $A$ in client role communicating to $B$ in server role. The key $k_{AB}$ is a unidirectional long-term key shared between $A$ and $B$ (should $B$ wish to play the client role, they would rely on a key $k_{BA}$, distinct from $k_{AB}$). The key $k_S$ is the session key freshly generated by $A$ and the payloads *request* and *response* are freshly generated. We assume that the server is always honest, but the client may be compromised. We write $bad(A)$ to mean that the client $A$ is compromised.

We aim to prove authentication and secrecy properties:

1. Authentication properties state that each principal can ensure that a received message was produced by the correct protocol participant. These properties are specified using event correspondences of the form:

$$
\begin{aligned}
&\mathit{server_reply}(A, B, req, resp) \implies \mathit{client_begin}(A, B, req) \vee bad(A) \\
&\mathit{client_accept}(A, B, req, resp) \implies \mathit{server_reply}(A, B, req, resp)
\end{aligned}
$$

   The first property states that, whenever *server_reply* happens, either *client_begin* has happened with corresponding parameters or the client is compromised. Similarly, the second property states that, whenever the event *client_accept* happens, the event *server_reply* has happened before. There is no compromise in this case as we assume the server to be honest.

2. Secrecy properties state that the attacker cannot learn values of payloads, where the secrecy of response is conditional on the client being honest.

## 1.2 Related Work

The project is situated at the intersection of two major research areas—*program verification* and *protocol verification*. The first deals with proving correctness of programs with respect to their specifications and the second deals with proving correctness of cryptographic protocols.

### 1.2.1 Program Verification

**Historical** Program verification has a very long history. The idea of using assertions in code to specify and prove its properties dates back to Turing [1949]. In the 1960s formalisms were proposed to precisely define meanings of programs, so that formal reasoning about programs became possible [Floyd, 1967; McCarthy, 1963]. The work of Hoare [1969] proposed the idea that the semantics of a programming language can actually be defined by stating axioms and proof rules about different constructs of the language. Hoare introduced the notation for writing such rules, which is now known as *Hoare triples*.

Using a formal definition of program semantics, it is possible to reduce the verification problem to a set of mathematical theorems, which can then be solved with an automated or manual theorem prover. Using this technique the first automated program verifier was built by King [1970].

A common obstacle in making verification fully automatic is the necessity to provide explicit loop invariants. Another difficulty is that reasoning about precise values of variables quickly becomes infeasible in practice. To address these problems, modern tools commonly use the technique of *abstract interpretation*: the sets of possible variable values are approximated by larger sets having simpler structure. For instance, instead of keeping track of an integer value, only its sign might be considered, or whether it belongs to a particular interval. The sets of values are typically propagated through the program until they saturate, i.e. do not change with further propagation. This incidentally establishes loop invariants which are sufficient for many purposes. The theoretical foundation for abstract interpretation was laid by Cousot and Cousot [1977].

**Static Analysis Tools** There is a wide range of tools that follow the approach, described above, of using theorem proving for program analysis. VCC [Cohen et al., 2009] is targeting concurrent system-level code, ASTRE [Blanchet et al., 2007] is aimed at verifying critical code in a restricted subset of C and Frama-C [Frama-C] is a plug-in based framework for program analysis. A much more comprehensive list can be found e.g. at List of Tools. The static analysis tool Calysto [Babić, 2008; Babić and Hu, 2008] uses a sophisticated symbolic execution algorithm to avoid combinatorial explosion when obtaining summaries for multiple paths. Also worth mentioning is the tool ESC/Java [Flanagan et al., 2002] that targets the Java language. Calysto and ESC/Java are not sound and provide no correctness guarantees. Since the term *verifier* is usually reserved for sound tools, Calysto and ESC/Java call themselves *extended static checkers* instead.

**Model Checking** An alternative approach to program analysis is systematic exploration of the program state space, known as *model checking*. It has been independently proposed by Clarke and Emerson [1982] and Queille and Sifakis [1982]. One of the first popular model checkers is SPIN [Holzmann, 1997], which has been applied to a range of critical software, for instance, at flood control gates or inside NASA space probes.

Model checking can produce very precise analysis, but at a very high cost: the size of the state space to be explored grows very rapidly with required precision. For this reason model checking is particularly useful for programs with bounded state space. An example of such application is the Microsoft SLAM tool [Ball et al., 2006] targeting device drivers. Modern tools often approximate the state space by retaining only partial information about the state, like it is done in *predicate abstraction*. Often this is combined with a feedback technique: if verification of an approximation fails, the results are used to construct a finer approximation. This is iterated until the verification succeeds or a feasible counterexample is found. Such technique is called *counterexample-guided abstraction refinement (CEGAR)* [Clarke et al., 2000] and is used, for instance, in the BLAST model checker [Beyer et al., 2007a].

Model checking and program analysis can be seen as two opposites on the precision-efficiency spectrum. Model checking is mostly concerned with precision, aiming to eliminate spurious counterexamples, while program analysis aims to give some performance guarantees. There are theoretical results suggesting that model checking and program analysis have much in common [Schmidt, 1998], so that the distinction between the two might blur in the future [Beyer et al., 2007b].

**Testing and Symbolic Execution** An active area of research is test generation by symbolic execution of programs: instead of supplying the normal inputs to programs, one supplies symbols representing arbitrary values. These symbols are propagated through the code, so that at the branching points we obtain terms describing the dependence of branching conditions on input values. This allows to formulate equations about the input so that solving them forces the branch to be taken in some particular direction. As a result, concrete inputs can be generated that force a desired execution path through the program thus increasing test coverage. This technique can be traced back to King [1976]. Modern tools like KLEE [Cadar et al., 2008] or SAGE [Godefroid et al., 2008] use concrete and symbolic execution in parallel to increase precision, this is often called *concolic* execution.

**Adoption and Future** Formal verification is now routinely used in firms like Microsoft [Ball et al., 2004] and Intel [Fix, 2008]. The *Verified Software Initiative* [VSI, 2007] is a commitment of leading computing researchers to producing a body of formally verified programs and tools that automate the verification. Constructing a verifying compiler has been proposed as a grand challenge in computing research [Hoare, 2003]. A more comprehensive survey of formal verification tools can be found in D'Silva et al. [2008].

### 1.2.2 Protocol Verification

**Security Properties** A foundational work in cryptographic protocol verification is Needham and Schroeder [1978]. This paper introduced the notion of the attacker, who is in full control of the network, which, by now, is a standard assumption in security verification. The property of *secrecy* guarantees that such an attacker cannot learn a certain value exchanged during the protocol. There is a distinction between *weak secrecy* in which the attacker cannot obtain full knowledge of the value, and *strong secrecy* in which the attacker cannot obtain any information about the value whatsoever.

Even though authentication protocols were being proposed since the 1970s, it was surprisingly hard to define what *authentication* exactly means. The definition that is widely accepted today is due to Woo and Lam [1993], where authentication is formulated using event correspondences, for instance, *"If $A$ believes that the message $m$ originates from $B$ then it must be the case that $B$ has indeed sent $m$"*.

**Computational Soundness Theorems** For a long time the divide between symbolic and computational security models has not been addressed. The first results linking the two approaches are due to Abadi and Rogaway [2000] and independently Pfitzmann et al. [2000]. This started a long series of *computational soundness theorems* that would map results obtained in the symbolic models to statements in computational model. The theorems usually differ according to the range of cryptographic primitives considered, the powers of the attacker (eavesdropping or active) and the definition of equivalence (trace-based or observational). For instance, Abadi and Jürjens [2001] provide a result for a passive (eavesdropping) attacker and observational equivalence, Backes et al. [2009] provide a result for the active attacker with trace-based equivalence, and Aizatulin et al. [2009] establish a result for strategy properties involved in analysis of contract-signing protocols.

**Verification of Formal Models** There is a range of tools for verification of protocol descriptions in a formal high-level language. ProVerif [Blanchet, 2009, 2014] and CryptoVerif [Blanchet, 2008] accept $\pi$-calculus models as inputs. The first tool works in the symbolic model and the second in the computational. AVISPA [Armando et al., 2005] and LySatool [Bodei et al., 2005] are further examples of well-developed protocol verification tools.

Many approaches to verifying formal specifications, such as Abadi [1999]; Focardi [2004]; Gordon and Jeffrey [2003] rely on typing. Their results state that a process in a certain formal language (such as the spi calculus used by Gordon and Jeffrey [2003]) is well-typed by certain rules then it is secure. The advantage of this approach is that typing can be easily automated. A disadvantage is that support for new types of cryptographic operations requires new typing rules.

CertiCrypt [Barthe et al., 2009] and EasyCrypt [Barthe et al., 2011] are frameworks for writing machine-checked cryptographic proofs. They work at a lower level than our verification and concentrate on proving properties of cryptographic primitives composed from other cryptographic primitives. They do not explicitly model protocol participant interaction in a network. Recently Almeida et al. [2013] added support for generating executable code from EasyCrypt models.

**Verification of Functional Implementations** Functional languages like ML have a formally defined clean semantics, which allows to translate them directly into $\pi$-calculus representations. This has been used for verification of protocol implementations in F# by translating them to ProVerif and CryptoVerif models [Bhargavan et al., 2006, 2008]. Unfortunately this approach is less suitable for low-level languages like C, because pointers and arithmetic have no direct representation inside $\pi$-calculus.

Another approach is using a type system for a dialect of F# such that type-correct programs can be proved secure [Bengtson et al., 2008]. This approach has recently been used by Bhargavan et al. [2013, 2014] to provide a fully verified reference implementation of the TLS protocol in F#. It is unclear, how well such an approach maps to pre-existing low-level implementations C, but we hope that some day verified functional implementations will become efficient and easy to write so that they will displace the C implementations.

**Verification of Java Implementations** Several projects concentrate on verification of protocols written in Java. One approach is translating Java programs into a set of Horn clauses that are fed directly into a general purpose theorem prover [Jürjens, 2006]. Another method is using static analysis to extract abstract models from Java programs that can subsequently be verified using LySatool [O'Shea, 2008]. These works can provide a lot of inspiration for C verification, in particular the second one follows a very similar strategy to what we are proposing. However, C language poses additional challenges, mostly due to its unrestricted memory access.

Recently there has been work on verifying indistinguishability properties for Java implementations in the universal composability framework Küsters et al. [2012, 2013, 2014]. If a Java program uses a particular library interface, it is possible to show that the concrete functionality provided by the library can be replaced by an "ideal" functionality that perfectly realizes the intended cryptographic goals. It is unclear how well this approach carries over to pre-existing Java implementations.

**Verification of C Implementations** One of the first attempts at cryptographic verification of C code is contained in Goubault-Larrecq and Parrennes [2005], where a C program is used to generate a set of Horn clauses that are then solved using a theorem prover. The method is implemented in the tool CSur. We improve upon CSur in two ways in particular.

First, we have an explicit attacker model with a standard computational attacker. The attacker in CSur is essentially symbolic—it is allowed to apply cryptographic operations, but cannot perform any arithmetic computations.

Second, we handle authentication properties in addition to secrecy properties. Adding authentication to CSur would be non-trivial, due to a rather coarse over-approximation of C code. For instance, the order of instructions in CSur is ignored, and writing a single byte into an array with unknown length is treated the same as overwriting the whole array. Authentication, however, crucially depends on the order of events in the execution trace as well as making sure that the authenticity of a whole message is preserved and not only of a single byte of it.

Jürjens [2005a,b]; Jürjens and Yampolskiy [2005] present an approach that provides a Dolev-Yao formalization in first-order logic starting from the program's control-flow graph, which can then be verified for security properties with automated theorem provers such as SPASS [Weidenbach et al., 2002]. There is a manual abstraction step between the (automated) generation of the control-flow graph and the (automated) generation of the logical representation from the control-flow graph.

ASPIER [Chaki and Datta, 2009] uses model checking to verify implementations of cryptographic protocols. The model checking operates on a protocol description language, which is rather more abstract than C; for instance, it does not contain pointers and cannot express variable message lengths. The translation from C to the protocol language is not described in the paper. Our method applies directly to C code with pointers, so that we expect it to provide much greater automation.

[Jeffrey and Ley-Wild, 2006] presents an approach for finding attacks on C code. One possible attack trace is chosen and encoded as a C program. The only thing left unresolved in the trace are the messages that the attacker sends at each point. Those are resolved by constructing an appropriate Prolog model. This approach requires the C code to use a specific API and places further restrictions on the code, such as the requirement that all variables should be assigned to only once.

Corin and Manzano [2011] report an extension of the KLEE test-generation tool [Cadar et al., 2008] that allows KLEE to be applied to cryptographic protocol implementations (but not to extract models, as in our work). They do not extend the class of properties that KLEE is able to test for; in particular, testing for trace properties is not yet supported. Similarly to our work, KLEE is based on symbolic execution; the main difference is that Corin and Manzano [2011] treat every byte in a memory buffer separately and thus only supports buffers of fixed length.

An advantage of the aforementioned tools (CSur, ASPIER, and KLEE) is that they do not share our limitation to single execution paths.

Dupressoir et al. [2010, 2014] show how to adapt a general-purpose verifier to security verification of C code. This approach does not have our restriction to non-branching code, on the other hand, it requires the code to be annotated (with about one line of annotation per line of code) and works in the symbolic model, requiring the pairing and projection operations to be treated as abstract.

**Code Generation** A very active recent trend is generating executable implementations from verified high-level models. This includes the work of Almeida et al. [2013] that generates executable implementations from EasyCrypt models. Further examples of this approach include generating OCaml code from verified CryptoVerif models [Cadé and Blanchet, 2012, 2013a,b] and the framework CAOVerif described by Almeida et al. [2014]; Barbosa et al. [2014].

**Manual Verification** Some protocols have been analysed manually using an interactive theorem prover. Examples include an analysis of the Bull-Otway protocol [Paulson, 1997] and TLS [Paulson, 1999] using Isabelle.

**Compliance Checking** Some work has been done on ensuring that an implementation of a protocol corresponds to the protocol definition in an RFC document [Udrea et al., 2008]. This is done by extracting correspondence rules from the document (for instance, *"Every time the server receives a request, it sends a reply."*) and annotating the code with them. Abstract interpretation is then used to check that the rules are indeed fulfilled. Such assurance is very useful, but it does not provide a formal guarantee of security, because only fragments of the specification are formalised and there is no proof that these fragments are actually enough for establishing security.

### 1.2.3 Reverse Engineering of Message Formats

Many tools aim to recover formats of protocol messages and are therefore related to our model extraction step. However, their intended purpose is very different—they are usually applied to compiled binaries such as malware, using some form of binary taint analysis, in an attempt to gain insights into its operation. As a result, they usually apply heuristics and do not aim for soundness—a message format they infer may happen to be incorrect. More importantly, because the tools work without access to source code, they do not analyse semantic information about the contents of the messages, such as whether a certain part of the message is encrypted.

Polyglot [Caballero et al., 2007] implements execution monitoring using the QEMU PC emulator enhanced to support dynamic taint analysis. It tries to identify length fields, field separators, keywords, and fixed-length fields. AutoFormat [Lin et al., 2008] is a similar tool based on Valgrind. Wondracek et al. [2008] uses a self-made dynamic taint analyser. The analysis first extracts format information for individual messages and then compares multiple messages to extract a more general hierarchical format specification. Tupni [Cui et al., 2008] uses Vigilante as its dynamic data flow engine and iDNA for capturing program traces. Antunes and Neves [2009] aim to reconstruct an automaton that accepts the protocol message formats.

Unlike the tools mentioned above, ReFormat [Wang et al., 2009] aims to identify buffers in memory that store encrypted information. It is based on the heuristic that cryptography typically uses much more arithmetic and bitwise instructions than the rest of the processing functions.

## 1.3 Basic Notation and Terminology

In order to match the semantics of C programs we only consider bitstrings that consist of an integral number of bytes: we define the set of all *bitstrings* to be $BS = \{0x00, \dots, 0xFF\}^*$. A more appropriate term would be "bytestring", but we choose to stick with "bitstring" to match the terminology of CryptoVerif used in chapter 4. For a bitstring $b$ we let $|b|$ denote the length of $b$ in bytes and we let $b[i]$ be the $i$th byte of $b$ counting from 0. The concatenation of two bitstrings $b_1$ and $b_2$ is written as $b_1|b_2$. For $a, b \in \mathbb{N}$ we define $\{a\}_b = \{a, \dots, a + b - 1\}$—this notation will be useful for selecting substrings from a bitstring.

Let *Var* be a countably infinite set of variables. We write $f \colon X \rightharpoonup Y$ to denote a partial function and let $\mathrm{dom}(f) \subseteq X$ be the set of $x$ for which $f(x)$ is defined. We use a special value $\bot$, which does not imply that a function is partial: if $f(x) = \bot$ for a given argument then we consider the function to be defined for $x$, that is, $x \in \mathrm{dom}(f)$. This value should therefore be understood as a None value in ML or Nothing in Haskell. Given a set $X$ we write $X_\bot$ as an abbreviation for $X \cup \{\bot\}$. We use the notation $f\{x \mapsto a\}$ to update functions.

When studying security properties of C implementations we shall be aiming to prove that the probability of an attack is a negligible function of the *security parameter*—an integer that relates to the lengths of inputs (most notably, keys) of cryptographic functions. A function $f \colon \mathbb{N} \to \mathbb{R}$ is called *negligible* if for every $c \in \mathbb{N}$ there exists $n_0 \in \mathbb{N}$ such that $f(n) < 1/n^c$ for all $n > n_0$. If we can show that the probability of an attack on an implementation is bounded by a negligible function, this means that we can make the implementation as secure as necessary by simply using longer keys—the probability of an attack will then scale with the inverse exponential of the key length. Another intuition for the use of negligible functions in the definition of security is that we are trying to prove that there is no better way to attack an implementation than exhaustively enumerating all the keys.

# Chapter 2

# Security Definitions

We define security for C programs by embedding them in a high-level process calculus. Our calculus IML, presented in section 2.1, is a version of the CryptoVerif calculus [Blanchet, 2008], but has a richer expression syntax which allows it to describe details of protocol message formats. We shall use IML both to represent models extracted from C programs and to describe the environment in which C programs and their models execute. The adversary is also modelled as an IML process and our security definition mirrors that of CryptoVerif.

The language CVM, presented in section 2.2, is a simple stack-based instruction language with memory access that we use to describe the behaviour of C programs. The semantics for CVM is given by describing the execution of a CVM process within an IML context. Compilation from C to CVM will be described informally in section 3.1.

Our security proofs will make heavy use of a simulation relation $\lesssim$ for mixed IML and CVM processes, developed in section 2.3. The main theoretical results of this chapter are theorem 2.2 saying that $Q \lesssim \tilde{Q}$ implies that $Q$ is at least as secure as $\tilde{Q}$ with respect to any trace property $\rho$, and theorem 2.3 saying that simulation is preserved by embedding—if $Q \lesssim \tilde{Q}$ then $C\{Q\} \lesssim C\{\tilde{Q}\}$ for any context $C$, which allows us to prove preservation of security properties with respect to all environments and adversaries.

## 2.1 The Language of Models—IML

This section presents the language *IML (Intermediate Model Language)* that we use to describe models extracted from protocol implementations. It is a process calculus inspired by the CryptoVerif tool [Blanchet, 2008] and can be seen as an intermediate stage between C programs and CryptoVerif models. On the one hand, IML does not operate on memory and pointers, on the other hand, it may still contain bitstring operations such as substring extraction or concatenation, both of which would be beyond CryptoVerif's reasoning abilities. Chapter 3 describes how we use symbolic execution to extract IML models that represent behaviours of C programs, and chapter 4 describes how we typecheck and simplify IML models to make them amenable to analysis with CryptoVerif.

The syntax of IML is shown in figure 2.1. Expressions are of three kinds: logical expressions (facts) that evaluate to boolean values, integer terms that evaluate to integers and binary terms that evaluate to bitstrings. Any expression can evaluate to a special value $\bot$ representing failure. Expressions evaluate with respect to an *environment* $\eta\colon Var \rightharpoonup BS_\bot$, a partial function that maps variables to bitstrings or $\bot$. The result of the evaluation will be denoted with $[\![e]\!]_\eta$. For an integer term $t$ we have $[\![t]\!]_\eta \in \mathbb{Z}_\bot$, and for a bitstring term $e$ we have $[\![e]\!]_\eta \in BS_\bot$. Following CryptoVerif, we identify boolean values with bitstrings, so that for a fact $\phi$ we have $[\![\phi]\!]_\eta \in BS_\bot$. We expect binary terms in boolean context to evaluate to a value from a set $Bool = \{\text{true}, \text{false}\} \subseteq BS$ with some arbitrary but fixed values of true and false.

The evaluation rules are shown in figure 2.2. If an expression $e$ cannot be evaluated by these rules for a particular environment $\eta$, we let $[\![e]\!]_\eta = \bot$.

Since our expressions will be used to represent values computed by C programs, they need to include a way to relate integer values and their bitstring representations. This is accomplished by viewing each C integer type as a partial function $\tau\colon BS \rightharpoonup \mathbb{Z}$ with an inverse $\tau^{-1}\colon \mathbb{Z} \rightharpoonup BS$ that encode and decode integers as bitstrings. We let $\mathbb{T}_I$ denote the set of all C types of interest. We shall assume that $\tau$ and $\tau^{-1}$ are inverse to each other on their respective domains: $\tau(\tau^{-1}(n)) = n$ for each $n \in \text{range}(\tau)$ and $\tau^{-1}(\tau(b)) = b$ for each $b \in \text{dom}(\tau)$. Throughout the thesis we shall be dealing with two's complement representations: a signed $l$-byte representation will be denoted with $\tau_{ls}$ and an unsigned representation with $\tau_{lu}$ with representable ranges $\text{range}(\tau_{lu}) = [0, 2^l - 1]$ and $\text{range}(\tau_{ls}) = [-2^{l-1}, 2^{l-1} - 1]$. We let $\text{len}(\tau)$ denote the length of the bitstrings of type $\tau$, that is, $\text{len}(\tau_{lu}) = \text{len}(\tau_{ls}) = l$. We are not concerned with whether the representation is big-endian or little-endian. Ones' complement representation could also be analysed, with minimal adjustments.

We let *Ops* be a set of *function symbols* such that to each $f \in Ops$ we associate an arity $n \in \mathbb{N}$ and an interpretation $I(f)\colon BS_\bot^n \to BS_\bot$. We require that $I(f)$ is computable in polynomial time in the length of the input and evaluate expressions of the form $f(e_1, \ldots, e_n)$ by applying $I(f)$ to the arguments. The set *Ops* is meant to contain both the primitive operations of the language (such as the arithmetic or comparison operators of C) and the cryptographic operations that are used by the implementation.

When analysing cryptographic protocol models, each cryptographic operation is usually a family of functions $f_k$, one for each security parameter $k \in \mathbb{N}$. At the same time cryptographic protocol implementations are written for a fixed value of the security parameter. Thus we fix an arbitrary value $k_0 \in \mathbb{N}$ and consider the interpretations $I(f)$ of functions symbols to be chosen with respect to that value. In chapter 4 we show how to link this view to CryptoVerif's definition that assigns a family of functions to each function symbol.

An expression $e\{p, l\}$ evaluates to the substring of $e$ starting at position $p$ of length $l$, and

| | |
|---|---|
| $\phi,\ \psi ::=$ | fact |
| $\quad \phi \circ \psi,\ \circ \in \{\wedge, \vee, \neg\}$ | logical operation |
| $\quad t \circ t',\ \circ \in \{=, \neq, <, >\}$ | integer comparison |
| $\quad e = e'$ | bitstring equality |
| $\quad e$ | bitstring fact |
| $t \in \mathit{ITerm} ::=$ | integer term |
| $\quad n \in \mathbb{Z}$ | integer |
| $\quad t \circ t',\ \circ \in \{+, -, *\}$ | integer operation |
| $\quad \mathrm{len}(e)$ | length |
| $\quad \tau(e),\ \tau \in \mathbb{T}_I$ | value |
| $e \in \mathit{IExp} ::=$ | bitstring term |
| $\quad x, i \in \mathit{Var}$ | variable |
| $\quad b \in \mathit{BS}$ | bitstring |
| $\quad f(e_1, \dots, e_n),\ f \in \mathit{Ops}$ | function application |
| $\quad \tau^{-1}(e),\ \tau \in \mathbb{T}_I$ | bitstring encoding |
| $\quad e\{t_p, t_l\}$ | substring |
| $\quad e \vert e'$ | concatenation |
| $Q \in \mathrm{IML}_I ::=$ | input process |
| $\quad 0$ | nil |
| $\quad Q \vert Q'$ | parallel composition |
| $\quad !^{i \le N} Q$ | replication $N$ times |
| $\quad \mathsf{in}(c[e_1, \dots, e_n], x);\ P$ | input |
| $P \in \mathrm{IML}_O ::=$ | output process |
| $\quad \mathsf{out}(c[e_1, \dots, e_n], e);\ Q$ | output |
| $\quad \mathsf{new}\ x\colon T;\ P$ | random number |
| $\quad \mathsf{let}\ x = e\ \mathsf{in}\ P$ | assignment |
| $\quad \mathsf{if}\ \phi\ \mathsf{then}\ P\ [\mathsf{else}\ P']$ | conditional |
| $\quad \mathsf{event}\ ev(e_1, \dots, e_n);\ P$ | event |
| $\quad \mathsf{assume}\ \phi;\ P$ | assumption |

Figure 2.1: The syntax of IML.

$$\frac{\circ \in \{\wedge, \vee, \neg\} \quad [\![\phi]\!]_\eta = \alpha \in \mathit{Bool} \quad [\![\psi]\!]_\eta = \beta \in \mathit{Bool}}{[\![\phi \circ \psi]\!]_\eta = \alpha \circ \beta}$$

$$\frac{\circ \in \{=, \neq, <, >\} \quad [\![t]\!]_\eta = n \in \mathbb{Z} \quad [\![t']\!]_\eta = n' \in \mathbb{Z}}{[\![t \circ t']\!]_\eta = n \circ n'}$$

$$\frac{\circ \in \{=, \neq, <, >\} \quad [\![t]\!]_\eta = \bot \text{ or } [\![t']\!]_\eta = \bot}{[\![t \circ t']\!]_\eta = \text{false}}$$

$$\frac{}{[\![e = e']\!]_\eta = ([\![e]\!]_\eta = [\![e']\!]_\eta)}$$

$$\frac{n \in \mathbb{Z}}{[\![n]\!]_\eta = n} \qquad \frac{b \in \mathit{BS}}{[\![b]\!]_\eta = b} \qquad \frac{x \in \text{dom}(\eta)}{[\![x]\!]_\eta = \eta(x)} \qquad \frac{[\![e]\!]_\eta = b \in \mathit{BS}}{[\![\text{len}(e)]\!]_\eta = |b|}$$

$$\frac{\circ \in \{+, -, *\} \quad [\![t]\!]_\eta = n \in \mathbb{Z} \quad [\![t']\!]_\eta = n' \in \mathbb{Z}}{[\![t \circ t']\!]_\eta = n \circ n'}$$

$$\frac{\tau \in \mathbb{T}_I \quad [\![e]\!]_\eta = b \in \text{dom}(\tau)}{[\![\tau(e)]\!]_\eta = \tau(b)} \qquad \frac{\tau \in \mathbb{T}_I \quad [\![t]\!]_\eta = n \in \text{range}(\tau)}{[\![\tau^{-1}(t)]\!]_\eta = \tau^{-1}(n)}$$

$$\frac{}{[\![f(e_1, \ldots, e_n)]\!]_\eta = I(f)([\![e_1]\!]_\eta, \ldots, [\![e_n]\!]_\eta)}$$

$$\frac{[\![e]\!]_\eta = b \in \mathit{BS} \quad [\![t_p]\!]_\eta = n_p \in \mathbb{N} \quad [\![t_l]\!]_\eta = n_l \in \mathbb{N} \quad n_p + n_l \leq |b|}{[\![e\{t_p, t_l\}]\!]_\eta = b[n_p] \ldots b[n_p + n_l - 1]}$$

$$\frac{[\![e]\!]_\eta = b \in \mathit{BS} \quad [\![e']\!]_\eta = b' \in \mathit{BS}}{[\![e|e']\!]_\eta = b|b'}$$

Figure 2.2: The evaluation of IML expressions.

an expression $e|e'$ evaluates to the concatenation of $e$ and $e'$.

There is an important difference between evaluation of bitstring and integer comparisons. For bitstring comparisons we follow CryptoVerif semantics in which a fact $e_1 = e_2$ evaluates to true if both $e_1$ and $e_2$ evaluate to $\bot$. On the other hand, integer comparisons are always used in context where the arguments are known to or required to be well-defined, thus a fact $t_1 = t_2$ evaluates to false if either $t_1$ or $t_2$ evaluates to $\bot$. This allows the rules of our rewriting solver presented in section 3.2 to be a lot simpler.

The syntax of IML is exactly the same as of the CryptoVerif calculus. We require that inputs and outputs alternate—this will guarantee that at any given time there is only one process ready to produce output, and all the others are waiting to receive input. We let $\text{IML} = \text{IML}_I \cup \text{IML}_O$ be the set of all input and output processes. The calculus uses *replication parameters*, denoted by $N$, to bound the number of process replications. The value of a parameter $N$ is written as $I(N) \in \mathbb{N}$. We let $[1, I(N)]$ be the set of integers from 1 to $I(N)$, represented as bitstrings

$$\{(\eta, S, 0)\} \uplus \mathcal{Q} \rightsquigarrow \mathcal{Q} \tag{INil}$$

$$\{(\eta, S, Q_1|Q_2)\} \uplus \mathcal{Q} \rightsquigarrow \{(\eta, S, Q_1), (\eta, S, Q_2)\} \uplus \mathcal{Q} \tag{IPar}$$

$$\{(\eta, S, !^{i \leq N} Q)\} \uplus \mathcal{Q} \rightsquigarrow \{(\eta\{i \mapsto b\}, S, Q) \mid b \in [1, I(N)]\} \uplus \mathcal{Q} \tag{IRepl}$$

reduce($\mathcal{Q}$) is the normal form of $\mathcal{Q}$ by $\rightsquigarrow$

Figure 2.3: The semantics of IML for input processes.

$$\frac{T = \mathtt{fixed}_n \text{ for some } n \in \mathbb{N} \quad |b| = n}{(\eta, S, \mathsf{new}\ x\colon T; P), \mathcal{Q} \rightarrow_{1/2^{8n}} (\eta\{x \mapsto b\}, S, P), \mathcal{Q}} \tag{INew}$$

$$\frac{[\![e]\!]_\eta = b \in BS_\bot}{(\eta, S, \mathsf{let}\ x = e\ \mathsf{in}\ P), \mathcal{Q} \rightarrow_1 (\eta\{x \mapsto b\}, S, P), \mathcal{Q}} \tag{ILet}$$

$$\frac{[\![\phi]\!]_\eta = \text{true}}{(\eta, S, \mathsf{if}\ \phi\ \mathsf{then}\ P\ \mathsf{else}\ P'), \mathcal{Q} \rightarrow_1 (\eta, S, P), \mathcal{Q}} \tag{IIfTrue}$$

$$\frac{[\![\phi]\!]_\eta = \text{false}}{(\eta, S, \mathsf{if}\ \phi\ \mathsf{then}\ P\ \mathsf{else}\ P'), \mathcal{Q} \rightarrow_1 (\eta, S, P'), \mathcal{Q}} \tag{IIfFalse}$$

$$\frac{[\![\phi]\!]_\eta = \text{true}}{(\eta, S, \mathsf{assume}\ \phi; P), \mathcal{Q} \rightarrow_1 (\eta, S, P), \mathcal{Q}} \tag{IAssume}$$

$$\frac{\begin{array}{l}[\![e]\!]_\eta = b \neq \bot \quad b' = \text{truncate}(b, I(\text{maxlen}(c))) \quad \forall i \leq n\colon [\![e_i]\!]_\eta = b_i \neq \bot \quad \mathcal{Q}' = \text{reduce}(\{(\eta, S, Q)\}) \\ \exists!(\eta', S', Q') \in \mathcal{Q}\colon Q' = \mathsf{in}(c[e'_1, \ldots, e'_n], x'); P' \wedge \forall i \leq n\colon [\![e'_i]\!]_{\eta'} = b_i\end{array}}{(\eta, S, \mathsf{out}(c[e_1, \ldots, e_n], e);\ Q), \mathcal{Q} \rightarrow_1 (\eta'\{x' \mapsto b'\}, S', P'), \mathcal{Q} \uplus \mathcal{Q}' \setminus \{(\eta', S', Q')\}} \tag{IOut}$$

$$\frac{\forall i \leq n\colon [\![e]\!]_\eta = b_i \neq \bot}{(\eta, S, \mathsf{event}\ ev(e_1, \ldots, e_n); P), \mathcal{Q} \xrightarrow{ev(b_1, \ldots, b_n)}_1 (\eta, S, P), \mathcal{Q}} \tag{IEvent}$$

Figure 2.4: The semantics of IML for output processes.

without leading zeros. We do not use C integer types for the representation, as we do not exclude the possibility that $I(N)$ will be greater than any value representable using a given C type. In IML the value of $I(N)$ is fixed, but in CryptoVerif it will vary depending on the security parameter. The calculus also assumes a countable set of channels. We denote channels by $c$ and for each channel $c$ let $I(\mathrm{maxlen}(c)) \in \mathbb{N}$ be the maximum length of the message that can be sent on that channel. Longer messages will be truncated—in figure 2.4 we use $\mathrm{truncate}(b, I(\mathrm{maxlen}(c)))$ to denote the result of the truncation. We make it explicit that the value $I(\mathrm{maxlen}(c))$ depends on the interpretation $I$ which in CryptoVerif calculus will vary depending on the security parameter and will make sure that the processes are executable in polynomial time.

The evaluation of IML processes is shown in figures 2.3 and 2.4. The basic unit of execution is an *executing process* of the form $(\eta, S, P)$, where $\eta$ is an environment, $P$ is an IML (input or output) process, and $S$ is a *process state*. In IML the process state is defined to be $\emptyset$ and is simply propagated through all the transformations. However, in section 2.2 we shall extend the semantics with processes that model the behaviour of C programs. Those processes will carry the memory and the execution stack in their state. We shall often write executing IML processes simply as $(\eta, P)$, meaning $(\eta, \emptyset, P)$.

The rules of figure 2.3 transform multisets of executing input processes. The process 0 does nothing, the process $P|P'$ executes $P$ and $P'$ in parallel, the process $!^{i \le N} P$ executes $I(N)$ copies of $P$ in parallel. The $n$th copy of $P$ has the value of $i$ set to $n$. The variable $i$ is called a *replication index* and may be used as a process identifier in input and output statements. The transformation relation $\rightsquigarrow$ is convergent [Blanchet, 2008], we denote the normal form of $Q$ by $\rightsquigarrow$ with $\mathrm{reduce}(Q)$.

Transition rules for output processes are shown in figure 2.4. These rules operate on *semantic configurations* of the form $\mathbb{C} = (\eta, S, P), \mathcal{Q}$, where $P$ is an output process and $\mathcal{Q}$ is a multiset of executing input processes. A rule $\mathbb{C} \xrightarrow{\mathcal{E}}_p \mathbb{C}'$ states that $\mathbb{C}$ gets transformed to $\mathbb{C}'$ with probability $p$ and an event sequence $\mathcal{E}$, which may be empty or contain a single event of the form $ev(b_1, \ldots, b_n)$, where $ev$ is an event symbol and $b_1, \ldots, b_n$ are bitstrings. We shall omit probability or event labels from the arrows when those are not relevant.

Communication is performed using the $\mathsf{in}$ and $\mathsf{out}$ constructs in which channel names are used with parameters, typically replication indices. When the semantic configuration is $(\eta, S, P), \mathcal{Q}$ and $P$ is ready to produce output $b$, we choose an input process $(\eta', S', Q)$ in $\mathcal{Q}$ that is ready to receive input into a variable $x$ on the same channel with the same parameters. The process $Q$ then reduces to an output process $P'$, the executing process $(\eta, S, P)$ reduces to a multiset of input processes $\mathcal{Q}'$, and the new semantic configuration is $(\eta'\{x \mapsto b\}, S', P'), \mathcal{Q} \uplus \mathcal{Q}' \setminus \{(\eta', S', Q)\}$. The channel parameters give the attacker the power to precisely specify the recipient of a message. Unlike in CryptoVerif, where there may be several matching recipients, here we require that there is precisely one. As we demonstrate below, this does not reduce the modelling power of the language, but it does make it easier to prove security theorems. The construct $\mathsf{new}\ x \colon T$ generates a value of type $T$ uniformly at random. For now we only consider types $T$ of the form $\mathtt{fixed}_n$ with $n \in \mathbb{N}$, representing the set of all bitstrings of length $n$. The construct $\mathsf{event}\ ev(e_1, \ldots, e_n)$ with an event symbol $ev$ is used to flag an event during the execution. The rest of the language is standard. A process of the form $\mathsf{if}\ \phi\ \mathsf{then}\ P$, without an else branch, is an abbreviation for $\mathsf{if}\ \phi\ \mathsf{then}\ P\ \mathsf{else}\ \mathsf{out}(yield[i_1, \ldots, i_n]; \varepsilon); 0$, where $yield$ is a distinguished channel name and $i_1, \ldots, i_n$ are the replication indices under which the process occurs.

Given an input process $Q$ the initial semantic configuration is defined to be $\mathrm{initConfig}(Q) = (\emptyset, \emptyset, \mathsf{out}(start, \varepsilon); 0), \mathrm{reduce}(\{(\emptyset, \emptyset, Q)\})$. The process $Q$ is thus expected to wait for a message on the $start$ channel. An IML *trace* is of the form $\mathcal{T} = \mathbb{C}_1 \rightarrow_{p_1} \ldots \rightarrow_{p_{n-1}} \mathbb{C}_n$, where $\mathbb{C}_1, \ldots, \mathbb{C}_n$ are semantic configurations such that for each $i < n$ there is a transition $\mathbb{C}_i \rightarrow_{p_i} \mathbb{C}_{i+1}$. We let $\mathrm{pr}(\mathcal{T}) = p_1 \ldots p_{n-1}$ be the probability of the trace, $\mathrm{events}(\mathcal{T})$ be the sequence of events on the transitions of the trace, and let $\mathrm{fst}(\mathcal{T}) = \mathbb{C}_1$ and $\mathrm{last}(\mathcal{T}) = \mathbb{C}_n$. We write $\mathbb{C}_1 \xrightarrow{\mathcal{E}}{}^*_p \mathbb{C}_n$ if a trace like the one above exists with $\mathcal{E} = \mathrm{events}(\mathcal{T})$ and $p = \mathrm{pr}(\mathcal{T})$.

Our goal is to investigate the probability that a process will generate a sequence of events with a certain property. For instance, authentication properties are formulated in the form *"If an event receive$(x)$ has been executed then an event send$(x)$ must have been executed before."* This motivates the following definition.

**Definition 2.1 (Trace Properties)** A *trace property* is a prefix-closed set of event sequences such that given an event sequence $\mathcal{E}$ it is decidable in polynomial time whether $\mathcal{E} \in \rho$. □

**Definition 2.2 (IML Security)** For an input IML process $Q$ and a trace property $\rho$ let

$$\mathrm{insec}(Q, \rho) = \sum_{\mathcal{T} \in \mathbb{T}} \mathrm{pr}(\mathcal{T}),$$

where $\mathbb{T}$ is the set of IML traces $\mathcal{T}$ such that $\mathrm{fst}(\mathcal{T}) = \mathrm{initConfig}(Q)$, $\mathrm{events}(\mathcal{T}) \notin \rho$, and $\mathcal{E} \in \rho$ for any proper prefix $\mathcal{E}$ of $\mathrm{events}(\mathcal{T})$. □

Given several processes $Q_1, \ldots, Q_n$ representing the protocol participants, we shall be studying their security in an environment that defines how they interact with each other (spawns multiple instances, distributes shared keys, etc) and with respect to an attacker. The environments are modelled by embedding the protocol process in a *context*.

**Definition 2.3 (Embedding)** An *evaluation context* is an IML process that contains a hole $[]$ in an input subprocess position. Given an evaluation context $C$ and an input process $Q$ we let $C[Q]$ be the process obtained from $C$ by replacing the hole by $Q$.

Given a process $Q$ and replication indices $i_1, \ldots, i_n$ we denote with $(i_1, \ldots, i_n).Q$ the process in which every channel expression of the form $c[e_1, \ldots, e_m]$ is replaced by $c[i_1, \ldots, i_n, e_1, \ldots, e_m]$.

Given a context $C$ and an input process $Q$ we define $C\{Q\} = C[(i_1, \ldots, i_n).Q]$, where $i_1, \ldots, i_n$ are the replication indices under which the hole occurs in $C$. The definition is easily generalised to multiple holes, we write $C\{Q_1, \ldots, Q_n\}$ in this case. □

When substituting $Q$ into $C\{Q\}$ under replication, we insert the replication indices to make sure that each instance of the process is uniquely identifiable. Given protocol participants $Q_1, \ldots, Q_n$, a trace property $\rho$, an environment context $C_E$, and an attacker process $Q_A$ we shall be interested in the value of $\mathrm{insec}(Q_A | C_E\{Q_1, \ldots, Q_n\}, \rho)$. We shall allow processes $Q_1, \ldots, Q_n$ to have free variables bound by $C_E$, which enables us to model shared secrets generated by the environment. However, we shall require that the full process is *well-formed*, as defined below.

**Definition 2.4 (Well-Formed Processes)** A variable is *defined* in an IML process in a let-binding, a new-binding, an input, or a replication. An IML process is *well-formed* if no variable is defined twice and every variable is defined before being used. □

The process syntax includes a way to add arbitrary assumptions that may be used by verification. We shall restrict our results to processes in which all the inline assumptions are always satisfied, as defined below:

**Definition 2.5 (Inline Assumptions)** We say that an IML semantic configuration $\mathbb{C}$ *satisfies inline assumptions* if every trace of the form

$$\mathbb{C} \longrightarrow^* (\eta, S, \mathsf{assume}\ \phi; P), \mathcal{Q}$$

satisfies $[\![\phi]\!]_\eta = \mathrm{true}$. An input IML process $Q$ *satisfies inline assumptions* if $\mathrm{initConfig}(Q)$ satisfies inline assumptions. □

**Example 2.1** We show how to set up an environment for verification of weak secrecy of one-time pad encryption, similar to our example in figure 1.1. Assume that our protocol has only one participant, either modelled by hand or extracted from implementation code, of the form

$$Q = \mathbf{in}(c_a[], ()); \mathbf{out}(c_b[], \mathit{XOR}(n, p)); 0$$

The participant simply sends a value of a nonce $n$ xored with a pad $p$, both of which are meant to be specified in the environment. Since this is more instructional, we will set up an environment in which the secrecy of the nonce is violated by reusing the same pad with several nonces:

$$\begin{aligned}
C_E = {} & (\mathbf{in}(c_1[], ()); \mathbf{new}\ p\colon \mathtt{fixed}_2; \mathbf{out}(c_2[], ()); \\
& \quad !^{i \le N}\ (\mathbf{in}(c_3[i], ()); \mathbf{new}\ n\colon \mathtt{fixed}_2; \mathbf{out}(c_4[i], ()); \\
& \qquad (\ [] \\
& \qquad |\ \mathbf{in}(c_5[i], ()); \mathbf{event}\ \mathit{leak}(n); \mathbf{out}(c_6[i], n); 0 \\
& \qquad |\ \mathbf{in}(c_7[i], x); \mathbf{if}\ x = \mathit{nonce}\ \mathbf{then\ event}\ \mathit{knows}(x); \mathbf{out}(c_8[], ()); 0)
\end{aligned}$$

The environment first generates a one-time pad and then spawns $I(N)$ sessions, each running with its own nonce (fur full generality one would usually put the environment itself under replication, but this is not necessary to demonstrate an attack). Each session consists of the participant $Q$ inserted in place of the hole [] while replacing $c_a[]$ and $c_b[]$ by $c_a[i]$ and $c_b[i]$, together with two processes required to formulate weak secrecy. One of these is a sentinel process that listens on the network and raises the event $knows(x)$ if the incoming message $x$ is the nonce, thus signalling that the attacker knows the nonce. Our model must include notion of compromise since recovery of a one time pad requires knowledge of at least one plaintext. This is achieved by the other process that is prepared to reveal the value $n$ of the nonce to the attacker, but only after raising the event $leak(n)$. The property of weak secrecy can now be formulated by letting the trace property $\rho$ be the set of traces in which every event $knows(x)$ is preceded by $leak(x)$. In a secure protocol we would expect this property to be violated with probability inversely proportional to the size of the type from which the plaintexts are drawn, that is, for every attacker process $Q_A$ we would expect that $\text{insec}(Q_A|C_E\{Q\}, \rho) = 2^{-16}$. This is of course not the case: the attacker can use a leaked nonce to recover the value of the pad which can then be used to retrieve a nonce from another session without having to compromise the participant. This attack is implemented by the following attacker process that always succeeds:

$$
\begin{aligned}
Q_A = {} & \mathbf{in}(start[], ()); \ \mathbf{out}(c_1[], ()); \ \mathbf{in}(c_2[], ()); \\
& \textit{(* find out the value of the pad by corrupting session 1 *)} \\
& \mathbf{out}(c_3[1], ()); \ \mathbf{in}(c_4[1], ()); \\
& \mathbf{out}(c_5[1], ()); \ \mathbf{in}(c_6[1], n_1); \\
& \mathbf{out}(c_a[1], ()); \ \mathbf{in}(c_b[1], msg_1); \\
& \mathbf{let}\ p = XOR(msg_1, n_1)\ \mathbf{in} \\
& \textit{(* use the value of the pad to recover the nonce in session 2 *)} \\
& \mathbf{out}(c_3[2], ()); \ \mathbf{in}(c_4[2], ()); \\
& \mathbf{out}(c_a[2], ()); \ \mathbf{in}(c_b[2], msg_2); \\
& \mathbf{let}\ n_2 = XOR(msg_2, p)\ \mathbf{in} \\
& \mathbf{out}(c_7[], n_2); \ 0
\end{aligned}
$$

Below we show the first several steps in one of the traces of $Q_A|C_E\{Q\}$, sufficient to demonstrate all of the important features of the calculus.

$$
\begin{aligned}
& \text{initConfig}(Q_A|C_E\{Q\}) \\
& = (\{\}, \mathsf{out}(start[], ()); 0), \text{reduce}(\{(\{\}, Q_A|C_E\{Q\})\}) \\
& = (\{\}, \mathsf{out}(start[], ()); 0), \ \{(\{\}, \mathsf{in}(start[], ()); \mathsf{out}(c_1[], ()); \ldots), \\
& \qquad\qquad (\{\}, \mathsf{in}(c_1[], ()); \ldots)\} \\
& \to (\{\}, \mathsf{out}(c_1[], ()); \mathsf{in}(c_2[], ()); \ldots), \ \{(\{\}, \mathsf{in}(c_1[], ()); \mathsf{new}\ p\colon \mathtt{fixed}_2; \ldots)\} \\
& \to (\{\}, \mathsf{new}\ p\colon \mathtt{fixed}_2; \mathsf{out}(c_2[], ()); \ldots), \ \{(\{\}, \mathsf{in}(c_2[], ()); \ldots)\} \\
& \to_{1/2^{16}} (\{p \mapsto 0x0201\}, \mathsf{out}(c_2[], ()); !^{i \le N}(\mathsf{in}(c_3[i], ()); \ldots)), \\
& \qquad \{(\{\}, \mathsf{in}(c_2[], ()); \mathsf{out}(c_3[1], ()); \ldots)\} \\
& \to (\{\}, \mathsf{out}(c_3[1], ()); \mathsf{in}(c_4[1], ()); \ldots), \ \{(\{p \mapsto 0x0201, i \mapsto 1\}, \mathsf{in}(c_3[i], ()); \ldots), \\
& \qquad\qquad (\{p \mapsto 0x0201, i \mapsto 2\}, \mathsf{in}(c_3[i], ()); \ldots), \ldots\}
\end{aligned}
$$

$$
\begin{aligned}
&\to (\{p \mapsto 0x0201, i \mapsto 1\}, \mathsf{new}\ n\colon \mathtt{fixed}_2; \mathsf{out}(c_4[i], ()); (\ldots)),\\
&\quad \{(\{\}, \mathsf{in}(c_4[1], ()); \ldots),\\
&\quad\ (\{p \mapsto 0x0201, i \mapsto 2\}, \mathsf{in}(c_3[i], ()); \ldots), \ldots\}\\
\to_{1/2^{16}}\ &(\{p \mapsto 0x0201, n \mapsto 0x0102, i \mapsto 1\},\\
&\quad\ \ \mathsf{out}(c_4[i], ()); (\mathsf{in}(c_a[i], ()); \ldots \,|\mathsf{in}(c_5[i], ()); \ldots \,|\mathsf{in}(c_7[i], ()); \ldots)),\\
&\quad \{(\{\}, \mathsf{in}(c_4[1], ()); \mathsf{out}(c_5[1], (); \ldots),\\
&\quad\ (\{p \mapsto 0x0201, i \mapsto 2\}, \mathsf{in}(c_3[i], (); \ldots), \ldots\}\\
&\to (\{\}, \mathsf{out}(c_5[1], (); \mathsf{in}(c_6[1], n_1); \ldots),\\
&\quad \{(\{p \mapsto 0x0201, n \mapsto 0x0102, i \mapsto 1\}, \mathsf{in}(c_a[i], ()); \ldots),\\
&\quad\ (\{p \mapsto 0x0201, n \mapsto 0x0102, i \mapsto 1\}, \mathsf{in}(c_5[i], ()); \mathsf{event}\ \mathit{leak}(n); \ldots),\\
&\quad\ (\{p \mapsto 0x0201, n \mapsto 0x0102, i \mapsto 1\}, \mathsf{in}(c_7[i], ()); \ldots),\\
&\quad\ (\{p \mapsto 0x0201, i \mapsto 2\}, \mathsf{in}(c_3[i], (); \ldots), \ldots\}\\
&\to (\{p \mapsto 0x0201, n \mapsto 0x0102, i \mapsto 1\}, \mathsf{event}\ \mathit{leak}(n); \mathsf{out}(c_6[i], n); \ldots),\\
&\quad \{(\{\}, \mathsf{in}(c_6[1], n_1); \mathsf{out}(c_a[1], ()); \ldots),\\
&\quad\ (\{p \mapsto 0x0201, n \mapsto 0x0102, i \mapsto 1\}, \mathsf{in}(c_a[i], ()); \ldots),\\
&\quad\ (\{p \mapsto 0x0201, n \mapsto 0x0102, i \mapsto 1\}, \mathsf{in}(c_7[i], ()); \ldots),\\
&\quad\ (\{p \mapsto 0x0201, i \mapsto 2\}, \mathsf{in}(c_3[i], (); \ldots), \ldots\}\\
\xrightarrow{\mathit{leak}(0x0102)}\ &(\{p \mapsto 0x0201, n \mapsto 0x0102, i \mapsto 1\}, \mathsf{out}(c_6[i], n); \ldots),\\
&\quad \{(\{\}, \mathsf{in}(c_6[1], n_1); \mathsf{out}(c_a[1], ()); \ldots),\\
&\quad\ (\{p \mapsto 0x0201, n \mapsto 0x0102, i \mapsto 1\}, \mathsf{in}(c_a[i], ()); \ldots),\\
&\quad\ (\{p \mapsto 0x0201, n \mapsto 0x0102, i \mapsto 1\}, \mathsf{in}(c_7[i], ()); \ldots),\\
&\quad\ (\{p \mapsto 0x0201, i \mapsto 2\}, \mathsf{in}(c_3[i], (); \ldots), \ldots\}\\
&\to \ldots
\end{aligned}
$$

This example highlights an important advantage of process calculi over Turing machines: it is feasible to explicitly write down a process, removing any ambiguities and making it much easier to reason about and amenable for formal verification. □

An important question is whether the process calculus is strong enough to model any efficient attacker. To show that this is the case, consider an attacker that is given by a pair of functions $f_{next}$ and $f_{output}$. The function $f_{next}$ takes a bitstring representation of an attacker state, say an encoding of a Turing machine tape, a message from the protocol, and a random value, and computes the representation of the attacker state that is ready to produce the next message to the protocol. The function $f_{output}$ takes a representation of an attacker state and returns a triple of values: an identifier $x_c$ of the channel on which to send the message to the protocol,

$Q'_E =$
$($
$\quad \mathbf{in}(c^1_{out}[x_1, \ldots, x_n],\ x_{in});$
$\quad \mathbf{new}\ r\colon \mathtt{fixed}_{maxr};$
$\quad \mathbf{out}(c_E[i+1],\ f_{next}(x_E, x_{in}, r));\ 0.$
$|$
$\quad \mathbf{in}(c^2_{out}[x_1, \ldots, x_n],\ x_{in});$
$\quad \mathbf{new}\ r\colon \mathtt{fixed}_{maxr};$
$\quad \mathbf{out}(c_E[i+1],\ f_{next}(x_E, x_{in}, r));\ 0.$
$)$

$P_E =$
$\mathbf{let}\ (x_c, (x_1, \ldots, x_n), x_{out}) = f_{output}(x_E)\ \mathbf{in}$
$\mathbf{if}\ x_c = 1$
$\mathbf{then}\ \mathbf{out}(c^1_{in}[x_1, \ldots, x_n],\ x_{out});\ Q'_E$
$\mathbf{else\ if}\ x_c = 2$
$\mathbf{then}\ \mathbf{out}(c^2_{in}[x_1, \ldots, x_n],\ x_{out});\ Q'_E$

$Q_E =$
$\mathbf{in}(start,\ _);\ \mathbf{out}(c_E[1],\ (E_0, \varepsilon));\ 0.$
$|\ !^{i \leq N_E}\ (\mathbf{in}(c_E[i],\ x_E);\ P_E)$

Figure 2.5: A universal attacker process for two input and two output channels.

the values $x_1, \ldots, x_n$ of the channel parameters, and the value $x_{out}$ of the message to send. Assume that the attacker is efficient, that is, there exists an upper bound on the execution time of $f_{next}$ and $f_{output}$, the attacker performs at most $maxn$ iterations and consumes at most $maxr$ bits of randomness per iteration (in a setting with an arbitrary security parameter $k$ these upper bounds would need to be polynomial in $k$, but in IML we deal with a fixed setting of the security parameter). Let $E_0$ be the initial attacker state.

Assume that a protocol process $Q$ is using two input channels $c^1_{in}, c^2_{in}$ and two output channels $c^1_{out}, c^2_{out}$. The interaction of the attacker with such a process can be modelled using the process $Q_E|Q$ with $Q_E$ as described in figure 2.5. First the attacker spawns $I(N_E)$ copies of the process $P_E$—we assume that $I(N_E) = maxn$. Each instance of $P_E$ represents a single iteration of the attacker. First $P_E$ computes the values of the channel $x_c$ and channel parameters $x_1, \ldots, x_n$ as well as the message $x_{out}$ to send to $Q$. Depending on the value of $x_c$ the message is sent either on $c^1_{in}$ or $c^2_{in}$. The attacker then uses parallel composition to wait for inputs on either $c^1_{out}$ or $c^2_{out}$. It then computes the next state using $f_{next}$ and passes it on to the next instance of $P_E$. The construction can be easily generalised to an arbitrary number of input and output channels of the protocol process.

Assume that every input and output in $Q$ that occurs under replication indices $i_1, \ldots, i_n$ uses all of the replication indices as channel parameters. Then it is easy to see that the restriction of the rule (IOut) that each output must have a uniquely determined recipient is satisfied. This restriction does not therefore impact our modelling capabilities.

## 2.2 The Source Language—CVM

This section describes the low-level source language *CVM (C Virtual Machine)* that we use to model the behaviour of C programs. The language is easy to formalise, while at the same time the operations of CVM are closely aligned with the operations performed by C programs, so that it is easy to translate from C to CVM. We shall discuss such a translation informally and present an example of a CVM program in section 3.1.

The model of execution of CVM is a stack-based machine with random memory access. All operations with values are performed on the stack, and values can be loaded from memory and stored back to memory. The language contains primitive operations that are necessary for implementing security protocols: reading values from the network or the execution environment, choosing random values, writing values to the network and signalling events. The only kind of conditional that CVM supports is a testing operation that checks a boolean value and aborts execution immediately if it is not true.

The fact that CVM permits no looping or recursion in the program allows us to inline all function calls, so that we do not need to add a call operation to the language itself. For simplicity of presentation we omit some aspects of the C language that are not essential for describing the approach, such as global variable initialisation and structures. Our implementation does not have these restrictions and deals with the full C language.

Modelling the behaviour of C programs is complicated by the fact that a lot of situations are left *undefined* by the C standard [ISO, 1999, Annex J2], notable examples being signed overflow or advancing a pointer outside the allocated range, except the special case when a pointer points one past the last element. Upon encountering an undefined situation the program is allowed to do absolutely anything, including divulging all the secrets to the attacker. Such a literal interpretation of the standard would leave unverifiable any protocol implementation that contains an undefined behaviour. A recent study [Dietz et al., 2012] found undefined integer overflows in many programs, including OpenSSL, Firefox, and GCC itself, which suggests that not allowing undefined behaviour at all may be too restrictive. Therefore we choose a more pragmatic approach and make additional assumptions about the execution of C programs, beyond what is provided by the standard. Specifically we assume that expressions resulting in undefined behaviour, such as unsigned integer overflow, advancing a pointer past the allocated memory range, or casting that truncates the argument, always evaluate to a certain value of the right length. For instance, we assume that signed addition of $\tau_{1s}^{-1}(1)$ and $\tau_{1s}^{-1}(127)$ will always return the same value of length 1, but we make no assumptions about what that value is. This matches how C programs behave in practice. Two other cases of undefined behaviour are more complicated, namely reading from and writing to memory that has not been allocated. A read operation will sometimes result in a crash, and sometimes return an unpredictable value, and a write operation may have unpredictable consequences due to memory corruption. In these cases we say that the program *goes bad*. Our symbolic execution will prove that this does not happen.

$c \in BS \cup \mathbb{N},\ v, x, i_1, \ldots, i_n \in Var,\ f \in Ops,\ n \in \mathbb{N}$

| | |
|---|---|
| $Q_0 \in \mathrm{CVM}_0$ ::= `Init`; $Q$ | initial process |
| $Q$ ::= | input process |
| `in`$(c[i_1, \ldots, i_n], x)$; $P$ | input |
| 0 | nil |
| $P$ ::= | output process |
| `Const` $c$; $P$ | constant value |
| `Ref` $v$; $P$ | pointer to variable |
| `Malloc`; $P$ | pointer to fresh memory |
| `Load`; $P$ | load from memory |
| `Store`; $P$ | write to memory |
| `New` $x$; $P$ | random value |
| `Apply` $f$; $P$ | operation |
| `Dup`; $P$ | duplicate |
| `Test`; $P$ | test a condition |
| `Assume`; $P$ | assume a fact |
| `ReadEnv` $x$; $P$ | read environment variable |
| `WriteEnv` $x$; $P$ | set environment variable |
| `Event` $ev\ n$; $P$ | event |
| `out`$(c[i_1, \ldots, i_n], x)$; $Q$ | output |

Figure 2.6: The syntax of CVM.

The syntax of CVM is shown in figure 2.6. Similarly to IML, we require that the inputs and the outputs alternate. This can be easily achieved for any CVM program by inserting dummy inputs and outputs. The semantics of CVM shall be described by embedding it within the IML calculus, producing a mixed language $\mathrm{IML}_C$, in which IML processes may spawn CVM programs. To describe the semantics of $\mathrm{IML}_C$ we shall only need to specify the additional rules that describe the execution of CVM processes. An advantage of this approach is that it reuses the definitions of IML for interprocess communication, relieving us of the burden of having to redefine it for CVM. A more important advantage is that $\mathrm{IML}_C$ provides a unified framework in which we can easily formulate and prove our security results. For instance, the soundness of our model extraction (theorem 3.2) can be formulated by saying that if CVM processes $Q_1, \ldots, Q_n$ yield IML models $\tilde{Q}_1, \ldots, \tilde{Q}_n$ then for any IML context $C_E$ the process $C_E\{Q_1, \ldots, Q_n\}$ is not less secure than $C_E\{\tilde{Q}_1, \ldots, \tilde{Q}_n\}$, thus reducing security verification for $\mathrm{IML}_C$ to much simpler security verification for pure IML.

**Definition 2.6 ($\mathbf{IML}_C$)** Let

$$\mathrm{IML}_C = \{C\{Q_1, \ldots, Q_n\} \mid C \text{ is an IML context, } Q_1, \ldots, Q_n \in \mathrm{CVM}_0 \cup \mathrm{IML}_I\} \qquad \square$$

The semantics of $\mathrm{IML}_C$ is given by combining the rules in figures 2.3 and 2.4 with the rules in figure 2.7. If at any point during the execution of the program there is no valid transition then the program *goes bad*. The *process states* of CVM are of the form $S = (\mathcal{A}, \mathcal{M}, \mathcal{S})$, where

- $\mathcal{M} \colon \mathbb{N} \rightharpoonup \{0x00, \ldots, 0xFF\}$ is a partial function that represents the memory.
- $\mathcal{A} \subseteq \mathbb{N}$ is the set of allocated memory addresses.
- $\mathcal{S}$ is a list of bitstrings and integers (including $\bot$) representing the execution stack.

The program represents pointers to memory as bitstrings of type $\tau_{\mathrm{ptr}}$ corresponding to C pointer types. We make no assumption about the set of valid pointers, that is, we make no assumptions about $\mathrm{range}(\tau_{\mathrm{ptr}})$ and $\mathrm{dom}(\tau_{\mathrm{ptr}})$ except that all pointers are of the same length $\mathrm{len}(\tau_{\mathrm{ptr}})$ and $0 \in \mathrm{range}(\tau_{\mathrm{ptr}})$. We use $\tau_{\mathrm{ptr}}^{-1}(0)$ to represent the null pointer in C programs, but we do not assume anything about its bitstring representation. Another distinguished type we use is $\tau_{\mathrm{size}}$ that corresponds to the C type `size_t` and is assumed to be unsigned.

Below we describe the effect that each CVM instruction has on the process state.

- When a CVM program gets spawned by an IML process, it inherits an empty process state from IML. The first instruction of each CVM program is `Init`, which results in the creation of an initial process state, ready to execute the program. In the initial state each referenced variable $v$ is allocated an address $addr(v) \in \mathbb{N}$ such that all allocations are non-overlapping. More precisely, let $\mathrm{var}(P)$ be the set of variables used in `Ref` instructions within the program $P$, for each $v \in \mathrm{var}(P)$ let $\mathrm{len}(v)$ be the size of the variable, and

$$\frac{\mathit{initmem}(Q) = (\bigcup_{v \in \mathrm{var}(P)} \{\mathit{addr}(v)\}_{\mathrm{len}(v)}}{\{(\eta,\ \emptyset,\ \texttt{Init};Q)\} \uplus \mathcal{Q} \rightsquigarrow \{(\eta,\ (\mathit{initmem}(Q),\ \emptyset,\ []),\ Q)\} \uplus \mathcal{Q}} \quad \text{(CInit)}$$

$$\frac{}{(\eta,\ (\mathcal{A},\ \mathcal{M},\ \mathcal{S}),\ \texttt{Const}\ c;P),\ \mathcal{Q} \rightarrow_1 (\eta,\ (\mathcal{A},\ \mathcal{M},\ c :: \mathcal{S}),\ P),\ \mathcal{Q}} \quad \text{(CConst)}$$

$$\frac{}{(\eta,\ (\mathcal{A},\ \mathcal{M},\ \mathcal{S}),\ \texttt{Ref}\ v;P),\ \mathcal{Q} \rightarrow_1 (\eta,\ (\mathcal{A},\ \mathcal{M},\ \tau_{\mathrm{ptr}}^{-1}(\mathit{addr}(v)) :: \mathcal{S}),\ P),\ \mathcal{Q}} \quad \text{(CRef)}$$

$$\frac{b_l \in \mathrm{dom}(\tau_{\mathrm{size}}) \quad l = \tau_{\mathrm{size}}(b_l) \quad p = \mathit{malloc}(\mathcal{A}, l) \quad \mathcal{A}' = \mathit{allocate}(\mathcal{A}, l)}{(\eta,\ (\mathcal{A},\ \mathcal{M},\ b_l :: \mathcal{S}),\ \texttt{Malloc};P),\ \mathcal{Q} \rightarrow_1 (\eta,\ (\mathcal{A}',\ \mathcal{M},\ \tau_{\mathrm{ptr}}^{-1}(p) :: \mathcal{S}),\ P),\ \mathcal{Q}} \quad \text{(CMalloc)}$$

$$\frac{\begin{array}{l} b_l \in \mathrm{dom}(\tau_{\mathrm{size}}) \quad l = \tau_{\mathrm{size}}(b_l) \quad b_p \in \mathrm{dom}(\tau_{\mathrm{ptr}}) \quad p = \tau_{\mathrm{ptr}}(b_p) \\ \{p\}_l \subseteq \mathrm{dom}(\mathcal{M}) \quad b = \mathcal{M}(p) \ldots \mathcal{M}(p + l - 1) \end{array}}{(\eta,\ (\mathcal{A},\ \mathcal{M},\ b_l :: b_p :: \mathcal{S}),\ \texttt{Load};P),\ \mathcal{Q} \rightarrow_1 (\eta,\ (\mathcal{A},\ \mathcal{M},\ b :: \mathcal{S}),\ P),\ \mathcal{Q}} \quad \text{(CLoad)}$$

$$\frac{b \neq \bot \quad b_p \in \mathrm{dom}(\tau_{\mathrm{ptr}}) \quad p = \tau_{\mathrm{ptr}}(b_p) \quad \{p\}_{|b|} \subseteq \mathcal{A} \quad \mathcal{M}' = \mathcal{M}\{p + i \mapsto b[i] \mid i < |b|\}}{(\eta,\ (\mathcal{A},\ \mathcal{M},\ b_p :: b :: \mathcal{S}),\ \texttt{Store};P),\ \mathcal{Q} \rightarrow_1 (\eta,\ (\mathcal{A},\ \mathcal{M}',\ \mathcal{S}),\ P),\ \mathcal{Q}} \quad \text{(CStore)}$$

$$\frac{b_l \in \mathrm{dom}(\tau_{\mathrm{size}}) \quad l = \tau_{\mathrm{size}}(b_l) \quad b \in \{0x00, \ldots, 0xFF\}^l}{(\eta,\ (\mathcal{A},\ \mathcal{M},\ b_l :: \mathcal{S}),\ \texttt{New}\ x;P),\ \mathcal{Q} \rightarrow_{1/2^{8l}} (\eta\{x \mapsto b\},\ (\mathcal{A},\ \mathcal{M},\ b :: \mathcal{S}),\ P),\ \mathcal{Q}} \quad \text{(CNew)}$$

$$\frac{f \text{ has arity } n \quad b = I(f)(b_1, \ldots, b_n)}{(\eta,\ (\mathcal{A},\ \mathcal{M},\ b_1 :: \ldots :: b_n :: \mathcal{S}),\ \texttt{Apply}\ f;P),\ \mathcal{Q} \rightarrow_1 (\eta,\ (\mathcal{A},\ \mathcal{M},\ b :: \mathcal{S}),\ P),\ \mathcal{Q}} \quad \text{(CApply)}$$

$$\frac{}{(\eta,\ (\mathcal{A},\ \mathcal{M},\ b :: \mathcal{S}),\ \texttt{Dup};P),\ \mathcal{Q} \rightarrow_1 (\eta,\ (\mathcal{A},\ \mathcal{M},\ b :: b :: \mathcal{S}),\ P),\ \mathcal{Q}} \quad \text{(CDup)}$$

$$\frac{}{(\eta,\ (\mathcal{A},\ \mathcal{M},\ \mathrm{true} :: \mathcal{S}),\ \texttt{Test};P),\ \mathcal{Q} \rightarrow_1 (\eta,\ (\mathcal{A},\ \mathcal{M},\ \mathcal{S}),\ P),\ \mathcal{Q}} \quad \text{(CTestTrue)}$$

$$\frac{}{(\eta,\ (\mathcal{A},\ \mathcal{M},\ \mathrm{false} :: \mathcal{S}),\ \texttt{Test};P),\ \mathcal{Q} \rightarrow_1 (\eta,\ \emptyset,\ \mathsf{out}(\mathit{yield}, \varepsilon); 0), \mathcal{Q}} \quad \text{(CTestFalse)}$$

$$\frac{}{(\eta,\ (\mathcal{A},\ \mathcal{M},\ \mathrm{true} :: \mathcal{S}),\ \texttt{Assume};P),\ \mathcal{Q} \rightarrow_1 (\eta,\ (\mathcal{A},\ \mathcal{M},\ \mathcal{S}),\ P),\ \mathcal{Q}} \quad \text{(CAssume)}$$

$$\frac{}{(\eta,\ (\mathcal{A},\ \mathcal{M},\ \mathcal{S}),\ \texttt{ReadEnv}\ x;P),\ \mathcal{Q} \rightarrow_1 (\eta,\ (\mathcal{A},\ \mathcal{M},\ [\![x]\!]_\eta :: \mathcal{S}),\ P),\ \mathcal{Q}} \quad \text{(CReadEnv)}$$

$$\frac{}{(\eta,\ (\mathcal{A},\ \mathcal{M},\ b :: \mathcal{S}),\ \texttt{WriteEnv}\ x;P),\ \mathcal{Q} \rightarrow_1 (\eta\{x \mapsto b\},\ (\mathcal{A},\ \mathcal{M},\ \mathcal{S}),\ P), \mathcal{Q}} \quad \text{(CWriteEnv)}$$

$$\frac{b_1, \ldots, b_n \neq \bot}{(\eta,\ (\mathcal{A},\ \mathcal{M},\ b_1 :: \ldots :: b_n :: \mathcal{S}),\ \texttt{Event}\ \mathit{ev}\ n;P),\ \mathcal{Q} \xrightarrow{\mathit{ev}(b_1, \ldots, b_n)}_1 (\eta,\ (\mathcal{A},\ \mathcal{M},\ \mathcal{S}),\ P), \mathcal{Q}} \quad \text{(CEvent)}$$

Figure 2.7: The semantics of CVM.

choose an allocation function $addr\colon \operatorname{var}(P) \to \mathbb{N}$. We require that the allocated memory ranges are valid and do not overlap, that is,

$$\{addr(v)\}_{\operatorname{len}(v)} \subseteq \operatorname{range}(\tau_{\text{ptr}}) \text{ for all } v \in \operatorname{var}(P),$$
$$\{addr(v)\}_{\operatorname{len}(v)} \cap \{addr(v')\}_{\operatorname{len}(v')} = \emptyset \text{ for all } v, v' \in \operatorname{var}(P) \text{ with } v \neq v'.$$

A process of the form $\mathsf{Init}\ ;\ P$ is an input process, therefore the reduction is an input reduction that gets added to the reductions in figure 2.3. All the other reductions of CVM are output reductions.

- $\mathsf{Const}\ c$ places an integer or bitstring $c$ on the stack.
- $\mathsf{Ref}\ v$ places $\tau_{\text{ptr}}^{-1}(addr(v))$ on the stack.
- $\mathsf{Malloc}$ takes a value $l = \tau_{\text{size}}(b_l)$ from the stack and uses the function *malloc* to allocate a block of memory $\{p\}_l$. The pointer $p$ to the beginning of the block is placed on the stack.

  The function $malloc\colon 2^{\mathbb{N}} \times \mathbb{N} \to \mathbb{N}$ models the memory manager: it takes a set $\mathcal{A} \subseteq \mathbb{N}$ of allocated memory addresses and a length $l \in \mathbb{N}$ and either returns 0 or an address $p \in \mathbb{N}$ such that $\{p\}_l \subseteq \operatorname{range}(\tau_{\text{ptr}}) \setminus \mathcal{A}$ (using the notation $\{p\}_l = \{p, \ldots, p+l-1\}$). We require that *malloc* runs in time polynomial in the size of the inputs.

  The allocation table is updated using the function *allocate* defined as follows:

$$allocate(\mathcal{A}, l) = \begin{cases} \mathcal{A}, & \text{if } malloc(\mathcal{A}, l) = 0, \\ \mathcal{A} \cup \{malloc(\mathcal{A}, l)\}_l\,, & \text{otherwise.} \end{cases}$$

- $\mathsf{Load}$ takes values $l$ and $p$ from the stack and returns the expression contained in the range $\{p\}_l$ in memory. If part of the range is not initialised then the program goes bad.
- $\mathsf{Store}$ is the reverse of $\mathsf{Load}$: it takes values $p$ and $b$ from the stack and writes $b$ into memory at position starting with $p$. If the memory to be written is not allocated then the program goes bad.
- $\mathsf{New}\ x$ takes a length $l$ from the stack and generates a random bitstring of length $l$. The bitstring is then both placed on the stack and recorded in the process environment $\eta$.
- $\mathsf{Apply}\ f$ where $f$ is a function symbol of arity $n$ applies $I(f)$ to $n$ values on the stack, replacing them by the result. As an abuse of notation we also allow instructions of the form $\mathsf{Apply}\ \tau$ and $\mathsf{Apply}\ \tau^{-1}$ for an integer type $\tau \in \mathbb{T}_I$ to convert bitstrings to and from integer values. The result of those operations is $\bot$ if the argument on the stack is not in the domain of the corresponding function.

- $\mathsf{Test}$ checks the bitstring at the top of the stack. If the bitstring is true the execution proceeds, if it is false the program terminates and yields to the evaluation context. Since C interpretation of boolean values may differ from IML interpretation, the C to CVM translation wraps every condition checked by the C program in a function $\mathit{truth}\colon BS_\bot \to \mathit{Bool}$ such that $\mathit{truth}(\bot) = \text{false}$ and $\mathit{truth}(b)$ returns true for $b \in BS$ if and only if the string $b$ evaluates to true according to C rules described in [ISO, 1999, 6.3.1.2].

- $\mathsf{ReadEnv}\ x$ and $\mathsf{WriteEnv}\ x$ are used by the program to access the environment $\eta$. When a program wants to receive network input, it does so in two steps. First the input is placed into the environment $\eta$ by the rule (IOut) of IML. Then the input is read from the environment and placed on the stack by the rule (CReadEnv). Conversely, when producing network output, first the rule (CWriteEnv) is used to set an environment variable, and then the rule (IOut) to actually send it. Appendix A shows examples of communication functions that demonstrate this behaviour.

- $\mathsf{Event}\ ev\ n$ raises an event with event label $ev$ and the top $n$ values on the stack as payload.

The security of $\text{IML}_C$ processes is defined in exactly the same way as for pure IML processes:

**Definition 2.7 ($\text{IML}_C$ Security)** For an input $\text{IML}_C$ process $Q$ and a trace property $\rho$ let

$$\text{insec}(Q, \rho) = \sum_{\mathcal{T} \in \mathbb{T}} \text{pr}(\mathcal{T}),$$

where $\mathbb{T}$ is the set of $\text{IML}_C$ traces $\mathcal{T}$ such that $\text{fst}(\mathcal{T}) = \text{initConfig}(Q)$, $\text{events}(\mathcal{T}) \notin \rho$, and $\mathcal{E} \in \rho$ for any proper prefix $\mathcal{E}$ of $\text{events}(\mathcal{T})$. □

By analogy with IML we define well-formed processes and processes that satisfy inline assumptions.

**Definition 2.8 (Well-Formed $\text{IML}_C$ Processes)** A variable in an $\text{IML}_C$ process is *defined* if it is used in a $\mathsf{New}$ or $\mathsf{WriteEnv}$ statement or defined by an IML construct as in definition 2.4. An $\text{IML}_C$ process is *well-formed* if no variable is defined twice and every variable is defined before being used. The well-formedness condition only speaks about the environment variables. There is no restriction on how program variables in $\mathsf{Ref}$ statements are used. □

**Definition 2.9 (Inline Assumptions)** We say that an $\text{IML}_C$ semantic configuration $\mathbb{C}$ *satisfies inline assumptions* if it satisfies IML inline assumptions as in definition 2.5 and every trace of the form

$$\mathbb{C} \longrightarrow^* (\eta, (\mathcal{A}, \mathcal{M}, b :: \mathcal{S}), \mathsf{Assume}; P), \mathcal{Q}$$

satisfies $b = \text{true}$. An input $\text{IML}_C$ process $Q$ *satisfies inline assumptions* if $\text{initConfig}(Q)$ satisfies inline assumptions. □

Our symbolic execution will prove that a CVM process never goes bad: there is no context that can force the process into a non-reducible state, except when there is no matching network communication partner. In order to make the link between a process and its descendant in the semantic configuration the following definition makes use of a process identifier stored in the process environment.

**Definition 2.10 (Going Bad)** An initial CVM process $Q_0 \in CVM_0$ *goes bad* if there exists a semantic configuration $\mathbb{C}_0 = P_0, \{(\eta_0, \emptyset, Q_0)\} \cup \mathcal{Q}_0$ such that $\mathbb{C}_0$ satisfies inline assumptions; there is a variable $id \in \mathrm{dom}(\eta_0)$ that does not occur in $P_0$ and $\mathcal{Q}_0$, and there is a trace

$$\mathbb{C}_0 \longrightarrow^* \mathbb{C} = (\eta, S, P), \mathcal{Q}$$

where $id \in \mathrm{dom}(\eta)$, $P$ does not start with an output, and $\mathbb{C}$ does not reduce. □

The use of the process identifier $id$ in the definition makes sure that the process $P$ is a descendant of the process $Q_0$. This allows us to exclude cases where the environment itself gets stuck even though the process does not.

Apart from the immediate guarantee that the process does not crash, not going bad is important from the modelling point of view: if a single process were to get stuck, this would halt the entire semantic configuration including the attacker and the other processes, which is, of course, not realistic.

## 2.3 The Simulation Relation

This section defines the main tool of our security proofs—a simulation relation $\lesssim$ between $\mathrm{IML}_C$ processes such that if $Q \lesssim \tilde{Q}$ then $Q$ is at least as secure as $\tilde{Q}$ with respect to any trace property (theorem 2.2). We also show that we can reduce the problem of checking simulation for the system of several participants in any environment to checking simulation for the participants in isolation (theorem 2.3).

Our definition for simulation requires that if an executing process $(\eta, S, Q)$ is simulated by $(\tilde{\eta}, \tilde{S}, \tilde{Q})$ then both executing processes reduce in a "similar" way to executing processes that again simulate each other. For processes that perform communication "similar" means "exactly the same" but for processes that perform computation we allow more freedom, so that many instructions in a CVM process can be simulated by a single statement in an IML model, as long as the probability of the transition and the executed events are the same.

Our definition of simulation applies to mixed IML and CVM processes, but there is nothing CVM-specific in it. It can therefore be readily applied to an embedding of any other programming language in a process calculus.

Given a relation $\lesssim$ between executing processes, we consider the *induced* relation on multisets of executing processes: For two such multisets $\mathcal{Q}$ and $\tilde{\mathcal{Q}}$ we write $\mathcal{Q} \lesssim \tilde{\mathcal{Q}}$ if there exists a pairwise matching between executing processes in $\mathcal{Q}$ and $\tilde{\mathcal{Q}}$ such that for each $(\eta, S, Q) \in \mathcal{Q}$ and its matching $(\tilde{\eta}, \tilde{S}, \tilde{Q}) \in \tilde{\mathcal{Q}}$ we have $(\eta, S, Q) \lesssim (\tilde{\eta}, \tilde{S}, \tilde{Q})$.

**Definition 2.11 (Simulation)** An IML$_C$ process $\tilde{Q}_0$ *simulates* an IML$_C$ process $Q_0$, denoted by $Q_0 \lesssim \tilde{Q}_0$, if there exists a relation $\lesssim$ between executing processes such that

1. For all $\eta$ we have $(\eta, \emptyset, Q_0) \lesssim (\eta, \emptyset, \tilde{Q}_0)$.
2. If $(\eta, S, \mathsf{in}(c[e_1, \ldots, e_n], x); P) \lesssim (\tilde{\eta}, \tilde{S}, \tilde{Q})$ then $\tilde{Q}$ is of the form $\tilde{Q} = \mathsf{in}(c[\tilde{e}_1, \ldots, \tilde{e}_n], \tilde{x}); \tilde{P}$ such that $[\![e_i]\!]_\eta = [\![\tilde{e}_i]\!]_{\tilde{\eta}}$ for all $i \leq n$ and $(\eta\{x \mapsto b\}, S, P) \lesssim (\tilde{\eta}\{\tilde{x} \mapsto b\}, \tilde{S}, \tilde{P})$ for all $b \in BS$.
3. If $(\eta, S, \mathsf{out}(c[e_1, \ldots, e_n], e); Q) \lesssim (\tilde{\eta}, \tilde{S}, \tilde{P})$ then $\tilde{P}$ is of the form $\tilde{P} = \mathsf{out}(c[\tilde{e}_1, \ldots, \tilde{e}_n], \tilde{e}); \tilde{Q}$ such that $[\![e]\!]_\eta = [\![\tilde{e}]\!]_{\tilde{\eta}}$ and $[\![e_i]\!]_\eta = [\![\tilde{e}_i]\!]_{\tilde{\eta}}$ for all $i \leq n$, and $(\eta, S, Q) \lesssim (\tilde{\eta}, \tilde{S}, \tilde{Q})$.
4. If $(\eta, S, Q) \lesssim (\tilde{\eta}, \tilde{S}, \tilde{Q})$ for some input process $Q$ then $\tilde{Q}$ is also an input process and $\mathrm{reduce}(\{(\eta, S, Q)\}) \lesssim \mathrm{reduce}(\{(\tilde{\eta}, \tilde{S}, \tilde{Q})\})$
5. If $(\eta, S, P) \lesssim (\tilde{\eta}, \tilde{S}, \tilde{P})$ for some output process $P$ and $(\eta, S, P), \mathcal{Q} \xrightarrow{\mathcal{E}}_p (\eta', S', P'), \mathcal{Q}$ with event sequence $\mathcal{E}$, probability $p$, and a multiset $\mathcal{Q}$ of input processes then $\tilde{P}$ is an output process and for any multiset $\mathcal{Q}'$ of input processes there exists a sequence of transitions of the form $(\tilde{\eta}, \tilde{S}, \tilde{P}), \mathcal{Q}' \xrightarrow{\mathcal{E}}{}^*_p (\tilde{\eta}', \tilde{S}', \tilde{P}'), \mathcal{Q}'$ such that $(\eta', S', P') \lesssim (\tilde{\eta}', \tilde{S}', \tilde{P}')$.
6. If $(\eta_1, S_1, P_1) \lesssim (\tilde{\eta}_1, \tilde{S}_1, \tilde{P}_1)$ and $(\eta_2, S_2, P_2) \lesssim (\tilde{\eta}_2, \tilde{S}_2, \tilde{P}_2)$ and $\eta_1(x) \neq \eta_2(x)$ for a variable $x$ that appears in a new-statement in $Q_0$ then there exists a variable $\tilde{x} \in \mathrm{dom}(\tilde{\eta}_1) \cap \mathrm{dom}(\tilde{\eta}_2)$ such that $\tilde{x}$ appears in a new-statement in $\tilde{Q}_0$ and $\tilde{\eta}_1(\tilde{x}) \neq \tilde{\eta}_2(\tilde{x})$. □

Condition 6 is necessary to prevent that several traces get simulated by a single trace, thus reducing the probability of the trace set. Without this condition we could prove $\mathsf{new}\ x \colon T;\ P \lesssim \mathsf{new}\ x \colon T;\ \mathsf{if}\ x = 0\ \mathsf{then}\ P\ \mathsf{else}\ Q$.

**Theorem 2.1** *The relation $\lesssim$ is transitive.* □

Proof Given two simulation relations $\lesssim_1$ and $\lesssim_2$ we can define a new relation $\lesssim$ by letting $(\eta, S, Q) \lesssim (\tilde{\eta}, \tilde{S}, \tilde{Q})$ if there exists an executing process $(\eta', S', Q')$ such that $(\eta, S, Q) \lesssim_1 (\eta', S', Q') \lesssim_2 (\tilde{\eta}, \tilde{S}, \tilde{Q})$. It is straightforward to check that $\lesssim$ is a simulation relation. ■

Next we give a proof that simulation is security-preserving (theorem 2.2). First we need to extend the simulation relation from single executing processes to semantic configurations: given a relation $\lesssim$ between executing processes we write $\mathbb{C} \lesssim \tilde{\mathbb{C}}$ for two semantic configurations $\mathbb{C} = (\eta, P), \mathcal{Q}$ and $\tilde{\mathbb{C}} = (\tilde{\eta}, \tilde{P}), \tilde{\mathcal{Q}}$ if $(\eta, P) \lesssim (\tilde{\eta}, \tilde{P})$ and $\mathcal{Q} \lesssim \tilde{\mathcal{Q}}$.

**Lemma 2.1** *Assume that $Q_0 \lesssim \tilde{Q}_0$ for two input IML$_C$ processes $Q_0$ and $\tilde{Q}_0$ and let $\lesssim$ be the corresponding simulation relation between executing processes. If $\mathbb{C} \lesssim \tilde{\mathbb{C}}$ and $\mathbb{C} \xrightarrow{\mathcal{E}}_p \mathbb{C}'$ for semantic configurations $\mathbb{C}$, $\mathbb{C}'$, and $\tilde{\mathbb{C}}$ then there exists a semantic configuration $\tilde{\mathbb{C}}'$ such that $\tilde{\mathbb{C}} \xrightarrow{\mathcal{E}}{}^*_p \tilde{\mathbb{C}}'$ and $\mathbb{C}' \lesssim \tilde{\mathbb{C}}'$.* □

PROOF If the transition from $\mathbb{C}$ to $\mathbb{C}'$ happens by any rule other than (IOut) then the statement follows immediately by condition 5 in definition 2.11. Assume that the transition happens by (IOut). Then $\mathbb{C}$ and $\mathbb{C}'$ are of the form

$$\mathbb{C} = (\eta, S, \mathsf{out}(c[e_1, \ldots, e_n], e);\ Q), \mathcal{Q},$$
$$\mathbb{C}' = (\eta'[x' \mapsto a], S', P'), \mathcal{Q} \uplus \mathrm{reduce}(\{(\eta, S, Q)\}) \setminus \{(\eta', S', Q')\},$$

where $a = [\![e]\!]_\eta$ and $(\eta', Q') \in \mathcal{Q}$ such that $Q'$ is of the form $Q' = \mathsf{in}(c[e'_1, \ldots, e'_n], x');\ P'$ and $[\![e_i]\!]_\eta = [\![e_i]\!]_{\eta'}$ for all $i \leq n$. By item 3 in definition 2.11 $\tilde{\mathbb{C}}$ is of the form

$$\tilde{\mathbb{C}} = (\tilde{\eta}, \tilde{S}, \mathsf{out}(c[\tilde{e}_1, \ldots, \tilde{e}_n], \tilde{e});\ \tilde{Q}), \tilde{\mathcal{Q}},$$

such that $(\eta, S, \mathcal{Q}) \lesssim (\tilde{\eta}, \tilde{S}, \tilde{\mathcal{Q}})$ and $[\![e]\!]_\eta = [\![\tilde{e}]\!]_{\tilde{\eta}}$ and $[\![e_i]\!]_\eta = [\![\tilde{e}_i]\!]_{\tilde{\eta}}$ for all $i \leq n$. Simulation provides us with $(\tilde{\eta}', \tilde{S}', \tilde{Q}') \in \tilde{\mathcal{Q}}$ such that $(\eta', S', Q') \lesssim (\tilde{\eta}', \tilde{S}', \tilde{Q}')$. By item 2 in definition 2.11 $\tilde{Q}'$ is of the form

$$\tilde{Q}' = \mathsf{in}(c[\tilde{e}'_1, \ldots, \tilde{e}'_n], \tilde{x}');\ \tilde{P}'$$

such that $(\eta'[x' \mapsto a], S', P') \lesssim (\tilde{\eta}'[\tilde{x} \mapsto a], \tilde{S}', \tilde{P}')$ and $[\![\tilde{e}'_i]\!]_{\tilde{\eta}'} = [\![e'_i]\!]_{\eta'}$ for all $i \leq n$. Let

$$\tilde{\mathbb{C}}' = (\tilde{\eta}'[\tilde{x} \mapsto a], \tilde{S}', \tilde{P}'), \tilde{\mathcal{Q}} \uplus \mathrm{reduce}(\{(\tilde{\eta}, \tilde{S}, \tilde{Q})\}) \setminus \{(\tilde{\eta}', \tilde{S}', \tilde{Q}')\}.$$

We show that $\tilde{\mathbb{C}}'$ satisfies the statement of the lemma. To prove $\tilde{\mathbb{C}} \to \tilde{\mathbb{C}}'$ the only thing left to check is the unique recipient condition: there exists no $(\tilde{\eta}'', \tilde{S}'', \tilde{Q}'') \in \tilde{\mathcal{Q}}$ with $(\tilde{\eta}'', \tilde{S}'', \tilde{Q}'') \neq (\tilde{\eta}', \tilde{S}', \tilde{Q}')$ such that

$$\tilde{Q}'' = \mathsf{in}(c[\tilde{e}''_1, \ldots, \tilde{e}''_n], \tilde{x}'');\ \tilde{P}'')$$

and $[\![\tilde{e}''_i]\!]_{\tilde{\eta}''} = [\![e'_i]\!]_{\eta'}$ for all $i \leq n$. Assume that such $(\tilde{\eta}'', \tilde{S}'', \tilde{Q}'')$ exists. By simulation we can find $(\eta'', S'', Q'') \in \mathcal{Q}$ such that $Q'' \neq Q'$ and $(\eta'', S'', Q'') \lesssim (\tilde{\eta}'', \tilde{S}'', \tilde{Q}'')$. All processes in $\mathcal{Q}$ start with an input, thus using item 2 in definition 2.11 we see that

$$Q'' = \mathsf{in}(c[e''_1, \ldots, e''_n], x'');\ P'')$$

such that $[\![e''_i]\!]_{\eta''} = [\![e'_i]\!]_{\eta'}$ for all $i \leq n$. This contradicts the unique recipient condition of the transition $\mathbb{C} \to \mathbb{C}'$.

To prove $\mathbb{C}' \lesssim \tilde{\mathbb{C}}'$ the only thing left to show is that $\mathrm{reduce}(\{(\eta, S, Q)\}) \lesssim \mathrm{reduce}(\{(\tilde{\eta}, \tilde{S}, \tilde{Q})\})$. This follows directly from $(\eta, S, \mathcal{Q}) \lesssim (\tilde{\eta}, \tilde{S}, \tilde{\mathcal{Q}})$ and condition 4 in definition 2.11. ∎

**Theorem 2.2 (Simulation Preserves Security)** *If $Q_0 \lesssim \tilde{Q}_0$ for two input $\mathrm{IML}_C$ processes $Q_0$ and $\tilde{Q}_0$ then $\mathrm{insec}(Q_0, \rho) \leq \mathrm{insec}(\tilde{Q}_0, \rho)$ for any correspondence property $\rho$.* □

PROOF Let $Q_0$ and $\tilde{Q}_0$ be as in the statement of the theorem. Obtain a relation $\lesssim$ between executing processes from definition 2.11. We shall define a function sim from traces to traces as follows. Consider a trace $\mathcal{T} = \mathbb{C}_1 \rightarrow \ldots \rightarrow \mathbb{C}_n$ where $\mathbb{C} = \mathrm{initConfig}(Q_0)$. We can extend $\lesssim$ by letting $(\emptyset, \emptyset, \mathsf{out}(\mathit{start}, \varepsilon); 0) \lesssim (\emptyset, \emptyset, \mathsf{out}(\mathit{start}, \varepsilon); 0)$ without breaking the properties of $\lesssim$. Then by conditions 1 and 4 of definition 2.11 $\mathrm{initConfig}(Q_0) \lesssim \mathrm{initConfig}(\tilde{Q}_0)$. Iterated application of lemma 2.1 yields a trace $\tilde{\mathcal{T}} = \tilde{\mathbb{C}}_1 \rightarrow^* \ldots \rightarrow^* \mathbb{C}_n$ such that $\tilde{\mathbb{C}}_1 = \mathrm{initConfig}(\tilde{Q}_0)$, $\mathrm{pr}(\mathcal{T}) = \mathrm{pr}(\tilde{\mathcal{T}})$, $\mathrm{events}(\mathcal{T}) = \mathrm{events}(\tilde{\mathcal{T}})$, and $\mathbb{C}_i \lesssim \tilde{\mathbb{C}}_i$ for all $i \leq n$. Let $\mathrm{sim}(\mathcal{T}) = \tilde{\mathcal{T}}$.

Consider a correspondence property $\rho$. Following definition 2.7 let $\mathbb{T}$ be the set of $\mathrm{IML}_C$ traces $\mathcal{T}$ such that $\mathrm{fst}(\mathcal{T}) = \mathrm{initConfig}(Q_0)$, $\mathrm{events}(\mathcal{T}) \notin \rho$, and $\mathcal{E} \in \rho$ for any proper prefix $\mathcal{E}$ of $\mathrm{events}(\mathcal{T})$. Then sim is injective on $\mathbb{T}$: Consider two traces $\mathcal{T} = \mathbb{C}_1 \rightarrow \ldots \rightarrow \mathbb{C}_n \in \mathbb{T}$ and $\mathcal{T}' = \mathbb{C}'_1 \rightarrow \ldots \rightarrow \mathbb{C}'_{n'} \in \mathbb{T}$ such that $\mathcal{T} \neq \mathcal{T}'$. By definition if a trace $\mathcal{T}$ is in $\mathbb{T}$ then no prefix of $\mathcal{T}$ is in $\mathbb{T}$. Given that $\mathbb{C}_1 = \mathbb{C}'_1$ there exists $i < \min(n, n')$ such that $\mathbb{C}_j = \mathbb{C}_j$ for all $j \leq i$ and $\mathbb{C}_{i+1} \neq \mathbb{C}'_{i+1}$. The only semantic rules that allow multiple transitions are (INew) and (CNew). Assume that the transition from $\mathbb{C}_i$ to $\mathbb{C}_{i+1}$ is taken by (INew). Then the semantic configurations have the form

$$\mathbb{C}_i = (\eta, \emptyset, \mathsf{new}\ x\colon T; P), \mathcal{Q} \qquad \mathbb{C}_{i+1} = (\eta\{x \mapsto a\}, \emptyset, P), \mathcal{Q} \qquad \mathbb{C}'_{i+1} = (\eta\{x \mapsto a'\}, \emptyset, P), \mathcal{Q}$$

for some $a \neq a'$. Let $\tilde{\mathbb{C}}_1 \rightarrow^* \ldots \rightarrow^* \tilde{\mathbb{C}}_n = \mathrm{sim}(\mathcal{T})$ and $\tilde{\mathbb{C}}'_1 \rightarrow^* \ldots \rightarrow^* \tilde{\mathbb{C}}'_{n'} = \mathrm{sim}(\mathcal{T}')$. By condition 6 of definition 2.11 it is easy to see that $\tilde{\mathbb{C}}_{i+1} \neq \tilde{\mathbb{C}}'_{i+1}$, thus $\mathrm{sim}(\mathcal{T}) \neq \mathrm{sim}(\mathcal{T}')$. The case that the transition from $\mathbb{C}_i$ to $\mathbb{C}_{i+1}$ is taken by (CNew) is similar.

Let $\tilde{\mathbb{T}}$ be the set of $\mathrm{IML}_C$ traces $\tilde{\mathcal{T}}$ such that $\mathrm{fst}(\tilde{\mathcal{T}}) = \mathrm{initConfig}(\tilde{Q}_0)$, $\mathrm{events}(\tilde{\mathcal{T}}) \notin \rho$, and $\mathcal{E} \in \rho$ for any proper prefix $\mathcal{E}$ of $\mathrm{events}(\tilde{\mathcal{T}})$. Then

$$\mathrm{insec}(Q_0, \rho) = \sum_{\mathcal{T} \in \mathbb{T}} \mathrm{pr}(\mathcal{T}) = \sum_{\mathcal{T} \in \mathbb{T}} \mathrm{pr}(\mathrm{sim}(\mathcal{T})) = \sum_{\mathcal{T} \in \mathrm{sim}(\mathbb{T})} \mathrm{pr}(\mathcal{T}) \leq \sum_{\tilde{\mathcal{T}} \in \tilde{\mathbb{T}}} \mathrm{pr}(\tilde{\mathcal{T}}) = \mathrm{insec}(\tilde{Q}_0, \rho).$$

The third equality is true by injectivity of sim and the inequality is true because by construction of sim we have $\mathrm{sim}(\mathbb{T}) \subseteq \tilde{\mathbb{T}}$. ■

We now prove that simulation is preserved by embedding of the form $C\{Q\}$ (theorem 2.3). This is done in two steps: first we show that simulation is preserved by prepending new indices to all channel arguments of a process, and then we show that simulation is preserved by simple embedding of the form $C[Q]$.

**Lemma 2.2** *Let $i_1, \ldots, i_n$ be replication indices and let $Q_0$ and $\tilde{Q}_0$ be $\mathrm{IML}_C$ processes such that $Q_0 \lesssim \tilde{Q}_0$ and $i_1, \ldots, i_n$ are defined neither in $Q_0$ nor in $\tilde{Q}_0$. Then $(i_1, \ldots, i_n).Q_0 \lesssim (i_1, \ldots, i_n).\tilde{Q}_0$.* □

PROOF Let $i_1, \ldots, i_n$, $Q_0$, and $\tilde{Q}_0$ be as in the statement of the lemma. Obtain the simulation relation $\lesssim$ between executing processes from definition 2.11. Define a new simulation relation $\lesssim^*$ as follows: let $(\eta, S, Q) \lesssim^* (\tilde{\eta}, \tilde{S}, \tilde{Q})$ if $\eta(i_j) = \tilde{\eta}(i_j)$ for all $j \leq n$ and there exist processes $Q'$ and $\tilde{Q}'$ such that $Q = (i_1, \ldots, i_n).Q$, $\tilde{Q} = (i_1, \ldots, i_n).\tilde{Q}$, and $(\eta, S, Q') \lesssim (\tilde{\eta}, \tilde{S}, \tilde{Q}')$. It is straightforward to check that $\lesssim^*$ defines a simulation between $(i_1, \ldots, i_n).Q_0$ and $(i_1, \ldots, i_n).\tilde{Q}_0$. ∎

**Theorem 2.3 (Simulation is Preserved by Embedding)** *Let $Q_1, \ldots, Q_n$ and $\tilde{Q}_1, \ldots, \tilde{Q}_n$ be* IML$_C$ *processes such that $Q_i \lesssim \tilde{Q}_i$ for each $i \leq n$. Then for any context $C$ with $n$ holes such that both $C\{Q_1, \ldots, Q_n\}$ and $C\{\tilde{Q}_1, \ldots, \tilde{Q}_n\}$ are well-formed we have*

$$C\{Q_1, \ldots, Q_n\} \lesssim C\{\tilde{Q}_1, \ldots, \tilde{Q}_n\}.$$ □

Proof Let $Q_1, \ldots, Q_n$, $\tilde{Q}_1, \ldots, \tilde{Q}_n$, and $C$ be as in the statement of the theorem. The condition that $C\{Q_1, \ldots, Q_n\}$ and $C\{\tilde{Q}_1, \ldots, \tilde{Q}_n\}$ are well-formed implies that neither $Q_i$ nor $\tilde{Q}_i$ for $i \leq n$ define any replication indices in $C$. Therefore by lemma 2.2 it suffices to show the statement for $C[Q_1, \ldots, Q_n]$ and $C[\tilde{Q}_1, \ldots, \tilde{Q}_n]$.

For each $i \leq n$ obtain a simulation relation $\lesssim_i$ between executing processes from definition 2.11 applied to $Q_i \lesssim \tilde{Q}_i$. Define a new simulation relation $\lesssim^*$ as follows: let $(\eta, S, Q) \lesssim^* (\tilde{\eta}, \tilde{S}, \tilde{Q})$ if either of the following is true:

- There exist indices $i_1, \ldots, i_k \leq n$ and a context $C'$ such that $Q = C'[Q_{i_1}, \ldots, Q_{i_k}]$ and $\tilde{Q} = C'[\tilde{Q}_{i_1}, \ldots, \tilde{Q}_{i_k}]$, and $\eta = \tilde{\eta}$, and $S = \tilde{S}$.
- $(\eta, S, Q) \lesssim_i (\tilde{\eta}, \tilde{S}, \tilde{Q})$ for some $i \leq n$.

It is straightforward to check that $\lesssim^*$ defines a simulation between the processes $C[Q_1, \ldots, Q_n]$ and $C[\tilde{Q}_1, \ldots, \tilde{Q}_n]$. ∎

# Chapter 3

# Model Extraction By Symbolic Execution

This chapter describes the path from a C program to its IML model. First the C program is compiled to a CVM process as explained in section 3.1. The translation does not descend into implementations of cryptographic primitives or system functions like memcpy, instead it replaces these functions by their models. For instance, a call to encrypt may be replaced by a single CVM instruction Apply *encrypt*. The models for cryptographic and system functions are provided by the user in the form of *proxy functions* that are themselves written in C.

To obtain an IML model of a CVM process we execute it with respect to symbolic semantics, in which instructions operate on expressions instead of concrete values, as outlined in section 3.5. These expressions over-approximate the set of all possible values that can occur at a given point. The main difference from existing symbolic execution algorithms, such as Cadar et al. [2008] or Godefroid et al. [2008], is that our symbolic variables represent bitstrings of potentially unknown length, whereas in previous algorithms a single variable corresponds to a single byte. We demonstrate symbolic execution on the example of RPC-enc (section 1.1) in section 3.6. The main theoretical results of this chapter are theorems 3.2 and 3.3 presented in section 3.7 that capture the correctness of symbolic execution. Theorem 3.2 states that model extraction by symbolic execution is sound: if $\tilde{Q}_1, \ldots, \tilde{Q}_n$ are IML models extracted from CVM programs $Q_1, \ldots, Q_n$ then for any execution context $C$ and trace property $\rho$ we have $\text{insec}(C\{Q_1, \ldots, Q_n\}, \rho) \leq \text{insec}(C\{\tilde{Q}_1, \ldots, \tilde{Q}_n\}, \rho)$. The proof proceeds by demonstrating a simulation relation $Q_i \lesssim \tilde{Q}_i$ (definition 2.11) and then applying theorems 2.2 and 2.3. Theorem 3.3 additionally proves that a CVM program that can be successfully symbolically executed never goes bad (definition 2.10). This means, in particular, that the CVM program is memory-safe. Symbolic execution treats bitstring arithmetic operations soundly with respect to overflow.

An important building block both for the symbolic execution and for the translation methods used in chapter 4 is the ability to check whether a certain fact implies another fact, where facts can contain a mixture of integer and binary operations. In section 3.2 we show how to deal with such facts in a sound way, using the SMT solver Yices [Dutertre and Moura, 2006] as our backend. Our solver is based on rewriting rules of the form $\phi \vdash e \rightsquigarrow e'$ meaning that $e$ and $e'$ evaluate to the same value whenever $\phi$ is true. The same set of rules can be used for simplifying expressions, as shown in section 3.3. Our solver is easily extensible—we can teach it to handle more symbolic expressions by simply adding more rewriting rules. This is demonstrated in section 3.4 where we add a new symbolic construct for representing C pointers during symbolic execution and add support for pointers to our solver simply by listing the relevant rewriting rules.

## 3.1 From C to CVM

Given a C program $P$ we use CIL [Necula et al., 2002] to modify it such that it outputs its own CVM representation when run. This means that we only analyse a single *main path* through the code, so our verification result applies to a restriction $\tilde{P}$ of $P$ that allows the execution to proceed only on the main path, and aborts immediately if another path is taken. All of the protocols that we analyse have linear structure—a participant stops execution if any of the cryptographic checks fails. It is therefore reasonable to assume that the restricted program $\tilde{P}$ still remains a functional implementation of the protocol, and so security of $\tilde{P}$ gives confidence in the security of $P$. It is important that, despite analysing a single path of the protocol, we still cover all possible executions that take that path. In section 7.1 we discuss how we could extend the architecture of our implementation to cover multiple branches.

The idea of instrumenting a program to emit a low-level set of instructions for symbolic execution at runtime as well as some initial implementation code were borrowed from the CREST symbolic execution tool [Burnim and Sen, 2008].

We do not formalise the compilation step from C to CVM and believe that prior work may be used to fill this gap. For instance, instead of CVM we could start our analysis from one of the intermediate stages of the CompCert compiler [Leroy, 2009] that has been formally proven correct. Another option would be to use the LLVM language [Lattner, 2002] as the starting point. LLVM is not formally verified, but using it would ensure that the low-level code we verify is the same code that gets executed.

As mentioned in chapter 1, we do not verify the source code of cryptographic functions, but instead trust that they implement the cryptographic algorithms correctly. Similarly, we would not be able to translate the source code of functions like `memcmp` into CVM directly, as these functions contain loops. Thus for the purpose of CVM translation we provide an abstraction for these functions. We do so by writing a *proxy function* `f_proxy` for each function `f` that needs to be abstracted. Whenever a call to `f` is encountered during the translation, it is replaced by the call to `f_proxy`. The proxy functions are part of the trusted base of the verification.

```
void LoadBuf(void * buf; size_t len) __attribute__((not_instrumented));
void LoadBuf(void * buf; size_t len){
  cvm("Load");
}

void Val(bool is_signed, size_t width){
  char cmd[100];
  sprintf(cmd, "Apply Val(%ld, %s)", width, is_signed ? "S" : "U");
  cvm(cmd);
}

void assume_len(const unsigned char * len, bool is_signed, size_t width){
  cvm("Dup");
  cvm("Len");
  LoadBuf(len, width);
  Val(is_signed, width);
  cvm("Apply EqInt/2");
  cvm("Assume");
}

void xor_proxy(unsigned char * buf, unsigned char * pad, size_t len){
  mute_cvm();
  xor(buf, pad, len);
  unmute_cvm();
  LoadBuf(buf, len);
  LoadBuf(pad, len);
  cvm("Apply xor/2");
  assume_len(&len, FALSE, sizeof(len));
  StoreBuf(buf);
}
```

Figure 3.1: Examples of proxy functions.

Examples of proxy functions are shown in figure 3.1. Each function f_proxy starts by calling the function f that is being proxied so that the concrete execution can proceed as usual. The proxy function then uses the function cvm to print to a file the sequence of instructions that represent in CVM the effect of calling f. The call to f itself is guarded by a pair of calls mute_cvm and unmute_cvm that silence the printing of CVM instructions from within f itself.

We shall now walk through the code of xor_proxy in detail. The calls to the helper function LoadBuf are meant to extract a value from memory and put it on the CVM execution stack. In order to understand how these calls work it is important to remember that more calls to cvm will be inserted by the automated instrumentation. In particular, when a function is called, the instrumentation inserts the necessary instructions to load the arguments from memory onto the execution stack (the called function is then expected to store the values from the stack into its local variables). Thus the call LoadBuf(buf, len) after instrumentation becomes

```
cvm("Ref buf"); cvm("Const 8"); cvm("Apply to_bitstring(8, U)/1"); cvm("Load");
cvm("Ref len"); cvm("Const 8"); cvm("Apply to_bitstring(8, U)/1"); cvm("Load");
LoadBuf(buf, len);
```

This instrumentation puts on the stack references to the variables buf and len, followed by the sizes of these variables (8 bytes, assuming a 64-bit architecture), and then uses the Load instructions to load the values of these variables onto the stack. The symbol " to_bitstring (8, U)" is the C string representation of $\tau_{U8}^{-1}$. When LoadBuf starts executing, the CVM stack already contains the correct arguments for the CVM Load instruction. Automatically generated instrumentation would store these values into the C argument variables. Instead we use a user-defined C attribute not_instrumented to signal to our instrumentation to keep LoadBuf unmodified, and we manually add the desired instruction Load.

After having loaded the arguments of the *xor* operation onto the stack xor_proxy uses the instruction Apply $xor/2$ which puts on the stack the result of the *xor* operation. It then uses the call to the helper function assume_len to communicate to the symbolic execution our assumption that the length of the result is equal to the length of the arguments. If $b$ is the result of *xor* and $b_l$ the bitstring contained in the len variable then assume_len constructs the fact $\text{len}(b) = \tau_{U8}(l)$ on the stack. It then uses the Assume instruction to make the symbolic execution assume that the value of the constructed fact is true in every possible run of the program.

The last step of xor_proxy is to use use the helper function StoreBuf (not shown here) that works similarly to LoadBuf and stores the result of *xor* from the stack into the memory pointed to by variable buf.

Proxy functions allow to model library functions that are only partially trusted. For instance, a proxy function can leak its arguments to the attacker, or make the result depend on a value that is read from the attacker. Appendix A contains more examples of proxy functions, including all the functions used by our example in figure 1.1.

Several aspects of our implementation are not mentioned elsewhere since they do not add to the core idea of the approach:

Init ; in(c, _); // *CVM initialisation*

*// unsigned char ∗ payload = malloc(PAYLOAD_LEN); // PAYLOAD_LEN = 20*
Const $\tau_{8u}^{-1}(20)$; Malloc; Ref payload; Store;

*// if (payload == NULL) exit(1);*
Ref payload; Const $\tau_{8u}^{-1}(8)$; Load; Apply null/0; Apply $==_{\tau_{\text{ptr}}}/2$; Apply $truth/1$; Apply $\neg/1$; Test;

*// RAND_bytes(payload, PAYLOAD_LEN);*
Const $\tau_{4s}^{-1}(20)$; New payload; Ref payload; Const $\tau_{8u}^{-1}(8)$; Load; Store;

*// size_t msg_len = PAYLOAD_LEN + 1;*
Const $\tau_{8u}^{-1}(20)$; Const $\tau_{8u}^{-1}(1)$; Apply $\bigoplus_{\tau_{8u}}/2$; Ref msg_len; Store;

*// unsigned char ∗ msg = malloc(msg_len);*
Ref msg_len; Const $\tau_{8u}^{-1}(8)$; Load; Malloc; Ref msg; Store;

*// if (msg == NULL) exit(1);*
Ref msg; Const $\tau_{8u}^{-1}(8)$; Load; Apply null/0; Apply $==_{\tau_{\text{ptr}}}/2$; Apply $truth/1$; Apply $\neg/1$; Test;

*// ∗msg = 0x01; // add the tag*
Const 0x01; Ref msg; Const $\tau_{8u}^{-1}(8)$; Load; Store;

*// memcpy(msg + 1, payload, PAYLOAD_LEN); // add the payload*
Ref payload; Const $\tau_{8u}^{-1}(8)$; Load; Const $\tau_{8u}^{-1}(20)$; Load;
Ref msg; Const $\tau_{8u}^{-1}(8)$; Load; Const $\tau_{8u}^{-1}(1)$; Apply $\bigoplus_{PI}^{\tau_{8u}}/2$; Store;

*// unsigned char ∗ pad = otp(msg_len);*
Ref msg_len; Const $\tau_{8u}^{-1}(8)$; Load; Malloc;
Dup; Apply null/0; Apply $!=_{\tau_{\text{ptr}}}/2$; Apply $truth/1$; Assume;
Ref pad; Const $\tau_{8u}^{-1}(8)$; Store;
ReadEnv pad;
Dup; Apply len/1; Apply $\tau_{8u}^{-1}$; Ref msg_len; Const $\tau_{8u}^{-1}(8)$; Load;
Apply $==_{\tau_{8u}}/2$; Apply $truth/1$; Assume;
Ref pad; Const $\tau_{8u}^{-1}(8)$; Load; Store;

*// xor(msg, pad, msg_len); // apply one−time pad*
Ref msg; Const $\tau_{8u}^{-1}(8)$; Load; Ref msg_len; Const $\tau_{8u}^{-1}(8)$; Load; Load;
Ref msg; Const $\tau_{8u}^{-1}(8)$; Load; Ref msg_len; Const $\tau_{8u}^{-1}(8)$; Load; Load;
Apply xor/2;
Dup; Apply len/1; Apply $\tau_{8u}^{-1}$; Ref msg_len; Const $\tau_{8u}^{-1}(8)$; Load;
Apply $==_{\tau_{8u}}/2$; Apply $truth/1$; Assume;
Ref msg; Const $\tau_{8u}^{-1}(8)$; Load; Store;

*// send(msg, msg_len);*
Ref msg; Const $\tau_{8u}^{-1}(8)$; Load; Ref msg_len; Const $\tau_{8u}^{-1}(8)$; Load; Load;
WriteEnv msg; out(c, msg); 0

Figure 3.2: Translation of the example C program (figure 1.1) into CVM.

- Our implementation supports C structures by representing field offsets using CVM functions. For instance, a pointer &(s.x) would be represented by taking the reference to s using Ref s and then adding to it an integer expression *field_offset_x*().

- Non-static variables in different calls to the same function are given different names. A CIL transformation pulls local variables in inner scopes to function scope, so we do not have to worry about locally scoped variables.

- Our verification works with respect to a particular machine architecture since it uses specific values for the sizes of variables. This could be improved by using symbolic expressions for these sizes (say, *sizeof_int* instead of 4) and making sure none of the messages that reach the network refer to such symbolic sizes in any way.

- We allow calls of functions via function pointers, as long as those can be statically resolved by the symbolic execution. For instance, we allow p = &f; p();, but not p = rand(); p();.

- We do not support variadic functions, but luckily the only variadic functions we came across are from the printf-family, and so can be removed from the code without affecting the implementation (assuming these calls are not observable by the attacker).

**Example 3.1** Figure 3.2 shows a CVM translation of the example program from figure 1.1, simplified to remove a level of indirection arising from function calls. We assume a 64-bit architecture, such that $\tau_{\text{size}} = \tau_{8u}$. An abbreviation Const $\tau(n)$ with an integer type $\tau$ and $n \in \mathbb{N}$ stands for the sequence of instructions Const $n$; Apply $\tau/1$. For convenience we write operation arguments of Apply together with their arities. Operations used by the program include the binary addition $\bigoplus_\tau$ and an addition of an integer to a pointer $\bigoplus_{PI}^{\tau}$, both parameterised by the type $\tau$ of their integer operands. The binary arithmetic operations are discussed in detail in section 3.2 and the pointer operations are discussed in detail in section 3.4. The function *truth* introduced in section 2.2 is used to convert C boolean values to CVM boolean values before applying the Test instructions.

The proxy function for otp loads the value from the environment variable *pad*—this will be a free variable in the extracted IML model. The user is expected to provide an environment, like the one in figure 1.1 that binds *pad* to a value. The program variable pad used in C code should not be confused with the environment variable *pad*.

The call to RAND_bytes generates a simpler sequence of instructions compared to the model shown in appendix A—here we assume that the function always succeeds. □

## 3.2 Facts and Implication

An important building block in all our algorithms is being able to decide whether a certain fact is implied by another fact. The problem is that we are dealing with facts of two types. On the one hand, we would like to prove facts about integers, say, to bound a length of a certain bitstring.

| | |
|---|---|
| $\phi, \psi ::=$ | fact |
| $\phi \circ \psi,\ \circ \in \{\wedge, \vee, \neg\}$ | logical operation |
| $t \circ t',\ \circ \in \{=, \neq, <, >\}$ | integer comparison |
| $e \circ_\tau e',\ \circ_\tau \in \{==_\tau, !=_\tau, <_\tau, >_\tau\},\ \tau \in \mathbb{T}_I$ | binary comparison |
| $e = e'$ | bitstring equality |
| $\mathrm{defined}(e)$ | $e$ is well-defined |
| $v \in \mathit{Var}$ | logical variable |
| $t, p, l ::=$ | integer term |
| $n \in \mathbb{Z}$ | integer |
| $t \circ t',\ \circ \in \{+, -, *\}$ | integer operation |
| $\mathrm{len}(e)$ | length |
| $\tau(e),\ \tau \in \mathbb{T}_I$ | value |
| $e ::=$ | bitstring term |
| $e \circ_\tau e',\ \circ_\tau \in \{\oplus_\tau, \ominus_\tau, \circledast_\tau\},\ \tau \in \mathbb{T}_I$ | binary operation |
| $\mathrm{cast}_{\tau_1 \to \tau_2}(e),\ \tau_1, \tau_2 \in \mathbb{T}_I$ | cast |
| $\mathit{cmp}(e_1, e_2)$ | C bitstring comparison |
| $\tau^{-1}(t),\ \tau \in \mathbb{T}_I$ | bitstring encoding |
| $x \in \mathit{Var}$ | variable |
| $b \in \mathit{BS}$ | bitstring |
| $e\{p, l\}$ | substring |
| $e \mid e'$ | concatenation |
| $f(e_1, \ldots, e_n)$ | other function |

Figure 3.3: The syntax of facts.

On the other hand, the C implementation provides us with facts about bitstring representation of integers. These representations can overflow or get truncated in a cast between a signed and an unsigned value. The matter is further complicated by the necessity to prove that bitstring expressions are well-defined (do not evaluate to $\bot$). This section describes how we use the SMT solver Yices [Dutertre and Moura, 2006] to solve these problems.

Yices contains both a theory of integers and a theory of fixed-size bitstrings, but it does not allow to mix them freely. For instance, there is no function corresponding to any of our $\tau \in \mathbb{T}_I$ that would allow to obtain an integer value of a bitstring. Since our expressions can contain unbounded integers the theory of fixed-width bitstrings alone is not sufficient to deal with them. Thus we chose to use integer arithmetic for all operations and to formulate side conditions to ensure that no overflows occur whenever we replace bitstring arithmetic operations with their integer counterparts. One advantage is efficiency: bitstring solving uses bit-blasting that may take exponential time in the size of the bitstrings, whereas integer solving uses the simplex method the runtime of which only depends logarithmically on the magnitude of the integers. Another advantage is that formulating side-conditions allows us to catch overflows early—if a side condition fails it is a good indication that something is wrong with the implementation being analysed. Bitstring solving treats overflows silently and does not provide such an indication. The downside of relying on the integer solver is that we cannot easily analyse a protocol that intentionally uses overflow for security-relevant purposes. Yices only supports linear arithmetic, but that turns out to be sufficient, as all our facts essentially deal with lengths of bitstrings and we have not yet seen an implementation that would multiply two lengths.

We shall be dealing with facts as defined in figure 2.1, but now we are specifically interested in some binary operations arising from the execution of C programs. For an integer type $\tau \in \mathbb{T}_I$ let $\oplus_\tau, \ominus_\tau, \circledast_\tau$ be binary addition, subtraction, and multiplication operators, and let $==_\tau, !=_\tau, <_\tau, >_\tau$ be binary comparison operators. For $\tau_1, \tau_2 \in \mathbb{T}_I$ let $\mathrm{cast}_{\tau_1 \to \tau_2}$ be the C cast function from type $\tau_1$ to $\tau_2$. Let *cmp* be the C bitstring comparison function (the application of *cmp* is generated by the proxy function for `memcmp`, appendix A). The syntax of facts in figure 3.3 makes these operations explicit—it is an instance of the syntax in figure 2.1, but now we split the function application rule into binary comparisons, binary arithmetic operations, and the remaining (non-arithmetic) function applications. We also make use of the function defined that returns false if its argument evaluates to $\bot$ and true otherwise.

Below we outline our assumptions regarding the binary operations. For an integer type $\tau \in \mathbb{T}_I$ let $\circ_\tau \in \{\oplus_\tau, \ominus_\tau, \circledast_\tau\}$ be a binary operator, and let $\circ' \in \{+, -, *\}$ be the corresponding integer operator. We assume that $\circ_\tau$ is only defined for two operands of type $\tau$, that is, $b_1 \circ_\tau b_2 = \bot$ whenever $b_1 \notin \mathrm{dom}(\tau)$ or $b_2 \notin \mathrm{dom}(\tau)$. Otherwise if $b_1, b_2 \in \mathrm{dom}(\tau)$ then the operation $\circ_\tau$ on bitstrings corresponds to the operation $\circ'$ on integer values as long as the result is within the representable range, that is,

$$\tau(b_1 \circ_\tau b_2) = \begin{cases} \tau(b_1) \circ' \tau(b_2), & \text{if } \tau(b_1) \circ' \tau(b_2) \in \mathrm{range}(\tau), \\ \text{arbitrary} & \text{otherwise.} \end{cases} \tag{3.21}$$

We do not assume a particular overflow behaviour, because such an assumption is provided by the C standard only for unsigned operations. Signed overflow is left undefined [ISO, 1999, Annex J2]. This means that a C program is free to do absolutely anything once a signed overflow occurs, including divulging secrets to the attacker. Such an assumption would make most protocols unverifiable, so we follow the behaviour of most compilers, and expect signed overflow to return a deterministic yet arbitrary result.

Let $\tau_1, \tau_2 \in \mathbb{T}_I$. We assume that $\mathrm{cast}_{\tau_1 \to \tau_2}$ is defined precisely for operands of type $\tau_1$:

$$\mathrm{cast}_{\tau_1 \to \tau_2}(b) \neq \bot \text{ iff } b \in \mathrm{dom}(\tau_1). \tag{3.22}$$

We also assume that $\mathrm{cast}_{\tau_1 \to \tau_2}$ does not change the value of the operand as long as it is representable in the new type $\tau_2$. More precisely, if $b \in \mathrm{dom}(\tau_1)$ and $\tau_1(b) \in \mathrm{range}(\tau_2)$ then

```
(define-type bitstringbot )

(define defined :: (-> bitstringbot bool))

;; integer types τ^y
(define value_y_unsigned :: (-> bitstringbot nat nat))
(define value_y_signed ::   (-> bitstringbot nat int))

;; inverse integer types τ^-1
(define bs_signed ::    (-> int bitstringbot ))
(define bs_unsigned ::  (-> nat bitstringbot))

(define len_y ::  (-> bitstringbot nat))
(define range :: (-> bitstringbot nat nat bitstringbot ))
(define concat :: (-> bitstringbot bitstringbot bitstringbot ))
```

Figure 3.4: The Yices preamble.

$\text{cast}_{\tau_1 \to \tau_2}(b) \in \text{dom}(\tau_2)$ and

$$\tau_2(\text{cast}_{\tau_1 \to \tau_2}(b)) = \tau_1(b). \tag{3.23}$$

This behaviour is provided by the C standard: *"When a value with integer type is converted to another integer type other than bool, if the value can be represented by the new type, it is unchanged."* [ISO, 1999, 6.3.1.3]. Some examples of invalid casts would include $\text{cast}_{\tau_{1s} \to \tau_{1u}}(0xFF)$ (the signed value of $0xFF$ is not representable as unsigned, and vice versa) or $\text{cast}_{\tau_{4u} \to \tau_{4s}}(\tau_{4u}^{-1}(1) \ominus_{\tau_{4u}} \tau_{4u}^{-1}(2))$. The last example highlights the fact that an operation x = x1 + x2; may not work as expected when x is signed and x1 and x2 are unsigned.

Similarly to binary arithmetic operators, binary comparison operators are assumed to be defined precisely for operands are of the right type. If both operands are of the right type, the comparison operators compare their values. This can be conveniently summarized as

$$truth(b_1 \circ_\tau b_2) = \tau(b_1) \circ' \tau(b_2) \tag{3.24}$$

for $b_1, b_2 \in BS$, $\tau \in \mathbb{T}_I$, $\circ_\tau \in \{==_\tau, !=_\tau, <_\tau, >_\tau\}$, and the corresponding integer operator $\circ' \in \{=, \neq, <, >\}$ (recall that $truth(\bot) = \text{false}$ and integer comparison operators return false if either operand evaluates to $\bot$)

Given an environment $\eta$ and a fact $\psi$ we write $\eta \models \psi$ if $[\![\psi]\!]_\eta = \text{true}$. Given two facts $\psi$ and $\psi'$ we write $\psi \models \psi'$ if for any environment $\eta$ such that $\eta \models \psi$ also $\eta \models \psi'$. This is the entailment relation that we are actually interested in. Below we show how given two facts $\psi$ and $\psi'$ we construct a Yices formula that implies $\psi \models \psi'$. It is important to point out that $\psi$ and $\psi'$ are unlikely to be valid Yices formulas themselves due to the mixing of integer and bitstring expressions that we discussed above.

Figure 3.4 shows the preamble that we give to Yices. Bitstring variables in our expressions map to Yices variables of type `bitstringbot` and logical variables map to Yices variables of type `bool`. Functions that return bitstrings, such as substring extractions or bitstring encodings of integers are represented as uninterpreted functions. Functions that return integers, namely len and $\tau$ for an integer type $\tau \in \mathbb{T}_I$ cannot be represented directly because they may return $\bot$ and Yices would not be able to perform arithmetic on such values. Instead we define two new functions, $\text{len}^y, \tau^y \colon BS \to \mathbb{N}$ that are defined exactly like len and $\tau$ except that they return an arbitrary integer instead of $\bot$. One of the goals of the translation that we describe below is to make sure that it is safe to convert len and $\tau$ into their Yices counterparts.

Let $e^y$ be the expression $e$ in which all occurrences of len are replaced by $\text{len}^y$ and all occurrences of $\tau$ for $\tau \in \mathbb{T}_I$ are replaced by $\tau^y$. We call $e$ *Yices-compatible* if $e = e^y$. Any Yices-compatible expression can be represented in Yices by replacing every bitstring subexpression $e'$ that is not an application of one of the functions in figure 3.4 by a bitstring variable $\mathit{opaque}[e']$. Given Yices-compatible facts $\psi$ and $\psi'$ we write $\psi \vdash^y \psi'$ if Yices claims that $\psi$ entails $\psi'$ which under the assumption of Yices correctness implies $\psi \models \psi'$.

The goal of our method is therefore to rewrite arbitrary $\psi$ and $\psi'$ into equivalent Yices-compatible formulas. Additionally we would like to convert as many as possible applications of binary arithmetic operations to integer operations (binary expressions that cannot be rewritten will have to be replaced by opaque variables). We achieve this by using rewriting rules of the form $\phi \vdash e \rightsquigarrow e'$ where $\phi$ is a fact and $e$ and $e'$ are arbitrary expressions. Such a rule means that $[\![\phi]\!]_\eta = \text{true}$ implies $[\![e]\!]_\eta = [\![e']\!]_\eta$. The idea of the method is based on the following self-evident theorem.

**Theorem 3.1 (Reduction of Facts to Yices-compatible Facts)** *Consider rewriting rules $\phi_y \vdash \psi \rightsquigarrow \psi_y$ and $\phi'_y \vdash \psi' \rightsquigarrow \psi'_y$ such that all of $\phi_y$, $\psi_y$, $\phi'_y$, and $\psi'_y$ are yices-compatible facts and $\psi$ and $\psi'$ are arbitrary facts. Then $\psi \models \psi'$ whenever $(\phi_y \Rightarrow \psi_y) \vdash^y (\phi'_y \wedge \psi'_y)$.* □

The difference in shape of the left- and the right-hand side of the conclusion can be understood as follows: both $\phi_y$ and $\phi'_y$ are conditions that need to be proved. Even though we assume $\psi$, we still need to prove $\phi_y$ in order to assume $\psi_y$.

Given two facts $\psi$ and $\psi'$ we rewrite them using the rules in figures 3.5 and 3.6. The first three rules allow to combine rewriting steps. Rule (R1) is the congruence rule: we can rewrite subexpressions in a larger expression. Rule (R2) shows that rule conditions themselves can be rewritten. This dictates the shape of our implementation: given a fact $\psi$ we first apply the rewriting rules to simplify it into a Yices-compatible fact $\psi^y$ and collect conditions $\phi_1, \ldots, \phi_n$ along the way. We then apply the same procedure recursively to rewrite $\phi_1, \ldots, \phi_n$ themselves. Rule (R3) captures the fact that adding a condition is not necessary when rewriting a conjunction if this condition is part of the conjunction already. The importance of this rule will be demonstrated in example 3.3. Rules (R26) and (R27) are responsible for making sure that the resulting expression is Yices-compatible.

The rule (R6) allows us to rewrite C logical conjunctions into their IML counterparts (the same kind of rule exists for disjunctions and negations). The rule (R8) is useful for modelling the

$$\frac{\phi \vdash e_1 \rightsquigarrow e_2}{\phi \vdash e[e_1] \rightsquigarrow e[e_2]}\ \text{(R1)} \qquad \frac{\phi \vdash e \rightsquigarrow e' \quad \psi \vdash \phi \rightsquigarrow \phi'}{\psi \wedge \phi' \vdash e \rightsquigarrow e'}\ \text{(R2)}$$

$$\frac{\phi_1 \vdash e_1 \rightsquigarrow (e_1' \wedge \phi_1') \quad (\phi_1' \wedge \phi_2) \vdash e_2 \rightsquigarrow e_2'}{(\phi_1 \wedge \phi_2) \vdash (e_1 \wedge e_2) \rightsquigarrow (e_1' \wedge e_2')}\ \text{(R3)}$$

$$\frac{\tau \in \mathbb{T}_I \quad (\circ_\tau, \circ') \in \{(==_\tau, =), (!=_\tau, \neq), (<_\tau, <), (>_\tau, >)\}}{\vdash \mathit{truth}(e \circ_\tau e') \rightsquigarrow \tau(e) \circ' \tau(e')}\ \text{(R4)}$$

$$\frac{\circ \in \{=, \neq, <, >\}}{\vdash t \circ t' \rightsquigarrow \text{defined}(t) \wedge \text{defined}(t') \wedge (t \circ t')}\ \text{(R5)} \qquad \frac{}{\vdash \mathit{truth}(e \,\&\&\, e') \rightsquigarrow \mathit{truth}(e) \wedge \mathit{truth}(e')}\ \text{(R6)}$$

$$\frac{\tau \in \mathbb{T}_I \quad (\circ_\tau, \circ') \in \{(\oplus_\tau, +), (\ominus_\tau, -), (\circledast_\tau, *)\}}{\text{defined}(e \circ_\tau e') \wedge (\tau(e) \circ' \tau(e') \in \text{range}(\tau)) \vdash \tau(e \circ_\tau e') \rightsquigarrow \tau(e) \circ' \tau(e')}\ \text{(R7)}$$

$$\frac{}{\text{defined}(e) \wedge \text{defined}(e') \vdash \tau_{\text{int}}(\mathit{cmp}(e, e')) = 0 \rightsquigarrow e = e'}\ \text{(R8)}$$

$$\frac{\tau \in \mathbb{T}_I}{\text{defined}(\tau^{-1}(t)) \vdash \tau(\tau^{-1}(t)) \rightsquigarrow t}\ \text{(R9)} \qquad \frac{\tau \in \mathbb{T}_I}{\text{defined}(\tau^{-1}(\tau(e))) \vdash \tau^{-1}(\tau(e)) \rightsquigarrow e}\ \text{(R10)}$$

$$\frac{\tau \in \mathbb{T}_I}{\text{defined}(\tau^{-1}(t)) \vdash \text{len}(\tau^{-1}(t)) \rightsquigarrow \text{width}(\tau)}\ \text{(R11)} \qquad \frac{b \in \mathit{BS}}{\vdash \text{len}(b) \rightsquigarrow |b|}\ \text{(R12)}$$

$$\frac{\tau \in \mathbb{T}_I \quad \circ_\tau \in \{\oplus_\tau, \ominus_\tau, \circledast_\tau\}}{\text{defined}(e \circ_\tau e') \vdash \text{len}(e \circ_\tau e') \rightsquigarrow \text{width}(\tau)}\ \text{(R13)}$$

$$\frac{}{\vdash \text{len}(e|e') \rightsquigarrow \text{len}(e) + \text{len}(e')}\ \text{(R14)} \qquad \frac{}{\text{defined}(e\{p, l\}) \vdash \text{len}(e\{p, l\}) \rightsquigarrow l}\ \text{(R15)}$$

$$\frac{}{\vdash \text{defined}(\text{defined}(e)) \rightsquigarrow \text{true}}\ \text{(R16)} \qquad \frac{\circ \in \{+, -, *\}}{\vdash \text{defined}(t \circ t') \rightsquigarrow \text{defined}(t) \wedge \text{defined}(t')}\ \text{(R17)}$$

$$\frac{}{\vdash \text{defined}(e\{p, l\}) \rightsquigarrow (p + l \leq \text{len}(e)) \wedge (p \geq 0) \wedge (l \geq 0)}\ \text{(R18)}$$

$$\frac{}{\vdash \text{defined}(e|e') \rightsquigarrow \text{defined}(e) \wedge \text{defined}(e')}\ \text{(R19)}$$

$$\frac{\tau \in \mathbb{T}_I}{\vdash \text{defined}(\tau^{-1}(t)) \rightsquigarrow t \in \text{range}(\tau)}\ \text{(R20)} \qquad \frac{\tau \in \mathbb{T}_I}{\vdash \text{defined}(\tau(e)) \rightsquigarrow e \in \text{dom}(\tau)}\ \text{(R21)}$$

$$\frac{\tau \in \mathbb{T}_I \quad \circ_\tau \in \{\oplus_\tau, \ominus_\tau, \circledast_\tau, ==_\tau, !=_\tau, <_\tau, >_\tau\}}{\vdash \text{defined}(e \circ_\tau e') \rightsquigarrow e, e' \in \text{dom}(\tau)}\ \text{(R22)} \qquad \frac{\circ \in \{=, \neq, <, >\}}{\vdash \text{defined}(e \circ e') \rightsquigarrow \text{true}}\ \text{(R23)}$$

$$\frac{}{\vdash \text{defined}(\text{len}(e)) \rightsquigarrow \text{defined}(e)}\ \text{(R24)} \qquad \frac{e = b \in \mathit{BS} \text{ or } e = n \in \mathbb{N}}{\vdash \text{defined}(e) \rightsquigarrow \text{true}}\ \text{(R25)}$$

$$\frac{\tau \in \mathbb{T}_I}{\text{defined}(e) \vdash \tau(e) \rightsquigarrow \tau^y(e)}\ \text{(R26)} \qquad \frac{}{\text{defined}(e) \,\&\&\, \text{len}(e) \rightsquigarrow \text{len}^y(e)}\ \text{(R27)}$$

Figure 3.5: Rewriting of expressions.

$$\frac{\tau, \tau' \in \mathbb{T}_I}{\vdash \text{defined}(\text{cast}_{\tau \to \tau'}(e)) \rightsquigarrow e \in \text{dom}(\tau)} \quad \text{(R28)}$$

$$\frac{\tau, \tau' \in \mathbb{T}_I}{\text{defined}(\text{cast}_{\tau \to \tau'}(e)) \vdash \text{len}(\text{cast}_{\tau \to \tau'}(e)) \rightsquigarrow \text{width}(\tau')} \quad \text{(R29)}$$

$$\frac{\tau, \tau' \in \mathbb{T}_I}{\text{defined}(\text{cast}_{\tau \to \tau'}(e)) \wedge (\tau(e) \in \text{range}(\tau')) \vdash \tau'(\text{cast}_{\tau \to \tau'}(e)) \rightsquigarrow \tau(e)} \quad \text{(R30)}$$

$$\frac{\tau_1, \tau_2 \in \mathbb{T}_I}{t \in \text{range}(\tau_1) \cap \text{range}(\tau_2) \vdash \text{cast}_{\tau_1 \to \tau_2}(\tau_2^{-1}(t)) \rightsquigarrow \tau_2^{-1}(t)} \quad \text{(R31)}$$

$$\frac{\tau_1, \tau_2, \tau_3 \in \mathbb{T}_I}{\tau_1(e) \in \text{range}(\tau_2) \cap \text{range}(\tau_3) \vdash \text{cast}_{\tau_2 \to \tau_3}(\text{cast}_{\tau_1 \to \tau_2}(e)) \rightsquigarrow \text{cast}_{\tau_1 \to \tau_3}(e)} \quad \text{(R32)}$$

$$\frac{\tau \in \mathbb{T}_I}{e \in \text{dom}(\tau) \vdash \text{cast}_{\tau \to \tau}(e) \rightsquigarrow e} \quad \text{(R33)}$$

Figure 3.6: Rewriting of cast expressions.

memcmp function, as shown in appendix A: it shows how the expression $cmp(e_1, e_2)$ generated by the proxy function for memcmp relates to the equality of $e_1$ and $e_2$.

Some rules mention facts of the form $t \in \text{range}(\tau)$ and $e \in \text{dom}(\tau)$ for an integer term $t$ or a bitstring term $e$ and a type $\tau \in \mathbb{T}_I$. These are easy to express for typical C types. For instance, for a signed byte type $\tau_{1s}$ we have $t \in \text{range}(\tau_{1s}) = (t \geq -128) \wedge (t \leq 127)$ and $e \in \text{dom}(\tau_{1s}) = (\text{len}(e) = 1)$.

Some rules show applications of $\text{defined}(\cdot)$ to integer formulas. This is not part of the syntax in figure 3.3, where $\text{defined}(\cdot)$ is meant to apply to bitstring expressions only. The meaning is obvious though and those applications are only transient and always get rewritten themselves.

We now prove that the rewriting rules in figures 3.5 and 3.6 are valid.

Proof (Rewriting rules) Rules (R4) and (R7) correspond directly to our assumptions (3.24) and (3.21) about binary operators. Rules (R29) and (R28) correspond to our assumptions (3.23) and (3.22). Next we prove (R31), (R32), and (R33).

(R31) Let $t$ be an integer expression such that $n = [\![t]\!]_\eta \in \text{range}(\tau_1) \cap \text{range}(\tau_2)$. Let $b = \tau_1^{-1}(n)$. By (3.23)

$$\text{cast}_{\tau_1 \to \tau_2}(\tau_2^{-1}(n)) = \text{cast}_{\tau_1 \to \tau_2}(b) = \tau_2^{-1}(\tau_2(\text{cast}_{\tau_1 \to \tau_2}(b))) = \tau_2^{-1}(\tau_1(b)) = \tau_2^{-1}(n).$$

(R32) Let $e$ be a bitstring expression such that $b = [\![e]\!]_\eta \in \text{dom}(\tau_1)$ and $n = \tau_1(b) \in \text{range}(\tau_2) \cap$

range($\tau_3$). By (R31)

$$\text{cast}_{\tau_2\to\tau_3}(\text{cast}_{\tau_1\to\tau_2}(b)) = \text{cast}_{\tau_2\to\tau_3}(\text{cast}_{\tau_1\to\tau_2}(\tau_1^{-1}(n))) = \text{cast}_{\tau_2\to\tau_3}(\tau_2^{-1}(n)) = \tau_3^{-1}(n)$$
$$= \text{cast}_{\tau_1\to\tau_3}(\tau_1^{-1}(n)) = \text{cast}_{\tau_1\to\tau_3}(b).$$

(R33) Let $e$ be a bitstring expression such that $b = [\![e]\!]_\eta \in \text{dom}(\tau)$. Let $n = \tau(b)$. By (R31)

$$\text{cast}_{\tau\to\tau}(b) = \text{cast}_{\tau\to\tau}(\tau^{-1}(n))) = \tau^{-1}(n)) = b.$$

All the other rules are straightforward. ■

**Example 3.2** We show how a single expression is simplified by rewriting. Suppose we would like to prove $\text{len}((x_1|x_2)\{5, \text{len}(x_2)\}) = \text{len}(x_2)$ for two bitstring variables $x_1$ and $x_2$. We make this expression number 1 below. Each rewriting step is labelled with some conditions and the conditions are themselves rewritten in the same way.

**1** $\quad \text{len}((x_1|x_2)\{5, \text{len}(x_2)\}) = \text{len}(x_2)$

(2) $\rightsquigarrow \text{len}(x_2) = \text{len}(x_2)$

(3) $\rightsquigarrow \text{len}^y(x_2) = \text{len}^y(x_2)$

**2** $\quad \text{defined}((x_1|x_2)\{5, \text{len}(x_2)\})$

$\rightsquigarrow (\text{len}(x_1|x_2) \geq (5 + \text{len}(x_2))) \wedge (5 \geq 0) \wedge (\text{len}(x_2) \geq 0)$

$\rightsquigarrow ((\text{len}(x_1) + \text{len}(x_2)) \geq (5 + \text{len}(x_2))) \wedge (\text{len}(x_2) \geq 0)$

(4) $\rightsquigarrow ((\text{len}^y(x_1) + \text{len}(x_2)) \geq (5 + \text{len}(x_2))) \wedge (\text{len}(x_2) \geq 0)$

(3) $\rightsquigarrow ((\text{len}^y(x_1) + \text{len}^y(x_2)) \geq (5 + \text{len}^y(x_2))) \wedge (\text{len}^y(x_2) \geq 0)$

**3** $\quad \text{defined}(x_2)$

**4** $\quad \text{defined}(x_1)$

For simplicity we remove the obvious condition $5 \geq 0$ from subsequent rewriting steps, but our implementation does not do this—all arithmetic facts are passed to Yices to deal with. In order to prove the original fact it is sufficient to prove the rewritten facts 1–4. Clearly this can be done under assumptions defined($x_1$), defined($x_2$), and $\text{len}^y(x_1) \geq 5$. Yices can discharge the condition $\text{len}^y(x_2) \geq 0$ on its own since it knows that lengths are non-negative since figure 3.4 defines `len_y` to have the return type `nat`. □

**Example 3.3** This example demonstrates how new assumptions, such as $\text{len}^y(x_1) \geq 5$ in the example above, are added to Yices. Suppose we would like Yices to know that $\tau_{8u}(x) \geq 200$ is true for some variable $x$. Applying our rewriting procedure we get

$$
\begin{aligned}
\mathbf{1} \quad & \tau_{8u}(x) \geq 200 \\
& \rightsquigarrow \text{defined}(\tau_{8u}(x)) \wedge \text{defined}(200) \wedge (\tau_{8u}(x) \geq 200) \\
& \rightsquigarrow (8 = \text{len}(x)) \wedge (\tau_{8u}(x) \geq 200) \\
& \rightsquigarrow \text{defined}(8) \wedge \text{defined}(\text{len}(x)) \wedge (8 = \text{len}(x)) \wedge (\tau_{8u}(x) \geq 200) \\
& \rightsquigarrow \text{defined}(\text{len}(x)) \wedge (8 = \text{len}(x)) \wedge (\tau_{8u}(x) \geq 200) \\
& \rightsquigarrow \text{defined}(x) \wedge (8 = \text{len}(x)) \wedge (\tau_{8u}(x) \geq 200) \\
(2) \; & \rightsquigarrow \text{defined}(x) \wedge (8 = \text{len}^y(x)) \wedge (\tau_{8u}(x) \geq 200) \\
(2) \; & \rightsquigarrow \text{defined}(x) \wedge (8 = \text{len}^y(x)) \wedge (\tau^y_{8u}(x) \geq 200)
\end{aligned}
$$

$$
\mathbf{2} \quad \text{defined}(x)
$$

In order to convert $\text{len}(x)$ into $\text{len}^y(x)$ and $\tau_{8u}(x)$ into $\tau^y_{8u}(x)$ we require the assumption $\text{defined}(x)$. However, this condition is already present in the expression that we are rewriting! This is where the rule (R3) comes in: it allows us to omit the condition in this case and just assert that $\text{defined}(x) \wedge (8 = \text{len}^y(x)) \wedge (\tau^y_{8u}(x) \geq 200)$ holds unconditionally. In general, when our implementation collects conditions while rewriting a conjunction, it drops all conditions that are already present in the conjunction itself. □

Appendix B contains an extended example with assumptions and queries that would be typical in our C program analysis.

## 3.3 Simplification

The rewriting system developed in the previous section can readily be used to simplify expressions: each rewriting rule $\psi \vdash e \rightsquigarrow e'$ can be seen as a simplification rule. In addition to that we add three rules for simplifying substring extraction expressions. Those rules are not part of the solver described in the previous section because the solver itself needs to be invoked to decide which instance of the rule to apply. We formalise this using the following simplification relation, parameterised by a fact set $\Phi$:

$$
\begin{aligned}
e &\rightsquigarrow_\Phi e', && \text{if } \psi \vdash e \rightsquigarrow e' \text{ and } \Phi \vdash \psi, \\
e\{p, l\} &\rightsquigarrow_\Phi \varepsilon, && \text{if } \Phi \vdash (l = 0), \\
(e_1| \dots |e_m)\{p, l\} &\rightsquigarrow_\Phi e_i| \dots |e_j, && \text{if } \Phi \vdash \text{defined}(e_1| \dots |e_m) \\
& && \quad \wedge (p = \text{len}(e_1| \dots |e_{i-1})) \wedge (l = \text{len}(e_i| \dots |e_j)) \\
e\{p, l\}\{p', l'\} &\rightsquigarrow_\Phi e\{p + p', l'\}, && \text{if } \Phi \vdash \{p + p'\}_{l'} \subseteq \{p\}_l \,.
\end{aligned}
$$

| | |
|---|---|
| $pb \in PBase ::=$ | pointer base |
| $addr(v)$ | stack pointer to variable $v$ |
| $malloc(e_{\mathcal{A}}, t_l)$ | heap pointer |
| $t \in ITerm ::=$ | integer term |
| $pb$ | pointer base |
| $\ldots$ | same as *ITerm* in figure 2.1 |
| $e \in IExp ::=$ | bitstring expression |
| $\mathrm{ptr}(pb, t_o)$ | pointer |
| $\ldots$ | same as *IExp* in figure 2.1 |

Figure 3.7: Symbolic expressions with pointers.

The second rule simplifies expressions that select an empty substring, the third rule simplifies substrings of concatenations, and the fourth rule simplifies repeated substring extractions. These rules are very important for symbolic execution that we shall describe in section 3.5 since without them the constructed symbolic expressions would quickly grow very large.

We exclude rules (R26) and (R27) from being used in simplification since the purpose of these rules is to make the expression Yices-compatible, but this is not the goal here.

Let $\mathrm{simplify}_\Phi(e)$ be recursively defined as follows: choose a rewriting rule $e \rightsquigarrow_\Phi e'$ and let $\mathrm{simplify}_\Phi(e) = \mathrm{simplify}_\Phi(e')$. If no such rule exists, let $\mathrm{simplify}_\Phi(e) = e$. Clearly this function is sound in the following sense:

**Lemma 3.1** *Given a fact set $\Phi$ and an environment $\eta$, if $\eta \models \Phi$ then* $[\![\mathrm{simplify}_\Phi(e)]\!]_\eta = [\![e]\!]_\eta$.$\square$

## 3.4 Symbolic Pointers

To track the values used as pointers during CVM execution we extend IML expressions with an additional construct as shown in figure 3.7. An expression of the form $\mathrm{ptr}(pb, t_o)$ represents a pointer into the memory location identified by the *pointer base* $pb$ with an *offset* $t_o$ relative to the beginning of the location. Pointer bases are of two kinds: a base of the form $addr(v)$ represents a pointer to the program variable $v$ and a base of the form $malloc(e_{\mathcal{A}}, t_l)$ represents the result of a Malloc. These two expressions are evaluated using functions *addr* and *malloc* defined in section 2.2. Given an environment $\eta$ we let $[\![addr(v)]\!]_\eta = addr(v)$ and $[\![malloc(e_{\mathcal{A}}, t_l)]\!]_\eta = malloc([\![e_{\mathcal{A}}]\!]_\eta, [\![t_l]\!]_\eta)$ whenever $[\![t_l]\!]_\eta \neq \bot$ and $[\![e_{\mathcal{A}}]\!]_\eta$ is an encoding of an allocation table suitable as argument to *malloc* (we let $[\![malloc(e_{\mathcal{A}}, t_l)]\!]_\eta = \bot$ otherwise). We allow to use a pointer base as a free-standing integer term.

Evaluation rules for pointer expressions are shown in figure 3.8. The rule (3.45) evaluates the null pointer—both the base and the offset must be zero in this case. Rules (3.46) and (3.47) evaluate stack and heap pointers. In each case we require that the offset points within the allocated memory area or one past the last element. This is specified by the C standard [ISO,

$$\frac{[\![pb]\!]_\eta = [\![t_o]\!] = 0}{[\![\mathrm{ptr}(pb, t_o)]\!]_\eta = \tau_{\mathrm{ptr}}^{-1}(0)} \tag{3.45}$$

$$\frac{\eta \models 0 \leq t_o \leq \mathrm{len}(v)}{[\![\mathrm{ptr}(addr(v), t_o)]\!]_\eta = [\![\tau_{\mathrm{ptr}}^{-1}(addr(v) + t_o)]\!]_\eta} \tag{3.46}$$

$$\frac{\eta \models malloc(e_{\mathcal{A}}, t_l) \neq 0 \quad \eta \models 0 \leq t_o \leq t_l}{[\![\mathrm{ptr}(malloc(e_{\mathcal{A}}, t_l), t_o)]\!]_\eta = [\![\tau_{\mathrm{ptr}}^{-1}(malloc(e_{\mathcal{A}}, t_l) + t_o)]\!]_\eta} \tag{3.47}$$

Figure 3.8: Pointer evaluation rules.

1999, 6.5.6 (8)] and creating (even without dereferencing) a pointer that fails these criteria results in undefined behaviour. This justifies the requirement $malloc(e_{\mathcal{A}}, t_l) \neq 0$ in (3.47) as we need to make sure that the memory area is successfully allocated before moving the offset. In both cases the evaluation results in the sum of the base and the offset converted to a bitstring of type $\tau_{\mathrm{ptr}}$.

In order to handle expressions with symbolic pointers we extend our solver from section 3.2 by the rules in figure 3.9. The first five rules are a direct consequence of the evaluation rules in figure 3.8. The last two rules encode our assumptions about the C operators $\oplus_{PI}^{\tau}$ and $\ominus_{PI}^{\tau}$ that are used to add or subtract an integer of type $\tau$ from a pointer, and $\ominus_{PP}$ that is used to subtract two pointers (the rule (R39) is only shown for $\oplus_{PI}^{\tau}$ but the same rule exists for $\ominus_{PI}^{\tau}$). The behaviour of $\ominus_{PP}$ is somewhat subtle, because the result of the operation is signed. According to [ISO, 1999, 6.5.6] *"When two pointers are subtracted, [...] the result is the difference of the subscripts of the two array elements. The size of the result is implementation-defined, and its type (a* signed *integer type) is* ptrdiff_t *[...].* The rule therefore subtracts the values of the offsets and then checks whether the result is representable as a bitstring of type $\tau_{\mathrm{ptrdiff}}$. This is not quite the same as using unsigned subtraction and then casting to a signed value (for instance, the result of $\mathrm{cast}_{\tau_{1u} \to \tau_{1s}}(\tau_{1u}^{-1}(1) \ominus_{\tau_{1u}} \tau_{1u}^{-1}(2))$ is undefined).

The above description of pointer operators assumes pointers of type void$*$. To deal with pointers to arbitrary underlying types we multiply the integer operand of $\oplus_{PI}^{\tau}$ and $\ominus_{PI}^{\tau}$, and divide the result of $\ominus_{PP}$ by the size of the underlying type.

The rule (R35) lists $pb = malloc(e_{\mathcal{A}}, t_l) \neq 0$ as one of its conditions. It is useful to see how such a condition will be proved in practice during symbolic execution. The symbolic execution of Malloc will generate an expression of the form $\mathrm{ptr}(pb, 0)$. The C program is expected to check that this value is not null before doing anything with it. Such a check would generate the fact

$$\frac{}{\vdash \text{defined}(\text{ptr}(\mathit{addr}(v), t_o)) \rightsquigarrow 0 \leq t_o \leq \text{len}(v)} \text{ (R34)}$$

$$\frac{}{\vdash \text{defined}(\text{ptr}(\mathit{malloc}(e_{\mathcal{A}}, t_l), t_o)) \rightsquigarrow (t_o = 0) \lor (\mathit{malloc}(e_{\mathcal{A}}, t_l) \neq 0 \land 0 \leq t_o \leq t_l)} \text{ (R35)}$$

$$\frac{}{\vdash \text{defined}(\tau_{\text{ptr}}^{-1}(0)) \rightsquigarrow \text{true}} \text{ (R36)}$$

$$\frac{}{\text{defined}(\text{ptr}(pb, t_o)) \vdash \tau_{\text{ptr}}(\text{ptr}(pb, t_o)) \rightsquigarrow pb + t_o} \text{ (R37)}$$

$$\frac{}{\text{defined}(\text{ptr}(pb, t_o)) \vdash \text{len}(\text{ptr}(pb, t_o)) \rightsquigarrow \text{len}(\tau_{\text{ptr}})} \text{ (R38)}$$

$$\frac{\tau \in \mathbb{T}_I}{\text{defined}(\text{ptr}(pb, t + \tau(e))) \vdash \text{ptr}(pb, t) \oplus_{PI}^{\tau} e \rightsquigarrow \text{ptr}(pb, t + \tau(e))} \text{ (R39)}$$

$$\frac{\tau \in \mathbb{T}_I}{\text{defined}(\tau_{\text{ptrdiff}}(t - t')) \vdash \text{ptr}(pb, t) \ominus_{PP} \text{ptr}(pb, t') \rightsquigarrow \tau_{\text{ptrdiff}}(t - t')} \text{ (R40)}$$

Figure 3.9: Rewriting of pointer expressions.

$\mathit{truth}(\text{ptr}(pb, 0) \mathrel{!=_{\tau_{\text{ptr}}}} \tau_{\text{ptr}}^{-1}(0))$. This fact would be rewritten by our solver as follows:

$$\begin{aligned}
& \mathit{truth}(\text{ptr}(pb, 0) \mathrel{!=_{\tau_{\text{ptr}}}} \tau_{\text{ptr}}^{-1}(0)) \\
\text{(R4)} \rightsquigarrow\ & \tau_{\text{ptr}}(\text{ptr}(pb, 0)) \neq \tau_{\text{ptr}}(\tau_{\text{ptr}}^{-1}(0)) \\
(1), \text{(R37)} \rightsquigarrow\ & pb \neq \tau_{\text{ptr}}(\tau_{\text{ptr}}^{-1}(0)) \\
(2), \text{(R9)} \rightsquigarrow\ & pb \neq 0
\end{aligned}$$

$$\begin{aligned}
\mathbf{1} \quad & \text{defined}(\mathit{ptr}(pb, 0)) \\
\text{(R34) or (R35)} \rightsquigarrow\ & \text{true}
\end{aligned}$$

$$\begin{aligned}
\mathbf{2} \quad & \text{defined}(\tau_{\text{ptr}}^{-1}(0)) \\
\text{(R36)} \rightsquigarrow\ & \text{true}
\end{aligned}$$

Despite having extended the class of symbolic expressions allowed in IML, the simulation relation (section 2.3) still holds since the proof does not rely on the particular shape of the expressions.

For convenience we define a function $\mathit{size}\colon \mathit{PBase} \to \mathit{ITerm}$ that gives the length of the allocated memory associated with a given pointer base, under the assumption that the pointer base is not null. We let

$$\mathit{size}(\mathit{addr}(v)) = \text{len}(v) \quad \text{and} \quad \mathit{size}(\mathit{malloc}(_, t_s)) = t_s$$

## 3.5 From CVM to IML: Symbolic Execution

We describe how to automatically extract an IML model from a CVM program. The model we extract will simulate the original program, as defined in section 2.3, and so every security property that we can prove on the model also holds for the original program. The key idea of the model extraction is to execute a CVM program in a symbolic semantics, where, instead of concrete bitstrings, memory locations contain IML expressions representing the set of all possible concrete values at a given execution point.

The algorithm for symbolic execution is determined by the set of rules presented in figure 3.10. The initial semantic configuration has the form $\emptyset, Q$ with an initial process $Q \in \mathrm{CVM}_0$. After executing the $\mathsf{Init}$ instruction at the start of $Q$ the semantic configurations take the form $\mathbb{S}, P = (\Phi, \mathcal{A}^s, \mathcal{M}^s, \mathcal{S}^s), P$, where

- $\Phi$ is a set of facts known to hold at a given execution point,
- $\mathcal{A}^s \in \mathit{IExp}$ is a symbolic expression representing the allocation table,
- $\mathcal{M}^s \colon \mathit{PBase} \rightharpoonup \mathit{IExp}$ is the symbolic memory,
- $\mathcal{S}^s$ is a list of symbolic (integer or bitstring) expressions representing the execution stack,
- $P \in \mathrm{CVM}$ is the executing program.

The symbolic execution rules essentially mimic the rules of the concrete execution. The algorithm makes heavy use of the implication relation $\vdash$ developed in section 3.2.

The crucial rules are (SLoad) and (SStore) that reflect the effect of storing and loading memory values on the symbolic level. The rule (SLoad) first checks that the dereferenced pointer is not null and is well-formed. The rule then constructs the expression $e = \mathcal{M}^s(pb)\{t_o, \tau_{\mathrm{size}}(e_l)\}$ that represents the value extracted from memory and checks that $\Phi \vdash \mathrm{defined}(e)$. In particular, $\mathrm{defined}(e)$ implies $\{t_o\}_{\tau_{\mathrm{size}}(e_l)} \subseteq \{0, \ldots, \mathrm{len}(\mathcal{M}^s(pb)) - 1\}$, which means that we are extracting from an initialised memory range.

The rule (SStore) first checks that the dereferenced pointer is not null and is well-formed, same as in (SLoad). It then distinguishes between two cases depending on how the expression $e$ to be stored is aligned with the expression $e_h$ that is already present in memory. If $e$ needs to be stored completely within the bounds of $e_h$ then we replace the contents of the memory location by $e'_h = e_h\{\ldots\}|e|e_h\{\ldots\}$ where the first and the last range expression represent the pieces of $e_h$ that are not covered by $e$. In case $e$ needs to be stored past the end of $e_h$, the new expression is of the form $e_h\{\ldots\}|e$. The rule still requires that the beginning of $e$ is positioned before the end of $e_h$, and hence it is impossible to write in the middle of an uninitialised memory location. This is for simplicity of presentation—the rule used in our implementation does not have this limitation (it creates an explicit “undefined” expression in these cases). Finally the rule (SStore) checks that the new expression fits within the allocated memory range, by testing the condition

$$\frac{\Phi_0 = \{\, addr(v) \neq 0 \mid v \in \mathrm{var}(Q)\} \quad \mathcal{M}^s_0 = \{\, addr(v) \mapsto \varepsilon \mid v \in \mathrm{var}(Q)\}}{\emptyset,\ \mathsf{Init}\,;Q \to (\Phi_0,\ initmem(Q),\ \mathcal{M}^s_0,\ []),\ Q} \quad \text{(SInit)}$$

$$\frac{x \text{ does not occur in } (\Phi,\ \mathcal{A}^s,\ \mathcal{M}^s,\ \mathcal{S}^s)}{(\Phi,\ \mathcal{A}^s,\ \mathcal{M}^s,\ \mathcal{S}^s),\ \mathtt{in}(c,x);P \xrightarrow{\mathtt{in}(c,x);} (\Phi \cup \{\mathrm{defined}(x)\},\ \mathcal{A}^s,\ \mathcal{M}^s,\ \mathcal{S}^s),\ P} \quad \text{(SIn)}$$

$$\frac{}{(\Phi,\ \mathcal{A}^s,\ \mathcal{M}^s,\ \mathcal{S}^s),\ \mathtt{out}(c,x);Q \xrightarrow{\mathtt{out}(c,x);} (\Phi,\ \mathcal{A}^s,\ \mathcal{M}^s,\ \mathcal{S}^s),\ Q} \quad \text{(SOut)}$$

$$\frac{}{(\Phi,\ \mathcal{A}^s,\ \mathcal{M}^s,\ \mathcal{S}^s),\ \mathtt{Const}\ c;P \to (\Phi,\ \mathcal{A}^s,\ \mathcal{M}^s,\ c :: \mathcal{S}^s),\ P} \quad \text{(SConst)}$$

$$\frac{}{(\Phi,\ \mathcal{A}^s,\ \mathcal{M}^s,\ \mathcal{S}^s),\ \mathtt{Ref}\ v;P \to (\Phi,\ \mathcal{A}^s,\ \mathcal{M}^s,\ \mathrm{ptr}(addr(v),\ 0) :: \mathcal{S}^s),\ P} \quad \text{(SRef)}$$

$$\frac{\Phi \vdash \mathrm{defined}(\tau_{\mathrm{size}}(e_l)) \quad \mathcal{A}^{s\prime} = allocate(\mathcal{A}^s, \tau_{\mathrm{size}}(e_l)) \quad pb = malloc(\mathcal{A}^s, \tau_{\mathrm{size}}(e_l))}{(\Phi,\ \mathcal{A}^s,\ \mathcal{M}^s,\ e_l :: \mathcal{S}^s),\ \mathtt{Malloc};P \to (\Phi',\ \mathcal{A}^{s\prime},\ \mathcal{M}^s\{pb \mapsto \varepsilon\},\ \mathrm{ptr}(pb,0) :: \mathcal{S}^s),\ P} \quad \text{(SMalloc)}$$

$$\frac{pb \in \mathrm{dom}(\mathcal{M}^s) \quad \Phi \vdash \mathrm{defined}(\mathrm{ptr}(pb,t_o)) \quad \Phi \vdash (pb \neq 0) \quad e = \mathcal{M}^s(pb)\{t_o, \tau_{\mathrm{size}}(e_l)\} \quad \Phi \vdash \mathrm{defined}(e)}{(\Phi,\ \mathcal{A}^s,\ \mathcal{M}^s,\ e_l :: \mathrm{ptr}(pb,t_o) :: \mathcal{S}^s),\ \mathtt{Load};P \to (\Phi,\ \mathcal{A}^s,\ \mathcal{M}^s,\ \mathrm{simplify}_\Phi(e) :: \mathcal{S}^s),\ P} \quad \text{(SLoad)}$$

$$\frac{\begin{array}{l} pb \in \mathrm{dom}(\mathcal{M}^s) \quad \Phi \vdash \mathrm{defined}(\mathrm{ptr}(pb,t_o)) \quad \Phi \vdash (pb \neq 0) \quad e_h = \mathcal{M}^s(pb) \\ \text{either } \Phi \vdash (t_o + \mathrm{len}(e) < \mathrm{len}(e_h)) \text{ and } e_h' = e_h\{0,t_o\}|e|e_h\{t_o + \mathrm{len}(e), \mathrm{len}(e_h) - (t_o + \mathrm{len}(e))\} \\ \quad \text{or } \Phi \vdash (t_o + \mathrm{len}(e) \geq \mathrm{len}(e_h)) \wedge (t_o \leq \mathrm{len}(e_h)) \text{ and } e_h' = e_h\{0,t_o\}|e \\ \Phi \vdash \mathrm{len}(e_h') \leq size(pb) \end{array}}{(\Phi,\ \mathcal{A}^s,\ \mathcal{M}^s,\ \mathrm{ptr}(pb,t_o) :: e :: \mathcal{S}^s),\ \mathtt{Store};P \to (\Phi,\ \mathcal{A}^s,\ \mathcal{M}^s\{pb \mapsto \mathrm{simplify}_\Phi(e_h')\},\ \mathcal{S}^s),\ P} \quad \text{(SStore)}$$

$$\frac{x \text{ does not occur in } (\Phi,\ \mathcal{A}^s,\ \mathcal{M}^s,\ \mathcal{S}^s) \quad n \in \mathbb{N} \quad T = \mathtt{fixed}_n}{(\Phi,\ \mathcal{A}^s,\ \mathcal{M}^s,\ \tau^{-1}_{\mathrm{size}}(n) :: \mathcal{S}^s),\ \mathtt{New}\ x;P \xrightarrow{\mathtt{new}\ x:\ T;} (\Phi \cup \{\mathrm{len}(x) = n\},\ \mathcal{A}^s,\ \mathcal{M}^s,\ x :: \mathcal{S}^s),\ P} \quad \text{(SNew)}$$

$$\frac{f \text{ has arity } n \quad e = f(e_1, \ldots, e_n)}{(\Phi,\ \mathcal{A}^s,\ \mathcal{M}^s,\ e_1 :: \ldots :: e_n :: \mathcal{S}^s),\ \mathtt{Apply}\ f;\ P \to (\Phi,\ \mathcal{A}^s,\ \mathcal{M}^s,\ e :: \mathcal{S}^s),\ P} \quad \text{(SApply)}$$

$$\frac{}{(\Phi,\ \mathcal{A}^s,\ \mathcal{M}^s,\ e :: \mathcal{S}^s),\ \mathtt{Dup};\ P \to (\Phi,\ \mathcal{A}^s,\ \mathcal{M}^s,\ e :: e :: \mathcal{S}^s),\ P} \quad \text{(SDup)}$$

$$\frac{\Phi \vdash \mathrm{defined}(e) \quad \phi = \mathrm{simplify}_\Phi(truth(e))}{(\Phi,\ \mathcal{A}^s,\ \mathcal{M}^s,\ truth(e) :: \mathcal{S}^s),\ \mathtt{Test};P) \xrightarrow{\mathsf{if}\ \phi\ \mathsf{then}} (\Phi \cup \{\phi\},\ \mathcal{A}^s,\ \mathcal{M}^s,\ \mathcal{S}^s),\ P} \quad \text{(STest)}$$

$$\frac{\phi \text{ is a fact}}{(\Phi,\ \mathcal{A}^s,\ \mathcal{M}^s,\ \phi :: \mathcal{S}^s),\ \mathtt{Assume};P) \xrightarrow{\mathsf{assume}\ \phi;} (\Phi \cup \{\phi\},\ \mathcal{A}^s,\ \mathcal{M}^s,\ \mathcal{S}^s),\ P} \quad \text{(SAssume)}$$

$$\frac{}{(\Phi,\ \mathcal{A}^s,\ \mathcal{M}^s,\ \mathcal{S}^s),\ \mathtt{ReadEnv}\ x;P \to (\Phi,\ \mathcal{A}^s,\ \mathcal{M}^s,\ x :: \mathcal{S}^s),\ P} \quad \text{(SReadEnv)}$$

$$\frac{x \text{ does not occur in } (\Phi,\ \mathcal{A}^s,\ \mathcal{M}^s,\ e :: \mathcal{S}^s)}{(\Phi,\ \mathcal{A}^s,\ \mathcal{M}^s,\ e :: \mathcal{S}^s),\ \mathtt{WriteEnv}\ x;P \xrightarrow{\mathsf{let}\ x = e\ \mathsf{in}} (\Phi \cup \{x = e\},\ \mathcal{A}^s,\ \mathcal{M}^s,\ \mathcal{S}^s),\ P} \quad \text{(SWriteEnv)}$$

$$\frac{\Phi \vdash \mathrm{defined}(e_1) \quad \ldots \quad \Phi \vdash \mathrm{defined}(e_n)}{(\Phi,\ \mathcal{A}^s,\ \mathcal{M}^s,\ e_1 :: \ldots :: e_n :: \mathcal{S}^s),\ \mathtt{Event}\ ev\ n;P \xrightarrow{\mathsf{event}\ ev(e_1, \ldots, e_n)} (\Phi,\ \mathcal{A}^s,\ \mathcal{M}^s,\ \mathcal{S}^s),\ P} \quad \text{(SEvent)}$$

Figure 3.10: The symbolic execution of CVM.

$\Phi \vdash \text{len}(e'_h) \leq size(pb)$. Both (SLoad) and (SStore) simplify the generated expressions using the function simplify described in section 3.3.

The symbolic allocation table $\mathcal{A}^s$ is initialised in rule (SInit) to the expression $initmem(Q)$ and updated using the function symbol *allocate* on every invocation of Malloc (we let the function symbols *initmem* and *allocate* be interpreted using the functions of the same name introduced in section 2.2). The rule (SMalloc) includes the current value of the allocation table in the pointer base expression that it constructs (this is the only place where the symbolic allocation table is used). As a result the symbolic expression generated by every Malloc explicitly depends on all preceding Malloc invocations. To see why this is necessary consider the following code:

```
void * ignored_ptr = malloc(secret);
void * checked_ptr = malloc(100);
if(checked_ptr == NULL) exit(1);
```

This program may leak information about secret since the success or failure of a malloc call may depend on how much memory has been previously allocated. Our symbolic execution makes this explicit by storing the following symbolic expression in checked_ptr:

$$\text{ptr}(malloc(allocate(initmem(Q), \tau_{\text{size}}(secret)), 100), 0).$$

The rule (STest) expects the condition to be wrapped in function *truth* that extracts an IML boolean value from a C bitstring. Our C to CVM translation described in section 3.1 inserts an application of *truth* before every Test instruction. The rule (SNew) requires that the length of the randomly generated value is a fixed integer $n$, so that the randomness generation can be represented using the IML new $x$: $\texttt{fixed}_n$ construct.

The information about lengths of expressions comes from three sources. First, the rule (SNew) adds information about the lengths of randomly generated values to the fact set $\Phi$. Second, the rewriting rules used by the solving procedure (section 3.2) encode knowledge about lengths of certain expressions. For instance, we rewrite expressions like $\text{len}(a \bigoplus_\tau b)$ to $\text{len}(\tau)$. Finally, explicit length tests and assumptions can be used to add knowledge about lengths via the (STest) and (SAssume) rules. This is demonstrated in the example of section 3.1 where the fact *"The result of XOR has the same length as the argument"* is added to the set $\Phi$ by an explicit assumption in the proxy function xor_proxy.

Whenever the CVM program performs a "model-relevant" action such as generating a random value or sending a value on the network, we record this action as a transition label. The extracted model is then simply the sequence of all the transition labels of the symbolic execution.

**Definition 3.1 (Model Extraction by Symbolic Execution)** Given a symbolic state $\mathbb{S}$ and a CVM process $P$, if there exists a symbolic trace $(\mathbb{S}, P) \xrightarrow{\lambda_1} \dots \xrightarrow{\lambda_n} (\mathbb{S}', 0)$ then we say that $(\mathbb{S}, P)$ *yields the model* $\tilde{Q} = \lambda_1 \dots \lambda_n 0 \in \text{IML}$. Given an initial process $Q \in \text{CVM}_0$ we say that

$Q_0$ *yields a model* $\tilde{Q}_0 \in \mathrm{IML}$ if $(\emptyset, Q_0)$ yields $\tilde{Q}_0$ □

The soundness of symbolic execution (theorem 3.2) which we prove in section 3.7 states that all the security properties that hold for the model also hold for the original program. Theorem 3.3 additionally states that successful symbolic execution proves that the program is safe—it never goes bad (definition 2.10).

## 3.6 Example: Symbolic Execution of RPC-enc

We walk through the symbolic execution of the function `send_request` of the client in the RPC-enc protocol described in section 1.1. The code of the function is shown in figure 3.11. The parameter of the function is the structure `ctx` that holds pointers to values relevant to the protocol execution: the identities of both the client and the server (fields `self` and `other`), the value of the request (field `request`), as well as fields `k_ab` and `k_s` pointing to a long-term key and the session key. The symbolic memory when entering the function looks as follows:

$$\begin{aligned}
&\mathit{addr}(\mathsf{ctx}) \mapsto \mathrm{ptr}(\mathit{pb}_1, 0)\\
&\mathit{pb}_1 \mapsto \{\, \mathsf{request} = \mathrm{ptr}(\mathit{pb}_2, 0),\ \mathsf{request_len} = \tau_{4u}^{-1}(\mathrm{len}(\mathit{request})),\\
&\qquad \mathsf{self} = \mathrm{ptr}(\mathit{pb}_3, 0),\ \mathsf{self_len} = \tau_{4u}^{-1}(\mathrm{len}(\mathit{clientID})),\\
&\qquad \mathsf{other} = \mathrm{ptr}(\mathit{pb}_4, 0),\ \mathsf{other_len} = \tau_{4u}^{-1}(\mathrm{len}(\mathit{serverID})),\\
&\qquad \mathsf{k_s} = \mathrm{ptr}(\mathit{pb}_5, 0),\ \mathsf{k_s_len} = \tau_{4u}^{-1}(\mathrm{len}(k_S)),\\
&\qquad \mathsf{k_ab} = \mathrm{ptr}(\mathit{pb}_6, 0),\ \mathsf{k_ab_len} = \tau_{4u}^{-1}(\mathrm{len}(\mathit{key}(\mathit{clientID}, \mathit{serverID})))\}\\
&\mathit{pb}_2 \mapsto \mathit{request},\ \mathit{pb}_3 \mapsto \mathit{clientID},\ \mathit{pb}_4 \mapsto \mathit{serverID},\\
&\mathit{pb}_5 \mapsto k_S,\ \mathit{pb}_6 \mapsto \mathit{key}(\mathit{clientID}, \mathit{serverID})
\end{aligned}$$

The pointer bases $\mathit{pb}_1, \ldots, \mathit{pb}_6$ are results of distinct calls to `malloc`. As described in section 3.1 such pointer bases refer to all preceding calls of `malloc` and can therefore grow quite large, so we do not spell them out explicitly. The values $\mathit{request}$, $\mathit{clientID}$, $\mathit{serverID}$, and $k_S$ are symbolic variables that have been created during symbolic execution of the preceding code, by a call to a random number generator in case of $k_S$ or by reading values from the environment in case of other variables. The value $\mathit{key}(\mathit{clientID}, \mathit{serverID})$ is a symbolic expression representing a long-term key. It is generated during symbolic execution of the call to `get_shared_key`.

We examine the symbolic execution of the function `send_request` line by line in figure 3.11, assuming that we start with the symbolic memory shown above. The left column shows the source code and the right column shows the corresponding updates to the symbolic memory as well as the generated IML expressions (in the last two lines).

In line 1, we compute the length of the encrypted part of the request message and store it in `m1_len`. The values $\mathrm{len}(k_S)$ and $\mathrm{len}(\mathit{request})$ are extracted from the fields of the `ctx` structure, the value 4 is the result of the sizeof operation.

Line 2 creates a new memory location $\mathit{pb}_7$ and stores the pointer to the beginning of the memory location (with offset 0) in `p` and `m1`. The length `m1_len` of the allocated area is stored in $pb_7$ and all future writes into $pb_7$ are checked to be within allocated bounds. We have removed the NULL checks after malloc for simplicity.

| | C line | symbolic execution steps |
|---|---|---|
| | `int send_request(RPCstate * ctx){` | |
| 1. | `uint32_t m1_len, m1_e_len, full_len;` `unsigned char * m1, * p, * m1_e;` `m1_len = 1 + ctx->k_s_len + sizeof(ctx->request_len) + ctx->request_len;` | $addr(\mathsf{m1_len}) \mapsto \tau_{4u}^{-1}(1) \oplus_{\tau_{4u}} \tau_{4u}^{-1}(\mathrm{len}(k_S)) \oplus_{\tau_{4u}} \tau_{4u}^{-1}(4) \oplus_{\tau_{4u}} \tau_{4u}^{-1}(\mathrm{len}(request))$ |
| 2. | `p = m1 = malloc(m1_len);` | $addr(\mathsf{p}) \mapsto \mathrm{ptr}(pb_7,\, 0)$ <br> $addr(\mathsf{m1}) \mapsto \mathrm{ptr}(pb_7,\, 0)$ <br> where $pb_7 = malloc(\ldots, 1 + \mathrm{len}(k_S) + 4 + \mathrm{len}(reqeust))$ |
| 3. | `memcpy(p, "p", 1);` | $pb_7 \mapsto \text{'p'}$ |
| 4. | `p += 1;` | $addr(\mathsf{p}) \mapsto \mathrm{ptr}(pb_7,\, 1)$ |
| 5. | `* (uint32_t *) p = ctx->request_len;` | $pb_7 \mapsto \text{'p'} \vert \tau_{4u}^{-1}(\mathrm{len}(request))$ |
| 6. | `p += sizeof(ctx->request_len);` | $addr(\mathsf{p}) \mapsto \mathrm{ptr}(pb_7,\, 5)$ |
| 7. | `memcpy(p, ctx->request, ctx->request_len);` | $pb_7 \mapsto \text{'p'} \vert \tau_{4u}^{-1}(\mathrm{len}(request)) \vert request$ |
| 8. | `p += ctx->request_len;` | $addr(\mathsf{p}) \mapsto \mathrm{ptr}(pb_7,\, 5 + \mathrm{len}(request))$ |
| 9. | `memcpy(p, ctx->k_s, ctx->k_s_len);` | $pb_7 \mapsto \text{'p'} \vert \tau_{4u}^{-1}(\mathrm{len}(request)) \vert request \vert k_S$ |
| 10. | `full_len = 1 + sizeof(ctx->self_len) + ctx->self_len + encrypt_len(ctx->k_ab, ctx->k_ab_len, m1, m1_len);` | $addr(\ \mathsf{full_len}\ ) \mapsto \tau_{4u}^{-1}(5) \oplus_{\tau_{4u}} \tau_{4u}^{-1}(\mathrm{len}(clientID)) \oplus_{\tau_{4u}} encrypt_len(msg1)$ <br> where $msg1 = \text{'p'} \vert \tau_{4u}^{-1}(\mathrm{len}(request)) \vert request \vert k_S$ |
| 11. | `p = m1_e = malloc(full_len);` | $addr(\mathsf{p}) \mapsto \mathrm{ptr}(pb_8,\, 0)$ <br> $addr(\mathsf{m1_e}) \mapsto \mathrm{ptr}(pb_8,\, 0)$ |
| 12. | `memcpy(p, "p", 1);` | $pb_8 \mapsto \text{'p'}$ |
| 13. | `p += 1;` | $addr(\mathsf{p}) \mapsto \mathrm{ptr}(pb_8,\, 1)$ |
| 14. | `* (uint32_t *) p = ctx->self_len;` | $pb_8 \mapsto \text{'p'} \vert \tau_{4u}^{-1}(\mathrm{len}(clientID))$ |
| 15. | `p += sizeof(ctx->self_len);` | $addr(\mathsf{p}) \mapsto \mathrm{ptr}(pb_8,\, 5)$ |
| 16. | `memcpy(p, ctx->self, ctx->self_len);` | $pb_8 \mapsto \text{'p'} \vert \tau_{4u}^{-1}(\mathrm{len}(clientID)) \vert clientID$ |
| 17. | `p += ctx->self_len;` | $addr(\mathsf{p}) \mapsto \mathrm{ptr}(pb_8,\, 5 + \mathrm{len}(clientID))$ |
| 18. | `m1_e_len = encrypt(ctx->k_ab, ctx->k_ab_len, m1, m1_len, p);` | $pb_8 \mapsto \text{'p'} \vert \tau_{4u}^{-1}(\mathrm{len}(clientID)) \vert clientID \vert cipher1$ <br> $addr(\mathsf{m1_e_len}) \mapsto \tau_{4u}^{-1}(\mathrm{len}(cipher1))$ <br> new fact: $\mathrm{len}(cipher1) \leq \tau_{4u}(encrypt_len(msg1))$ <br> $cipher1 = E(key(clientID, serverID), msg1)$ |
| 19. | `full_len = 1 + sizeof(ctx->self_len) + ctx->self_len + m1_e_len;` | $addr(\ \mathsf{full_len}\ ) \mapsto \tau_{4u}^{-1}(5) \oplus_{\tau_{4u}} \tau_{4u}^{-1}(\mathrm{len}(clientID)) \oplus_{\tau_{4u}} \tau_{4u}^{-1}(\mathrm{len}(cipher1))$ |
| 20. | `send(&(ctx->bio), &full_len, sizeof(full_len));` | $\mathsf{out}(c, \tau_{4u}^{-1}(5) \oplus_{\tau_{4u}} \tau_{4u}^{-1}(\mathrm{len}(clientID)) \oplus_{\tau_{4u}} \tau_{4u}^{-1}(\mathrm{len}(cipher1)));$ |
| 21. | `send(&(ctx->bio), m1_e, full_len); }` | $\mathsf{out}(c,\ \text{'p'} \vert \tau_{4u}^{-1}(\mathrm{len}(clientID)) \vert clientID \vert cipher1);$ |

Figure 3.11: Symbolic execution of the send_request function.

The call to function memcpy in line 3 invokes a proxy function (section 3.1 and appendix A) that copies the literal bitstring 'p' into the memory location $pb_7$ stored in p. Line 4 increments p by 1 so that the value stored in p becomes $\mathrm{ptr}(pb_7, 1)$.

Line 5 stores the value $\mathrm{len}(request)$ from ctx→request_len into p. Now that p points to the offset 1 from the beginning of $pb_7$ the value $\mathrm{len}(request)$ is written just past the value 'p' that is already in $pb_7$, so that the memory location now contains a concatenation. Line 6 increments p to point just past the end of the concatenation. Lines 7–9 proceed similarly and append *request* and $k_S$ to the contents of the memory location.

Line 10 computes an upper bound of the length of the request message. The function encrypt_len gives an upper bound of the length of an encryption based on the key and the plaintext. In this case the plaintext is the concatenation in $pb_7$. We are not verifying correctness of cryptographic implementations and so the call to encrypt_len is replaced by a call to a proxy function that constructs the symbolic expression $\mathit{encrypt_len}(\mathit{msg1})$, where $\mathit{msg1}$ is the plaintext.

Lines 11–17 work similarly to lines 2–9 and build up the request message in the memory location $pb_8$. In line 18 we add the ciphertext to the message by calling encrypt. The call is redirected to a proxy function which generates a new symbolic expression $\mathit{cipher1} = E(\mathit{key}(\mathit{clientID}, \mathit{serverID}), \mathit{msg1})$ that represents the ciphertext and writes that expression through the pointer passed as argument. The function returns the expression $\mathrm{len}(\mathit{cipher1})$. It additionally checks that $\mathrm{len}(\mathit{cipher1}) \leq \mathit{encrypt_len}(\mathit{msg1})$. This fact is added to the set of known facts and is used to prove that the length of the contents of $pb_8$ is still less than or equal to the allocated size of $pb_8$, that is, the encryption function does not write past the end of the allocated buffer.

Line 19 computes the full length of the request message. Lines 20 and 21 send this length followed by the actual message. The call to send redirects to a proxy function that generates the IML statement $\mathsf{out}(c, e)$, where $e$ is the symbolic expression contained in the buffer passed to send.

Section 4.2 shows the complete IML model extracted from the RPC-enc implementation and the steps that we take to turn it into a CryptoVerif model.

## 3.7 Symbolic Execution Soundness

This section proves the main result of this chapter—the soundness of symbolic execution (theorem 3.2). In a nutshell, the soundness theorem states that if a CVM process $Q$ yields an IML model $\tilde{Q}$ then for any acceptable context $C$ the process $C\{Q\}$ is not less secure than $C\{\tilde{Q}\}$. The idea of the proof is to demonstrate a simulation $Q \lesssim \tilde{Q}$ (lemma 3.3) after which the statement of soundness follows from theorems 2.2 and 2.3. Another important result that we prove is theorem 3.3 which states that if the symbolic execution of $Q$ succeeds then $Q$ never goes bad.

The main tool in the proof is a relation $S \sim_\eta \mathbb{S}$ between a concrete process state $S$ and a symbolic state $\mathbb{S}$ with respect to an environment $\eta$. The relation holds whenever evaluating all symbolic expressions in $\mathbb{S}$ with respect to $\eta$ would yield corresponding bitstrings in $S$. The relation also imposes a consistency condition on $\mathbb{S}$: all the symbolic facts should hold with respect to $\eta$, the symbolic expressions in memory should be contained within allocated regions, and the allocated regions corresponding to different pointer bases should not overlap. The following definitions make this precise.

**Definition 3.2 ($\eta$-consistent Symbolic States)** For a symbolic memory $\mathcal{M}^s \colon \mathit{PBase} \rightharpoonup \mathit{IExp}$ and an environment $\eta$ let

$$\mathrm{base}_\eta(\mathcal{M}^s) = \{pb \in \mathrm{dom}(\mathcal{M}^s) \mid [\![pb]\!]_\eta \neq 0\}\,.$$

$$\begin{array}{ccccc}
(\eta_1, S_1, P_1) & \longrightarrow & \dots & \longrightarrow & (\eta_n, S_n, P_n) \\
\Uparrow S_1 \sim_{\eta_1} \mathbb{S}_1 & & & & \Uparrow S_n \sim_{\eta_n} \mathbb{S}_n \\
(\mathbb{S}_1, P_1) & \longrightarrow & \dots & \longrightarrow & (\mathbb{S}_n, P_n)
\end{array}$$

Figure 3.12: The correspondence between the concrete and the symbolic trace.

Given a symbolic state $\mathbb{S} = (\Phi, \mathcal{A}^s, \mathcal{M}^s, \mathcal{S}^s)$ and an environment $\eta$ we say that $\mathbb{S}$ is *$\eta$-consistent* when the following conditions hold:

1. $\eta \models \Phi$.
2. For all $pb \in \text{base}_\eta(\mathcal{M}^s)$: $\eta \models (\text{len}(\mathcal{M}^s(pb)) \leq \mathit{size}(pb))$. In particular, $[\![\mathcal{M}^s(pb)]\!]_\eta \neq \bot$ and $[\![pb]\!]_\eta \neq \bot$.
3. For all $pb$, $pb' \in \text{base}_\eta(\mathcal{M}^s)$ with $pb \neq pb'$:
$$\{[\![pb]\!]_\eta\}_{[\![\mathit{size}(pb)]\!]_\eta} \cap \{[\![pb']\!]_\eta\}_{[\![\mathit{size}(pb')]\!]_\eta} = \emptyset.$$
4. $[\![\mathcal{A}^s]\!]_\eta = \bigcup \left\{ \{[\![pb]\!]_\eta\}_{[\![\mathit{size}(pb)]\!]_\eta} \;\middle|\; pb \in \text{base}_\eta(\mathcal{M}^s) \right\}.$

The initial symbolic state $\emptyset$ is defined to be $\eta$-consistent for any environment $\eta$. □

**Definition 3.3 (State Correspondence)** Given an environment $\eta$ we say that an $\eta$-consistent symbolic state $\mathbb{S} = (\Phi, \mathcal{A}^s, \mathcal{M}^s, \mathcal{S}^s)$ *corresponds* to a concrete process state $S = (\mathcal{A}, \mathcal{M}, \mathcal{S})$ with respect to $\eta$, writing $S \sim_\eta \mathbb{S}$, if $\mathcal{S}$ is obtained from $\mathcal{S}^s$ by applying $[\![\cdot]\!]_\eta$ to each element, $\mathcal{A} = [\![\mathcal{A}^s]\!]_\eta$, and for each $pb \in \text{base}_\eta(\mathcal{M}^s)$

$$[\![\mathcal{M}^s(pb)]\!]_\eta = \mathcal{M}(p) \dots \mathcal{M}(p + l - 1), \text{ where } p = [\![pb]\!]_\eta \text{ and } l = |[\![\mathcal{M}^s(pb)]\!]_\eta|. \tag{3.78}$$

Additionally we let $\emptyset \sim_\eta \emptyset$ for every environment $\eta$. □

The main idea of the proof is illustrated in figure 3.12. Given a concrete trace with executing processes $(\eta_1, S_1, P_1), \dots, (\eta_n, S_n, P_n)$ and a symbolic trace $(\mathbb{S}_1, P_1), \dots, (\mathbb{S}_n, P_n)$ we shall show that $S_i \sim_{\eta_i} \mathbb{S}_i$ for each $i \leq n$. We shall then define a simulation relation $\lesssim$ between executing CVM states and executing IML states in such a way that for each $i \leq n$ we have $(\eta_i, S_i, P_i) \lesssim (\eta_i, \tilde{P}_i)$, whenever $(\mathbb{S}_i, P_i)$ yields the model $\tilde{P}_i$. We shall then prove that $\lesssim$ satisfies the conditions in definition 2.11.

We start by proving a lemma that provides the inductive step necessary to establish the correspondence in figure 3.12 for output processes. In the rest of the section $\mathcal{Q}$ is an arbitrary multiset of executing input processes.

**Lemma 3.2** *Assume that there is a concrete transition of the form* $(\eta, S, P), \mathcal{Q} \to (\eta', S', P'), \mathcal{Q}$ *with output CVM processes* $P$ *and* $P'$*. Further assume that there is a symbolic transition of the form* $(\mathbb{S}, P) \to (\mathbb{S}', P')$ *such that* $\mathbb{S}$ *is* $\eta$*-consistent, and* $S \sim_\eta \mathbb{S}$*. Then* $\mathbb{S}'$ *is* $\eta'$*-consistent, and* $S' \sim_{\eta'} \mathbb{S}'$*.* □

Proof We prove the statement by a case split based on the first instruction in $P$. In the following $\mathcal{M}, \ldots$ and $\mathcal{M}', \ldots$ refer to components of $S$ and $S'$, and $\mathcal{M}^s, \ldots$ and $\mathcal{M}^{s\prime}, \ldots$ refer to components of $\mathbb{S}$ and $\mathbb{S}'$. For the purpose of this proof the labels of the concrete and the symbolic transitions are not relevant.

- Const $c$

  Both the concrete transition (CConst) and the symbolic transition (SConst) have the effect of putting the same bitstring or integer $c$ onto the stack. Thus both the $\eta$-consistency and the state correspondence are preserved and the lemma holds with $\eta' = \eta$.

- Ref $v$

  The concrete transition (CRef) puts $\tau_{\text{ptr}}^{-1}(addr(v))$ on the stack and the symbolic transition (SRef) puts $\text{ptr}(addr(v), 0)$ on the stack. By equation (3.46) (section 3.4)

  $$[\![\text{ptr}(addr(v), 0)]\!]_\eta = [\![\tau_{\text{ptr}}^{-1}(addr(v) + 0)]\!]_\eta = \tau_{\text{ptr}}^{-1}(addr(v))$$

  and the lemma holds with $\eta' = \eta$.

- Malloc

  Let $b_l$, $l$, and $p$ be defined as in rule (CMalloc), and let $e_l$ and $pb$ be defined as in rule (SMalloc). According to (CMalloc) $\eta' = \eta$. First we prove that $\mathbb{S}'$ is $\eta$-consistent.

  1 The condition $\eta \models \Phi'$ holds trivially since $\Phi' = \Phi$.

  2 The symbolic rule adds $pb$ to $\text{base}_\eta(\mathcal{M}^s)$. According to (SRef) $[\![\tau_{\text{size}}(e_l)]\!]_\eta \neq \bot$. Together with the assumption that $\tau_{\text{size}}$ is unsigned this gives

  $$[\![size(pb)]\!]_\eta = [\![\tau_{\text{size}}(e_l)]\!]_\eta \geq 0 = |\varepsilon| = [\![\text{len}(\mathcal{M}^s(pb))]\!]_\eta.$$

  Thus $\eta \models (\text{len}(\mathcal{M}^s(pb)) \leq size(pb))$.

  3 Choose $pb' \in \text{base}_\eta(\mathcal{M}^s)$ such that $pb \neq pb'$. By condition 4 of definition 3.2 applied to $\eta$-consistency of $\mathbb{S}$ we have $\{[\![pb']\!]_\eta\}_{[\![size(pb')]\!]_\eta} \subseteq [\![\mathcal{A}^s]\!]_\eta$. On the other hand,

  $$[\![pb]\!]_\eta = [\![malloc(\mathcal{A}^s, \tau_{\text{size}}(e_l))]\!]_\eta = malloc([\![\mathcal{A}^s]\!]_\eta, [\![size(pb)]\!]_\eta)$$

  Under the assumption of correctness of *malloc* (section 2.2) we have

  $$\{[\![pb]\!]_\eta\}_{[\![size(pb)]\!]_\eta} \cap \{[\![pb']\!]_\eta\}_{[\![size(pb')]\!]_\eta} \subseteq \{[\![pb]\!]_\eta\}_{[\![size(pb)]\!]_\eta} \cap [\![\mathcal{A}^s]\!]_\eta = \emptyset.$$

4 This condition follows directly from the definition of *allocate* (section 2.2).

Now we prove that $S' \sim_{\eta'} \mathbb{S}'$. By the assumption $S \sim_\eta \mathbb{S}$ we have $\mathcal{A} = [\![\mathcal{A}^s]\!]_\eta$ and by comparing the symbolic and the concrete rule we see that $\mathcal{A}' = [\![\mathcal{A}^{s\prime}]\!]_\eta$. It is left to prove that $[\![\mathrm{ptr}(pb, 0)]\!]_\eta = \tau_{\mathrm{ptr}}^{-1}(p)$. By (R37) we have $[\![\mathrm{ptr}(pb, 0)]\!]_\eta = \tau_{\mathrm{ptr}}^{-1}([\![pb]\!]_\eta)$. By the assumption $S \sim_\eta \mathbb{S}$ we have $[\![e_l]\!]_\eta = b_l$ and thus $[\![\tau_{\mathrm{size}}(e_l)]\!]_\eta = l$. Therefore

$$\begin{aligned}[\![\mathrm{ptr}(pb, 0)]\!]_\eta &= \tau_{\mathrm{ptr}}^{-1}([\![pb]\!]_\eta) = \tau_{\mathrm{ptr}}^{-1}([\![\mathit{malloc}(\mathcal{A}^s, \tau_{\mathrm{size}}(e_l))]\!]_\eta) \\ &= \tau_{\mathrm{ptr}}^{-1}(\mathit{malloc}(\mathcal{A}, [\![\tau_{\mathrm{size}}(e_l)]\!]_\eta)) = \tau_{\mathrm{ptr}}^{-1}(\mathit{malloc}(\mathcal{A}, l)) = \tau_{\mathrm{ptr}}^{-1}(p).\end{aligned}$$

- Load

Both the concrete rule (CLoad) and the symbolic rule (SLoad) have the effect of replacing two values on the stack with a new value. In the concrete transition the new value is $b = \mathcal{M}(p) \ldots \mathcal{M}(p + l - 1)$, where $p$ and $l$ are the pointer and the length value taken from the stack. In the symbolic transition the new value is $e' = \mathrm{simplify}_\Phi(e)$ with $e = \mathcal{M}^s(pb)\{t_o, \tau_{\mathrm{size}}(e_l)\}$, where $pb$ and $t_o$ are the base and offset of a symbolic pointer taken from the stack and and $e_l$ is the length expression taken from the stack. We shall prove that $[\![e']\!]_\eta = b$, so that $\eta' = \eta$ satisfies the requirements. Let

$$p_b = [\![pb]\!]_\eta \qquad n_o = [\![t_o]\!]_\eta \qquad l_h = |[\![\mathcal{M}^s(pb)]\!]_\eta|$$

All of these are well-defined due to the conditions checked by the rule (SLoad). One of these conditions implies $p_b \neq 0$, and so by condition (3.78) of the assumption $S \sim_\eta \mathbb{S}$

$$[\![\mathcal{M}^s(pb)]\!]_\eta = \mathcal{M}(p_b) \ldots \mathcal{M}(p_b + l_h - 1).$$

By $S \sim_\eta \mathbb{S}$ we know $l = [\![\tau_{\mathrm{size}}(e_l)]\!]_\eta$. By condition $\Phi \vdash \mathrm{defined}(e)$ of the rule (SLoad) the following expression is well-defined:

$$[\![e]\!]_\eta = [\![\mathcal{M}^s(pb)\{t_o, \tau_{\mathrm{size}}(e_l)\}]\!]_\eta = \mathcal{M}(p_b + n_o) \ldots \mathcal{M}(p_b + n_o + l - 1).$$

Using the assumption $S \sim_\eta \mathbb{S}$, the condition $\Phi \vdash \mathrm{defined}(\mathrm{ptr}(pb, t_o))$ of the rule (SLoad), and the pointer evaluation rule (R37) we get

$$p = [\![\tau_{\mathrm{ptr}}(\mathrm{ptr}(pb, t_o))]\!]_\eta = \tau_{\mathrm{ptr}}(\tau_{\mathrm{ptr}}^{-1}([\![pb]\!]_\eta + [\![t_o]\!]_\eta)) = p_b + n_o,$$

and therefore

$$[\![e]\!]_\eta = \mathcal{M}(p) \ldots \mathcal{M}(p + 1 - 1) = b.$$

By the soundness of simplification $[\![e']\!]_\eta = [\![\mathrm{simplify}_\Phi(e)]\!]_\eta = [\![e]\!]_\eta$.

- Store

Let $p$ and $b$ be defined as in rule (CStore), and let $pb$, $t_o$, $e_h$, and $e'_h$ be defined as in rule (SStore). Both the concrete transition (CStore) and the symbolic transition (SStore) perform a memory update. These updates are

$$\mathcal{M}' = \mathcal{M}\left\{p + i \mapsto b[i] \mid i < |b|\right\}, \tag{3.79}$$
$$\mathcal{M}^{s\prime} = \mathcal{M}^{s}\left\{pb \mapsto \text{simplify}_{\Phi}(e'_h)\right\}, \tag{3.710}$$

We shall prove that the lemma is satisfied with $\eta' = \eta$. Let

$$l_h = |[\![e_h]\!]_\eta| \qquad l'_h = |[\![e'_h]\!]_\eta| \qquad n_o = [\![t_o]\!]_\eta$$
$$b_h = [\![e_h]\!]_\eta \qquad l = |[\![e]\!]_\eta| \qquad p_b = [\![pb]\!]_\eta$$

All of these values are well-defined: $[\![e_h]\!]_\eta \neq \bot$ because $\mathbb{S}$ is $\eta$-consistent, $[\![e'_h]\!]_\eta \neq \bot$ because of the condition $\Phi \vdash \text{len}(e'_h) \leq \mathit{size}(pb)$ checked by the rule (SStore), and $[\![t_o]\!]_\eta \neq \bot$ and $[\![pb]\!]_\eta \neq \bot$ because of the condition $\Phi \vdash \text{defined}(\text{ptr}(pb, t_o))$.

It is easy to see that $S'$ is $\eta$-consistent: as the transition only updates the memory, we only need to check that $\eta \models \text{len}(\mathcal{M}^{s\prime}(pb)) \leq \mathit{size}(pb)$, but this is exatly the condition checked by the rule (SStore).

To show $S' \sim_\eta \mathbb{S}'$ we only need to prove (3.78) with $\mathcal{M}'$ and $\mathcal{M}^{s\prime}$. By assumption $S \sim_\eta \mathbb{S}$ we have $b = [\![e]\!]_\eta$ and $p = [\![\tau_{\text{ptr}}(\text{ptr}(pb, t_o))]\!]_\eta$. Using the same argument as in the proof for Load we get $p = p_b + n_o$. To show (3.78) for $pb$ we shall consider two cases, as in the rule (SStore). The first case is

$$\Phi \vdash (t_o + \text{len}(e) < \text{len}(e_h))$$
$$e'_h = e_h\{0, t_o\} \,|\, e \,|\, e_h\{t_o + \text{len}(e), \text{len}(e_h) - (t_o + \text{len}(e))\}.$$

Applying (3.78) with $S \sim_\eta \mathbb{S}$ and using (3.79) we get

$$\begin{aligned}[\![e'_h]\!]_\eta &= b_h[0]\dots b_h[n_o - 1]b[0]\dots b[l-1]b_h[n_o + l]\dots b_h[l_h - 1] \\ &= \mathcal{M}(p_b)\dots\mathcal{M}(p_b + n_o - 1)\mathcal{M}'(p)\dots\mathcal{M}'(p + l - 1) \\ &\qquad \mathcal{M}(p_b + n_o + l)\dots\mathcal{M}(p_b + l_h - 1).\end{aligned}$$

Using $p = p_b + n_o$ we conclude

$$\begin{aligned}[\![e'_h]\!]_\eta &= \mathcal{M}'(p_b)\dots\mathcal{M}'(p-1)\mathcal{M}'(p)\dots\mathcal{M}'(p+l-1)\mathcal{M}'(p+l)\dots\mathcal{M}'(p_b + l_h - 1) \\ &= \mathcal{M}'(p_b)\dots\mathcal{M}'(p_b + l_h - 1).\end{aligned}$$

By the soundness of simplification $[\![\text{simplify}_\Phi(e'_h)]\!]_\eta = [\![e'_h]\!]_\eta$, thus by applying (3.710) we see that (3.78) is satisfied for $\mathcal{M}'$ and $\mathcal{M}^{s\prime}$ with $pb$.

The second case is

$$\Phi \vdash (t_o + \mathrm{len}(e) \geq \mathrm{len}(e_h)) \wedge (t_o \leq \mathrm{len}(e_h)) \qquad \text{and} \qquad e_h' = e_h\{0, t_o\}|e$$

and it is proved the same way as the case above.

It remains to show that none of the other memory regions get overwritten, that is, we need to show that (3.78) still holds for every $pb' \in \mathrm{base}_\eta(\mathcal{M}^{s\prime})$ with $pb' \neq pb$. Condition $\Phi \vdash \mathrm{defined}(\mathrm{ptr}(pb, t_o))$ of the rule (SStore) implies that $n_o \geq 0$. In particular, $p = p_b + n_o \geq p_b$. At the same time both cases above imply $\Phi \vdash t_o + \mathrm{len}(e) \leq \mathrm{len}(e_h')$, so that $n_o + l \leq l_h' \leq [\![size(pb)]\!]_\eta$. Therefore

$$\{p\}_l = \{p_b + n_o\}_l \subseteq \{p_b\}_{[\![size(pb)]\!]_\eta} = \{[\![pb]\!]_\eta\}_{[\![size(pb)]\!]_\eta}.$$

Considering (3.79) this means that $\mathcal{M}'(i) = \mathcal{M}(i)$ for all $i \notin \{[\![pb]\!]_\eta\}_{[\![size(pb)]\!]_\eta}$. By conditions 2 and 3 of $\eta$-consistency of $\mathbb{S}$ we see that (3.78) is satisfied for $pb'$.

- New $x$

  The concrete rule (CNew) places on the stack a value $b$ of length $l = \tau_{\text{size}}(b_l)$, where $b_l$ is taken from the stack and updates $\eta' = \eta\{x \mapsto b\}$. The symbolic rule (SNew) takes $\tau_{\text{size}}^{-1}(n)$ for some $n \in \mathbb{N}$ from the stack, places $x$ on the stack, and adds a fact $\mathrm{len}(x) = n$. It is straightforward to check that the lemma is satisfied with $\eta'$. It is important that (SNew) checks that $x$ does not occur in $\mathbb{S}$, so that $\eta$-consistency is not disrupted.

- Apply $f$

  The rule (CApply) places on the stack the bitstring $b = I(f)(b_1, \ldots, b_n)$, whereby $b_1, \ldots, b_n$ are taken from the stack. The rule (SApply) places on the stack the value $e = f(e_1, \ldots, e_n)$, whereby $e_1, \ldots, e_n$ are taken from the stack. By $S \sim_\eta \mathbb{S}$ we have $[\![e_i]\!]_\eta = b_i$ for all $i$, therefore $b = [\![e]\!]_\eta$ and the lemma holds with $\eta' = \eta$.

- Dup with rules (CDup) and (SDup)

  The lemma is satisfied with $\eta' = \eta$.

- Test with rules (CTestTrue) and (STest)

  The lemma is satisfied with $\eta' = \eta$. We do not deal with the rule (CTestFalse) here since it violates the premise of the lemma that the resulting state of the concrete transition is the same as the resulting state of the symbolic transition. We shall cover (CTestFalse) separately in the proof of lemma 3.3.

- Assume with rules (CAssume) and (SAssume)

  The rule (SAssume) adds a fact $\phi$ to $\Phi$. The condition of the concrete transition implies that $[\![\phi]\!]_\eta = \mathrm{true}$. Thus $\mathbb{S}'$ is $\eta$-consistent and the lemma is satisfied with $\eta' = \eta$.

- $\mathsf{ReadEnv}\ x$ with rules (CReadEnv) and (SReadEnv)

  The lemma is satisfied with $\eta' = \eta$.

- $\mathsf{WriteEnv}\ x$ with rules (CWriteEnv) and (SWriteEnv)

  The lemma is satisfied with $\eta' = \eta\{x \mapsto b\}$.

- $\mathsf{Event}\ ev\ n$ with rules (CEvent) and (SEvent)

  The lemma is satisfied with $\eta' = \eta$. ■

We are now ready to define a simulation relation between a CVM process and its model obtained by symbolic execution.

**Lemma 3.3** *Let $Q_0$ be a CVM process that defines every variable at most once and satisfies all inline assumptions. If $Q_0$ successfully yields an IML model $\tilde{Q}_0$ then $Q_0 \lesssim \tilde{Q}_0$.* □

PROOF Define the simulation relation between executing states as follows: let $(\eta, S, P) \lesssim (\tilde{\eta}, \tilde{P})$ if $\eta = \tilde{\eta}$ and there exists a symbolic state $\mathbb{S}$ such that $\mathbb{S}$ is $\eta$-consistent, $S \sim_\eta \mathbb{S}$, and $(\mathbb{S}, P)$ successfully yields the model $\tilde{P}$. In order to cover reduction by rule (CTestFalse) we also let $(\eta, \mathsf{out}(\mathit{yield}, \varepsilon); 0) \lesssim (\eta, \mathsf{out}(\mathit{yield}, \varepsilon); 0)$ and $(\eta, 0) \lesssim (\eta, 0)$ for each environment $\eta$.

We shall now check that $\lesssim$ satisfies the conditions of definition 2.11.

1. To satisfy condition 1 in definition 2.11 we need to show that $(\eta, \emptyset, Q_0) \lesssim (\eta, \tilde{Q}_0)$ for any environment $\eta$. Choose $\mathbb{S} = \emptyset$. By definition $\emptyset \sim_\eta \emptyset$, and by assumption of the lemma $(\emptyset, Q_0)$ successfully yields $\tilde{Q}_0$.

2. Assume that $(\eta, S, Q) \lesssim (\eta, \tilde{Q})$ and $Q = \mathsf{in}(c, x); P$. To satisfy condition 2 in definition 2.11 we show that $\tilde{Q} = \mathsf{in}(c, x); \tilde{P}$ and $(\eta', S, P) \lesssim (\eta', \tilde{P})$ for all $b \in BS$ and $\eta' = \eta\{x \mapsto b\}$.

   Obtain the symbolic state $\mathbb{S}$ from the definition of the simulation relation $\lesssim$. Then by rule (SIn) $(\mathbb{S}, Q) \xrightarrow{\mathtt{in}(c,x);} (\mathbb{S}', P)$ and therefore $\tilde{Q} = \mathsf{in}(c, x); \tilde{P}$ such that $(\mathbb{S}', P)$ yields the model $\tilde{P}$. Choose $b \in BS$ and let $\eta' = \eta\{x \mapsto b\}$. The configuration $\mathbb{S}'$ differs from $\mathbb{S}$ by an addition of the fact $\mathrm{defined}(x)$ which is true with respect to $\eta'$. Thus $\mathbb{S}'$ is $\eta'$-consistent and $S \sim_{\eta'} \mathbb{S}'$, and so we see that $(\eta', S, P) \lesssim (\eta', \tilde{P})$.

3. Condition 3 in definition 2.11 is proved in the same way as condition 2.

4. Assume that $(\eta, S, Q) \lesssim (\eta, \tilde{Q})$ for an input CVM process $Q$ and an input IML process $\tilde{Q}$. To satisfy condition 4 in definition 2.11 we need to show $\mathrm{reduce}(\{(\eta, S, Q)\}) \lesssim \mathrm{reduce}(\{(\eta, \tilde{Q})\})$.

   First consider the case $\tilde{Q} = 0$. Then clearly $Q = 0$ as well, and so $\mathrm{reduce}(\{(\eta, S, Q)\}) = \mathrm{reduce}(\{(\eta, \tilde{Q})\}) = \emptyset$.

   Now assume $\tilde{Q} \neq 0$ and thus $Q \neq 0$. The symbolic execution does not produce any replications or parallel compositions, thus $\mathrm{reduce}(\{(\eta, \tilde{Q})\}) = \{(\eta, \tilde{Q})\}$. The only non-nil CVM process that performs a non-trivial reduction by $\rightsquigarrow$ is of the form $Q = \mathsf{Init}\,; Q'$ (figure 2.7). By (CInit) $S = \emptyset$ and $\mathrm{reduce}(\{(\eta, S, Q)\}) = \{(\eta, S_0, Q')\}$ with $S_0 = (\mathit{initmem}(Q), \emptyset, [])$. Let $\mathbb{S}_0$ be the symbolic state such that $(\emptyset, Q) \to (\mathbb{S}_0, Q')$ by (SInit). We shall show that $(\eta, S_0, Q') \lesssim (\eta, \tilde{Q})$ with $\mathbb{S}_0$. Clearly $(\mathbb{S}_0, Q')$ yields $\tilde{Q}$. Condition 2 of $\eta$-consistency of $\mathbb{S}_0$ is trivial. Conditions 1, 3 and 4 follow by the choice of *addr* and *initmem* functions (section 2.2). Finally, it is straightforward to check that $S_0 \sim_\eta \mathbb{S}_0$.

5. Assume that $(\eta, S, P) \lesssim (\eta, \tilde{P})$ for an output CVM process $P$ and an output IML process $\tilde{P}$ and there is a transition of the form $(\eta, S, P), \mathcal{Q} \xrightarrow{\mathcal{E}}_p (\eta', S', P'), \mathcal{Q}$. To satisfy condition 5 in definition 2.11 we need to find an IML process $\tilde{P}'$ such that there is a transition $(\eta, \tilde{P}), \mathcal{Q} \xrightarrow{\mathcal{E}}{}^{*}_{p} (\eta', \tilde{P}'), \mathcal{Q}$ and $(\eta', S', P') \lesssim (\eta', \tilde{P}')$.

   Obtain $\mathbb{S}$ from the definition of $(\eta, S, P) \lesssim (\eta, \tilde{P})$. First consider the case that $P = \mathsf{Test}; P^*$ and the transition happens by (CTestFalse) so that $P' = \mathsf{out}(\mathit{yield}, \varepsilon); 0$. From $S \sim_\eta \mathbb{S}$ it follows that $\tilde{P}$ is of the form $\tilde{P} = \mathsf{if}\ \phi\ \mathsf{then}\ \ldots$ with $[\![\phi]\!]_\eta = b = \mathrm{false}$, where $b$ is the top of stack in $S$. By (IIfFalse) $(\eta, \tilde{P})$ also reduces to $\mathsf{out}(\mathit{yield}, \varepsilon); 0$.

   Now assume that the concrete transition happens by a rule other than (CTestFalse). In that case the resulting process $P'$ is the same as in the next step of the symbolic execution, that is, there exists a symbolic state $\mathbb{S}'$ such that $(\mathbb{S}, P) \xrightarrow{\lambda} (\mathbb{S}', P')$ and $(\mathbb{S}', P')$ successfully yields a process $\tilde{P}'$ such that $\tilde{P} = \lambda\tilde{P}'$. By lemma 3.2 $\mathbb{S}'$ is $\eta'$-consistent and $S' \sim_{\eta'} \mathbb{S}'$. Thus $(\eta', S', P') \lesssim (\eta', \tilde{P}')$.

   To show $(\eta, \tilde{P}), \mathcal{Q} \xrightarrow{\mathcal{E}}{}^{*}_{p} (\eta', \tilde{P}'), \mathcal{Q}$ we make a case split on the value of $\lambda$.

   - $\lambda = \varepsilon$ with rules (SConst), (SRef), (SMalloc), (SStore), (SApply), and (SReadEnv). In all these cases $\mathcal{E} = \varepsilon$ and $p = 1$, so that $\tilde{P} = \tilde{P}'$ and an empty trace satisfies the conditions.
   - $\lambda = \mathsf{new}\ x \colon T;$ with rule (SNew).

     Assume that the CVM rule (CNew) executes with $\eta' = \eta\{x \mapsto b\}$ for some $b \in BS$ with $|b| = n$ and $p = 1/2^n$. By (SNew) $T = \mathtt{fixed}_n$, so that by (INew) there is a transition $(\eta, \mathsf{new}\ x \colon \mathtt{fixed}_n; \tilde{P}'), \mathcal{Q} \to_p (\eta\{x \mapsto b\}, \tilde{P}'), \mathcal{Q}$.
   - The statement for the rules (STest), (SWriteEnv), and (SEvent) is proved similarly.

6. The condition 6 in definition 2.11 is trivially satisfied since by definition $\eta = \tilde{\eta}$ for any $(\eta, S, P) \lesssim (\tilde{\eta}, \tilde{P})$. ∎

**Theorem 3.2 (Symbolic Execution is Sound)** *Let $Q_1, \ldots, Q_n$ be CVM processes that successfully yield IML models $\tilde{Q}_1, \ldots, \tilde{Q}_n$. Then for any context $C$ such that both $C\{Q_1, \ldots, Q_n\}$ and $C\{\tilde{Q}_1, \ldots, \tilde{Q}_n\}$ are well-formed and any trace property $\rho$*

$$\mathrm{insec}(C\{Q_1, \ldots, Q_n\}, \rho) \le \mathrm{insec}(C\{\tilde{Q}_1, \ldots, \tilde{Q}_n\}, \rho).$$ □

Proof Let $Q_1, \ldots, Q_n$, $\tilde{Q}_1, \ldots, \tilde{Q}_n$, $C$, and $\rho$ be defined as in the statement of the theorem. By lemma 3.3 $Q_i \lesssim \tilde{Q}_i$ for each $i \leq n$. By theorem 2.3 $C\{Q_1, \ldots, Q_n\} \lesssim C\{\tilde{Q}_1, \ldots, \tilde{Q}_n\}$. By theorem 2.2 $\text{insec}(C\{Q_1, \ldots, Q_n\}, \rho) \leq \text{insec}(C\{\tilde{Q}_1, \ldots, \tilde{Q}_n\}, \rho)$. ■

**Theorem 3.3 (Symbolic Execution Proves Safety)** *If a CVM process $Q_0$ yields an IML model $\tilde{Q}_0$ then $Q_0$ does not go bad.* □

Proof Following definition 2.10 consider a semantic configuration $\mathbb{C}_0 = P_0, \{(\eta_0, \emptyset, Q_0)\} \cup \mathcal{Q}_0$ such that $\mathbb{C}_0$ satisfies inline assumptions and there is a variable $id \in \text{dom}(\eta_0)$ that does not occur in $P_0$ and $\mathcal{Q}_0$. Let there be a trace

$$\mathbb{C}_0 \longrightarrow^* \mathbb{C} = (\eta, S, P), \mathcal{Q}$$

such that $id \in \text{dom}(\eta)$ and $P$ does not start with an output. We need to show that there exists a transition $\mathbb{C} \to \mathbb{C}'$ for some semantic configuration $\mathbb{C}'$.

Let $\tilde{\mathbb{C}}_0 = P_0, \{(\eta_0, \emptyset, \tilde{Q}_0)\} \cup \mathcal{Q}_0$. Define a relation between executing states as follows: let $(\eta, S, P) \lesssim (\tilde{\eta}, \tilde{S}, \tilde{P})$ if $\eta = \tilde{\eta}$ and one of the following conditions hold:

- $id \in \text{dom}(\eta)$ and $\lesssim$ is defined as in the proof of lemma 3.3.
- $id \notin \text{dom}(\eta)$, $S = \tilde{S}$, and $P = \tilde{P}$.

Following the proof of lemma 3.3 it is easy to see that $\lesssim$ is a simulation relation. Furthermore, $\mathbb{C}_0 \lesssim \tilde{\mathbb{C}}_0$. Applying lemma 2.1 to the trace $\mathbb{C}_0 \to^* \mathbb{C}$ we obtain a semantic configuration $\tilde{\mathbb{C}} = (\tilde{\eta}, \tilde{S}, \tilde{P}), \tilde{\mathcal{Q}}$ such that $\tilde{\mathbb{C}}_0 \to^* \tilde{\mathbb{C}}$ and $\mathbb{C} \lesssim \tilde{\mathbb{C}}$, in particular $(\eta, S, P) \lesssim (\tilde{\eta}, \tilde{S}, \tilde{P})$. Since $id \in \text{dom}(\eta)$ and $P$ is an output process that does not start with an output, by definition of $\lesssim$ in lemma 3.3 there exists a symbolic state $\mathbb{S}$ such that $\mathbb{S}$ is $\eta$-consistent, $S \sim_\eta \mathbb{S}$, and $(\mathbb{S}, P)$ successfully yields the model $\tilde{P}$. In particular, since $P$ is an output process and thus $P \neq 0$, there exists a symbolic state $\mathbb{S}'$ such that $(\mathbb{S}, P) \to (\mathbb{S}', P')$. It is straightforward to verify that the conditions checked by the symbolic execution imply that there is also a reduction of $(\eta, S, P)$ (and thus also of $\mathbb{C}$) in the concrete semantics. The only exception is the case when $P$ starts with an Assume statement, in which case the existence of a reduction follows from our assumption that $\mathbb{C}_0$ satisfies inline assumptions. ■

# Chapter 4

# Model Verification in Computational Setting

This chapter describes the path from an IML model extracted from a C program to a model that is ready for verification with CryptoVerif. We start by reviewing the calculus of CryptoVerif in section 4.1. Most important difference between CryptoVerif and IML is that bitstring-manipulating expressions are no longer available in CryptoVerif. The goal of our translation, described in section 4.2 will thus be to abstract these expressions away.

A crucial role in the translation is played by typechecking. One reason is that CryptoVerif uses typing information to derive security properties. For instance, if an encryption may reveal the length of a payload then the payload must be of a fixed length in order to establish secrecy. As another example, two random values may need to come from a large type in order to eliminate collisions between them. Typing is also important due to specifics of dealing with an implementation, as opposed to a model. We would like to supply CryptoVerif with facts about our parsing functions. For instance, if $\mathit{conc}(x, y) = \mathtt{0x01}|\tau_{4u}^{-1}(\mathrm{len}(x_1))|x_1|x_2$ and $\mathit{parse}(x) = x\{5, \tau_{4u}(x\{1, 4\})\}$ then we would like CryptoVerif to know that $\mathit{parse}(\mathit{conc}(x, y)) = x$. Unfortunately, such an equation does not hold for all values of $x$ and $y$—if $x$ is too long then the length stored by the implementation overflows—$\tau_{4u}^{-1}(\mathrm{len}(x))$ cannot be evaluated, and so $\mathit{conc}(x, y)$ evaluates to $\bot$, making the equation invalid. It will therefore be important to give $\mathit{conc}$ a type $T_x \times T_y \to T$ such that the fact $\mathit{parse}(\mathit{conc}(x, y)) = x$ is true for all $x \in I(T_x)$ and $y \in I(T_y)$. Once types for all such formatting functions have been established, we typecheck the whole process to make sure that a formatting function will always be supplied with arguments of the right type.

The translation that we develop transforms an IML process $Q$ to a CryptoVerif *model* $(\tilde{Q}, \Gamma, \Phi)$ with a typing environment $\Gamma$ and a set of facts $\Phi$. The main result of this chapter is theorem 4.3 (section 4.11) that captures the soundness of this translation: First, $Q \lesssim \tilde{Q}$, and so $Q$ is at least as secure as $\tilde{Q}$ against any trace property $\rho$. Second, if CryptoVerif verifies the model $(\tilde{Q}, \Gamma, \Phi)$ against a property $\rho$ then $\tilde{Q}$ is (asymptotically) secure against $\rho$.

| | |
|---|---|
| $\phi ::=$ | facts |
| $e = e'$ | bitstring equality |
| $e$ | bitstring fact |
| $e ::=$ | bitstring term |
| $x, i \in \mathit{Var}$ | variable |
| $f(e_1, \ldots, e_n),\ f \in \mathit{Ops}$ | function application |
| $Q \in \mathrm{CV} ::=$ | input process |
| $0$ | nil |
| $Q \vert Q'$ | parallel composition |
| $!^{i \leq N} Q$ | replication $N$ times |
| $\mathsf{in}(c[e_1, \ldots, e_n], x);\ P$ | input |
| $P ::=$ | output process |
| $\mathsf{out}(c[e_1, \ldots, e_n], e);\ Q$ | output |
| $\mathsf{new}\ x \colon T;\ P$ | random number |
| $\mathsf{let}\ x = e\ \mathsf{in}\ P$ | assignment |
| $\mathsf{if}\ \phi\ \mathsf{then}\ P\ [\mathsf{else}\ P']$ | conditional |
| $\mathsf{event}\ ev(e_1, \ldots, e_n);\ P$ | event |

Figure 4.1: The syntax of CryptoVerif processes.

An interesting challenge arises from the fact that CryptoVerif assumes for each function symbol $f$ a family of functions, $\tilde{I}_k(f)$, one for each security parameter $k \in \mathbb{N}$, but a C implementation $I(f)$ is written for a single value of the security parameter, an issue that is also discussed by Küsters et al. [2012]. We therefore need to define generalisations for our formatting functions to arbitrary security parameters in such a way that the set of facts that we supply to CryptoVerif remains true for the generalisations. This is described in section 4.10.

We start by reviewing the CryptoVerif calculus and semantics in section 4.1. We then present a high-level overview of the whole translation procedure in section 4.2. The remaining sections of this chapter fill in the details.

## 4.1 Review—CryptoVerif

This section reviews the input language of the CryptoVerif tool [Blanchet, 2008] that we use to verify our models. It is a subset of our IML calculus—the process syntax is exactly the same (except for the lack of inline assumptions), but expressions are restricted to variables and cryptographic function applications as well as bitstring equality tests, as shown in figure 4.1. In particular, CryptoVerif cannot reason directly about substring extraction and concatenation or about arithmetic facts, so we shall aim to replace those by sound abstractions.

Apart from the lack of reasoning about low-level primitives a big difference between CryptoVerif and IML is that CryptoVerif processes are typed. This is necessary because some properties of cryptographic functions only hold with respect to bitstring of bounded or fixed lengths, a property that can be established by typing. The interpretation $I$ defined in section 2.1 for function symbols will now be extended to types: for every type $T$ we consider its interpretation $I(T) \subseteq BS_\perp$. A *typing environment* $\Gamma$ assigns types $T$ to variables and *function types* $T_1 \times \ldots \times T_n \to T$ to function symbols. We shall assume that we are given a typing environment $\Gamma_0$ that contains types for all the cryptographic function symbols used by the process. In our implementation this typing environment is provided by the user in the CryptoVerif template file. We do not assume that $\Gamma_0$ contains types for any formatting symbols that we shall introduce in translation—those types will be inferred.

**Definition 4.1** An *interpretation* is a function $I$ that maps function symbols $f$ of arity $n$ to functions $I(f)\colon BS_\perp^n \to BS_\perp$, types $T$ to sets $I(T) \subseteq BS_\perp$, replication parameters $N$ to values $I(N) \in \mathbb{N}$ and for every channel $c$ maps the term $\mathrm{maxlen}(c)$ to a value $I(\mathrm{maxlen}(c)) \in \mathbb{N}$. □

In the rest of this chapter we shall make use of the following type-safety judgements.

**Definition 4.2 (Typing Relations)** For a function symbol $f$, types $T$, $T_1, \ldots, T_n$, an environment $\eta$, a typing environment $\Gamma$, and an IML process $Q$ let

- $I, f \models T_1 \times \ldots \times T_n \to T$ if $I(f)(b_1, \ldots, b_n) \in I(T)$ for all $b_1 \in I(T_1), \ldots, b_n \in I(T_n)$.
- $I, f \models \Gamma$ if $f \in \mathrm{dom}(\Gamma)$ and $I, f \models \Gamma(f)$.
- $I|_Q \models \Gamma$ if $I, f \models \Gamma$ for all $f$ used by $Q$.
- $\eta \models \Gamma$ if for all $x \in \mathrm{dom}(\eta)$ we have $x \in \mathrm{dom}(\Gamma)$ and $\eta(x) \in I(\Gamma(x))$.

A process $Q$ is *well-typed* with respect to a typing environment $\Gamma$ that contains only function symbols, if $\Gamma \vdash Q$ according to the rules in figure 4.2. □

The semantics of the calculus as described in Blanchet [2008] and implemented in the CryptoVerif tool is shown in figures 4.3 to 4.5. The semantics is parameterized by an interpretation $I$ and a typing environment $\Gamma$. The only difference from the IML semantics described in figures 2.3 and 2.4 is that the evaluation of expressions takes the types of functions into account: if a function is ever applied to an argument of a wrong type, the evaluation fails and the execution of the process stops. The judgement $\eta, e \Downarrow b$ means that the expression $e$ evaluates to the bitstring $b$ in the environment $\eta$. The rules of expression evaluation are shown in figure 4.3. The semantic rules for processes are shown in figures 4.4 and 4.5. They are the same as the rules for IML except that now we evaluate expressions using $\eta, e \Downarrow b$ instead of $[\![e]\!]_\eta$. We define an initial configuration $\mathrm{initConfig}(Q)$ and traces the same way as for IML, except that now the trace additionally depends on the typing environment.

$$\frac{\Gamma(x) = T}{\Gamma \vdash x\colon T} \tag{TVar}$$

$$\frac{\Gamma(f) = T_1 \times \ldots \times T_n \to T \quad \forall i \leq n\colon \Gamma \vdash e_i\colon T_i}{\Gamma \vdash f(e_1, \ldots, e_n)\colon T} \tag{TFun}$$

$$\frac{\Gamma \vdash e\colon T \quad \Gamma \vdash e'\colon T' \quad T = T'}{\Gamma \vdash e = e'\colon \texttt{bool}} \tag{TEq}$$

$$\Gamma \vdash 0 \tag{TNil}$$

$$\frac{\Gamma \vdash Q \quad \Gamma \vdash Q'}{\Gamma \vdash Q|Q'} \tag{TPar}$$

$$\frac{\Gamma\{i \mapsto \texttt{bitstring}\} \vdash Q}{\Gamma \vdash !^{i \leq n} Q} \tag{TRepl}$$

$$\frac{\forall i \leq n\colon \Gamma \vdash e_i\colon T_i \quad \Gamma\{x \mapsto \texttt{bitstring}\} \vdash P}{\Gamma \vdash \mathsf{in}(c[e_1, \ldots, e_n], x); P} \tag{TIn}$$

$$\frac{\forall i \leq n\colon \Gamma \vdash e_i\colon T_i \quad \Gamma \vdash e\colon T \quad \Gamma \vdash Q}{\Gamma \vdash \mathsf{out}(c[e_1, \ldots, e_n], e); Q} \tag{TOut}$$

$$\frac{T = \texttt{fixed}_n \quad \Gamma\{x \mapsto T\} \vdash P}{\Gamma \vdash \mathsf{new}\ x\colon T; P} \tag{TNew}$$

$$\frac{\Gamma \vdash e\colon T \quad \Gamma\{x \mapsto T\} \vdash P}{\Gamma \vdash \mathsf{let}\ x = e\ \mathsf{in}\ P} \tag{TLet}$$

$$\frac{\Gamma \vdash e\colon \texttt{bool} \quad \Gamma \vdash P \quad \Gamma \vdash P'}{\Gamma \vdash \mathsf{if}\ e\ \mathsf{then}\ P\ \mathsf{else}\ P'} \tag{TIf}$$

$$\frac{\forall i \leq n\colon \Gamma \vdash e_i\colon T_i \quad \Gamma \vdash P}{\Gamma \vdash \mathsf{event}\ ev(e_1, \ldots, e_n); P} \tag{TEvent}$$

Figure 4.2: Typing rules for CryptoVerif processes.

$$\frac{x \in \mathrm{dom}(\eta)}{\eta, x \Downarrow \eta(x)}$$

$$\frac{f \in \mathrm{dom}(\Gamma) \quad \Gamma(f) = T_1 \times \ldots \times T_n \to T \quad \forall i \leq n\colon \eta, e_i \Downarrow a_i \wedge a_i \in I(T_i)}{\eta, f(e_1, \ldots, e_n) \Downarrow I(f)(a_1, \ldots, a_n)}$$

$$\frac{\eta, e \Downarrow a \quad \eta, e' \Downarrow a'}{\eta, e = e' \Downarrow a = a'}$$

Figure 4.3: The evaluation of CryptoVerif expressions.

$$\{(\eta, 0)\} \uplus \mathcal{Q} \rightsquigarrow \mathcal{Q} \tag{CVNil}$$

$$\{(\eta, Q_1|Q_2)\} \uplus \mathcal{Q} \rightsquigarrow \{(\eta, Q_1), (\eta, Q_2)\} \uplus \mathcal{Q} \tag{CVPar}$$

$$\{(\eta, !^{i \leq N} Q)\} \uplus \mathcal{Q} \rightsquigarrow \{(\eta\{i \mapsto a\}, Q) \mid a \in [1, I(N)]\} \uplus \mathcal{Q} \tag{CVRepl}$$

reduce($\mathcal{Q}$) is the normal form of $\mathcal{Q}$ by $\rightsquigarrow$

Figure 4.4: The semantics of the CryptoVerif calculus for input processes.

$$\frac{T = \texttt{fixed}_n \text{ for some } n \in \mathbb{N} \quad a \in I(T)}{(\eta, \mathsf{new}\ x\colon T;\ P), \mathcal{Q} \to_{1/|I(T)|} (\eta\{x \mapsto a\}, P), \mathcal{Q}} \tag{CVNew}$$

$$\frac{\eta, e \Downarrow a}{(\eta, \mathsf{let}\ x = e\ \mathsf{in}\ P), \mathcal{Q} \to_1 (\eta\{x \mapsto a\}, P), \mathcal{Q}} \tag{CVLet}$$

$$\frac{\eta, e \Downarrow \mathrm{true}}{(\eta, \mathsf{if}\ e\ \mathsf{then}\ P\ \mathsf{else}\ P'), \mathcal{Q} \to_1 (\eta, P), \mathcal{Q}} \tag{CVIfTrue}$$

$$\frac{\eta, e \Downarrow \mathrm{false}}{(\eta, \mathsf{if}\ e\ \mathsf{then}\ P\ \mathsf{else}\ P'), \mathcal{Q} \to_1 (\eta, P'), \mathcal{Q}} \tag{CVIfFalse}$$

$$\frac{\begin{array}{l}\eta, e \Downarrow a \in BS \quad a' = a\{0, I(\mathrm{maxlen}(c))\} \quad \forall i \leq n\colon \eta, e_i \Downarrow a_i \quad \mathcal{Q}' = \mathrm{reduce}(\{(\eta, Q)\}) \\ \exists!(\eta', Q') \in \mathcal{Q}\colon Q' = \mathsf{in}(c[e'_1, \ldots, e'_n], x'); P' \wedge \forall i \leq n\colon \eta', e'_i \Downarrow a_i\end{array}}{(\eta, \mathsf{out}(c[e_1, \ldots, e_n], e);\ Q), \mathcal{Q} \to_1 (\eta'\{x' \mapsto a'\}, P'), \mathcal{Q} \uplus \mathcal{Q}' \setminus \{(\eta', Q')\}} \tag{CVOut}$$

$$\frac{\eta, e_i \Downarrow a_i \text{ for all } i \leq n}{(\eta, \mathsf{event}\ ev(e_1, \ldots, e_n);\ P), \mathcal{Q} \xrightarrow{ev(a_1, \ldots, a_n)}_1 (\eta, P), \mathcal{Q}} \tag{CVEvent}$$

Figure 4.5: The semantics of the CryptoVerif calculus for output processes.

In the CryptoVerif security definition we shall make it explicit that the execution depends on the interpretation $I$ (definition 4.1) that provides implementations of the function symbols, interpretations of types, values of the replication parameters and the maximum lengths of messages that can be sent on a channel. Making the interpretations explicit will allow us to vary them depending on the security parameter.

**Definition 4.3 (CryptoVerif Security)** For an input CryptoVerif process $Q$, a trace property $\rho$, a typing environment $\Gamma$, and an interpretation $I$ let

$$\mathrm{cvinsec}(Q, I, \Gamma, \rho) = \sum_{\mathcal{T} \in \mathbb{T}} \mathrm{pr}(\mathcal{T}),$$

where $\mathbb{T}$ is the set of CryptoVerif traces $\mathcal{T}$ with respect to $\Gamma$ and $I$ such that $\mathrm{fst}(\mathcal{T}) = \mathrm{initConfig}(Q)$, $\mathrm{events}(\mathcal{T}) \notin \rho$, and $\mathcal{E} \in \rho$ for any proper prefix $\mathcal{E}$ of $\mathrm{events}(\mathcal{T})$. □

CryptoVerif checks *correspondence properties* of the form "If an event has been executed then a certain other event must have been executed before". We refer the reader to Blanchet [2008] for the exact definition. In our presentation we only rely on the fact that correspondence properties are trace properties, and so are accounted for in our treatment. CryptoVerif inputs a *model* of the form $(Q, \Gamma, \Phi)$ where $\Gamma$ is a typing environment, $Q$ is a well-formed process (definition 2.4) that is well-typed with respect to $\Gamma$, and $\Phi$ is a set of facts. In addition to facts defined in figure 4.1 CryptoVerif allows facts containing quantifiers and logical connectives, with obvious evaluation rules. Given a set of facts $\Phi$ we write $I \models \Phi$ if $\emptyset, \phi \Downarrow$ true for each $\phi \in \Phi$.

CryptoVerif makes a statement about the process $Q_A|Q$ where $Q_A$ is an arbitrary *attacker process* that must be well-formed and well-typed and is not allowed to contain events. CryptoVerif constructs a series of processes $Q_1, \ldots, Q_n$ such that $Q_1 = Q$ and for each $i$ the processes $Q_i$ and $Q_{i+1}$ are indistinguishable in any context. In particular, for any attacker $Q_A$ the processes $Q_A|Q_i$ and $Q_A|Q_{i+1}$ produce a trace belonging to $\rho$ with almost the same probability. Verification succeeds if the final process $Q_n$ trivially satisfies the security property. For instance, $Q_n$ preserves the secrecy of a particular bitstring $x$ if $x$ does not occur in $Q_n$ at all. We can then conclude that $Q$ also satisfies the property with overwhelming probability.

In order to formally define what "almost the same" and "overwhelming" probabilities mean CryptoVerif considers a family of interpretations $(\tilde{I}_k)_{k \in \mathbb{N}}$. A successful verification result implies that the probability of breaking the security property is negligible in $k$. In order for the semantics to be polynomial time executable the family of interpretations must be efficiently computable as defined below.

**Definition 4.4 (Efficient Interpretations)** A family of interpretations $(\tilde{I}_k)_{k \in \mathbb{N}}$ is called *efficient* if the following conditions are satisfied:

- For each function symbol $f$ the function $\tilde{I}_k(f)$ is computable in time polynomial in $k$ and the length of the inputs.

- For each type $T$ the set $\tilde{I}_k(T)$ is recognizable in time polynomial in $k$.
- For each replication parameter $N$ and channel $c$, both $\tilde{I}_k(N)$ and $\tilde{I}_k(\text{maxlen}(c))$ are polynomially bounded and efficiently computable functions of $k$. $\square$

In the input template the user can mark the cryptographic functions as satisfying certain standard properties, say, IND-CPA (indistinguishability with respect to chosen-plaintext attack) for encryption functions or UF-CMA (unforgeability against chosen-message attack) for message authentication codes. These properties enable CryptoVerif to perform certain kinds of transformations. For instance, if *enc* is an IND-CPA encryption function, *keygen* is the corresponding key generation function, and $Z$ a function that replaces every bit of the argument with 0, then an expression of the form $enc(x, key, seed)$ can be rewritten to $enc(Z(x), key, seed)$ provided *seed* is a randomly generated encryption seed that is not used anywhere else and *key* is a key generated using *keygen* and not used anywhere else.

We let $\tilde{I}_k^c$ for $k \in \mathbb{N}$ denote the family of interpretations of cryptographic functions used by our implementation together with interpretations of all types in $\Gamma_0$. We shall assume that these interpretations are sound with respect to the typing environment $\Gamma_0$ provided by the user, that is, $\tilde{I}_k^c, f \models \Gamma_0$ for each cryptographic function $f \in \text{dom}(\tilde{I}_k^c)$ and $k \in \mathbb{N}$.

The following theorem summarizes the requirements that CryptoVerif places on its input and the guarantee that it provides.

**Theorem 4.1 (Correctness of CryptoVerif [Blanchet, 2008])** *Let $(Q, \Gamma, \Phi)$ be a CryptoVerif model with a typing environment $\Gamma$, a set of facts $\Phi$ and a well-formed input process $Q \in \text{CV}$ that is well-typed with respect to $\Gamma$. Let $(\tilde{I}_k)_{k\in\mathbb{N}}$ be an efficient family of interpretations such that for each $k \in \mathbb{N}$ the interpretation $\tilde{I}_k$ agrees with $\tilde{I}_k^c$ for all cryptographic functions, $\tilde{I}_k|_Q \models \Gamma$ and $\tilde{I}_k \models \Phi$.*

*If CryptoVerif verifies the model $(Q, \Gamma, \Phi)$ with respect to a correspondence property $\rho$ then for any well-formed process $Q_A$ that is well-typed with respect to $\Gamma$ and does not contain events the function $\text{cvinsec}(Q_A|Q, \tilde{I}_k, \Gamma, \rho)$ is negligible in $k$.* $\square$

We can reformulate correctness of CryptoVerif in such a way that the verification result does not depend on the shape of the non-cryptographic functions used by the process. The idea is to take the "least secure" implementation of non-cryptographic functions for each security parameter.

**Definition 4.5** Given a CryptoVerif model $(Q, \Gamma, \Phi)$, a correspondence property $\rho$, a polyno-mial $p$, and a security parameter $k \in \mathbb{N}$ let

$$\begin{aligned} \mathrm{cvbound}_p(Q, \Gamma, \Phi, \rho, k) = \sup\{ \ & \mathrm{cvinsec}(Q, I, \Gamma, \rho) \mid \\ & I \text{ agrees with } \tilde{I}_k^c \text{ for cryptographic functions}, \\ & I|_Q \models \Gamma, \ I \models \Phi, \\ & \mathrm{runtime}(I(f)(b_1, \ldots, b_n)) \leq p(b_1 + \ldots + b_n + k) \\ & \quad \text{for all } f \text{ in } Q \text{ and } b_1, \ldots, b_n \in \mathit{BS} \ \}, \end{aligned}$$

where $\mathrm{runtime}(I(f)(b_1, \ldots, b_n))$ for a function $f$ of arity $n$ denotes the time it takes to compute $I(f)(b_1, \ldots, b_n)$. □

Using the above definition we can restate theorem 4.1 as follows.

**Lemma 4.1** *Let $(Q, \Gamma, \Phi)$ be a CryptoVerif model with a typing environment $\Gamma$, a set of facts $\Phi$ and a well-formed input process $Q \in \mathrm{CV}$ that is well-typed with respect to $\Gamma$.*

*If CryptoVerif verifies the model $(Q, \Gamma, \Phi)$ with respect to a correspondence property $\rho$ then for any well-formed process $Q_A$ that is well-typed with respect to $\Gamma$ and does not contain events and for any polynomial $p$ the function $\mathrm{cvbound}_p(Q_A|Q, \Gamma, \Phi, \rho, k)$ is negligible in $k$.* □

Proof Assume that the conditions of the lemma are satisfied but $\mathrm{cvbound}_p(Q_A|Q, \Gamma, \Phi, \rho, k)$ is not negligible in $k$. For each $k \in \mathbb{N}$ choose an interpretation $\tilde{I}_k$ provided by definition 4.5 such that

$$\mathrm{cvinsec}(Q_A|Q, \tilde{I}_k, \Gamma, \rho) \geq a * \mathrm{cvbound}_p(Q_A|Q, \Gamma, \Phi, \rho, k)$$

for an arbitrary constant $a \in (0, 1)$. The family of interpretations constructed in this way satisfies the conditions of theorem 4.1 (in particular, it is efficient since the runtime of all the functions is bounded by $p$), but the function $\mathrm{cvinsec}(Q_A|Q, \tilde{I}_k, \Gamma, \rho)$ is not negligible in $k$, yielding a contradiction. ■

Lemma 4.1 sets the goal of this chapter: to verify an IML process $Q$ obtained by symbolic execution of a C program we must translate it to a model $(\tilde{Q}, \Gamma, \Phi)$ that satisfies the conditions of the lemma. Since C programs are only written for a single value of the security parameter $k_0$, we first obtain a model that satisfies the conditions for $k_0$ only: the model will be evaluated with respect to a single interpretation $I$ obtained by combining the cryptographic functions in $\tilde{I}_{k_0}^c$ with new functions for constructing and parsing messages that we add to the process. In section 4.10 we show how we can generalize $I$ to a family of interpretations $(\tilde{I}_k)_{k \in \mathbb{N}}$ to prove that $\mathrm{cvbound}_p(Q, \Gamma, \Phi, \rho, k)$ is well-defined for all $k \in \mathbb{N}$ with a sufficiently large polynomial $p$.

The CryptoVerif implementation goes further than the asymptotic statement in lemma 4.1 and actually provides a concrete estimate of the value $\mathrm{cvbound}_p(Q_A|Q, \Gamma, \Phi, \rho, k)$ in terms of the security parameter, the runtime of the attacker, and the time it takes to break the cryptographic functions. This would remove the need for a generalization step, but alas this feature of CryptoVerif has not been formalised.

We made some changes with respect to the original version of the calculus in Blanchet [2008]. These changes are as follows:

- In CryptoVerif bitstring equality is implemented using a family of functions $=_T : T \times T \rightarrow$ `bool`. In our presentation we lift bitstring equality to a separate syntactic form. The semantics of both versions are obviously equivalent—our evaluation rule for bitstring equality can be viewed as an instance of function application with a function $=_T$.

- The semantics in [Blanchet, 2008] uses a single global environment $\eta$ where each variable is an array, and each process uses a single cell to store its local value. This allows a process to read the variables of another process by using the special construct `find`. As we don't use `find` in our modelling, we simplify the semantics, and bring it closer to the IML semantics by giving each process a local environment.

- Our version of the rule (CVOut) in figure 4.5 requires that there is only one matching recipient of a message. The original version does not have this restriction. This is fine since we are dealing with trace properties and every trace in the restricted semantics is also a trace in the original semantics.

- CryptoVerif definitions use "proper" bitstrings—sequences of 0 and 1, in contrast to sequences of bytes used in this dissertation. The input statement in CryptoVerif allows to specify the expected type of the message, so that a participant can refuse to accept messages that do not contain an integral number of bytes.

## 4.2 From IML to CryptoVerif: Summary

This section describes how we generate a CryptoVerif model $(\tilde{Q}, \Gamma, \Phi)$ from an IML model $Q$. We aim to provide motivation and a complete high-level overview of the method, while the remaining sections of this chapter provide the details.

Since CryptoVerif cannot reason about string manipulation and arithmetic expressions, we shall aim to replace these expressions by higher-level abstractions. The crucial observation is that these expressions are used to implement tupling and projection operations, and so we replace them by application of fresh function symbols for tupling and projection.

**Definition 4.6 (Encoders and Parsers)** We call an expression $e$ with variables $x_1, \dots, x_n$ an *encoding expression* (or simply *encoder*) when $e = e_1 | \dots | e_m$ and each $e_i$ is either a constant, a variable $x_j$ for some $j \leq n$, or has the form $\tau^{-1}(\mathrm{len}(x_j))$ for some integer type $\tau \in \mathbb{T}_I$ and $j \leq n$. We call an expression $e$ with a single variable $x$ a *parsing expression* (or simply *parser*) when $e = x\{t_o, t_l\}$ and $t_o$ and $t_l$ are arithmetic expressions that can contain integer constants,

len$(x)$, or expressions of the form $\tau(e')$ where $e'$ is itself a parsing expression with the variable $x$ and $\tau \in \mathbb{T}_I$. Encoders and parsers are collectively called *formatting functions*. □

An example of an encoder is $conc(x, y) = \tau_{4u}^{-1}(\text{len}(x))|\texttt{0x01}|x|y$ and an example of a parser is $parse(x) = x\{5, \tau_{4u}(x\{0, 4\})\}$. The definition does not consider an expression of the form $\tau_{4u}^{-1}(\text{len}(x|y))|\texttt{0x01}|x|y$ to be an encoder, instead such an expression can be viewed as a nested application of two encoders: $x|y$ and $\tau_{4u}^{-1}(\text{len}(x))|\texttt{0x01}|x$. Message formats in cryptographic protocols like SSH or TLS use length fields to delimit parts of the message, so that these formats are included in our definition. Protocols that use field separators in their messages (like the newline character used to separate the fields in the HTTP protocol) are currently not covered by our implementation.

Encoders and parsers are the right level of abstraction because they capture the intent of the protocol designer and as such possess useful properties that are crucial for the verification of the protocol. For instance, given appropriate types $T_x$ and $T_y$ it is true that $parse(conc(x, y)) = x$ for each $x \in I(T_x)$ and $y \in I(T_y)$ and the functions *parse* and *conc* defined as above. No such property can be formulated for the naked string concatenation operator $|$ since it is not injective. As another example, if $conc'(x, y) = \tau_{4u}^{-1}(\text{len}(x))|\texttt{0x02}|x|y$ then *conc* and *conc'* have disjoint ranges, which is often crucial for proving that there are no injection attacks. Facts like these make up the set $\Phi$ that is part of the CryptoVerif model we generate.

Appropriate choice of the typing environment $\Gamma$ will allow CryptoVerif to prove that certain bitstring have types of bounded or fixed length. One example where this is important is the use of encryption that does not hide the length of the plaintext. Secrecy under such encryption can only be proved if the plaintext is known to be of fixed length. Typing is also important to make sure that formatting functions are only ever called with values of appropriate types, such as $T_x$ and $T_y$ above.

Being well-typed is a necessary condition for the soundness of our extracted model. The only difference between the CryptoVerif and the IML semantics is that execution in CryptoVerif stops whenever inputs to a function are not of the right type. In lemma 4.9 we show that this never happens for well-typed processes, and so the notions of IML and CryptoVerif security coincide. Together with the fact that the extracted CryptoVerif model simulates the IML model ($Q \lesssim \tilde{Q}$) this establishes the soundness of our translation procedure (theorem 4.3) which is the main result of this chapter.

**Example 4.1** To describe our transformation steps, we use the RPC-enc protocol (section 1.1) as a running example: figure 4.6 shows the initial IML model, while figure 4.7 shows the resulting CryptoVerif model (some of the symbolic execution that generates the IML model is shown in section 3.6). Due to space constraints we only show the most important parts of the typing environment $\Gamma$ and the fact set $\Phi$ and also omit some condition checks in the CryptoVerif process $\tilde{Q}$. Appendix C shows the complete CryptoVerif model without omissions.

The extracted IML process contains free variables which are bound in the user-provided environment, as shown in detail in appendix C. The intended meaning of the variables is as

```
  let client =
    if clientID = xClient then
    let key1 = lookup(clientID, serverID, db) in
    new kS_seed1: fixed_16;
    let key2 = kgen(kS_seed1) in
    event client_begin(clientID, serverID, request);
    let msg1 = "p"|τ4u^-1(len(request))|request|key2 in          | ⇝ let msg1 = conc1(request, key2) in
    new nonce1: fixed_16;
    let cipher1 = E(msg1, key1, nonce1) in
    let msg3 = "p"|τ4u^-1(len(clientID))|clientID|cipher1 in     | ⇝ let msg3 = conc2(clientID, cipher1) in
    out(c_out, msg3);
    in(c_in, msg4);
    assume len(msg4) = (4);
▷   if τ4u(msg4) = 1056 then                                     | ⇝ if cond1(msg4) then
    in(c_in, cipher2);
    assume len(cipher2) = τ4u(msg4);
    let msg6 = D(cipher2, key2) in
    let injbot(msg7) = msg6 in
▷   if len(msg7) = 1024 then                                     | ⇝ if cond2(msg7) then
    event client_accept(clientID, serverID, request, msg7);

  let server =
    in(c_in, msg8);
    assume len(msg8) = 4;
▷   if (τ4u(msg8) ≥ 1082) ∧ (τ4u(msg8) ≤ 2106) then                | ⇝ if cond3(msg8) then
    in(c_in, msg9);
    assume len(msg9) = τ4u(msg8);
▷   if "p" = msg9{0, 1} then                                      | ⇝ if cond4(msg9) then
▷   if τ4u(msg9{1, 4}) ≤ 1024 then                                | ⇝ if cond5(msg9) then
▷   if τ4u(msg9{1, 4}) = len(xClient) then                        | ⇝ if cond6(msg9, xClient) then
▷   if τ4u(msg8) − (5 + τ4u(msg9{1, 4})) ≤ 1077 then              | ⇝ if cond7(msg8, msg9) then
    let client2 = msg9{5, τ4u(msg9{1, 4})} in                     | ⇝ let client2 = parse1(msg9) in
    if client2 = xClient then
    let key3 = lookup(client2, serverID, db) in
    let cipher3 = msg9{5 + τ4u(msg9{1, 4}),                       | ⇝ let cipher3 = parse2(msg9) in
                       len(msg9) − (5 + τ4u(msg9{1, 4}))} in
    let msg10 = D(cipher3, key3) in
    let injbot(msg11) = msg10 in
▷   if len(msg11) ≤ (τ4u(msg8) − (5 + τ4u(msg9{1, 4}))) − 32 then | ⇝ if cond8(msg8, msg9) then
▷   if len(msg11) ≥ 5 then                                        | ⇝ if cond9(msg11) then
▷   if τ4u(msg11{1, 4}) = 1024 then                               | ⇝ if cond10(msg11) then
▷   if len(msg11) > 5 + τ4u(msg11{1, 4}) then                     | ⇝ if cond11(msg11) then
▷   if "p" = msg11{0, 1} then                                     | ⇝ if cond12(msg11) then
▷   if len(response) = 1024 then                                  | ⇝ if cond13(response) then
▷   if len(msg11)) = (5 + τ4u(msg11{1, 4})) = 16 then             | ⇝ if cond14(msg11) then
    let var10 = msg11{5, τ4u(msg11{1, 4}))} in                    | ⇝ let var10 = parse3(msg11) in
    event server_reply(client2, serverID, var10, response);
    let key4 = msg11{5 + τ4u(msg11{1, 4}),                        | ⇝ let key4 = parse4(msg11) in
                     len(msg11) − (5 + τ4u(msg11{1, 4}))} in
    new nonce2: fixed_16;
    let msg14 = E(response, key4, nonce2) in
    out(c_out, msg14);
```

Figure 4.6: The IML model of RPC-enc extracted from the C code. Formatting and auxiliary test substitutions are shown on the right. Auxiliary tests are marked with ▷.

```
(* Types from the template Γ0 *)
clientID: bounded1024
E: bounded1045 × fixed16 × fixed16 → bounded1077
...

(* Inferred types Γ *)
conc1: fixed1024 × fixed16 → bounded1045
conc2: bounded1024 × bounded1077 → bitstring
...

(* Inferred facts Φ *)
forall x1 : fixed1024, x2 : fixed16;
    Z_bounded1045(conc1(x1, x2)) = Z_bounded1045(conc1'(Z_fixed1024(x1), Z_fixed16(x2))).
forall x1 : fixed1024, x2 : fixed16;
    cond14(conc1(x1, x2)) = cond14'(conc1(Z_fixed1024(x1), Z_fixed16(x2))).
forall x : fixed16; Z_fixed16(x) = zero_fixed16.
forall x : fixed1024; Z_fixed1024(x) = zero_fixed16.
...

(* The model Q̃ *)
let client =
  in(c_in, ());
  if clientID = xClient then
  let key1 = lookup(clientID, serverID, db) in
  new kS_seed1: fixed16;
  let key2 = kgen(kS_seed1) in
  event client_begin(clientID, serverID, request);
  let msg1 = conc1(request, key2) in
  new nonce1: fixed16;
  let cipher1 = E(msg1, key1, nonce1) in
  let msg3 = conc2(clientID, cipher1) in
  out(c_out, msg3);
  in(c_in, (msg4: bitstring, cipher2: bitstring));
  let msg6 = D(tcast_bitstring→bounded1077(cipher2), key2) in
  let injbot(msg7) = msg6 in
  if cond2(msg7) then
  event client_accept(clientID, serverID, request, tcast_bounded1045→fixed1024(msg7));
  yield .

let server =
  in(c_in, (msg8: bitstring, msg9: bitstring));
  let conc2(client2, cipher3) = msg9 in
  if client2 = xClient then
  let key3 = lookup(client2, serverID, db) in
  let msg10 = D(cipher3, key3) in
  let injbot(msg11) = msg10 in
  if cond14(msg11) then
  let conc1(var10, key4) = msg11 in
  event server_reply(client2, serverID, var10, response);
  new nonce2: fixed16;
  let msg14 = E(tcast_fixed1024→bounded1045(response), key4, nonce2) in
  out(c_out, msg14); 0 .
```

Figure 4.7: The CryptoVerif model of RPC-enc. Some auxiliary tests are omitted.

follows: *clientID* and *serverID* are global constants containing the names of an honest client and an honest server, *xClient* is the attacker-chosen name of the client that the server should communicate with, *request* and *response* are randomly chosen bitstrings, and *db* is the key database used to look up shared keys (some of which may be compromised). The client request message is sent in line 11 and received in line 26. The server response message is sent in line 52 and received in line 15.

We use uint32_t for all integer variables in the C code. Consistent use of an unsigned type makes sure that there are no casts from signed to unsigned and back—all the operations in the example are unsigned. Conditions checked by the C code are simplified by the symbolic execution, converting them from bitstring facts to integer facts where possible. For instance, in line 38 the condition checked by the C code is

$$\textbf{if } \mathrm{msg8} \ominus_{\tau_{4u}} (\tau_{4u}^{-1}(5) \oplus_{\tau_{4u}} \mathrm{msg9}\{1, 4\}) \leq_{\tau_{4u}} \tau_{4u}^{-1}(1077) \textbf{ then}$$

During symbolic execution the rule (STest) (chapter 3) uses the function simplify (section 3.3) to turn this into

$$\textbf{if } \tau_{4u}(\mathrm{msg8}) - (5 + \tau_{4u}(\mathrm{msg9}\{1, 4\})) \leq 1077 \textbf{ then}$$

Lines 18 and 38 contain pattern matching of the form $\mathsf{let}\ \mathrm{injbot}(\mathrm{msg7}) = \mathrm{msg6}\ \mathsf{in}\ P$. This is CryptoVerif abbreviation for $\mathsf{let}\ \mathrm{msg7} = \mathrm{injbot}^{-1}(\mathrm{msg6})\ \mathsf{in}\ \mathsf{if}\ \mathrm{injbot}(\mathrm{msg7}) = \mathrm{msg6}\ \mathsf{then}\ P$. In line 18 msg6 is a result of a decryption and injbot is the natural injection from the type $T$ of plaintexts into `bitstringbot`. The function $\mathrm{injbot}^{-1}\colon \mathtt{bitstringbot} \to T$ is the identity function for arguments of type $T$, and returns an arbitrary value otherwise. Therefore the construct $\mathsf{let}\ \mathrm{injbot}(\mathrm{msg7}) = \mathrm{msg6}\ \mathsf{in}\ P$ checks that the result of the decryption is of the correct type. In particular, the result is not $\perp$, which means that the encryption succeeded.

Part of the model that is not shown in this example is the sending of msg2 by the client that contains the length of msg3:

$$\begin{aligned}&\textbf{let } \mathrm{msg2} = \tau_{4u}^{-1}(5) \oplus_{\tau_{4u}} \tau_{4u}^{-1}(\mathrm{len}(\mathrm{clientID})) \oplus_{\tau_{4u}} \tau_{4u}^{-1}(\mathrm{len}(\mathrm{cipher1})) \ \ \textbf{in}\\ &\textbf{out}(\mathrm{c}, \mathrm{msg2});\end{aligned}$$

In our implementation we change all such arithmetic expressions to opaque expressions of the form *arithmetic*(*clientID*, *cipher1*) of type `bitstring`.

Compared to the example in section 3.6 the model here does not make abstract the function encrypt_len that computes the length of the ciphertext based on the length of the plaintext. Instead, to make the model more concise, we symbolically execute the actual code of the function and capture what it does concretely, namely add 32 (the encryption overhead) to the length of the plaintext. This explains why the type of the ciphertext is $\mathtt{bounded}_{1077}$ given that the type of the plaintext is $\mathtt{bounded}_{1045}$.

The example omits dummy inputs and outputs that are necessary to make inputs and outputs alternate as well as conditionals originating from null pointer checks. The full model in appendix C contains all these details □

The conditions checked by the implementation are of two kinds that need to be treated differently. Arithmetic conditions like the one in line 30 are unlikely to be useful for CryptoVerif, but will be necessary during typechecking to prove that formatting functions are applied correctly. The expressions in these conditions can be replaced by applications of opaque functions in the final CryptoVerif model. On the other hand, bitstring comparisons such as the check of the client identity by the server in line 33 are likely to be crucial for CryptoVerif verification and thus need to be preserved in the model. We use the following heuristic to distinguish the two kinds.

**Definition 4.7 (Cryptographic and Auxiliary Facts)** A *cryptographic fact* is a fact of the form $f(e_1, \ldots, e_n)$ where $f$ is a cryptographic function or of the form $e = e'$ where both $e$ and $e'$ are bitstring terms of one of the following kinds: a variable, a cryptographic function application, a substring or a concatenation expression. A fact that is not cryptographic is called *auxiliary.*□

The heuristic is formulated in such a way that the checking of a message tag as in line 28 is considered an auxiliary test since it involves a literal bitstring on one side. Figure 4.6 marks auxiliary tests in the process with $\triangleright$. In the CryptoVerif model all auxiliary facts are replaced by opaque function applications of the form $cond_i(x_1, \ldots, x_n)$ with variables $x_1, \ldots, x_n$ and $i \in \mathbb{N}$. As we shall see below, in order for the verification to succeed it may be necessary to prove additional properties of these functions. For instance, in order to prove secrecy of some of the $x_i$ it is usually necessary to make sure that the condition only depends on the length of the $x_i$, but not on their actual contents.

**Formatting Abstraction** The first step in our transformation pipeline is bringing the process into *formatting-normal form* in which every parsing and encoding expression occurs in its own let-binding, except in auxiliary statements which are left unchanged. This step is described in detail in section 4.3. Our example in figure 4.6 is already in formatting-normal form. The main purpose of this form is to simplify the presentation (and the implementation) of the other transformations. Now that expressions of different kinds occur separately, it is straightforward to abstract away applications of formatting functions and auxiliary statements, as show on the right of figure 4.6. This rewriting step is formalized in section 4.4.

**Type Inference** Next we need to decide upon appropriate types for the new function symbols. It turns out that we can infer most of the types automatically since we already have types for cryptographic functions from the user-provided template. Consider an application of a concatenation function of a form $\mathsf{let}\ x = f(x_1, \ldots, x_n)\ \mathsf{in}\ P$. If $x_1, \ldots, x_n$ are results of cryptographic functions with result types $T_1, \ldots, T_n$ and $x$ is used as an argument of a cryptographic function with argument type $T$ then $f$ can be given type $f\colon T_1 \times \ldots \times T_n \to T$. If $x$ is an argument of another formatting function, we cannot decide what type it should have. In these cases the user can provide an explicit type annotation in C code by inserting a special CVM instruction, but so far we did not have to use any type annotations for our models. Section 4.5 describes in detail how we infer the typing environment $\Gamma$ that we submit to CryptoVerif as part of the model. The soundness of the method does not depend on this inference step since the process is typechecked separately during the next step.

In addition to the types `bitstring`, `bitstringbot`, and `bool` that are predefined in CryptoVerif we use types $\texttt{fixed}_n$ and $\texttt{bounded}_n$ of bitstrings with length exactly or at most $n$. It is worth considering the reasons for this particular choice—why don't we use weaker types (say, only `bitstring`) or stronger types (say, bitstrings described by arbitrary logical formulas or concatenation expressions in the spirit of refinement types of Bengtson et al. [2008])? First, the types should be strong enough to prove that certain bitstrings are of bounded or fixed length. For instance, to prove the secrecy of the request and the response we will need to use the fact that they both have fixed length since our encryption does not hide the length of the message. On the other hand, more sophisticated types would not be preserved by application of cryptographic functions. Additionally we need to choose types that are easy to generalise to arbitrary security parameters so that we can appeal to the asymptotic security result for models verified with CryptoVerif.

One problem with our choice of types is that they are too weak to give a sound type for parsers since parsers rely on the internal structure of the input. We deal with this problem separately in section 4.9 where we replace the parsers by strengthened versions that always return a value from the return type instead of failing. We verify that the conditions checked by the implementation are sufficient for such a replacement to be sound. We also argue that all our facts $\Phi$ still hold with respect to the strengthened parsers.

**Typechecking** During the next step we typecheck the model against the typing environment $\Gamma$ that we just inferred. In order to ensure that the process is well-typed it may be necessary to insert type casts such as the cast of *msg7* at the end of the client model in figure 4.7. Unfortunately, it is not always possible to eliminate explicit typecasts by choosing the right types for our functions since membership of a certain type is typically checked explicitly in the code. For instance, we cannot give the decryption function $D$ the result type $\texttt{fixed}_{1024}$ since the decryption function does not check the length of its result. The length of *msg7* is only checked in line 19 of the IML model (*cond2* in the CryptoVerif model), and so the typecast would not be sound above that line. In section 4.6 we describe a transformation $\Gamma, Q \rightsquigarrow \tilde{Q}$ that adds typecasts into the model in such a way that $Q \lesssim \tilde{Q}$ and $\tilde{Q}$ is well-typed with respect to $\Gamma$.

**Fact Inference** The final ingredient of our model is the fact set $\Phi$. Equation facts are interpreted by CryptoVerif as rewriting rules—whenever an expressions matches the left hand side, it is rewritten according to the equation. An inequality fact gives rise to the rewriting rule that allows to rewrite an equality check to false. Section 4.7 provides details of how the set of facts is computed. In summary, we add facts of the following kinds:

- Parsing equations of the form $\forall(x_1 : T_1, \ldots, x_n : T_n) \colon f_p(f_c(x_1, \ldots, x_n)) = x_i$.

- Injectivity equations of the form

$$\forall(x_1 : T_1, \dots, x_n : T_n, y_1 : T_1, \dots, y_n : T_n) : f_c(x_1, \dots, x_n) = f_c(y_1, \dots, y_n) \Rightarrow \bigwedge_{i \le n} x_i = y_i.$$

- Disjointness facts of the form

$$\forall(x_1 : T_1, \dots, x_n : T_n, y_1 : T'_1, \dots, y_m : T'_m) : f_c(x_1, \dots, x_n) \neq f'_c(y_1, \dots, y_m).$$

- Equations that allow to erase arguments of concatenation expressions. For instance, when CryptoVerif applies the security definition of the symmetric encryption to our model, it replaces the plaintext $\mathit{msg1} = \mathit{conc1}(\mathit{request}, \mathit{key2})$ by $Z_{\mathtt{bounded}_{1045}}(\mathit{msg1})$, where $Z_{\mathtt{bounded}_{1045}}$ is the function that sets all bits of the input to 0. We provide CryptoVerif with equations that allow to push zero functions further down the expression and to prove that $Z_{\mathtt{bounded}_{1045}}(\mathit{msg1})$ does not depend on the values of *request* and *key2*, and so those remain secret:

$$\begin{aligned}
&\forall(x_1 : \mathtt{fixed}_{1024}, x_2 : \mathtt{fixed}_{16}) : \\
&\quad Z_{\mathtt{bounded}_{1045}}(\mathit{conc1}(x_1, x_2)) = Z_{\mathtt{bounded}_{1045}}(\mathit{conc1}'(Z_{\mathtt{fixed}_{1024}}(x_1), Z_{\mathtt{fixed}_{16}}(x_2))). \\
&\forall(x : \mathtt{fixed}_{16}) : Z_{\mathtt{fixed}_{16}}(x) = 0_{\mathtt{fixed}_{16}}. \\
&\forall(x : \mathtt{fixed}_{1024}) : Z_{\mathtt{fixed}_{1024}}(x) = 0_{\mathtt{fixed}_{1024}}.
\end{aligned}$$

  We use $\mathit{conc1}'$ on the right hand side of the equation to prevent CryptoVerif from infinitely looping applying the same rule over and over again.

- Equations that allow to erase arguments of conditional expressions. For instance, we can prove that *cond14* when applied to $\mathit{conc1}(x_1, x_2)$ depends only on the lengths of $x_1$ and $x_2$, and so we can replace $x_1$ and $x_2$ by zero bitstrings of corresponding lengths. This is captured using the following fact:

$$\begin{aligned}
&\forall(x_1 : \mathtt{fixed}_{1024}, x_2 : \mathtt{fixed}_{16}) : \\
&\quad \mathit{cond14}(\mathit{conc1}(x_1, x_2)) = \mathit{cond14}'(\mathit{conc1}(Z_{\mathtt{fixed}_{1024}}(x_1), Z_{\mathtt{fixed}_{16}}(x_2))).
\end{aligned}$$

  Using the fact that our encryption is authenticated (IND-CTXT) CryptoVerif is able to rewrite the argument *msg11* of *cond14* to $\mathit{conc1}(\mathit{var10}, \mathit{key4})$. Together with the rewriting rules above this allows us to prove that $\mathit{cond14}(\mathit{msg11})$ does not depend on values of *var10* and *key4* and thus does not interfere with their secrecy.

**Parsing Safety** Our next transformation strengthens applications of parsers where possible. It is considered good practice to fully check the format of incoming messages before making use of any of the fields (one needs to be careful though not to reveal to the attacker too much information if the message is malformed, as shown by Albrecht et al. [2009]). CryptoVerif has a special pattern matching form that allows to represent this kind of check: given a poly-injective function $f$ a statement of the form let $f(x_1, \ldots, x_n) = e$ in $P$ is an abbreviation for

$$\textbf{let } x_1 = f_1^{-1}(e) \textbf{ in } \ldots \textbf{ let } x_n = f_n^{-1}(e) \textbf{ in if } f(x_1, \ldots, x_n) = e \textbf{ then } P$$

where $f_1^{-1}, \ldots, f_n^{-1}$ are the partial inverses of $f$ provided by injectivity. For each parser application let $x = f_p(e)$ in $P$ we check whether the facts that hold at the point of the application imply that $e$ is in the range of a particular encoder $f_c$ and if so, we replace the parser application by let $f_c(\ldots, x, \ldots) = e$ in $P$, where $x$ stands at position $i$ if $f_p$ extracts the $i$th field of $f_c$. The key ingredient of the transformation is being able to determine whether a given expression is in the range of a particular encoder. Section 4.8 shows how this can be done and provides further details of the transformation.

Parsing safety is necessary to prevent collision attacks in some protocols. Consider the Needham-Schroeder-Lowe protocol where first $A$ sends a challenge $c_1(A, N_A)$ to $B$, encrypted with $B$'s public key, $B$ responds with $c_2(B, N_A, N_B)$, and finally $A$ responds with $c_3(N_B)$. It is important to make sure that $B$ never confuses the first and the third message. Suppose we would implement the third message naively by sending the naked nonce: $c_3(N_B) = N_B$. The attacker could then replay $N_B$ as the first message in a different session. The parsing error when trying to parse $N_B$ as $c_1(A, N_A)$ may reveal information about $N_B$ to the attacker.

The attack suggests that we must tag both the first and the third message and check the complete message format, including the tag, before proceeding. The process that receives the first message must be of the form

$$\begin{aligned}
&\textbf{let } d = D(\mathit{cipher}, \mathit{key}) \textbf{ in}\\
&\textbf{let } \mathit{injbot}(m) = d \textbf{ in}\\
&\textbf{let } c_1(A, N_A) = m \textbf{ in}
\end{aligned}$$

When CryptoVerif applies the IND-CCA2 property of the encryption, it will prove that $m$ can be an application of $c_1$, or an application of $c_3$, or a message that does not contain sensitive information. Assuming that $\Phi$ contains a disjointness fact $c_1(\ldots) \neq c_3(\ldots)$ and using the fact $m = c_1(A, N_A)$ obtained from pattern matching CryptoVerif will be able to prove that $m$ cannot be an application of $c_3$.

Even when the security of the protocol does not depend on parsing safety, CryptoVerif often works better when we are able to replace applications of parsers by pattern matches. Consider, for instance, the following fragment of the server model:

$$\begin{aligned}
&\textbf{let } \mathrm{conc2}(\mathrm{client2}, \mathrm{cipher3}) = \mathrm{msg9} \textbf{ in}\\
&\textbf{if } \mathrm{client2} = \mathrm{xClient} \textbf{ then}\\
&\textbf{let } \mathrm{key3} = \mathrm{lookup}(\mathrm{client2}, \mathrm{serverID}, \mathrm{db}) \textbf{ in}
\end{aligned}$$

CryptoVerif will distinguish two cases when verifying the protocol: $\mathit{xClient} = \mathit{clientID}$ and $\mathit{xClient} \neq \mathit{clientID}$. In the first case CryptoVerif will be able to rewrite $\mathit{lookup}(\mathit{client2}, \ldots)$ into $k_{AB}$, the honest shared key, using the rewriting rules for the key database provided by the template (appendix C). This works because the condition check introduces a rewriting rule $\mathit{client2} \rightsquigarrow \mathit{xClient}$. If $\mathit{client2}$ were defined by a parser application, $\mathit{client2} = \mathit{parse1}(\mathit{msg9})$, this would introduce another rewriting rule, and so CryptoVerif would effectively not know which rewriting rule to apply. As a result, the application of the lookup function does not get simplified and CryptoVerif cannot verify the protocol.

Our transformation merges pattern matches of the same expression. For instance, if a pattern match of the form $f(x, _) = e$ is followed by a pattern match $f(_, y) = e$ later on in the process then we replace the first occurrence by $f(x, y) = e$.

**Soundness** Now our model $(\tilde{Q}, \Gamma, \Phi)$ is ready for verification with CryptoVerif. The soundness theorem 4.3 in section 4.11 states that $\mathrm{insec}(Q_A|Q, \rho) \leq \mathrm{cvbound}_p(Q_A|\tilde{Q}, \Gamma, \Phi, \rho, k_0)$ for any attacker $Q_A$, trace property $\rho$, and a sufficiently large polynomial $p$. The CryptoVerif implementation provides a concrete bound on the security of $\tilde{Q}$ in terms of the runtime of $Q_A$ and the constants associated with cryptographic assumptions, which immediately gives a concrete bound on the security of $Q$. Unfortunately, this feature of CryptoVerif has not been formalized and Blanchet [2008] only makes an asymptotic statement: the function $\mathrm{cvbound}_p(Q_A|\tilde{Q}, \Gamma, \Phi, \rho, k)$ is negligible in $k$ (lemma 4.1). In order to appeal to this statement we need to show that this function is well-defined for all $k \in \mathbb{N}$: we must construct at least one interpretation $\tilde{I}_k$ for each $k \in \mathbb{N}$ such that $\tilde{I}_k|_{\tilde{Q}} \models \Gamma$ and $\tilde{I}_k \models \Gamma$. This construction is described in section 4.10.

All the transformations in this chapter are presented for a single well-formed process. This process will consist of the IML models of the participants (*client* and *server* in our example) embedded into a user-provided environment, given as a CryptoVerif template—an input file with missing subprocesses. In the implementation we only need to perform the transformations of the participant processes since the template is already well-typed and does not contain any formatting expressions. We only use the template to provide types for the cryptographic functions and free variables of the participant processes (such as *clientID* in our example).

## 4.3 Formatting-Normal Form

Our first transformation brings the process into a form that makes the following transformations easier to formulate and implement. The goal of the transformation is to move each formatting expression into its own let-binding.

**Definition 4.8** A process is in *formatting-normal form* if

- Formatting expressions occur only in let-bindings.
- There are no cryptographic function applications in auxiliary conditions.
- The expression $e$ in every statement of the form $\mathsf{let}\ x = e\ \mathsf{in}\ P$ is of one of the following

kinds: a variable, an encoding expression, a parsing expression, or an expression containing only cryptographic function applications. □

We transform the process into formatting-normal form by pulling out formatting subexpressions into separate let-bindings, except in auxiliary test conditions, which we leave unchanged.

**Example 4.2** Consider the following process (without distinction between integer and binary terms):

**if** len(x) $\geq$ 4 + x{0, 4} **then**
**let** msg = hash(x{4, x{0, 4}}, keygen(keyseed)) **in**
**if** x{4 + x{0, 4}, len(x) − (4 + x{0, 4})} = msg **then**
**out**(c, len(msg)|msg);

The first test gets classified as auxiliary, the second as cryptographic. The process gets transformed into the following formatting-normal form:

**if** len(x) $\geq$ 4 + x{0, 4} **then**
**let** v1 = x{4, x{0, 4}} **in**
**let** msg = hash(v1, leygen(keyseed)) **in**
**let** v2 = x{4 + x{0, 4}, len(x) − (4 + x{0, 4})} **in**
**if** v2 = msg **then**
**let** v3 = len(msg)|msg **in**
**out**(c, v3);

The concatenation expression len($msg$)|$msg$ is pulled out of the output statement to satisfy the requirement that formatting expressions occur only in let-bindings. □

To justify the correctness of the transformation we show that the basic step (pulling an expression into a separate let binding) is sound.

**Lemma 4.2** *For an output process $P_1$ let $x'$ be a variable that does not occur in $P_1$ and let*

$$P_0 = \textsf{let } x = e[e'/x'] \textsf{ in } P_1,$$
$$\tilde{P}_0 = \textsf{let } x' = e' \textsf{ in let } x = e \textsf{ in } P_1.$$

*Then $P_0 \lesssim \tilde{P}_0$.* □

Proof Let $P_1$, $P_0$, and $\tilde{P}_0$ be as in the statement of the lemma. For two executing processes $(\eta, P)$ and $(\tilde{\eta}, \tilde{P})$ define $(\eta, P) \lesssim (\tilde{\eta}, \tilde{P})$ if either $\tilde{\eta} = \eta$ and $P = P_0$ and $\tilde{P} = \tilde{P}_0$ or $\tilde{\eta} = \eta\{x' \mapsto b\}$ for some $b \in BS$ and $P = \tilde{P} = P'$, where $P'$ is a subprocess of $P_1$. It is straightforward to check that $\lesssim$ satisfies the conditions of definition 2.11. ■

## 4.4 Formatting Extraction

We replace every auxiliary condition and every application of a formatting expression by an application of a fresh function symbol. This transformation is formalised by the rules in fig-

$$\frac{e \text{ is a formatting expression with variables } x_1,\ldots,x_n \qquad f \text{ is a fresh symbol with } I(f) = [\![e]\!] \qquad P \rightsquigarrow \tilde{P}}{\mathsf{let}\ x = e\ \mathsf{in}\ P \rightsquigarrow \mathsf{let}\ x = f(x_1,\ldots,x_n)\ \mathsf{in}\ \tilde{P}}$$

$$\frac{\phi \text{ is an auxiliary fact with variables } x_1,\ldots,x_n \qquad f \text{ is a fresh symbol with } I(f) = [\![\phi]\!] \qquad P \rightsquigarrow \tilde{P}}{\mathsf{if}\ \phi\ \mathsf{then}\ P \rightsquigarrow \mathsf{if}\ f(x_1,\ldots,x_n)\ \mathsf{then}\ \tilde{P}}$$

$$\frac{P \rightsquigarrow \tilde{P}}{\lambda P \rightsquigarrow \lambda \tilde{P}} \qquad \frac{}{0 \rightsquigarrow 0}$$

Figure 4.8: Formatting extraction rules.

ure 4.8. The rules are shown in order of priority so that the last two rules are defaults that apply the transformation to the subprocess. For an expression $e$ with variables $x_1,\ldots,x_n$ let $[\![e]\!]\colon BS_\bot^n \to BS_\bot$ be the function defined by $[\![e]\!](b_1,\ldots,b_n) = [\![e]\!]_{\{x_1\mapsto b_1,\ldots,x_n\mapsto b_n\}}$.

**Example 4.3** In our example in figure 4.6 the following lines introduce new formatting functions:

7 **let** msg1 = conc1(request, key2) **in** ...
10 **let** msg3 = conc2(clientID, cipher1) **in** ...

32 **let** client2 = parse1(msg9) **in** ...
36 **let** cipher3 = parse2(msg9) **in** ...
46 **let** var10 = parse3(msg11) **in** ...
48 **let** key4 = parse4(msg11) **in** ...

with new formatting functions defined as follows:

$$\begin{aligned}
&\mathit{conc1}(x,y) = \mathit{conc2}(x,y) = \text{"p"}|\tau_{4u}^{-1}(\mathit{len}(x))|x|y \\
&\mathit{parse1}(x) = \mathit{parse3}(x) = x\{5, \tau_{4u}(x\{0,4\})\} \\
&\mathit{parse2}(x) = \mathit{parse4}(x) = x\{5+\tau_{4u}(x\{1,4\}), \mathrm{len}(x) - (5+\tau_{4u}(x\{1,4\}))\}
\end{aligned}$$

It is important that every occurrence of a formatting expression gets assigned a distinct function symbol, even if the expressions are the same, as in the case of *conc1* and *conc2*. The reason is that *conc1* and *conc2* will be assigned different types by the type inference. □

If $f(x_1,\ldots,x_n)$ it is a result of the substitution of an auxiliary condition $\phi$, we shall keep calling it an auxiliary condition in the remaining sections. When adding $f(x_1,\ldots,x_n)$ as a fact to the knowledge of our prover, we expand it first, replacing it by $\phi$.

## 4.5 Type Inference

Type inference aims to generate types for the new function symbols added in the previous transformation. This is done in two stages. First we collect types for the variables in the process based on how they are generated or used. For instance, if a variable is a result of a cryptographic function with return type $T$ (as specified in the typing environment $\Gamma_0$ from the user-provided template) or bound in a statement $\mathsf{new}\ x\colon T$ then the variable can be assigned type $T$. Alternatively a variable can be given type $T$ if it is used as an argument of a function with argument type $T$. Since our input process is in a formatting-normal form, every occurrence of a formatting function $f$ is of the form $\mathsf{let}\ x = f(x_1, \ldots, x_n)$. Knowledge of the types of $x, x_1, \ldots, x_n$ allows us to give a type to $f$. The soundness result does not depend on the correctness of type inference, as all the types are checked during the next stage (section 4.6).

We shall use the following types:

- for each $n \in \mathbb{N}$ the type $\mathtt{fixed}_n$ with $I(\mathtt{fixed}_n) = \{0x00, \ldots, 0xFF\}^n$,
- for each $n \in \mathbb{N}$ the type $\mathtt{bounded}_n$ with $I(\mathtt{bounded}_n) = \{0x00, \ldots, 0xFF\}^{\leq n}$,
- the type $\mathtt{bitstring}$ with $I(\mathtt{bitstring}) = BS$,
- the type $\mathtt{bitstringbot}$ with $I(\mathtt{bitstringbot}) = BS_\perp$,
- the type $\mathtt{bool}$ with $I(\mathtt{bool}) = Bool$.

When type inference collects multiple suggestions for the type of a variable, we consolidate those by taking the intersection of the types. To define the intersection we first define a subtyping relation $\sqsubseteq$ as follows: we let $T \sqsubseteq T$ for every type $T$ and additionally

$$T \sqsubseteq \mathtt{bitstringbot} \text{ for any } T \qquad T \sqsubseteq \mathtt{bitstring} \text{ for any } T \neq \mathtt{bitstringbot}$$

$$\mathtt{bounded}_n \sqsubseteq \mathtt{bounded}_m \text{ iff } n \leq m \qquad \mathtt{fixed}_n \sqsubseteq \mathtt{bounded}_m \text{ iff } n \leq m$$

$$\mathtt{bounded}_n \sqsubseteq \mathtt{fixed}_m \text{ iff } n = m = 0$$

Clearly $T \sqsubseteq T'$ implies $I(T) \subseteq I(T')$. A type $T$ is said to be an *intersection* of two types $T_1$ and $T_2$ if $T = T_1 \sqsubseteq T_2$ or $T = T_2 \sqsubseteq T_1$ (an intersection need not necessarily exist). Given two typing environments $\Gamma$ and $\Gamma'$ that contain only variables we can unify the information contained in them by taking type intersection for variables that are present in both $\Gamma$ and $\Gamma'$. Formally, we define

$$\begin{aligned} \Gamma \sqcup \Gamma' = \quad & \{x \mapsto \Gamma(x) \mid x \in \mathrm{dom}(\Gamma) \setminus \mathrm{dom}(\Gamma')\} \\ & \cup \{x \mapsto \Gamma'(x) \mid x \in \mathrm{dom}(\Gamma') \setminus \mathrm{dom}(\Gamma)\} \\ & \cup \{x \mapsto \Gamma(x) \sqcap \Gamma'(x) \mid x \in \mathrm{dom}(\Gamma) \cap \mathrm{dom}(\Gamma')\} . \end{aligned}$$

IF the type intersection for one of the variables does not exist, we fail.

$$\frac{T \neq \bot}{x,\, T \rightsquigarrow \{x \mapsto T\}} \qquad \frac{}{x,\, \bot \rightsquigarrow \emptyset}$$

$$\frac{e,\, T \rightsquigarrow \Gamma}{(e\colon T),\, _ \rightsquigarrow \Gamma}$$

$$\frac{\Gamma_0(f) = T_1 \times \ldots \times T_n \to T \quad e_i,\, T_i \rightsquigarrow \Gamma_i}{f(e_1, \ldots, e_n),\, _ \rightsquigarrow \bigsqcup_i \Gamma_i}$$

$$\frac{f \notin \mathrm{dom}(\Gamma_0) \quad e_i,\, \bot \rightsquigarrow \Gamma_i}{f(e_1, \ldots, e_n),\, _ \rightsquigarrow \bigsqcup_i \Gamma_i}$$

$$\frac{e,\, \bot \rightsquigarrow \Gamma \quad e',\, \bot \rightsquigarrow \Gamma'}{(e = e'),\, _ \rightsquigarrow \Gamma \sqcup \Gamma'}$$

Figure 4.9: Type inference rules for expressions.

$$\frac{\mathrm{type}_{\Gamma_0}(e) \neq \bot \quad e,\, \bot \rightsquigarrow \Gamma \quad P \rightsquigarrow \Gamma'}{\mathsf{let}\ x = e\ \mathsf{in}\ P \rightsquigarrow (\Gamma \sqcup \Gamma')\{x \mapsto \mathrm{type}_{\Gamma_0}(e)\}}$$

$$\frac{\mathrm{type}_{\Gamma_0}(e) = \bot \quad e,\, \bot \rightsquigarrow \Gamma \quad P \rightsquigarrow \Gamma'}{\mathsf{let}\ x = e\ \mathsf{in}\ P \rightsquigarrow \Gamma \sqcup \Gamma'}$$

$$\frac{P \rightsquigarrow \Gamma}{\mathsf{new}\ x\colon T; P \rightsquigarrow \Gamma\{x \mapsto T\}}$$

$$\frac{P \rightsquigarrow \Gamma}{\mathsf{in}(c[e_1, \ldots, e_n], x); P \rightsquigarrow \Gamma\{x \mapsto \mathtt{bitstring}\}}$$

$$\frac{P \rightsquigarrow \Gamma}{\mathsf{assume}\ \phi;\ P \rightsquigarrow \Gamma}$$

$$\frac{\exp(\lambda) = e_1, \ldots, e_n \quad e_i, \bot \rightsquigarrow \Gamma_i \quad P \rightsquigarrow \Gamma}{\lambda P \rightsquigarrow \Gamma \sqcup \Gamma_1 \sqcup \ldots \sqcup \Gamma_n}$$

Figure 4.10: Type inference rules for processes.

We allow the user to provide explicit type annotations on expressions, that is, we consider a new form $e\colon T$ of expressions, with a type $T$. The annotations are added by a call in the C program. For instance, a call `typehint(buf, len, "bounded_1024");` will assign type $\mathtt{bounded}_{1024}$ to the bitstring contained in buffer `buf` of length `len`. Type hints are picked up and propagated by symbolic execution. Annotations are only used for type inference, and are erased after that. In particular, typechecking of the next section does not rely on annotations.

We shall write $\mathrm{type}_\Gamma(e)$ to represent the type of an expression $e$ in typing environment $\Gamma$. This is computed by the following rules, in order of precedence:

$$\begin{aligned}
\mathrm{type}_\Gamma(e\colon T) &= T, \\
\mathrm{type}_\Gamma(x) &= \Gamma(x) \text{ if } x \in \mathrm{dom}(\Gamma), \\
\mathrm{type}_\Gamma(f(\ldots)) &= T \text{ if } f \in \mathrm{dom}(\Gamma) \text{ and } \Gamma(f) = T_1 \times \ldots \times T_n \to T, \\
\mathrm{type}_\Gamma(e) &= \bot \text{ otherwise.}
\end{aligned}$$

The rules of type inference for processes (figure 4.10) are of the form $P \rightsquigarrow \Gamma$ meaning that a process $P$ yields a typing environment $\Gamma$ that contains types for variables in $P$. The rules for expressions (figure 4.9) are of the form $e, T \rightsquigarrow \Gamma$, meaning that an expression $e$ yields an environment $\Gamma$ under the assumption that the type of $e$ is $T$. In the default rule in figure 4.10 we use $\exp(\lambda)$ to denote the expressions of a syntactic form $\lambda$. Typing information comes from several sources: explicit type annotations, applications of cryptographic functions with known types (given by $\Gamma_0$), use of new $x\colon T$, and inputs. We do not derive any type constraints from bitstring equality tests, and we do not add types for replication indices—those will be assigned the default type `bitstring`. Assumptions are ignored for the purpose of type inference.

**Definition 4.9 (Type Inference)** We say that a typing environment $\Gamma$ is the *result of type inference of function types* from a process $P$ if it is constructed as follows: First obtain a typing environment $\Gamma_{var}$ by $P \rightsquigarrow \Gamma_{var}$ and complete it by setting $\Gamma_{var}(x) = \texttt{bitstring}$ for every variable $x$ of $P$ with $x \notin \mathrm{dom}(\Gamma_{var})$. Next generate the typing environment $\Gamma_{fun}$ with types of formatting and auxiliary test functions as follows: For each formatting function $f$ that occurs in a statement let $x = f(x_1, \ldots, x_n)$ set $\Gamma_{fun}(f) = \Gamma_{var}(x_1) \times \ldots \times \Gamma_{var}(x_n) \to \Gamma_{var}(x)$. For each auxiliary test function that occurs in a statement if $f(x_1, \ldots, x_n)$ then set $\Gamma_{fun}(f) = \Gamma_{var}(x_1) \times \ldots \times \Gamma_{var}(x_n) \to \texttt{bool}$. Finally, let $\Gamma = \Gamma_0 \sqcup \Gamma_{fun}$, where $\Gamma_0$ is the typing environment provided by the user that contains types of the cryptographic functions. □

**Example 4.4** Consider the application of *conc2* in line 10 of figure 4.6: let msg3 = conc2( clientID, cipher1). The variable *clientID* is bound in the user-provided template and declared to be of type $\texttt{bounded}_{1024}$. The variable *cipher1* is the result of an application of the encryption function $E$ which in the user template has return type $\texttt{bounded}_{1077}$ (see below for the origin of the number 1077). Finally, the result *msg3* of *conc2* is output in line 11, and so is assigned type `bitstring` for lack of more precise information. This gives us the type *conc2* : $\texttt{bounded}_{1024} \times \texttt{bounded}_{1077} \to \texttt{bitstring}$. The types of the other functions are derived in the same way (user-provided type annotations are not necessary for RPC-enc). The types we infer for our formatting functions are

$$
\begin{aligned}
&\mathit{conc1} : \texttt{fixed}_{1024} \times \texttt{fixed}_{16} \to \texttt{bounded}_{1045}\\
&\mathit{conc2} : \texttt{bounded}_{1024} \times \texttt{bounded}_{1077} \to \texttt{bitstring}\\
&\mathit{parse1} : \texttt{bitstring} \to \texttt{bounded}_{1024}\\
&\mathit{parse2} : \texttt{bitstring} \to \texttt{bounded}_{1077}\\
&\mathit{parse3} : \texttt{bounded}_{1045} \to \texttt{fixed}_{1024}\\
&\mathit{parse4} : \texttt{bounded}_{1045} \to \texttt{fixed}_{16}
\end{aligned}
$$

The type $\texttt{fixed}_{16}$ is the type of keys used by the process and the type $\texttt{fixed}_{1024}$ is the type of payloads. It is important that both payloads and keys have fixed types to preserve their secrecy. The type $\texttt{bounded}_{1045}$ is the type of concatenations of payload and key, which results from adding the concatenation overhead of 5 bytes to the sum of payload length of 1024 and key length of 16. The type $\texttt{bounded}_{1077}$ is obtained by taking the type $\texttt{bounded}_{1045}$ and adding 32 bytes of encryption overhead. □

The types we give to our formatting functions are not necessarily sound. In particular, parsers expect input with a certain internal structure since they use length fields to extract the concatenation arguments. We shall deal with this problem in section 4.9 where we check that the types of encoders are in fact sound and that the parsers can be safely replaced by strengthened versions that are sound with respect to their types.

## 4.6 Typechecking

The purpose of the next transformation step is to insert typecasts where necessary to ensure that the process is well-typed with respect to the typing environment $\Gamma$ generated by the type inference. We use the facts that are known to be true at a particular point in order to establish that this transformation is sound. Facts come from the conditions that are checked by the process as well as from the types of expressions that have been checked so far. We use facts of the form $\mathrm{intype}(e, T)$ to record that an expression $e$ has type $T$. These are defined as follows:

$$\begin{aligned} \mathrm{intype}(e, \mathtt{fixed}_n) &= (\mathrm{len}(e) = n), \\ \mathrm{intype}(e, \mathtt{bounded}_n) &= (\mathrm{len}(e) \le n), \\ \mathrm{intype}(e, \mathtt{bitstring}) &= \mathrm{defined}(e), \\ \mathrm{intype}(e, \mathtt{bitstringbot}) &= \mathrm{true}\,. \end{aligned}$$

We never form intype expressions for boolean types. Given a typing environment $\Gamma$ we let

$$\mathrm{typefacts}(\Gamma) = \{\mathrm{intype}(x, \Gamma(x)) \mid x \in \mathrm{dom}(\Gamma)\}\,.$$

The typechecking rules use two typing judgements, presented in figure 4.11. The judgement $\Gamma, \Phi \vdash P \rightsquigarrow \tilde{P}$ means that under the assumption that both $\Phi$ and $\mathrm{typefacts}(\Gamma)$ are true the process $P$ can be rewritten to a process $\tilde{P}$ that is well-typed with respect to $\Gamma$. The judgement $\Gamma, \Phi, T \vdash e \rightsquigarrow \tilde{e}$ means that under the assumption that both $\Phi$ and $\mathrm{typefacts}(\Gamma)$ are true the expression $e$ can be rewritten to a well-typed expression $\tilde{e}$ of type $T$. Most of the rules are straightforward and simply propagate the typing information and facts through the process, inserting typecasts where necessary and justified. A typecast from a type $T$ to a type $T'$ is performed by a function $\mathrm{tcast}_{T \to T'} : T \to T'$ such that $\mathrm{tcast}_{T \to T'}(x) = x$ for all $x \in I(T) \cap I(T')$ and for all $x \notin I(T)$, and $\mathrm{tcast}_{T \to T'}(x)$ is an arbitrary but fixed value of type $T'$ otherwise. We require the value to be fixed since later we make use of the fact that casts are length-regular. The definition is chosen in such a way that $\mathrm{tcast}_{T \to T}$ is the identity function, and so can be omitted.

The rules that rewrite an expression $e$ into $\mathrm{tcast}_{T \to T'}(e)$ check that $e$ already belongs to type $T'$ as implied by the context, and so the the resulting expressions evaluates the same as $e$. Most of the work is done by the rules (TCVar) and (TCFormat) that prove that variables and results of formatting function applications belong to the required type. We do not cast the results of formatting function applications since in formatting-normal form those results

$$\frac{\Gamma(x) = T_x \quad \Phi \cup \text{typefacts}(\Gamma) \vdash \text{intype}(x, T)}{\Gamma, \Phi, T \vdash x \rightsquigarrow \text{tcast}_{T_x \to T}(x)} \tag{TCVar}$$

$$\frac{f \text{ is a formatting function} \quad \Gamma(f) = T_1 \times \ldots \times T_n \to T \\ \Gamma, \Phi, T_i \vdash e_i \rightsquigarrow \tilde{e}_i \quad \Phi \cup \text{typefacts}(\Gamma) \vdash \text{intype}(\text{expand}(f(e_1, \ldots, e_n)), T)}{\Gamma, \Phi, T \vdash f(e_1, \ldots, e_n) \rightsquigarrow f(\tilde{e}_1, \ldots, \tilde{e}_n)} \tag{TCFormat}$$

$$\frac{f \text{ is a cryptographic or auxiliary test function} \\ \Gamma(f) = T_1 \times \ldots \times T_n \to T_f \quad \Gamma, \Phi, T_i \vdash e_i \rightsquigarrow \tilde{e}_i \quad T_f \sqsubseteq T}{\Gamma, \Phi, T \vdash f(e_1, \ldots, e_n) \rightsquigarrow \text{tcast}_{T_f \to T}(f(\tilde{e}_1, \ldots, \tilde{e}_n))} \tag{TCCrypto}$$

$$\frac{T = \text{type}_\Gamma(e) \sqcup \text{type}_\Gamma(e') \quad \Gamma, \Phi, T \vdash e \rightsquigarrow \tilde{e} \quad \Gamma, \Phi, T \vdash e' \rightsquigarrow \tilde{e}'}{\Gamma, \Phi, \texttt{bool} \vdash (e = e') \rightsquigarrow (\tilde{e} = \tilde{e}')} \tag{TCEq}$$

$$\Gamma, \Phi \vdash 0 \rightsquigarrow 0 \tag{TCNil}$$

$$\frac{\Gamma, \Phi \vdash Q \rightsquigarrow \tilde{Q} \quad \Gamma, \Phi \vdash Q' \rightsquigarrow \tilde{Q}'}{\Gamma, \Phi \vdash Q | Q' \rightsquigarrow \tilde{Q} | \tilde{Q}'} \tag{TCPar}$$

$$\frac{\Gamma\{i \mapsto \texttt{bitstring}\}, \Phi \vdash Q \rightsquigarrow \tilde{Q}}{\Gamma, \Phi \vdash !^{i \le N} Q \rightsquigarrow !^{i \le N} \tilde{Q}} \tag{TCRepl}$$

$$\frac{\Gamma, \Phi, \text{type}_\Gamma(e) \vdash e \rightsquigarrow \tilde{e} \quad \Gamma\{x \mapsto \text{type}_\Gamma(e)\}, \Phi \vdash P \rightsquigarrow \tilde{P}}{\Gamma, \Phi \vdash \mathsf{let}\ x = e\ \mathsf{in}\ P \rightsquigarrow \mathsf{let}\ x = \tilde{e}\ \mathsf{in}\ \tilde{P}} \tag{TCLet}$$

$$\frac{\Gamma, \Phi, \texttt{bool} \vdash \phi \rightsquigarrow \tilde{\phi} \quad \Gamma, \Phi \cup \{\phi\} \vdash P \rightsquigarrow \tilde{P}}{\Gamma, \Phi \vdash \mathsf{if}\ \phi\ \mathsf{then}\ P \rightsquigarrow \mathsf{if}\ \tilde{\phi}\ \mathsf{then}\ \tilde{P}} \tag{TCIf}$$

$$\frac{\Gamma, \Phi \cup \{\phi\} \vdash P \rightsquigarrow \tilde{P}}{\Gamma, \Phi \vdash \mathsf{assume}\ \phi;\ P \rightsquigarrow \mathsf{assume}\ \phi;\ \tilde{P}} \tag{TCAssume}$$

$$\frac{\forall i \le n\colon \Gamma, \Phi, \text{type}_\Gamma(e_i) \vdash e_i \rightsquigarrow \tilde{e}_i \quad \Gamma\{x \mapsto \texttt{bitstring}\}, \Phi \vdash P \rightsquigarrow \tilde{P}}{\Gamma, \Phi \vdash \mathsf{in}(c[e_1, \ldots, e_n], x); P \rightsquigarrow \mathsf{in}(c[\tilde{e}_1, \ldots, \tilde{e}_n], x); \tilde{P}} \tag{TCIn}$$

$$\frac{T = \texttt{fixed}_n \quad \Gamma\{x \mapsto T\}, \Phi \vdash P \rightsquigarrow \tilde{P}}{\Gamma, \Phi \vdash \mathsf{new}\ x\colon T;\ P \rightsquigarrow \mathsf{new}\ x\colon T;\ \tilde{P}} \tag{TCNew}$$

$$\frac{\forall i \le n\colon \Gamma, \Phi, \text{type}_\Gamma(e_i) \vdash e_i \rightsquigarrow \tilde{e}_i \quad \Gamma, \Phi, \text{type}_\Gamma(e) \vdash e \rightsquigarrow \tilde{e} \quad \Gamma, \Phi \vdash Q \rightsquigarrow \tilde{Q}}{\Gamma, \Phi \vdash \mathsf{out}(c[e_1, \ldots, e_n], e);\ Q \rightsquigarrow \mathsf{out}(c[\tilde{e}_1, \ldots, \tilde{e}_n], \tilde{e}); \tilde{Q}} \tag{TCOut}$$

$$\frac{\forall i \le n\colon \Gamma, \Phi, \text{type}_\Gamma(e_i) \vdash e_i \rightsquigarrow \tilde{e}_i \quad \Gamma, \Phi \vdash P \rightsquigarrow \tilde{P}}{\Gamma, \Phi \vdash \mathsf{event}\ ev(e_1, \ldots, e_n); P \rightsquigarrow \mathsf{event}\ ev(\tilde{e}_1, \ldots, \tilde{e}_n);\ \tilde{P}} \tag{TCEvent}$$

Figure 4.11: Typechecking rules for expressions and processes.

are always assigned to variables that can be cast later as needed. Since types inferred for the formatting functions need not necessarily be sound, as mentioned in the previous section, we do not trust these types. Instead we expand the definitions of the formatting functions and check the resulting expressions explicitly: for a formatting function $f = [\![e]\!]$ where $e$ is a formatting expression with variables $x_1, \ldots, x_n$ we let $\text{expand}(f(e_1, \ldots, e_n)) = e[e_1/x_1, \ldots, e_n/x_n]$. When adding the fact $\phi$ to the set $\Phi$ the rule (TCIf) expands applications of auxiliary test functions in the same way.

The rule (TCEq) casts both operands of an equality test to the union of their types. Given two types $T$ and $T'$ we define $T \sqcup T'$ in the obvious way, similarly to the way $T \sqcap T'$ is defined in section 4.5.

**Example 4.5** The inferred type of *parse2* in our example is $\texttt{bitstring} \to \texttt{bounded}_{1077}$. The facts that are checked before applying *parse2* to *msg9* in line 36 include $\Phi = \{\phi_1, \phi_2, \phi_3, \phi_4\}$, where

$$\begin{aligned}
\phi_1 &= (\tau_{4u}(\mathit{msg8}) \geq 1082) \wedge (\tau_{4u}(\mathit{msg8}) \leq 2106),\\
\phi_2 &= \text{len}(\mathit{msg9}) = \tau_{4u}(\mathit{msg8}),\\
\phi_3 &= \tau_{4u}(\mathit{msg9}\{1,4\}) \leq 1024,\\
\phi_4 &= \tau_{4u}(\mathit{msg8}) - (5 + \tau_{4u}(\mathit{msg9}\{1,4\})) \leq 1077.
\end{aligned}$$

To prove that the result of parsing is of the right type the rule (TCFormat) will check that $\Phi$ implies

$$\begin{aligned}
&\text{intype}(\text{expand}(\mathit{parse2}(\mathit{msg9})), \texttt{bounded}_{1077})\\
&\quad = \text{len}(\mathit{msg9}\{5 + \tau_{4u}(\mathit{msg9}\{1,4\}), \text{len}(\mathit{msg9}) - (5 + \tau_{4u}(\mathit{msg9}\{1,4\}))\}) \leq 1077.
\end{aligned}$$

The fact $\phi_1$ is used to show that the substring extraction is within correct bounds, $\phi_3$ is necessary to show that there is no overflow, and $\phi_2$ and $\phi_4$ together imply the necessary arithmetic condition. □

We now state several results that capture the soundness of typechecking. In the following we assume that $\Gamma$ agrees with $\Gamma_0$ (the template environment with types for cryptographic functions) and contains types for all the typecast functions used in the process. The following lemma is obvious.

**Lemma 4.3 (Typechecking of Expressions is Sound)** *Assume that* $\Gamma, \Phi, T \vdash e \rightsquigarrow \tilde{e}$. *If* $\eta \models \Gamma$ *and* $\eta \models \Phi$ *for an environment* $\eta$ *then* $[\![e]\!]_\eta = [\![\tilde{e}]\!]_\eta \in I(T)$. □

**Lemma 4.4 (Typechecking of Processes is Sound)** *Assume that* $\Gamma, \emptyset \vdash Q \rightsquigarrow \tilde{Q}$ *for a typing environment* $\Gamma$ *that does not contain variables. Then* $Q \lesssim \tilde{Q}$. □

PROOF For two executing processes $(\eta, P)$ and $(\tilde{\eta}, \tilde{P})$ let $(\eta, P) \lesssim (\tilde{\eta}, \tilde{P})$ if $\eta = \tilde{\eta}$ and there exists a set of facts $\Phi$ and a typing environment $\Gamma$ such that $\eta \models \Phi$ and $\eta \models \Gamma$ and $\Gamma, \Phi \vdash P \rightsquigarrow \tilde{P}$. We show that $\lesssim$ satisfies the conditions of definition 2.11. Item 1 is fulfilled by assumption $\Gamma, \emptyset, Q \rightsquigarrow \tilde{Q}$ of the lemma. Here it is important that $\Gamma$ does not contain variables, so that $\eta \models \Gamma$ for any environment $\eta$. Items 2 to 5 can be proved by induction on the size of the process. We shall show the proof for the case of let-expressions, the other cases follow in the same way.

Assume that $(\eta, P) \lesssim (\eta, \tilde{P})$ and $P = \mathsf{let}\ x = e\ \mathsf{in}\ P'$. By definition of $\lesssim$ the rule (TCLet) must apply, so that $\tilde{P} = \mathsf{let}\ x = \tilde{e}\ \mathsf{in}\ \tilde{P}'$. We therefore need to check that item 5 in definition 2.11 is satisfied: both executing processes reduce together and still simulate each other after reduction, that is, $(\eta', P') \lesssim (\tilde{\eta}', \tilde{P}')$, where $\eta' = \eta\{x \mapsto [\![e]\!]_\eta\}$ and $\tilde{\eta}' = \eta\{x \mapsto [\![\tilde{e}]\!]_\eta\}$. By lemma 4.3 applied to the premise of the rule (TCLet) we have $[\![e]\!]_\eta = [\![\tilde{e}]\!]_\eta \in I(\mathrm{type}_\Gamma(e))$ so that $\eta' = \tilde{\eta}'$ and $\eta' \models \Gamma\{x \mapsto \mathrm{type}_\Gamma(e)\}$. By the rule (TCLet) $\Gamma\{x \mapsto \mathrm{type}(e, \Gamma)\}, \Phi \vdash P' \rightsquigarrow \tilde{P}'$, and so $(\eta', P') \lesssim (\tilde{\eta}', \tilde{P}')$.

Finally, condition 6 in definition 2.11 follows from the fact that $\eta = \tilde{\eta}$ when $(\eta, P) \lesssim (\tilde{\eta}, \tilde{P})$. ∎

**Lemma 4.5 (Typechecked Processes are Well-Typed)** *If $\Gamma, \emptyset \vdash Q \rightsquigarrow \tilde{Q}$ and $\tilde{Q}'$ is the result of removing inline assumptions from $\tilde{Q}$ then $\Gamma \vdash \tilde{Q}'$.* □

PROOF The statement follows by structural induction on $Q$ and direct comparison of rules in figure 4.2 with rules in figure 4.11. As an auxiliary statement we prove for expressions: if $\Gamma, \Phi, T \vdash e \rightsquigarrow \tilde{e}$ then $\Gamma \vdash \tilde{e} \colon T$. ∎

## 4.7 Fact Inference

This section describes the set of facts $\Phi$ that we supply to CryptoVerif as part of the generated model. The facts encode the properties of our formatting functions and include parsing equations, disjointness of ranges, injectivity, and length-regularity (the length of the output depends only on the lengths of the inputs). We shall make use of quantifiers of the form $\forall (x \colon T) \colon \phi$, with the obvious meaning. For the rest of the section we fix a process $Q$ and a typing environment $\Gamma$ and write $f \colon T_1 \times \ldots \times T_n \to T$ to mean $\Gamma(f) = T_1 \times \ldots \times T_n \to T$. The environment $\Gamma$ is assumed to contain types for all functions used in $Q$.

**Parsing Equations** Consider an encoder $f_c \colon T_1 \times \ldots \times T_n \to T$ defined by an encoding expression $e_c$ with variables $x_1, \ldots, x_n$ and a parser $f_p \colon T \to T_i$ for some $i \leq n$ defined by a parsing expression $e_p$ with a variable $x$. We can check the result of applying the parser to the encoder by forming the expression $e_p[e_c/x]$ and checking whether we can simplify it to $x_i$ using the function simplify defined in section 3.3. Let $\Phi_{arg} = \{\mathrm{intype}(x_1, T_1), \ldots, \mathrm{intype}(x_n, T_n)\}$. If $\mathrm{simplify}_{\Phi_{arg}}(e_p[e_c/x]) = x_i$ for some $i \leq n$ then we add the following fact to $\Phi$:

$$parse_\Gamma(f_p, f_c, i) = \forall (x_1 \colon T_1, \ldots, x_n \colon T_n) \colon f_p(f_c(x_1, \ldots, x_n)) = x_i.$$

**Example 4.6** In our RPC-enc example we add the following parsing equations:

$$\forall(x\colon \mathtt{fixed}_{1024}, y\colon \mathtt{fixed}_{16})\colon \mathit{parse3}(\mathit{conc1}(x,y)) = x,$$
$$\forall(x\colon \mathtt{fixed}_{1024}, y\colon \mathtt{fixed}_{16})\colon \mathit{parse4}(\mathit{conc1}(x,y)) = y,$$
$$\forall(x\colon \mathtt{bounded}_{1024}, y\colon \mathtt{bounded}_{1077})\colon \mathit{parse1}(\mathit{conc2}(x,y)) = x,$$
$$\forall(x\colon \mathtt{bounded}_{1024}, y\colon \mathtt{bounded}_{1077})\colon \mathit{parse2}(\mathit{conc2}(x,y)) = y.$$

We do not add equations for *parse3* applied to *conc2* and the other way round because the types do not match. □

**Injectivity of Encoders** We check which encoders are injective, taking into account the types of the arguments. A sufficient condition for the encoder to be injective is when there is at most one argument which is not preceded by its length in the encoding and is not of a fixed-length type. We record this property for future use.

**Definition 4.10** An encoder $f\colon T_1 \times \ldots \times T_n \to T$ defined by an expression $e = e_1|\ldots|e_m$ with variables $x_1, \ldots, x_n$ is called *robustly injective* when there is at most one $i$ such that $e_i = x_j$ for some $j \leq n$, there is no expression $\mathrm{len}(x_j)$ preceding $e_i$ in $e$, and $T_j$ is not a fixed-length type. □

If $f$ is robustly injective, we add the following fact to the set $\Phi$:

$$\mathit{inj}_\Gamma(f) = \forall(x_1\colon T_1, \ldots, x_n\colon T_n, y_1\colon T_1, \ldots, y_n\colon T_n)\colon$$
$$f(x_1, \ldots, x_n) = f(y_1, \ldots, y_n) \Rightarrow \bigwedge_{i \leq n} x_i = y_i.$$

**Example 4.7** Applying our criterion, we see that both *conc1* and *conc2* in RPC-enc are injective: there is only one argument variable (the second argument) that is not preceded by its length. □

CryptoVerif makes use of a stronger notion called *poly-injectivity*: a function is poly-injective if it is injective and its inverses are efficiently computable. This is necessary to use the pattern-matching construct in section 4.8. For our encoders injectivity clearly implies poly-injectivity.

**Disjointness of Encoders** We establish disjointness of encoders by finding constant tags in the formatting that are always aligned and have different values. We record this criterion for future use.

**Definition 4.11** Let $f\colon T_1 \times \ldots \times T_n \to T$ and $f'\colon T'_1 \times \ldots \times T'_{n'} \to T$ be two encoders given by expressions $e_1|\ldots|e_m$ and $e'_1|\ldots e'_{m'}$. We say that $f$ and $f'$ are *robustly disjoint* if there exists $k \leq \min(m, m')$ such that $e_k$ and $e'_k$ are constant tag expressions such that neither is a prefix of the other and $e_i = e'_i$ for all $i < k$. □

For every two encoders $f$ and $f'$ that are robustly disjoint, we add following fact to the set $\Phi$:

$$\mathit{disj}_\Gamma(f, f') = \forall(x_1\colon T_1, \ldots, x_n\colon T_n, y_1\colon T'_1, \ldots, y_m\colon T'_m)\colon f(x_1, \ldots, x_n) \neq f'(y_1, \ldots, y_m).$$

In our example the two encoders are not disjoint since they are identical, but disjointness will become very important when proving computational soundness conditions in chapter 5.

**Equality of Encoders** If two encoders $f, f'$ have identical types and defining expressions, we add the following fact to $\Phi$:

$$\mathit{eq}_\Gamma(f, f') = \forall(x_1\colon T_1, \ldots, x_n\colon T_n)\colon f(x_1, \ldots, x_n) = f'(x_1, \ldots, x_n).$$

This is useful for protocols where the same message is computed in different places. An example would be a contract-signing protocol in which two participants each compute the same message (the contract), sign the message with their private keys, and exchange only the signatures.

**Cast Equations** For every two types $T$, $T'$ such that $T \sqsubseteq T'$ and $Q$ contains both $\mathrm{tcast}_{T \to T'}$ and $\mathrm{tcast}_{T' \to T}$ we add the following fact to $\Phi$:

$$\mathit{cast}_\Gamma(T, T') = \forall(x\colon T)\colon\ \mathrm{tcast}_{T' \to T}(\mathrm{tcast}_{T \to T'}(x)) = x.$$

**Length-Regularity of Encoders** By construction all our encoders are *length-regular*: the length of their result depends only on the length of the arguments. This property is important to establish the security of RPC-enc, where the client sends the encryption of $\mathit{msg1} = \mathit{conc1}(\mathit{request}, \mathit{key2})$ and we need to prove that this encryption does not leak any information about $\mathit{key2}$. We assume that the encryption only leaks the length of the plaintext, therefore it is important that the length of $\mathit{msg1}$ does not depend on the value of $\mathit{key2}$.

When CryptoVerif applies the security definition of the symmetric encryption, it replaces the plaintext $\mathit{msg1}$ by $Z_{\mathtt{bounded}_{1045}}(\mathit{msg1})$ where $Z_{\mathtt{bounded}_{1045}}\colon \mathtt{bounded}_{1045} \to \mathtt{bounded}_{1045}$ is the function that sets all bits of its input to 0. We need to give enough facts to CryptoVerif so that it can rewrite $Z_{\mathtt{bounded}_{1045}}(\mathit{conc1}(\mathit{request}, \mathit{key2}))$ to a value that does not depend on $\mathit{key2}$. This is done as follows: for each encoder $f\colon T_1 \times \ldots \times T_n \to T$ we add the following fact:

$$\mathit{reg}_\Gamma(f) = \forall(x_1\colon T_1, \ldots, x_n\colon T_n)\colon Z_T(f(x_1, \ldots, x_n)) = Z_T(f'(Z_{T_1}(x_1), \ldots, Z_{T_n}(x_n))).$$

We use a fresh symbol $f'$ to prevent CryptoVerif from looping when applying the rewriting rule. For every fixed-length type $T$ we also add the fact

$$\mathit{zero}_\Gamma(T) = \forall(x\colon T)\colon Z_T(x) = 0_T,$$

where $0_T$ is a constant of type $T$. Our cast functions are defined in such a way that they are also length-regular (section 4.6), and so for every function $\mathrm{cast}_{T \to T'}$ used in our model we add the fact $reg_\Gamma(\mathrm{cast}_{T \to T'})$ as above.

**Example 4.8** In RPC-enc we add the equations

$$\begin{aligned}
&\forall (x_1 \colon \mathtt{fixed}_{1024}, x_2 \colon \mathtt{fixed}_{16})\colon \\
&\quad Z_{\mathtt{bounded}_{1045}}(conc1(x_1, x_2)) = Z_{\mathtt{bounded}_{1045}}(conc1'(Z_{\mathtt{fixed}_{1024}}(x_1), Z_{\mathtt{fixed}_{16}}(x_2))). \\
&\forall (x \colon \mathtt{fixed}_{16})\colon Z_{\mathtt{fixed}_{16}}(x) = 0_{\mathtt{fixed}_{16}}. \\
&\forall (x \colon \mathtt{fixed}_{1024})\colon Z_{\mathtt{fixed}_{1024}}(x) = 0_{\mathtt{fixed}_{1024}}.
\end{aligned}$$

□

In order for the functions $Z_T$ to be type-safe it is important that our types themselves are length-regular: if $b \in I(T)$ then also $b' \in I(T)$ for every bitstring $b'$ such that $|b| = |b'|$. In particular, $Z_T(b) \in I(T)$ (an exception is the type `bool` which is not assumed to be length-regular, but that is not a problem since `bool` never occurs as a result or argument type of any formatting or cast functions in our models). This partially motivates our choice of types—they are simple enough to be length-regular.

**Length-Regularity of Auxiliary Tests** Consider a condition of the form `if` $f(x_1, \dots, x_n)$ `then` $P$ where $f \colon T_1 \times \ldots \times T_n \to T$ is a function defined by an expression $e_f$. Without further help CryptoVerif would not be able to prove secrecy of the variables $x_i$ since the control flow may depend on the value of those variables. However, often the shape of $e_f$ is such that it contains only expressions of the form $\mathrm{len}(x_i)$ for some $i$. Alternatively, during CryptoVerif rewriting an $x_i$ may be instantiated to a more structured expression, say, $c(y_1, y_2)$ with some concatenation symbol $c$, and it may be the case that $f$ destructs $c(y_1, y_2)$ in such a way that the value of $y_2$ is never accessed. To enable CryptoVerif to prove secrecy in these cases we generate some additional facts, similar to the facts for length-regularity of encoders described above.

We generate two types of facts, corresponding to the two cases described above. Let us start with a simpler case where the value of $f$ depends only on lengths of some of the $x_i$ or does not depend on them at all. In order to check this for one of the arguments, say $x_1$, we can introduce two fresh variables $x$ and $x'$ and simply use our solver to check whether $\mathrm{len}(x) = \mathrm{len}(x')$ implies $e_f[x, x_2, \dots, x_n] = e_f[x', x_2, \dots, x_n]$. In practice, an even simpler check is enough: we introduce new variables $l$, $x$, and $x'$, obtain $e'_f$ from $e_f$ by replacing every occurrence of $\mathrm{len}(x_1)$ with $l$ and check whether the expressions $e'_f[x/x_1]$ and $e'_f[x'/x_1]$ are syntactically equal. We repeat this check for every argument $x_i$. If it succeeds, say, for $x_1, \dots, x_k$ then we generate the fact

$$reg_\Gamma(f) = \forall (x_1 \colon T_1, \dots, x_n \colon T_n)\colon f(x_1, \dots, x_n) = f'(Z_{T_1}(x_1), \dots, Z_{T_k}(x_k), x_{k+1}, \dots, x_n),$$

where the functions $Z_T$ returns a string of zeroes of equal length to the length of the argument. The function $f'$ is defined exactly like $f$ in the template, its sole purpose is to prevent CryptoVerif from looping.

The second case is that we cannot erase one of the arguments $x_i$, but we can erase a substring of $x_i$ if we assume that $x_i$ is a result of an encoding operation. During verification CryptoVerif will often be able to rewrite $f(x_1, \ldots, x_n)$ into, say, $f(c(y_1, \ldots, y_m), x_2, \ldots, x_n)$, where $c\colon T'_1 \times \ldots \times T'_m \to T_1$ is a concatenation operator defined by an expression $e_c$. We can now check whether we can erase one of the arguments of $c$, say, $y_1$ in a way similar to above. The only additional step is that we simplify the expression. Let

$$\Phi = \{\text{intype}(y_1, T'_1), \ldots, \text{intype}(y_m, T'_m)\}$$

and let $e = \text{simplify}_\Phi(e_f[e_c[y_1, \ldots, y_m], x_2, \ldots, x_n])$ where simplify is our simplification function defined in section 3.3. Now we can proceed exactly as above—in order to check that we can erase $y_1$ replace every occurrence of $\text{len}(y_1)$ in $e$ by a fresh variable $l$, and let $e'$ be the resulting expression. Now check whether for two fresh variables $y$ and $y'$ the expressions $e'[y/y_1]$ and $e'[y'/y_1]$ are syntactically equal. Repeat for each $y_i$. If the check succeeds for $y_1, \ldots, y_k$ then we generate the fact

$$\begin{aligned} &\mathit{reg}'_\Gamma(f, c) = \forall (y_1\colon T'_1, \ldots, y_m\colon T'_m, x_2\colon T_2, \ldots, x_n\colon T_n)\colon \\ &\quad f'(c(y_1, \ldots, y_m), x_2, \ldots, x_n) = f'(c'(Z_{T'_1}(y_1), \ldots, Z_{T'_k}(y_k), y_{k+1}, \ldots, y_m), x_2, \ldots, x_n). \end{aligned}$$

**Example 4.9** Suppose we have identified two concatenation operators $c_1, c_2\colon \texttt{bounded}_{10} \times \texttt{bounded}_{20} \to \texttt{bounded}_{100}$ defined as

$$\begin{aligned} c_1(y_1, y_2) &= 0x00|\tau_{4u}^{-1}(\text{len}(y_1))|y_1|y_2, \\ c_2(y_1, y_2) &= 0x01|\tau_{4u}^{-1}(\text{len}(y_1))|y_1|y_2, \end{aligned}$$

and consider a condition $\texttt{if}\ f(x_1, x_2)\ \texttt{then}\ P$ where $f\colon \texttt{bitstring} \times \texttt{bitstring} \to \texttt{bool}$ is

$$f(x_1, x_2) = (x_1\{5, x_1\{1, 4\}\} = x_2\{5, x_2\{1, 4\}\}).$$

It is clear that neither $x_1$ nor $x_2$ can be erased. However, consider what happens when we substitute $x_1$ by $c_1(y_1, y_2)$. The resulting condition becomes

$$\begin{aligned} &f(c_1(y_1, y_2)), x_2) = \\ &\quad ((0x00|\tau_{4u}^{-1}(\text{len}(y_1))|y_1|y_2)\{5, (0x00|\tau_{4u}^{-1}(\text{len}(y_1))|y_1|y_2)\{1, 4\}\}) = x_1\{5, x_1\{1, 4\}\}). \end{aligned}$$

This is simplified as follows:

$$\text{simplify}_\Phi(f(c_1(y_1, y_2), x_2)) = y_1, \quad \text{where } \Phi = \{\text{intype}(y_1, \texttt{bounded}_{10}), \text{intype}(y_2, \texttt{bounded}_{20})\}$$

This expression does not depend on $y_2$ at all, so we can add a fact

$$\forall(y_1 \colon \texttt{bounded}_{10}, y_2 \colon \texttt{bounded}_{20}, x_2 \colon \texttt{bitstring}) \colon$$
$$f(c_1(y_1, y_2), x_2) = f(c_1(y_1, Z_{\texttt{bounded}_{20}}(y_2)), x_2).$$

Altogether we generate 4 such facts—with both $c_1$ and $c_2$ in each of the positions $x_1$ and $x_2$.$\square$

## 4.8 Parsing Safety

An application of a parser $f_p(e)$ is called *safe* if the facts in the context imply that $e$ is in the range of some encoder $f$. In such cases, whenever $f$ is injective and $f_p$ matches $f$ at position $i$, we can replace $\mathsf{let}\ x_i = f_p(e)\ \mathsf{in}\ P$ by $\mathsf{let}\ f(x_1, \ldots, x_n) = e\ \mathsf{in}\ P$, which is a CryptoVerif abbreviation for

$$\textbf{let } x_1 = f_1^{-1}(e) \textbf{ in} \ldots \textbf{let } x_n = f_n^{-1}(e) \textbf{ in if } f(x_1, \ldots, x_n) = e \textbf{ then } P$$

where $f_i^{-1}$ are the partial inverses of $f$ provided by injectivity: if $f$ has type $T_1 \ldots T_n \to T$ then each $f_i^{-1}$ is a function from $I(T)$ to $I(T_i)$ such that $f_i(f(b_1, \ldots, b_n)) = b_i$ for all $b_1 \in I(T_1), \ldots, b_n \in I(T_n)$. As discussed in section 4.2, the use of pattern matching allows CryptoVerif to succeed more often. This section describes how we check parsing safety and transform the process to use pattern matching.

We define the following condition that encodes the fact that the value of an expression $e$ that does not contain variables $x_1, \ldots, x_n$ comes from the range of a function $f \colon T_1 \times \ldots \times T_n \to T$:

$$\mathit{inrange}(e, f) = \exists(x_1 \colon T_1, \ldots, x_n \colon T_n) \colon e = f(x_1, \ldots, x_n).$$

We say that an application $f_p(e)$ of a parser $f_p \colon T \to T'$ to an expression $e$ *safely matches* an encoder $f \colon T_1 \times \ldots \times T_n \to T$ at position $i$ in context $\Phi$, if $f$ is injective, $T' = T_i$, and $\Phi \vdash \mathit{inrange}(e, f) \wedge \mathit{parse}_\Gamma(f_p, f, i)$. The condition of injectivity takes the types of the arguments into account: $f$ only needs to be injective for the arguments of types $T_1, \ldots, T_n$. CryptoVerif actually requires poly-injectivity, that is, the inverse functions of $f$ need to be efficiently computable. Clearly all our encoders are poly-injective whenever they are injective.

The rules in Figure 4.12 show the transformation that we apply to the process. It may happen that several encoders are safely matched by a parsing expression if they have overlapping ranges and equal return types. In that case it is safe to choose one encoder arbitrarily, but our implementation terminates with an error to alert the user that more distinctive typing of encoders is necessary. If we cannot show parsing safety for the application of a particular parser, we leave the parsing unchanged and proceed with the rest of the process.

In section 4.7 we described how to check whether $f$ is injective and whether $\mathit{parse}_\Gamma(f_p, f, i)$ is satisfied. We now show how to check the *inrange* condition. Given a bitstring $b$, to check $\mathit{inrange}(b, f)$ we shall construct an explicit parser for every field of $f$ and check that applications

$$
\frac{\begin{array}{l} f_p(e) \text{ safely matches } f \text{ at position } i \text{ in context } \Phi \\ x_j \text{ unused in } P \text{ for } j \neq i \quad \Phi \vdash P \rightsquigarrow \tilde{P} \end{array}}{\Phi \vdash \mathsf{let}\ x_i = f_p(e)\ \mathsf{in}\ P \rightsquigarrow \mathsf{let}\ f(x_1, \ldots, x_n) = e\ \mathsf{in}\ \tilde{P}}
$$

$$
\frac{\Phi \cup \{\phi\} \vdash P \rightsquigarrow \tilde{P}}{\Phi \vdash \mathsf{if}\ \phi\ \mathsf{then}\ P \rightsquigarrow \mathsf{if}\ \phi\ \mathsf{then}\ \tilde{P}} \qquad \frac{\Phi \cup \{\phi\} \vdash P \rightsquigarrow \tilde{P}}{\Phi \vdash \mathsf{assume}\ \phi;\ P \rightsquigarrow \mathsf{assume}\ \phi;\ \tilde{P}}
$$

$$
\frac{\Phi \vdash P \rightsquigarrow \tilde{P}}{\Phi \vdash \lambda P \rightsquigarrow \lambda \tilde{P}}
$$

Figure 4.12: Parsing safety rules.

of those parsers to $b$ are well-defined, that the tag fields in $b$ have expected values, and that the argument fields have expected types. We formalise this idea below.

Given a concatenation function symbol $f\colon T_1 \times \ldots \times T_n \to T$ with $I(f) = [\![e]\!]$ for an encoding expression $e$ with variables $x_1, \ldots, x_n$, we check the fact *inrange*$(b, f)$ as follows. First check that $e$ is *well-formed*: it is required to be of a form $e_1 | \ldots | e_m$, where each of the $e_i$ is either a constant tag, one of the variables $x_1, \ldots, x_n$ or a length of one of these variables. More precisely, $\{1, \ldots, m\} = I_x \cup I_l \cup I_t$ such that

- for all $i \in I_x$ it is $e_i = x_k$ for some $k \leq n$,
- for all $i \in I_l$ it is $e_i = \tau_i^{-1}(\mathrm{len}(x_k))$ for some $k \leq n$ and $\tau_i \in \mathbb{T}_I$,
- for all $i \in I_t$ it is $e_i = t_i$ for some constant bitstring $t_i$.

We require that each variable occurs at most once among the arguments and at most once among the lengths, and that

$$
I_x = \{i \mid e_i = x_j \text{ s.t. } e \text{ contains } \mathrm{len}(x_j)\} \uplus \{m\},
$$

that is, the expression $e$ contains lengths for all parameters except the last one—the missing length can then be derived from knowing the total length of the concatenation. Each field length must precede the field itself: if $e_i = \tau_i^{-1}(\mathrm{len}(x_k))$ and $e_j = x_k$ for some $k$ then $i < j$.

Next we inductively define integer terms $l_1(x), \ldots, l_m(x)$ and bitstring terms $p_1(x), \ldots, p_m(x)$ that would extract field lengths and fields when applied to an expression in the range of $f$. We then apply them to $b$ and check that the results behave as expected. Assume that $p_1, \ldots, p_{i-1}$ and $l_1, \ldots, l_{i-1}$ are defined. Let

$$l_i(x) = \begin{cases} |t_i|, & \text{if } e_i = t_i \in BS, \\ \text{len}(\tau_i), & \text{if } e_i = \tau_i^{-1}(\text{len}(x_k)), \\ \tau_j(p_j(x)), & \text{if } e_i = x_k \text{ and } e_j = \tau_j^{-1}(\text{len}(x_k)), \\ \text{len}(x) - \sum_{j=1}^{m-1} l_j(x), & \text{if } i = m \end{cases}$$

$$p_i(x) = x\left\{\sum_{j=1}^{i-1} l_j(x),\, l_i(x)\right\}.$$

Using these constructs we can check that the fields of the argument are well-defined. We also check that the argument fields are of the correct type and that the tag fields have correct value. Formally, we define

$$\begin{aligned} \mathit{fields}(x, f) &= \bigwedge_{i=1}^{m} \text{defined}(p_i(x)), \\ \mathit{types}(x, f) &= \bigwedge_{i \in I_x} \text{intype}(p_i(x),\, T_j), \text{ when } e_i = x_j, \\ \mathit{tags}(x, f) &= \bigwedge_{i \in I_t} p_i(x) = t_i. \end{aligned}$$

**Theorem 4.2 (Parsing Safety)** *For any well-formed encoder $f$ and bitstring $b$*

$$\mathit{fields}(b, f) \wedge \mathit{types}(b, f) \wedge \mathit{tags}(b, f) \iff \mathit{inrange}(b, f). \qquad \square$$

Proof Assume that $\mathit{fields}(b, f)$, $\mathit{types}(b, f)$, and $\mathit{tags}(b, f)$ are true (since we use a constant $b$, these facts do not contain variables, and so can be evaluated independently of an environment). We show $\mathit{inrange}(b, f)$ by finding $b_1 \in I(T_1), \ldots, b_n \in I(T_n)$ such that $b = [\![e]\!](b_1, \ldots, b_n)$.

For $i = 1, \ldots, m$ let $a_i = [\![p_i(b)]\!]$. By $\mathit{fields}(b, f)$ each $a_i$ is well-defined. By construction of the parsers $p_i$ it is easy to see that $b = a_1|\ldots|a_m$.

For $i \in I_x$ such that $e_i = x_j$ let $b_j = a_i$. By $\mathit{types}(b, f)$ we have $b_1 \in I(T_1), \ldots, b_n \in I(T_n)$. Let $\eta = \{x_1 \mapsto b_1, \ldots, x_n \mapsto b_n\}$. We show that $[\![e]\!]_\eta = a_1|\ldots|a_m$ by showing $[\![e_i]\!]_\eta = a_i$ for all $i \leq m$. Consider the following cases:

- $i \in I_t$ and $e_i = t_i$ for some constant bitstring $t_i$. By $\mathit{tags}(b, f)$ we have $a_i = t_i$ as well.

- $i \in I_x$ and $e_i = x_j$. Then $\eta(x_j) = b_j = a_i$.

- $i \in I_l$ and $e_i = \tau_i^{-1}(\text{len}(x_k))$ for some $k \leq n$. Let $j \in I_x$ such that $e_j = x_k$. Expanding

the definitions of $p_j$ and $l_j$ we see that $p_j$ is of the form

$$p_j(x) = x\left\{\sum_{k=1}^{j-1} l_k(x),\ \tau_i(p_i(x))\right\},$$

so that

$$\begin{aligned}\llbracket e_i \rrbracket_{\eta_x} =& \tau_i^{-1}(|\eta(x_k)|) = \tau_i^{-1}(|a_j|)\\ =& \tau_i^{-1}(|\llbracket p_j(b)\rrbracket|) = \tau_i^{-1}(\llbracket \tau_i(p_i(b))\rrbracket) = \tau_i^{-1}(\tau_i(\llbracket p_i(b)\rrbracket)) = a_i\end{aligned}$$

The reverse implication is straightforward. ■

**Example 4.10** In our RPC-enc example all parser applications are safe. For instance, consider application of *parse4* in line 48 in figure 4.6. The only encoder that matches the type of *parse4* is $conc1 : \mathtt{fixed}_{1024} \times \mathtt{fixed}_{16} \to \mathtt{bounded}_{1045}$. To check that the argument *msg11* of *parse4* comes from the range of *conc1* we apply the construction above to the defining expression $\text{"}\mathtt{p}\text{"}|\tau_{4u}^{-1}(\mathrm{len}(x))|x|y$ of *conc1* and check that the context at the call site implies the following conditions:

$$\begin{aligned}
&\mathit{fields}(\mathit{msg11}, \mathit{conc1}) =\\
&\qquad \mathrm{defined}(\mathit{msg11}\{0,1\}) \wedge \mathrm{defined}(\mathit{msg11}\{1,4\})\\
&\quad \wedge \mathrm{defined}(\mathit{msg11}\{5, \tau_{4u}(\mathit{msg11}\{1,4\})\})\\
&\quad \wedge \mathrm{defined}(\mathit{msg11}\{5 + \tau_{4u}(\mathit{msg11}\{1,4\}), \mathrm{len}(\mathit{msg}) - (5 + \tau_{4u}(\mathit{msg11}\{1,4\}))\}),\\
&\mathit{types}(\mathit{msg11}, \mathit{conc1}) =\\
&\qquad \mathrm{intype}(\mathit{msg11}\{5, \tau_{4u}(\mathit{msg11}\{1,4\})\}, \mathtt{fixed}_{1024})\\
&\quad \wedge \mathrm{intype}(\mathit{msg11}\{5 + \tau_{4u}(\mathit{msg11}\{1,4\}), \mathrm{len}(\mathit{msg}) - (5 + \tau_{4u}(\mathit{msg11}\{1,4\}))\}, \mathtt{fixed}_{16}),\\
&\mathit{tags}(\mathit{msg11}, \mathit{conc1}) = (\mathit{msg11}\{0,1\} = \text{"}\mathtt{p}\text{"}).
\end{aligned}$$

□

It is easy to see that a similar construction can be applied to an arbitrary robustly injective encoder (definition 4.11). We shall make use of this fact in section 4.10 where we shall create generalized parsers that always check whether their argument is in the range of some robustly injective encoder.

We now formalise the soundness of parsing safety analysis.

**Lemma 4.6** *If $\emptyset, Q \rightsquigarrow \tilde{Q}$ by the rules in figure 4.12 then $Q \lesssim \tilde{Q}$.* □

Proof For two executing processes $(\eta, P)$ and $(\tilde{\eta}, \tilde{P})$ let $(\eta, P) \lesssim (\tilde{\eta}, \tilde{P})$ if $\tilde{\eta} = \eta \cup \eta_*$ where all variables in $\eta_*$ do not occur in $\tilde{P}$ and there exists a fact set $\Phi$ such that $\eta \models \Phi$ (and therefore $\tilde{\eta} \models \Phi$) and $\Phi \vdash P \rightsquigarrow \tilde{P}$. We show that $\lesssim$ satisfies definition 2.11. The only non-trivial case

$$\frac{f \text{ is a formatting function} \quad \Gamma(f) = T_1 \times \ldots \times T_n \to T \qquad \Gamma, \Phi, T_i \vdash e_i \rightsquigarrow \tilde{e}_i \quad \Phi \cup \text{typefacts}(\Gamma) \vdash \text{intype}(\text{expand}(f(e_1, \ldots, e_n)), T)}{\Gamma, \Phi, T \vdash f(e_1, \ldots, e_n) \rightsquigarrow \text{tcast}_{\texttt{bitstringbot} \to T}(f(\tilde{e}_1, \ldots, \tilde{e}_n))} \quad (\text{TCFormat'})$$

Figure 4.13: A modified typechecking rule for formatting functions.

corresponds to the rule for let bindings in which

$$\begin{aligned} P &= \mathsf{let}\ x_i = f_p(e)\ \mathsf{in}\ P', \\ \tilde{P} &= \mathsf{let}\ f(x_1, \ldots, x_n) = e\ \mathsf{in}\ \tilde{P}' \\ &= \mathsf{let}\ x_1 = f_1^{-1}(e)\ \mathsf{in}\ \ldots\ \mathsf{let}\ x_n = f_n^{-1}(e)\ \mathsf{in}\ \mathsf{if}\ f(x_1, \ldots, x_n) = e\ \mathsf{then}\ \tilde{P}'. \end{aligned}$$

The executing process $(\eta, P)$ reduces to $(\eta', P')$ with $\eta' = \eta\{x_i \mapsto [\![f_p(e)]\!]_\eta\}$. By the premise of the rule we have $\tilde{\eta} \models \mathit{inrange}(e, f)$, so that the test condition in $\tilde{P}$ is true and $\tilde{P}$ reduces to $(\tilde{\eta}', \tilde{P}')$ with

$$\tilde{\eta}' = \tilde{\eta}\{x_1 \mapsto [\![f_1^{-1}(e)]\!]_{\tilde{\eta}}, \ldots, x_n \mapsto [\![f_n^{-1}(e)]\!]_{\tilde{\eta}}\}.$$

By the premise of the rule $\eta \models \mathit{parse}_\Gamma(f_p, f, e)$ so that $[\![f_p(e)]\!]_\eta = [\![f_1^{-1}(e)]\!]_\eta = [\![f_1^{-1}(e)]\!]_{\tilde{\eta}}$. Thus

$$\tilde{\eta}' = \eta' \cup \eta_* \cup \{x_j \mapsto [\![f_j^{-1}(e)]\!]_{\tilde{\eta}} \mid j \leq n,\ j \neq i\}\,.$$

By the premise of the rule $\Phi \vdash P' \rightsquigarrow \tilde{P}'$, so that $(\eta', P') \lesssim (\tilde{\eta}', \tilde{P}')$. ■

## 4.9 Type-Safety of Formatting functions

We come back to the problem that we left unresolved in previous sections—making sure that our formatting functions are type-safe—given inputs of the right type they should always produce outputs of the right type, as specified by the typing environment $\Gamma$ inferred in section 4.5. Using the notation introduced in definition 4.2 we would like to make sure that $I, f \models \Gamma$ for each formatting function $f$. We need to follow different approaches for parsers and encoders. Encoders are expected to be type-safe without any further modification, so we simply check that they are. To check that $I, f \models T_1 \times \ldots \times T_n \to T$ we ask our solver to prove that

$$\{\text{intype}(x_1, T_1), \ldots, \text{intype}(x_n, T_n)\} \vdash \text{intype}(\text{expand}(f(x_1, \ldots, x_n)), T),$$

where expand replaces the application of $f$ by its definition, as explained in section 4.6.

We cannot expect parsers to be type-safe since they rely on the internal structure of the bitstring that they parse and this structure is not captured by the types we use. Therefore we replace parsers by strengthened versions that coerce their result to the right type. This is justified by the typechecking we do in section 4.6. Consider the typechecking rules in figure 4.11 with one slight modification to the rule (TCFormat), shown in figure 4.13: when we check an application $f(e_1, \ldots, e_n)$ of a formatting function $f\colon T_1 \times \ldots \times T_n \to T$, we wrap it in a typecast to become $\mathrm{tcast}_{\texttt{bitstringbot} \to T}(f(\tilde{e}_1, \ldots, \tilde{e}_n))$. It is clear that this change does not affect the behaviour of the process—the rule checks that $f(e_1, \ldots, e_n)$ already belongs to the type $T$, and so the typecast leaves the argument unchanged. Thus it is safe replace every parser $f\colon T_1 \times \ldots \times T_n \to T$ by a fresh parser $f'$ such that $I(f') = \mathrm{tcast}_{\texttt{bitstringbot} \to T} \circ I(f)$. Such a strengthening makes all our parsers type-safe.

In addition to preserving the behaviour of the process parser strengthening also preserves the validity of our fact set $\Phi$ inferred in section 4.7. The only facts that refer to parsers are the parsing equations, and it is easy to see that those hold for the modified parsers as well. The need to preserve the validity of $\Phi$ further justifies different treatment of encoders and parsers: if we were to strengthen the encoders in the same way, instead of explicitly checking their type-safety, the facts in $\Phi$ would in general not be true any more.

## 4.10 Generalized Formatting

The verification result of CryptoVerif (theorem 4.1) is formulated with respect to a family of interpretations $\tilde{I}_k$, one for for each value $k \in \mathbb{N}$ of the security parameter. In this section we show that given the model $(Q, \Gamma, \Phi)$ and the interpretation $I$ that we constructed so far, we can generalize $I$ to such a family $(\tilde{I}_k)_{k \in \mathbb{N}}$ in a way that satisfies the conditions of theorem 4.1: for each $k \in \mathbb{N}$ the interpretation $\tilde{I}_k$ agrees with the interpretation $\tilde{I}^c_k$ of the cryptographic functions and additionally $\tilde{I}_k|_Q \models \Gamma$ and $\tilde{I}_k \models \Phi$ (lemma 4.7).

In order for the cryptographic functions in $\tilde{I}^c_k$ to satisfy their asymptotic security properties it must be the case that the sizes of their arguments and outputs grow with $k$. We shall therefore make an assumption that for each $n \in \mathbb{N}$ it is $\tilde{I}^c_k(\texttt{fixed}_n) = \{0x00, \ldots, 0xFF\}^{kn}$ and $\tilde{I}^c_k(\texttt{bounded}_n) = \{0x00, \ldots, 0xFF\}^{\leq kn}$, the sets of bitstrings with lengths equal or at most $kn$. The interpretations of types `bitstring`, `bitstringbot`, and `bool` will stay the same across all values of $k$.

We start by summarizing the shape of the interpretation that we have constructed in the previous sections. We shall make one additional assumption that all our encoders are robustly injective (definition 4.10). This is a reasonable assumption insofar as an encoder that is not robustly injective should be considered ill-designed. All the encoders we have encountered so far satisfy this property.

**Definition 4.12 (Admissible Interpretations)** Given an interpretation $I$ and a CryptoVerif model $(Q, \Gamma, \Phi)$ we call the tuple $(I, Q, \Gamma, \Phi)$ a *constrained interpretation* if $I|_Q \models \Gamma$, and $I \models \Phi$. A constrained interpretation $(I, Q, \Gamma, \Phi)$ is *admissible* if

- All types in $\Gamma$ are `bool`, `bitstringbot`, `bitstring`, $\texttt{fixed}_n$, or $\texttt{bounded}_n$ for $n \in \mathbb{N}$.

- Each function $f \in \mathrm{dom}(\Gamma)$ is of one of the following kinds:
  - A cryptographic function.
  - A robustly injective encoder $f_c$ with $\Gamma(f_c) = T_1 \times \ldots \times T_n \to T$ where none of the types $T$, $T_1, \ldots, T_n$ is `bool` or `bitstringbot`.
  - A parser $f_p$ with $\Gamma(f_p) = T \to T'$ where neither of the types $T$ or $T'$ is `bool` or `bitstringbot`.
  - An auxiliary test function $f$ such that $\Gamma(f) = T_1 \times \ldots \times T_n \to$ `bool` for some types $T_1, \ldots, T_n$.
  - A cast function $\mathrm{tcast}_{T \to T'}$ with $\Gamma(\mathrm{tcast}_{T \to T'}) = T \to T'$.
  - A constant $0_T$ with $\Gamma(0_T) = T$ for some fixed type $T$.
- $\Phi$ only contains facts of the form described in section 4.7, that is, facts of the form $\mathit{parse}_\Gamma(f_c, f_p, i)$, $\mathit{inj}_\Gamma(f_c)$, $\mathit{disj}_\Gamma(f_c, f'_c)$, $\mathit{eq}_\Gamma(f_c, f'_c)$, $\mathit{reg}_\Gamma(f_c)$, $\mathit{reg}_\Gamma(f)$, $\mathit{reg}'_\Gamma(f, f_c)$ $\mathit{zero}_\Gamma(T)$, and $\mathit{cast}_\Gamma(T, T')$ with a parser $f_p$, an encoder $f_c$, an auxiliary test function $f$, and types $T$ and $T'$. If $\mathit{disj}_\Gamma(f_c, f'_c) \in \Phi$ then $f_c$ and $f'_c$ must be robustly disjoint (definition 4.11). If $\mathit{eq}_\Gamma(c, c') \in \Phi$ then $f_c$ and $f'_c$ must have equal defining expressions. □

There are two reasons why the interpretation $I$ is not suitable for arbitrary security parameters as is. The first reason is that with growing input types the sizes of length fields in encoders may become too small to hold the length of the inputs. The other reasons is that an encoder may need to have a fixed-length return type. Since the return type is growing with the security parameter, so must the length of the encoding. We solve both of these issues by widening each encoder such that for a security parameter $k$ every tag field is replaced by $k$ copies of itself and every length field is made $k$ times wider.

**Definition 4.13 (Generalized Interpretations)** Consider an admissible constrained interpretation $(I, \Gamma, \Phi)$. We define the *generalized interpretation* $\tilde{I}_k$ for each $k \in \mathbb{N}$ as follows:

- $\tilde{I}_k(\mathtt{fixed}_n) = \{0x00, \ldots, 0xFF\}^{kn}$ and $\tilde{I}_k(\mathtt{bounded}_n) = \{0x00, \ldots, 0xFF\}^{\leq kn}$. The interpretations of the types `bitstring`, `bitstringbot`, and `bool` will stay the same as their interpretation by $I$.
- For every cryptographic function $f$ let $\tilde{I}_k(f) = \tilde{I}^c_k(f)$.
- For every encoder $f_c$ where $I(f_c) = [\![e_1|\ldots|e_n]\!]$ we let $\tilde{I}_k(f_c) = [\![\tilde{e}_1|\ldots|\tilde{e}_n]\!]$ where for each $i$ the field $\tilde{e}_i$ is defined as follows:
  - If $e_i = t \in BS$ then $\tilde{e}_i = t|\ldots|t$ ($k$ times).
  - If $e_i = \tau^{-1}(\mathrm{len}(x))$ for a variable $x$ where $\tau$ is a (signed or unsigned) integer type of length $l$ then $\tilde{e}_i = \tilde{\tau}^{-1}(\mathrm{len}(x))$ where $\tilde{\tau}$ is an unsigned type of length $kl$.
  - If $e_i = x$ for a variable $x$ then $\tilde{e}_i = x$ as well.

- For every parser $f_p$ with $\Gamma(f_p) = T \to T'$ let

$$\tilde{I}_k(f_p) = \begin{cases} b_i, & \text{if } b = c(b_1, \ldots, b_n) \text{ for some encoder } c \text{ and } \mathit{parse}_\Gamma(f_c, f_p, i) \in \Phi, \\ 0_{T'}, & \text{otherwise.} \end{cases}$$

  The check whether $b$ is in the range of an encoder $f_c$ can be performed efficiently in the way described in section 4.8 since we assumed that all the encoders are robustly injective.

- For every auxiliary test function $f$ let $\tilde{I}_k(f)$ be the function that returns true for every input.

- The cast functions and zeroes are updated according to the change in the interpretation of their types:

$$\tilde{I}_k(\mathrm{tcast}_{T \to T'}) = \mathrm{tcast}_{\tilde{I}_k(T) \to \tilde{I}_k(T')} \qquad \tilde{I}_k(0_T) = 0_{\tilde{I}_k(T)} \qquad \square$$

**Example 4.11** Consider an encoder $f_c$ with $I(f_c)(x, y) = \tau_{4u}^{-1}(\mathrm{len}(x))|\texttt{0x01}|x|y$. The generalization of $f_c$ for $k = 4$ would be $\tilde{I}_4(f_c)(x, y) = \tau_{16u}^{-1}(\mathrm{len}(x))|\texttt{0x01010101}|x|y$. □

It is straightforward to check that the generalized interpretations still satisfy $\Phi$ and $\Gamma$, and so the following lemma holds.

**Lemma 4.7** *Let $(I, Q, \Gamma, \Phi)$ be an admissible constrained interpretation and let $(\tilde{I}_k)_{k \in \mathbb{N}}$ be the family of generalized interpretations as in definition 4.13. Then $(\tilde{I}_k)_{k \in \mathbb{N}}$ is an efficient family such that for each $k \in \mathbb{N}$ the interpretation $\tilde{I}_k$ agrees with $\tilde{I}_k^c$ for cryptographic functions and $\tilde{I}_k \models \Phi$ as well as $\tilde{I}_k|_Q \models \Gamma$.* □

In definition 4.12 we make some assumptions about the structure of the encoders beyond what is provided by the fact set $\Phi$. For instance, when two encoders are disjoint according to $\Phi$, we require that they are robustly disjoint (have tag fields that are always aligned and neither tag is a prefix of the other—definition 4.11). This is a necessity since our construction does not preserve disjointness as such. For instance, consider two encoders $f_c$ and $f_c'$ (without arguments) given by

$$I(f_c)(x) = \tau_{2u}^{-1}(\mathrm{len}(x))|x \qquad I(f_c') = \texttt{0b11000000}.$$

It is clear that these two encoders are disjoint: there is no $x$ such that $f_c(x) = f_c'$. However if we apply our construction in definition 4.13 then for $k = 2$ we get

$$\tilde{I}_2(f_c)(x) = \tau_{4u}^{-1}(\mathrm{len}(x))|x \qquad \tilde{I}_2(f_c') = \texttt{0b1100000011000000}$$

and we see that $\tilde{I}_2(f_c)(\texttt{0b000011000000}) = \tilde{I}_2(f_c')$.

Definition 4.12 does not require the parsers to be of any particular shape. Instead we define the behaviour of the parsers based on the $parse_\Gamma$ facts in $\Phi$. This means that the generalization procedure is fully applicable to the interpretation obtained by strengthening the parsers in section 4.9.

## 4.11 Summary and Soundness

We are now ready to prove soundness of our translation from IML processes to CryptoVerif models. We start by summarizing the translation procedure.

**Definition 4.14 (IML to CryptoVerif Translation)** Consider a well-formed input IML process $Q$. We say that $Q$ *yields a CryptoVerif model* $(\tilde{Q}, \Gamma, \Phi)$ if there exist processes $Q_1, \ldots, Q_4$ such that

1. $Q_1$ is the formatting-normal form of $Q$ (section 4.3).
2. $Q_2$ is the formatting abstraction of $Q_1$, that is, $Q_1 \rightsquigarrow Q_2$ by the rules in figure 4.8 (section 4.4).
3. $\Gamma$ is the result of type inference of function types from $Q$ (section 4.5).
4. $Q_3$ is the result of typechecking $Q_2$ with respect to $\Gamma$, that is, $\Gamma, \emptyset \vdash Q_2 \rightsquigarrow Q_3$ by the rules in figure 4.11 (section 4.6).
5. $\Phi$ is the result of fact inference (section 4.7).
6. $Q_4$ is the result of parsing safety analysis of $Q_3$, that is, $\emptyset \vdash Q_3 \rightsquigarrow Q_4$ by the rules in figure 4.12 (section 4.8).
7. $I, f \models \Gamma$ for every encoder $f$ (section 4.9).
8. $\tilde{Q}$ is obtained from $Q_4$ by removing inline assumptions. □

We need to provide a relation between IML security (definition 2.2) and CryptoVerif security (definition 4.3). The only difference between the IML and the CryptoVerif semantics, as pointed out in section 4.1 is the fact that CryptoVerif processes are executed with respect to a typing environment and the execution stops whenever a function is passed an argument of a wrong type. Thus we shall be able to prove that the CryptoVerif semantics coincides with the IML semantics for a process $Q$ whenever the process is well-typed ($\Gamma \vdash Q$) and the interpretation is type-safe ($I|_Q \models \Gamma$) as in definition 4.2. We start by defining a semantic consequence of these two conditions: a process is *type-safe* if no function is ever called with input of a wrong type.

**Definition 4.15 (Type-Safe Processes)** An executing process $(\eta, Q)$ is *type-safe* with respect to a typing environment $\Gamma$ if for each top-level expression $f(e_1, \ldots, e_n)$ in $Q$:

- $f \in \text{dom}(\Gamma)$.
- If $\Gamma(f) = T_1 \times \ldots \times T_n \to T$ then $[\![e_i]\!]_\eta \in I(T_i)$ for all $i \leq n$.

An input process $Q_0$ is *type-safe* with respect to a typing environment $\Gamma$ if for each semantic configuration $(\eta, P), \Omega$ reachable from $\text{initConfig}(Q_0)$ both $(\eta, P)$ and each $(\eta, Q) \in \Omega$ are type-safe with respect to $\Gamma$. □

**Lemma 4.8** *If a well-formed input CryptoVerif process $Q$ is well-typed with respect to a typing environment $\Gamma$ and $I|_Q \models \Gamma$ then $Q$ is type-safe with respect to $\Gamma$.* □

PROOF The proof is straightforward by structural induction on the derivation $\Gamma \vdash Q$, similar to Blanchet [2008, Appendix B2]. ■

**Lemma 4.9** *If a well-formed input CryptoVerif process $Q$ is well-typed with respect to a typing environment $\Gamma$ and $I|_Q \models \Gamma$ then for any trace property $\rho$ we have* $\text{insec}(Q, \rho) = \text{cvinsec}(Q, I, \Gamma, \rho)$. □

PROOF Let $Q$ and $\Gamma$ be as in the statement of the lemma. By lemma 4.8 $Q$ is type-safe with respect to $\Gamma$. Let $\mathbb{C}$ be a reachable semantic configuration with respect to IML semantics, let $(\eta, P)$ be an executing process in $\mathbb{C}$ (either the output process or one of the input processes) and $e$ an expression in $P$. The only difference in the CryptoVerif and the IML semantics are the rules for expression evaluation. It therefore suffices to show that $e, \eta \Downarrow [\![e]\!]_\eta$—then obviously any reduction of $\mathbb{C}$ by the IML semantics will also be a reduction by the CryptoVerif semantics.

We prove the statement by structural induction on $e$. The case $e = x \in \mathit{Var}$ is trivial. Let $e = f(e_1, \ldots, e_n)$. Assume that $\Gamma(f) = T_1 \times \ldots \times T_n \to T$. By type-safety of $Q$ we have $[\![e_i]\!]_\eta \in I(T_i)$ for each $i \leq n$. Applying the induction hypothesis we conclude $e, \eta \Downarrow [\![e]\!]_\eta$. ■

We are now ready to present the main result of this chapter.

**Theorem 4.3 (IML to CryptoVerif Translation is Sound)** *Consider a well-formed input IML process $Q$ that satisfies inline assumptions. Assume that $Q$ yields a CryptoVerif model $(\tilde{Q}, \Gamma, \Phi)$ and CryptoVerif verifies this model against a correspondence property $\rho$. Assume that $I$ agrees with $\tilde{I}^c_{k_0}$ for cryptographic functions. Then*

1. *For any attacker $Q_A$ (a well-formed process that is well-typed with respect to $\Gamma$ and does not contain events) there is a sufficiently large polynomial $p$ such that*

$$\text{insec}(Q_A|Q, \rho) \leq \text{cvbound}_p(Q_A|\tilde{Q}, \Gamma, \Phi, \rho, k_0).$$

2. *The function $\text{cvbound}_p(Q_A|\tilde{Q}, \Gamma, \Phi, \rho, k)$ is well-defined for all $k \in \mathbb{N}$ and is negligible in $k$ for any attacker $Q_A$ and a sufficiently large polynomial $p$.* □

Proof Obtain processes $Q_1, \ldots, Q_4$ from definition 4.14 and choose a trace property $\rho$ and an attacker $Q_A$. Conversion to formatting-normal form is sound (lemma 4.2) and clearly formatting abstraction is sound as well, thus $Q \lesssim Q_1 \lesssim Q_2$. By lemmas 4.4 and 4.6 $Q_2 \lesssim Q_3 \lesssim Q_4$ and finally, since $Q$ satisfies inline assumptions, $Q_4 \lesssim \tilde{Q}$. Thus overall $Q \lesssim \tilde{Q}$. By theorems 2.2 and 2.3

$$\text{insec}(Q_A|Q, \rho) \leq \text{insec}(Q_A|\tilde{Q}, \rho). \tag{4.37}$$

Condition 7 in the translation makes sure that all encoders $f$ are type-safe: $I, f \models \Gamma$. As described in section 4.9, we can strengthen the parsers to make them type-safe as well. Let $(\tilde{Q}^*, \Gamma^*, \Phi^*)$ be the model in which every parser has been replaced by the strengthened version. Then $I|_{Q^*} \models \Gamma^*$ and $I \models \Phi^*$ and the behaviour of the process is preserved:

$$\text{insec}(Q_A|\tilde{Q}, \rho) = \text{insec}(Q_A|\tilde{Q}^*, \rho). \tag{4.38}$$

By lemma 4.5 $\tilde{Q}$ is well-typed with respect to $\Gamma$, therefore $\tilde{Q}^*$ is well-typed with respect to $\Gamma^*$. The attacker $Q_A$ is assumed to be well-formed and well-typed with respect to $\Gamma$. We can assume that $Q_A$ does not use the parsing symbols used by $\tilde{Q}$, therefore $Q_A$ is well-typed with respect to $\Gamma^*$. Applying lemma 4.9 we get

$$\text{insec}(Q_A|\tilde{Q}^*, \rho) = \text{cvinsec}(Q_A|\tilde{Q}^*, I, \Gamma^*, \rho). \tag{4.39}$$

Since the interpretation $I$ is efficient (we assume efficiency for cryptographic functions and all our formatting functions are efficient by construction) there exists a polynomial $p$ such that

$$\text{cvinsec}(Q_A|\tilde{Q}^*, I, \Gamma^*, \rho) \leq \text{cvbound}_p(Q_A|\tilde{Q}^*, \Gamma^*, \Phi^*, \rho, k_0). \tag{4.40}$$

Finally, since the models $(\tilde{Q}, \Gamma, \Phi)$ and $(\tilde{Q}^*, \Gamma^*, \Phi^*)$ are the same up to renaming of symbols and use the same cryptographic functions, we have

$$\text{cvbound}_p(Q_A|\tilde{Q}^*, \Gamma^*, \Phi^*, \rho, k_0) = \text{cvbound}_p(Q_A|\tilde{Q}, \Gamma, \Phi, \rho, k_0). \tag{4.41}$$

Equations (4.37) to (4.41) together prove statement (1) of the theorem.

In order to show that $\text{cvbound}_p(Q_A|\tilde{Q}, \Gamma, \Phi, \rho, k)$ is defined for all $k \in \mathbb{N}$ we need to prove that there is at least one interpretation for each $k \in \mathbb{N}$ that satisfies the definition of cvbound. The tuple $(I, \tilde{Q}^*, \Gamma^*, \Phi^*)$ forms an admissible constrained interpretation (definition 4.12). Applying lemma 4.7 we obtain a family of interpretations $(\tilde{I}_k)_{k \in \mathbb{N}}$ such that the interpretation $\tilde{I}_k$ satisfies the definition of cvbound for $k$. Our construction is efficient, so we can choose the polynomial $p$ large enough to allow each function in $\tilde{I}_k$ enough time to execute. Statement (2) of the theorem now follows by lemma 4.1. ■

Theorem 4.3 can be summarized as follows: Assume that an IML process $Q$ translates to a CryptoVerif model $(\tilde{Q}, \Gamma, \Phi)$ and CryptoVerif successfully verifies $(\tilde{Q}, \Gamma, \Phi)$ against a trace property $\rho$. Then $Q$ is at least as secure as $\tilde{Q}$ with respect to $\rho$ and $\tilde{Q}$ can be made as secure as necessary simply by increasing $k_0$. The shape of the formatting functions is not relevant as long as they satisfy $\Gamma$ and $\Phi$. Section 4.10 shows one way how such formatting functions can be constructed for every value of the security parameter.

# Chapter 5

# Model Verification in Symbolic Setting

*Symbolic models of cryptography* view cryptographic messages as abstract terms that possess no properties except the desired ones [Dolev and Yao, 1983]. For instance, the only way to extract information from an encrypted message would be to decrypt it, having a valid key. In other words, symbolic models assume unbreakable cryptography.

ProVerif [Blanchet, 2009, 2014] is a tool that verifies security of processes with respect to a symbolic model. It inputs a process $Q$, the desired trace property $\rho$, and a set $R$ of rewriting rules that model cryptography in the formal model, such as $\mathit{decrypt}(\mathit{encrypt}(m, k), k) = m$. It then checks whether $Q$ *symbolically satisfies* $\rho$ with respect to $R$.

We can always take a process resulting from the CryptoVerif translation in chapter 4 and analyse this process in the symbolic model with ProVerif. What such an analysis implies about the computational security of the process is a matter of active research. Under some circumstances it is possible to make a precise computational statement based on a symbolic result. This chapter shows how to apply a set of *computational soundness conditions* derived by Backes et al. [2009] in their *CoSP* framework in order to obtain such a statement.

We start by reviewing the result of Backes et al. [2009] in section 5.1. Their work defines a notion of computational execution, called *computational pi-execution*, which we review in section 5.1.1. This notion is different from ours in that the attacker is not a process, but an external machine that interacts with the executing protocol process. The computational soundness result itself is presented in section 5.1.3. It applies to a set of primitives that include an IND-CCA2 encryption operation. More interestingly for us, the result lists several properties that all encoders and parsers in the process must satisfy: the encoders must be defined for all inputs, the parsers must check correctness of the message format before parsing, and the ranges of all encoders must be pairwise disjoint.

There are three things that we need to accomplish in order to formulate a security result for an IML process. First, we need to transform the process into a pi process that satisfies some syntactic conditions required by the result of Backes et al. [2009]. This is done by the transformations described in sections 5.3 and 5.4.

Second, we need to make sure that the operations used by the process satisfy the encryption soundness conditions, so we can apply the soundness result and obtain a security statement for the computational pi execution. This is described in section 5.5. The condition that the encoders must be defined for bitstrings of arbitrary length is the most problematic one, as it cannot be satisfied by implementations that use a fixed-size length field. To deal with this we specify a relaxed set of soundness conditions that only need to be satisfied for well-typed inputs. In order to check these conditions, we can make use of a fact set $\Phi$ obtained as in section 4.7. We can generalize the set of primitives that satisfies the relaxed conditions to a set of primitives that satisfies full conditions in such a way that the generalized primitives agree with the primitives used by the process for well-typed inputs. We then typecheck the process (using the same method as in chapter 4) to make sure that formatting inputs in all runs will indeed be well-typed. Thus if $Q$ is a process that represents our implementation, we show that we can generalise and strengthen the primitives in $Q$ to yield a process $Q^*$ that satisfies the computational soundness conditions such that $Q \lesssim Q^*$.

Finally, we need to show that security in the pi model implies security in the IML model. We prove this in section 5.6 by demonstrating how to take any attack represented by an IML process and turn it into an attack against the computational pi execution.

The full translation from IML to ProVerif is summarized in section 5.2 and the remaining sections fill in the details. The main theoretical result of this chapter is theorem 5.2 (section 5.7). The theorem states that if an IML process $Q$ successfully translates to a ProVerif model $\tilde{Q}$ and certain soundness conditions are satisfied then there exists a function $f$ of the security parameter $k$ such that the insecurity of $Q$ can be bounded from above by $f(k_0)$ and $f$ is negligible in $k$.

The computational result from the ProVerif analysis is both weaker and harder to obtain than the result we get from using CryptoVerif directly. We are yet to see a protocol that would support a computational soundness result and not be amenable to direct CryptoVerif verification. The value of the work presented in this chapter is therefore less practical that in chapter 4, but we see this work as a validation for applicability of our verification approach: our process transformations could be adapted to a new setting almost without change.

We feel that ProVerif is more useful in protocol design, where it can be used to rapidly check ideas for logical flaws. Mature protocols with existing implementations would profit more from direct verification with CryptoVerif using our method in chapter 4.

$b \in BS,\ x, i \in Var,\ f \in Ops$

| | |
|---|---|
| $e \in PExp ::=$ | expression |
| $x$ | variable |
| $f(e_1, \ldots, e_n)$ | constructor/destructor |
| $Q ::=$ | input process |
| $0$ | nil |
| $!^{i \le N} Q$ | replication |
| $Q \vert Q'$ | parallel composition |
| $\mathsf{in}(c[e_1, \ldots, e_n], x);\ P$ | input |
| $P ::=$ | output process |
| $\mathsf{new}\ x \colon \mathtt{nonce};\ P$ | randomness |
| $\mathsf{out}(c[e_1, \ldots, e_n], e);\ Q$ | output |
| $\mathsf{event}\ ev();\ P$ | event |
| $\mathsf{if}\ e =_\bot e'\ \mathsf{then}\ P\ [\mathsf{else}\ P']$ | conditional |
| $\mathsf{let}_\bot\ x = e\ \mathsf{in}\ P\ [\mathsf{else}\ P']$ | evaluation |

Figure 5.1: The syntax of the applied pi calculus.

## 5.1 Review—CoSP

### 5.1.1 The Applied Pi Calculus

We start by reviewing the calculus used by Backes et al. [2009] to formalize their computational soundness result. It is a variant of the applied pi calculus [Blanchet et al., 2008] understood by the ProVerif tool [Blanchet, 2009]. The calculus is not a syntactic subset of IML, but in section 5.2 we shall show how we can embed it within IML by introducing appropriate syntactic sugar.

The syntax of the calculus is shown in figure 5.1. The main difference from IML is the changed let-binding rule. The binding is now allowed to have an else-branch which is taken whenever the expression evaluates to $\bot$. We write $\mathsf{let}_\bot$ to distinguish this construct (we shall call it a *strong let*) from a normal IML let. As usual, the missing else-branch in either if or let is an abbreviation for else $\mathsf{out}(yield[i_1, \ldots, i_n]; \varepsilon)$; 0 where $i_1, \ldots, i_n$ are the replication indices under which the process occurs. The only form of the conditional expression is $e =_\bot e'$ that evaluates to $\bot$ if either $e$ or $e'$ evaluates to $\bot$. We shall call $=_\bot$ a *strong equality check*.

Compared to IML, the following restrictions are necessary to support the computational soundness result in section 5.1.3:

- We only allow nonces of one special type, `nonce`. We shall later assume, following Backes et al. [2009], that the set of nonces is disjoint from the ranges of all the functions in *Ops*. Thus in the implementation nonces need to be tagged.

$$\frac{x \in \mathrm{dom}(\eta)}{[\![x]\!]^{\pi}_{I,\eta} = \eta(x)} \qquad \frac{[\![e_1]\!]^{\pi}_{I,\eta} = b_1 \in BS \quad \ldots \quad [\![e_n]\!]^{\pi}_{I,\eta} = b_n \in BS}{[\![f(e_1, \ldots, e_n)]\!]^{\pi}_{I,\eta} = I(f)(b_1, \ldots, b_n)}$$

Figure 5.2: The evaluation of pi expressions with respect to an interpretation $I$. We let $[\![e]\!]^{\pi}_{I,\eta} = \bot$ if neither rule applies.

- Events cannot contain parameters. This is a limitation of the result of Backes et al. [2009]—events there are fixed bitstrings.

The result of evaluating a pi expression $e$ with respect to an interpretation $I$ and an environment $\eta$ will be denoted with $[\![e]\!]^{\pi}_{I,\eta}$. The evaluation rules are shown in figure 5.2. We make explicit the dependence on the interpretation since we shall allow the interpretation to vary depending on the security parameter. The main difference from the IML evaluation rules (figure 2.2) is that failures shortcut—the whole expression evaluates to $\bot$ whenever any subexpression evaluates to $\bot$.

The execution of pi calculus is defined with respect to a security parameter $k \in \mathbb{N}$ and assumes a family of interpretations $(\tilde{I}_k)_{k\in\mathbb{N}}$ (often shortened to $\tilde{I}$ in the following). The evaluation is carried out by a single machine that is supposed to work across all security parameters. This places additional efficiency conditions on $(\tilde{I}_k)_{k\in\mathbb{N}}$—it must be implementable by a single machine for all values of $k$. This is captured by the following definition.

**Definition 5.1 (Uniformly Efficient Interpretations)** A family of interpretations $(\tilde{I}_k)_{k\in\mathbb{N}}$ is *uniformly efficient* if it is efficient in the sense of definition 4.4 and additionally for each symbol $f \in \mathit{Ops}$ of arity $n$ there exists an efficient algorithm $A_f$ that computes $\tilde{I}_k(f)(b_1, \ldots, b_n)$ given $k, b_1, \ldots, b_n$ as inputs. □

The evaluation of processes is defined differently from our IML evaluation. Instead of representing the attacker as a process that runs in parallel with the protocol process, Backes et al. [2009] makes the attacker external—it is an arbitrary machine that interacts with the executing process (called an *interactive machine*). Another difference is that pi evaluation uses a single global environment and relies on renaming of variables to avoid clashes. The definition is as follows.

**Definition 5.2 (Computational Pi-execution [Backes et al., 2009, Definition 18])** Let $Q_0$ be a well-formed pi process, $\tilde{I}$ a uniformly efficient family of interpretations, and let $A$ be an interactive machine called the adversary. We define the computational $\pi$-execution with respect to $\tilde{I}$ as an interactive machine $\mathrm{Exec}_{\tilde{I},Q_0}(1^k)$ that takes a security parameter $k$ as argument and interacts with $A$:

- Start: Let $Q^* := Q_0$ (where we rename all bound variables and names such that they are pairwise distinct and distinct from all unbound ones). Let $\eta$ be a totally undefined environment.

- Main Loop: Send $P$ to the adversary and expect an evaluation context $C$ with one or two holes from the adversary. Distinguish the following cases:
  - $Q^* = C[\mathsf{in}(c[\ldots], x); Q']$: Request a channel name $c'$ and a bitstring $b$ from the adversary and if $c' = c$, set $\eta := \eta\{x \mapsto b\}$ and $Q^* := C[Q']$. Ignore channel parameters.
  - $Q^* = C[\mathsf{new}\ x\colon \mathtt{nonce}; Q']$. Pick $r \in I(\mathtt{nonce})$ at random and set $Q^* := C[Q']$ and $\eta := \eta\{x \mapsto r\}$.
  - $Q^* = C[\mathsf{out}(c[\ldots], e); Q_1][\mathsf{in}(c[\ldots], x); Q_2]$: If $b = [\![e]\!]^{\pi}_{I_k,\eta} \neq \bot$ then set $Q^* := C[Q_1][Q_2]$ and $\eta := \eta\{x \mapsto b\}$. Ignore channel parameters.
  - $Q^* = C[\mathsf{let}\ x = e\ \mathsf{in}\ Q_1\ \mathsf{else}\ Q_2]$: If $b = [\![e]\!]^{\pi}_{I_k,\eta} \neq \bot$ then set $Q^* := C[Q_1]$ and $\eta := \eta\{x \mapsto b\}$. Else set $Q^* := C[Q_2]$.
  - $Q^* = C[\mathsf{if}\ e = e'\ \mathsf{then}\ Q_1\ \mathsf{else}\ Q_2]$: Let $b = [\![e]\!]^{\pi}_{I_k,\eta}$ and $b' = [\![e]\!]^{\pi}_{I_k,\eta}$. If $b = b' \in BS$ then set $Q := C[Q_1]$. Else if $b \in BS$ and $b \in BS$ but $b \neq b'$ set $Q^* := C[Q_2]$.
  - $Q^* = C[\mathsf{event}\ ev();\ Q']$: Set $Q^* := C[Q']$ and raise event $ev$.
  - $Q^* = C[!^{i \leq N} Q']$: Rename all bound variables of $Q'$ such that they are pairwise distinct and distinct from all variables and names in $Q$ and in the domain of $\eta$, yielding a process $Q''$. Set $Q^* = C[Q''|!^{i \leq N} Q']$. Ignore the replication index.
  - $Q^* = C[\mathsf{out}(c[\ldots], e); Q']$: If $b = [\![e]\!]^{\pi}_{I_k,\eta} \neq \bot$ request a channel name $c'$ from the adversary and if $c' = c$, send $b$ to the adversary and set $Q^* := C[Q']$. Ignore channel parameters.
  - In all other cases do nothing. □

The above leads directly to the definition of security for pi processes.

**Definition 5.3 (Pi Security [Backes et al., 2009, Definition 19])** Given a uniformly efficient family of interpretations $\tilde{I}$, a polynomial-time interactive machine $A$, a closed pi process $Q$, and a polynomial $p$, let $\mathrm{Events}(Q, A, \tilde{I}, p, k)$ be the list of events raised within the first $p(k)$ computation steps of $A(1^k)$ interacting with $\mathrm{Exec}_{\tilde{I},Q}(1^k)$ (jointly counted for $A(1^k)$ and $\mathrm{Exec}_{\tilde{I},Q}(1^k)$). For a trace property $\rho$ let $\mathrm{insec}_{\pi}(Q, A, \tilde{I}, \rho, p, k)$ be the probability that $\mathrm{Events}(Q, A, \tilde{I}, p, k) \notin \rho$. □

**Differences from Original Sources** The calculus shown in figure 5.1 is a restricted version of the pi calculus presented in Backes et al. [2009], as we do not need the full generality used there. Our restrictions are as follows:

- The processes are required to be in alternating input-output form, to match IML syntax.
- The original calculus makes a distinction between variables (results of network inputs) and names (results of random number generation). We drop this distinction since both behave identically for the purpose of the computational execution.

- The original calculus divides the operations in $Ops$ into constructors and destructors. This is necessary to define the symbolic semantics of the calculus where constructors create new symbolic terms and destructors are used to rewrite the terms. We drop the distinction between constructors and destructors since we do not formalize the symbolic semantics.

- Conditional expressions are not included in the original calculus. However, let-expressions of the form $\mathsf{let}_\perp\ x = e\ \mathsf{in}\ P\ \mathsf{else}\ P'$ can be used to conditionally choose based on equality of bitstrings by assuming that there exists an operation $eq \in Ops$ such that $I(eq)(b, b) = b$ and $I(eq)(b, b') = \perp$ for all $b \neq b'$.

- We keep replication indices and channel parameters in the syntax, but they are only there to make our correspondence proof easier to formulate. The computational pi execution (definition 5.2) makes no use of replication indices or channel parameters.

### 5.1.2 ProVerif

ProVerif [Blanchet, 2009, 2014] verifies the security of a pi process with respect to symbolic semantics in which processes create and rewrite terms based on a given set of constructors and rewriting rules. The rewriting rules also specify the powers of the attacker. Verification proceeds by transforming the process into a set of Horn clauses that include atoms such as $\mathrm{attacker}(t)$ meaning that the attacker can derive the term $t$ and $\mathrm{event}(ev(t))$ meaning that event $ev$ has been executed with payload $t$. Example clauses could have the form $\mathrm{attacker}(t_1) \wedge \ldots \wedge \mathrm{attacker}(t_n) \Rightarrow \mathrm{event}(ev(t))$ meaning that the attacker can force event $ev(t)$ to be executed if it is in possession of $t_1, \ldots, t_n$ or $\mathrm{event}(ev(t)) \Rightarrow \mathrm{attacker}(t')$ meaning that the attacker can obtain $t'$ but only after $ev(t)$ has been executed (all of the $t$s may contain variables). ProVerif makes a distinction between the events that may have been executed (those used on the right-hand side of the clauses) and the events that must have been executed (used on the left-hand side), but we gloss over this distinction.

To verify a correspondence property $ev(t) \Rightarrow ev'(t')$, meaning that every event $ev'(t')$ must be preceded by an event $ev(t)$ ProVerif runs a resolution algorithm on the set of clauses unifying clauses with each other until saturation (which may lead to non-termination). If after resolution each clause with $ev'(t')$ in the conclusion has $ev(t)$ among the hypotheses then we have proved the required correspondence.

We do not formalize the symbolic semantics of the calculus. We treat ProVerif together with the symbolic soundness result described in the next section as a black box—the soundness result states that if the process symbolically verifies the property under a certain set of rewriting rules (as verified by ProVerif) then it computationally satisfies the property with respect to definition 5.3. We refer the reader to Backes et al. [2009]; Blanchet [2009] for a formal definition of symbolic process semantics and symbolic security properties.

### 5.1.3 Computational Soundness For Public-Key Encryption

Backes et al. [2009] provide an example of a set of operations $Ops$ and a set of *soundness conditions* restricting their implementations that are sufficient to establish computational soundness. The set $Ops$ contains a public key encryption operation that is required to be IND-CCA2 secure. The soundness result is established for the class of *key-safe* processes that always use fresh randomness for encryption and key generation, only use honestly generated decryption keys and never send decryption keys. We start by listing the function symbols that the soundness result applies to, splitting them into two groups: cryptographic operations and formatting operations.

**Definition 5.4 (Operations for Encryption-Soundness)** Let

$$Ops_E = \{E/3, ek/1, dk/1, D/2, isenc/1, isek/1, ekof/1, garbage/1, garbageE/2\}$$

be the set of operations that include encryption and decryption functions $E$ and $D$, functions $ek$ and $dk$ for generating encryption and decryption keys, a function $isenc$ for testing whether a bitstring is an encryption, a function $isek$ for testing whether whether a bitstring is an encryption key, and a function $ekof$ for extracting the encryption key from an encryption. The functions $garbage$ and $garbageE$ are not meant to be used by the process—their purpose is to allow additional abilities to the attacker in the symbolic model.

A set of function symbols $Ops_F$ is called a *formatting set* if $Ops_F = Ops_C \cup Ops_P$ such that for each symbol $c \in Ops_C$ with arity $n$ the set $Ops_P$ contains the symbols $c_i^{-1}$ for $i = 1, \ldots, n$. □

The result of Backes et al. [2009] includes soundness for signatures, but we omit them to simplify the presentation (our NSL example in appendix D does use signatures though). The result of Backes et al. [2009] is presented for a single pairing operation, but in practice protocols will use a variety of message formats, so we use the obvious generalisation to an arbitrary set of formatting operations.

Next we outline the soundness conditions that the operations in $Ops_E \cup Ops_F$ need to satisfy.

**Definition 5.5 (Encryption-Soundness Conditions)** Given a uniformly efficient family of interpretations $\tilde{I}$ and a formatting set $Ops_F = Ops_C \cup Ops_P$ we say that $\tilde{I}$ is *encryption-sound* for $Ops_E \cup Ops_F$ if for each $k \in \mathbb{N}$:

1. For each $c \in Ops_C$ with arity $n$ and $b_1, \ldots, b_n \in BS$ and for each $i \leq n$ we have $\tilde{I}_k(c_i^{-1})(\tilde{I}_k(c)(b_1, \ldots, b_n)) = b_i$. For every $b \notin \text{range}(\tilde{I}_k(c))$ and each $i \leq n$ we have $\tilde{I}_k(c_i^{-1})(b) = \bot$.

2. The sets $\tilde{I}_k(\texttt{nonce})$, $\text{range}(\tilde{I}_k(E))$, $\text{range}(\tilde{I}_k(ek))$, $\text{range}(\tilde{I}_k(dk))$, and $\text{range}(\tilde{I}_k(c))$ for $c \in Ops_C$ are disjoint and efficiently recognizable.

3. The functions $\tilde{I}_k(E)$, $\tilde{I}_k(ek)$, $\tilde{I}_k(dk)$, and $\tilde{I}_k(c)$ for $c \in Ops_C$ are *length-regular*—the length of their result depends only on the lengths of their parameters. All $m \in \tilde{I}_k(\texttt{nonce})$ have the same length.

4. $\mathrm{range}(\tilde{I}_k(ekof)) \subseteq \mathrm{range}(\tilde{I}_k(ek))$.

5. $\tilde{I}_k(ekof)(\tilde{I}_k(E)(p,x,y)) = p$ for all $p \in \mathrm{range}(\tilde{I}_k(ek))$, $x \in BS$, and $y \in \tilde{I}_k(\texttt{nonce})$. $\tilde{I}_k(ekof)(e) \neq \bot$ for any $e \in \mathrm{range}(\tilde{I}_k(E))$ and $\tilde{I}_k(ekof)(e) = \bot$ for any $e \notin \mathrm{range}(\tilde{I}_k(E))$.

6. $\tilde{I}_k(E)(p,m,y) = \bot$ if $p \notin \mathrm{range}(\tilde{I}_k(ek))$.

7. $\tilde{I}_k(D)(\tilde{I}_k(dk)(r),m) = \bot$ if $r \in \tilde{I}_k(\texttt{nonce})$ and $\tilde{I}_k(ekof)(m) \neq \tilde{I}_k(ek)(r)$.

8. $\tilde{I}_k(D)(\tilde{I}_k(dk)(r), \tilde{I}_k(E)(\tilde{I}_k(ek)(r), m, r')) = m$ for all $r, r' \in \tilde{I}_k(\texttt{nonce})$.

9. $\tilde{I}_k(isek)(x) = x$ for any $x \in \tilde{I}_k(ek)$. $\tilde{I}_k(isek)(x) = \bot$ for any $x \notin \tilde{I}_k(ek)$.

10. $\tilde{I}_k(isenc)(x) = x$ for any $x \in \tilde{I}_k(E)$. $\tilde{I}_k(isenc)(x) = \bot$ for any $x \notin \tilde{I}_k(ek)$.

Additionally the following two conditions need to be satisfied:

11. We define an encryption scheme $(KeyGen, Enc, Dec)$ as follows: $KeyGen$ picks a random $r$ in $\tilde{I}_k(\texttt{nonce})$ and returns $(\tilde{I}_k(ek)(r), \tilde{I}_k(dk)(r))$. $Enc(p,m)$ picks a random $r$ in $\tilde{I}_k(\texttt{nonce})$ and returns $\tilde{I}_k(E)(p,m,r)$. $Dec(k,c)$ returns $\tilde{I}_k(D)(k,c)$. We require that the defined encryption scheme is IND-CCA2 secure with respect to the security parameter $k$.

12. For all $e$ of type encryption key and $m \in BS$ the probability that $\tilde{I}_k(E)(e,m,r) = \tilde{I}_k(E)(e,m,r')$ for uniformly chosen $r, r' \in \tilde{I}_k(\texttt{nonce})$ is negligible in $k$.

No restrictions are put on the implementations of *garbage* and *garbageE* since those symbols are not used by the process. □

As noted by Backes et al. [2009] any IND-CCA2 secure encryption scheme can be transformed into a set of operations satisfying the above conditions by suitable tagging and padding. It is worth noting that the encryption operation can only be realised using hybrid encryption, as it has to produce output for all $x \in BS$.

The soundness result of Backes et al. [2009] is proved for a class of *key-safe* processes. In a nutshell, key-safe processes always use fresh randomness for encryption and key generation and only use honestly generated (that is, through key generation) decryption keys for decryption. Decryption keys may not be sent around (in particular, this avoids the key-cycle problems). The grammar of key-safe processes is summarised in figure 5.3. We let $x$, $x_d$, $k_s$, and $r$ stand for different sets of variables: general purpose, decryption key, signing key, and randomness variables.

$$
\begin{aligned}
c &\in \mathit{Ops}_C \text{ with arity } n \\
m &::= x \mid c(m_1, \ldots, m_n) \\
e &::= m \ \mid c_i^{-1}(e) \text{ for } i \le n \mid \mathit{isek}(e) \mid \mathit{isenc}(e) \mid D(x_d, e) \mid \mathit{ekof}(e) \\
P,\, P' &::= \mathsf{out}(c[\ldots], x);\ P \mid \mathsf{in}(c[\ldots], x);\ P \mid 0 \mid !^{i \le N} P \mid (P|P') \\
&\mid \mathsf{new}\ x\colon \mathtt{nonce};\ P \\
&\mid \mathsf{event}\ ev();\ P \\
&\mid \mathsf{let}_\perp\ x = e\ \mathsf{in}\ P\ [\mathsf{else}\ P'] \\
&\mid \mathsf{if}\ e =_\perp e\ \mathsf{then}\ P\ [\mathsf{else}\ P'] \\
&\mid \mathsf{new}\ r\colon \mathtt{nonce};\ \mathsf{let}_\perp\ x = ek(r)\ \mathsf{in}\ \mathsf{let}_\perp\ x_d = dk(r)\ \mathsf{in}\ P \\
&\mid \mathsf{new}\ r\colon \mathtt{nonce};\ \mathsf{let}_\perp\ x = E(\mathit{isek}(e_1), e_2, r)\ \mathsf{in}\ P
\end{aligned}
$$

Figure 5.3: The syntax of key-safe processes.

**Definition 5.6 (Symbolic Rewriting Rules for Encryption-Soundness)** The *symbolic rewriting rules* for $\mathit{Ops}_E \cup \mathit{Ops}_F$ with a formatting set $\mathit{Ops}_F = \mathit{Ops}_C \cup \mathit{Ops}_P$ are

$$
\begin{aligned}
D(dk(t_1), E(ek(t_1), m, t_2)) &= m, \\
\mathit{isenc}(E(ek(t_1), t_2, t_3)) &= E(ek(t_1), t_2, t_3), \\
\mathit{isenc}(\mathit{garbageE}(t_1, t_2)) &= \mathit{garbageE}(t_1, t_2) \\
\mathit{isek}(ek(t)) &= ek(t), \\
\mathit{ekof}(E(ek(t_1), m, t_2)) &= ek(t_1), \\
\mathit{ekof}(\mathit{garbageE}(t_1, t_2)) &= t_1, \\
c_i^{-1}(c(x_1, \ldots, x_n)) &= x_i,\ \text{ for } c \in \mathit{Ops}_C\ i \le n = \mathrm{arity}(c).
\end{aligned}
$$
□

**Theorem 5.1 (Computational Soundness [Backes et al., 2009, Theorem 4])** *Assume that a well-formed key-safe pi process $Q$ uses only operations in $\mathit{Ops}_E \cup \mathit{Ops}_F$ for some formatting set $\mathit{Ops}_F$ and there exists a uniformly efficient family of interpretations $\tilde{I}$ that is encryption-sound for $\mathit{Ops}_E \cup \mathit{Ops}_F$. If $Q$ symbolically satisfies a trace property $\rho$ with respect to symbolic rewriting rules for $\mathit{Ops}_E \cup \mathit{Ops}_F$ (definition 5.6) then for all polynomial-time interactive machines $A$ and all polynomials $p$ the function* $\mathrm{insec}_\pi(Q, A, \tilde{I}, \rho, p, k)$ *is negligible in $k$.* □

## 5.2 From IML to ProVerif: Summary

Our translation from an IML to a ProVerif model reuses many of the steps from our IML to CryptoVerif translation (chapter 4). Transforming the process into a formatting-normal form, formatting abstraction, type inference, and type checking will remain exactly the same. This section gives a high-level overview of the steps that differ and the remaining sections of this chapter fill in the details.

**Process Strengthening** The first step in the translation is making sure that our process is a valid pi process. Pi calculus uses strong lets that terminate whenever the evaluation of an expression results in $\bot$. These can be added to IML as syntactic sugar: we define $\mathsf{let}_\bot\ x = e\ \mathsf{in}\ P$ to be an abbreviation for $\mathsf{if}$ defined($e$) $\mathsf{then}\ \mathsf{let}\ x = e\ \mathsf{in}\ P$. Every let-binding that evaluates an expression $e$ can be turned into a strong let-binding if we can prove the fact defined($e$) from the context. This is done in section 5.3.

An additional transformation that is performed in section 5.3 is strengthening of the parsers. Encryption-soundness conditions require that the parsers are *safe*: they must evaluate to $\bot$ whenever their arguments are not in the range of the encoder that they are intended to match. To achieve this we can reuse our parsing safety analysis developed in section 4.8. Now instead of parsing safety being optional, it is required at every parsing site, which allows us to replace all the parsers by their safe versions.

**Erasing Conditionals** IML models extracted from cryptographic implementations contain a lot of auxiliary condition checks (definition 4.7) resulting from if-statements in the C code. The syntax of key-safe processes (figure 5.3) does not allow for arbitrary condition checks, so they need to be removed. This is done in section 5.4 by a transformation that is in a way dual to the strengthening of let-bindings described above. Given a process of the form

$$\textbf{if}\ \phi\ \textbf{then}\ \textbf{let}_\bot\ x = f_p(e)\ \textbf{in}\ P$$

with a safe parser $f_p$, we can remove the condition check if we can prove that the condition $\phi$ is implied by $\mathit{inrange}(e, f_c)$, where $f_c$ is the encoder matched by $f_p$ (see section 4.8 for a discussion of *inrange* facts). This transformation, described in section 5.4, is able to remove all auxiliary condition checks that are used for parsing safety. Unfortunately, most C programs also perform other types of checks. For instance, any check for the `NULL` result of a `malloc` will introduce an if-statement into the model that cannot be removed. Thus in order to apply the computational soundness result we need to make an assumption that the program executes with unbounded memory and `malloc` calls (as well as any other system calls used by the program) never fail. We review the extracted model and manually remove the condition checks for which we can reasonably assume that they always succeed.

**Encryption-Soundness** In section 5.5 we make sure that our model uses functions that satisfy the encryption-soundness conditions (definition 5.5). The conditions can be split into two groups: those speaking about the cryptographic functions $\mathit{Ops}_E$ will need to be assumed since we are not verifying the implementations of the cryptographic functions. On the other hand, the conditions speaking about the formatting functions $\mathit{Ops}_F$ can be checked by our verification tools. One of the conditions in definition 5.5 is particularly problematic, namely the requirement that encoding functions should be defined for arguments of arbitrary lengths. This is unlikely to be satisfied by any practical protocol implementation since message formats typically use length fields of a fixed length. In order to overcome this problem we show that our formatting functions can be strengthened to handle arguments of arbitrary types yielding a new formatting set $Ops_F^*$ that agrees with $Ops_F$ for well-typed arguments (lemma 5.3). We then argue that if our process is well-typed then we can replace functions from $Ops_F$ by their strengthened versions from $Ops_F^*$—the behaviour of the process will not change since it never applies formatting functions to arguments of a wrong type.

Our experience suggests that cryptographic security results should always include some concept of typing since that makes it much easier to apply those results to existing cryptographic implementations. (Incidentally, Backes et al. [2009] do speak about types, but those refer to the types of terms in the symbolic model, not to the bitstring types.)

**From Pi Security to IML Security** Next we prove that computational security in the pi model implies security in the IML model. To do so we show that any IML attacker process $Q_A$ can be transformed into an adversary $A$ (an interactive machine) that interacts with $\mathrm{Exec} = \mathrm{Exec}_{\tilde{I},Q_0}(1^{k_0})$ for a process $Q_0$ in the same way as $Q_A$ interacts with $Q_0$ by the IML rules. When the IML attacker sends a message on a channel $c$ with channel parameters $n_1, \ldots, n_k$, we need to translate the *address* $c, n_1, \ldots, n_k$ into a context $C$ that $A$ should send to Exec. The context $C$ is supposed to select the same subprocess in the process $Q^*$ maintained by Exec as would be matched by the address $c, n_1, \ldots, n_k$ in the IML semantic configuration. Section 5.6 provides details of the construction.

**Failure Propagation and Typing** The encryption-soundness conditions do not allow arbitrary condition checks and instead require that functions (most notably, parsers) should signal failure by returning $\bot$. This stands in the way of type-safety: now we cannot require that $I|_Q \models \Gamma$ for a process $Q$ and a typing environment $\Gamma$ since any function can return $\bot$. Instead we define a weaker notion of type-safety that discards failure.

**Definition 5.7 (Failure-Insensitive Type-Safety)** Given a function symbol $f$ of arity $n$ and a typing environment $\Gamma$ we say that $f$ is *type-safe up to failure* with respect to $\Gamma$, writing $I, f \models_\bot \Gamma$, if the following holds: Let $\Gamma(f) = T_1 \times \ldots \times T_n \to T$ be the type of $f$ in $\Gamma$ (in particular, $f \in \mathrm{dom}(\Gamma)$). Then for all $b_1 \in I(T_1), \ldots, b_n \in I(T_n)$ and $b = I(f)(b_1, \ldots, b_n)$ either $b = \bot$ or $b \in I(T)$.

Given an IML process $Q$ and a typing environment $\Gamma$ we say that *functions in $Q$ are type-safe up to failure* with respect to $\Gamma$, writing $I|_Q \models_\bot \Gamma$, if $I, f \models_\bot \Gamma$ for every $f$ in $Q$.

A function $f$ is called *$\bot$-propagating* if $I(f)$ returns $\bot$ when any of the arguments is $\bot$. □

Given a pi process $Q$ that is well-typed with respect to $\Gamma$, it turns out that if all functions in $Q$ are type-safe up to failure and $\bot$-propagating then that is enough to establish type-safety of $Q$: every executing process $(\eta, P)$ in a semantic configuration that is reachable from $\mathrm{initConfig}(Q)$ will satisfy $\eta \models \Gamma$. This is intuitively clear since any failure will be propagated to the top level by $\bot$-propagating functions, at which point the process will be stopped if it only uses strong lets. Thus we never update the environment with a value of the wrong type. We shall use an argument of this kind in the proof lemma 5.4.

By construction all the functions used by processes in this chapter will be $\bot$-propagating. When we construct a function for arguments $b_1, \ldots, b_n \in BS$, we imply that it returns $\bot$ when one of the arguments is $\bot$.

**Interpretation of the Result** The soundness result (theorem 5.2) that we prove in section 5.7 states that if the translation from an IML process $Q$ to a pi model $\tilde{Q}$ is successful and the cryptographic functions satisfy certain conditions then we can bound $\text{insec}(Q_A|Q, \rho)$ for any attacker $Q_A$ and trace property $\rho$ by a function of the form $\text{insec}_\pi(\tilde{Q}^*, A, \tilde{I}^*, \rho, p, k)$ that is negligible in $k$. The process $\tilde{Q}^*$ is obtained from $\tilde{Q}$ by strengthening of the formatting functions as described in section 5.5, the adversary $A$ is constructed from $Q_A$ as described in section 5.6, $\tilde{I}^*$ is a suitable family of interpretations that extends the functions used by the implementation to arbitrary security parameters, and $p$ is some polynomial large enough to give $A$ enough time to execute any attack performed by $Q_A$.

This result is less convincing that the corresponding result in theorem 4.3 that we obtained for CryptoVerif. The negligible function that provides the upper bound explicitly depends on the implementations of the functions used by the process ($\tilde{I}^*$), and we cannot exclude that this function has a particularly high value for $k_0$, since it uses the formatting functions provided by the implementation for $k = k_0$ and uses "ideal" formatting functions for all values of $k \neq k_0$.

The difference between theorem 4.3 and theorem 5.2 is down to the fact that the pi security definition (definition 5.3) requires *uniform* efficiency of the interpretation family that is used for defining the computational execution. For every function $f$ there must exist a single machine that is able to execute $\tilde{I}_k{}^*(f)$ for any value of $k$. This prevents us from using the same argument as in the proof of lemma 4.1 to define a function that is independent of a particular interpretation. There we took the least secure interpretation for each security parameter and combined all those interpretations into a new family, arguing that the insecurity with respect to that family is still negligible. Now we cannot do so any more since a family constructed in this way is not necessarily uniformly efficient.

The above suggests that one should prefer non-uniform security definitions. This corresponds well with practice since there is often no expectation that the cryptographic algorithms should be executable for arbitrary security parameters. For instance, symmetric block ciphers are very specific to a particular length of the key.

We can still make an informal argument that the function $\text{insec}_\pi(\tilde{Q}^*, A, \tilde{I}^*, \rho, p, k)$ is a good bound since it is more likely to depend on the properties of the cryptographic functions than on the properties of the formatting functions.

$$\frac{\begin{array}{l} f_p(e) \text{ safely matches } f_c \text{ in context } \Phi \\ \tilde{f}_p \text{ is a fresh symbol with } I(\tilde{f}_p) = I(f_p)|_{\mathrm{range}(f_c)} \quad \Phi \vdash P \rightsquigarrow \tilde{P} \end{array}}{\Phi \vdash \mathsf{let}\ x = f_p(e)\ \mathsf{in}\ P \rightsquigarrow \mathsf{let}_\bot\ x = \tilde{f}_p(e)\ \mathsf{in}\ \tilde{P}}$$

$$\frac{\Phi \vdash \mathrm{defined}(e) \quad \Phi \vdash P \rightsquigarrow \tilde{P}}{\Phi \vdash \mathsf{let}\ x = e\ \mathsf{in}\ P \rightsquigarrow \mathsf{let}_\bot\ x = e\ \mathsf{in}\ \tilde{P}} \qquad \frac{\Phi \vdash \mathrm{defined}(e) \wedge \mathrm{defined}(e') \quad \Phi \vdash P \rightsquigarrow \tilde{P}}{\Phi \vdash \mathsf{if}\ e = e\ \mathsf{then}\ P \rightsquigarrow \mathsf{if}\ e =_\bot e\ \mathsf{then}\ \tilde{P}}$$

$$\frac{\Phi \cup \{\phi\} \vdash P \rightsquigarrow \tilde{P}}{\Phi \vdash \mathsf{if}\ \phi\ \mathsf{then}\ P \rightsquigarrow \mathsf{if}\ \phi\ \mathsf{then}\ \tilde{P}} \qquad \frac{\Phi \cup \{\phi\} \vdash P \rightsquigarrow \tilde{P}}{\Phi \vdash \mathsf{assume}\ \phi;\ P \rightsquigarrow \mathsf{assume}\ \phi;\ \tilde{P}}$$

$$\frac{\Phi \vdash P \rightsquigarrow \tilde{P}}{\Phi \vdash \lambda P \rightsquigarrow \lambda \tilde{P}}$$

Figure 5.4: Process strengthening rules.

## 5.3 Process Strengthening

Unlike CryptoVerif where parsing safety is optional, the soundness conditions in definition 5.5 require that every parser $f_p$ is *safe*, that is, $I(f_p)(b) = \bot$ whenever $b \notin \mathrm{range}(I(f_c))$ for the encoder $f_c$ matched by $f_p$. We modify our parsing safety analysis from section 4.8 such that it fails whenever parsing safety is not satisfied. Then we know that at each parsing site the parsing input will always be in the range of the corresponding encoding function, which allows us to safely replace every parser $f_p$ in the process by a parser $\tilde{f}_p$ that fails whenever the input is outside the range. We use the notation $f|_A$ to denote a function that returns $f(x)$ for all $x \in A$ and $\bot$ otherwise.

In addition to strengthening the parsers we replace let-bindings by strong let-bindings and equality checks by strong equality checks whenever possible. The transformation rules are formalised in figure 5.4. The transformation obviously preserves the behaviour of the process, as stated below.

**Lemma 5.1** *If $\emptyset, Q \rightsquigarrow \tilde{Q}$ by the rules in figure 5.4 then $Q \lesssim \tilde{Q}$.* □

## 5.4 Erasing Conditionals

Since the syntax for key-safe processes does not allow auxiliary conditions, we erase those conditions when possible by comparing them with the conditions checked by the strong let-bindings. More precisely, we use the transformations shown in figure 5.5. The first transformation allows to erase a condition that is implied by the fact defined$(e)$ since that fact is implicitly checked by the strong let that follows. The second transformation works in the same way, but now it also uses the fact that a safe parser ensures that its argument is in the range of the matching encoder. Therefore the condition can be erased when it is implied by *inrange*$(e, f_c)$ (the fact

$$\frac{\text{defined}(e) \vdash \phi}{\mathsf{if}\ \phi\ \mathsf{then}\ \mathsf{let}_\perp\ x = e\ \mathsf{in}\ P \rightsquigarrow \mathsf{let}_\perp\ x = e\ \mathsf{in}\ \tilde{P}}$$

$$\frac{\text{safe parser } f_p \text{ matches encoder } f_c \quad \mathit{inrange}(e, f_c) \vdash \phi}{\mathsf{if}\ \phi\ \mathsf{then}\ \mathsf{let}_\perp\ x = f_p(e)\ \mathsf{in}\ P \rightsquigarrow \mathsf{let}_\perp\ x = f_p(e)\ \mathsf{in}\ \tilde{P}}$$

$$\frac{\lambda \text{ is not an output}}{\mathsf{if}\ \phi\ \mathsf{then}\ \lambda P \rightsquigarrow \lambda\ (\mathsf{if}\ \phi\ \mathsf{then}\ P)} \qquad \frac{P \rightsquigarrow \tilde{P}}{C[P] \rightsquigarrow C[\tilde{P}]}$$

Figure 5.5: Rules for erasing conditionals.

$\mathit{inrange}(e, f_c)$ can be expanded using theorem 4.2).

In order to enable the erasing described above we move condition checks down when necessary as shown by the third rule. We have to be careful not to push a condition check past an output since both an output and a condition failure are observable by the attacker and it can therefore detect which happened first. The soundness of the transformation relies on the fact that both the failures of strong let-bindings and the failures of condition checks produce output on the same dedicated channel *yield* such that the attacker cannot tell which condition failed first. The transformations clearly preserve the behaviour of the process, as stated below.

**Lemma 5.2** *If $Q \rightsquigarrow \tilde{Q}$ by the rules in figure 5.5 then $Q \lesssim \tilde{Q}$.* □

**Example 5.1** Consider the following IML fragment:

1 **if** $\text{defined}(D(\mathit{msg}, \mathit{key}))$ **then**
2 **let**$_\perp$ $d = D(\mathit{msg}, \mathit{key})$ **in**
3 **if** $\text{len}(d) \geq 5$ **then**
4 **if** $\text{ptr}(\mathit{malloc}(\ldots), 0) \neq \tau_{\text{ptr}}^{-1}(0)$ **then** $\ldots$
5 **let**$_\perp$ $x_1 = f_p(d)$ **in**

The first condition is likely to be added by the proxy function for decryption, the second check is part of ensuring parsing safety by the implementation, and the third check comes from checking a result of a malloc against being NULL. Assume that the parser $f_p$ matches an encoder $f_c$ of the form $f_c(x_1, x_2) = \texttt{0x01}|\tau_{4u}^{-1}(\text{len}(x_1))|x_1|x_2$. Then $\mathit{inrange}(d, f_c)$ will have the form shown in example 4.10 (section 4.8) with $d$ in place of *msg11*. In particular, $\mathit{inrange}(d, f_c)$ will imply the condition $\text{len}(d) \geq 5$ checked above the parsing site.

Using our transformations we can erase condition checks in lines 1 and 3. There is nothing we can do with the check in line 4 except making the assumption that the program runs with unbounded memory and removing the check manually. □

## 5.5 Local Encryption-Soundness

The encryption soundness conditions in definition 5.5 form two groups: conditions 11 and 12 are global—they speak about the security of the encryption function evaluated across all values of the security parameter. These conditions will have to be assumed for the particular functions used by our implementation, since no encryption function has been proven IND-CCA2 secure so far. In contrast, conditions 1 to 10 are local in the sense that they are evaluated separately for each value of the security parameter. We shall now discuss how we make sure that our implementation satisfies the local conditions. We start by giving the local encryption soundness conditions their own definition.

**Definition 5.8 (Local Encryption-Soundness)** Given a formatting set $Ops_F$ we say that $Ops_E \cup Ops_F$ is *locally encryption-sound* if it satisfies conditions 1-10 of definition 5.5 with the interpretation $I$ in place of $\tilde{I}_k$. □

The local encryption soundness conditions can themselves be split into three groups: those that speak about $Ops_E$, about $Ops_F$, and about the interaction of $Ops_E$ and $Ops_F$. Conditions 2-10 require that $Ops_E$ on its own is locally encryption sound. Condition 2 requires that certain functions in $Ops_E$ have disjoint ranges from encoders in $Ops_F$. These two condition sets will have to be assumed, just like conditions 11 and 12 since we do not verify the implementations of the cryptographic functions.

Conditions 1 to 3 contain the required properties of the formatting set $Ops_F$: all the encoding functions must be length-regular and have disjoint ranges and the parsing functions must correctly extract the arguments of the encoding functions and must fail (return $\bot$) whenever the argument is not a valid encoding. These conditions can be explicitly checked during verification—we shall make use of the fact set $\Phi$ generated by our fact inference (section 4.7).

An unfortunate property of the soundness conditions is that the parsing equations must be true for arguments of arbitrary length. This will of course not be satisfied in practice since implementations use fixed-length fields in message formats. In order to overcome this obstacle we first formulate a relaxed set of conditions that need to be satisfied only for inputs of the right type, and then show how we can extend our formatting functions to inputs of all types.

**Definition 5.9 (Regular Formatting Sets)** A finite formatting set $Ops_F = Ops_C \cup Ops_P$ is called *regular* with respect to a typing environment $\Gamma$ if for every encoder $c \in Ops_C$ such that $\Gamma(c) = T_1 \times \ldots \times T_n \to T$ and for each $b_1 \in I(T_1), \ldots, b_n \in I(T_n)$

1. $I(c)(b_1, \ldots, b_n) \in I(T)$ and $T$ is a bounded type.
2. For each $i \leq n$ we have $I(c_i^{-1})(I(c)(b_1, \ldots, b_n)) = b_i$.
3. For each $b \notin \text{range}(I(c))$ and $i \leq n$ we have $I(c_i^{-1})(b) = \bot$.
4. For each $c' \in Ops_C$ such that $c \neq c'$ the ranges of functions $I(c)$ and $I(c')$ are disjoint.

5. $I(c)$ is length-regular. □

It is easy to check regularity using the fact set $\Phi$ generated by the method in section 4.7. Condition (2) is satisfied if for each encoder $c \in Ops_C$ and each parser $c_i^{-1}$ for $i \leq \text{arity}(c)$ we have $parse_\Gamma(c, p, i) \in \Phi$. Condition (4) is satisfied if for every two encoders $c, c' \in Ops_C$ we have $disj_\Gamma(c, c') \in \Phi$. Condition (1) requires that every encoder $c$ is type-safe ($I, c \models \Gamma$) and can be checked as described in section 4.9. Condition (3) is enforced by the parser strengthening that we perform in section 5.3. Finally, condition (5) is satisfied since all our encoders are length-regular by construction.

Given a regular formatting set $Ops_F$ we shall aim to replace it with a formatting set $Ops_F^*$ that satisfies the parsing equations for inputs of arbitrary lengths. The crucial property that we would like to capture is that $Ops_F$ and $Ops_F^*$ agree for inputs of the correct type. This motivates the following definition.

**Definition 5.10 (Consistent Formatting Sets)** Let $\Gamma$ be a typing environment. Two functions $f$ and $f^*$ with $\Gamma(f) = T_1 \times \ldots \times T_n \to T$ are called *consistent* with respect to $\Gamma$, denoted by $f \sim_\Gamma f^*$, if $I(f)(b_1, \ldots, b_n) = I(f^*)(b_1, \ldots, b_n)$ for all $b_1 \in I(T_1), \ldots, b_n \in I(T_n)$.

Two formatting sets $Ops_F$ and $Ops_F^*$ are called *consistent* with respect to $\Gamma$, denoted $Ops_F \sim_\Gamma Ops_F^*$, if each $f \in Ops_F$ maps one-to-one to $f^* \in Ops_F^*$ such that $f \sim_\Gamma f^*$. □

We now show how to generalise a regular formatting set $Ops_F$ to a formatting set $Ops_F^*$ that satisfies local encryption soundness conditions. We shall make use of "ideal" formatting functions $(\cdot)_c$ for each $c \in Ops_C$ and use $(b_1, \ldots, b_n)_c$ whenever $c(b_1, \ldots, b_n)$ is not defined. For parsing a value we shall need to decide whether it was created using $(\cdot)_c$ or $c$. In order to do so we make use of the assumption that the return types of all the encoders are bounded. Since $Ops_F$ is finite, there exists some maximal length $\mathcal{L}$ of the output returned by any encoder. We shall therefore make $(\cdot)_c$ return outputs longer than $\mathcal{L}$, which makes parsing unambiguous. We start by listing the properties that we require of the ideal formatting functions.

**Definition 5.11** Let $Ops_F = Ops_C \cup Ops_P$ be a formatting set and $\Gamma$ a typing environment. A *strong encoder family* for $Ops_E \cup Ops_F$ with respect to $\Gamma$ and $\mathcal{L} \in \mathbb{N}$ is a family of functions $(\cdot)_c \colon BS^n \to BS$ for each function $c \in Ops_C$ of arity $n$ such that

- $(b_1, \ldots, b_n)_c$ is computable in polynomial time in $\sum_{i \leq n} |b_i|$.
- $(\cdot)_c$ is injective with inverses $\pi_{c,1}, \ldots, \pi_{c,n}$. Each inverse is computable in polynomial time in the length of the input.
- For each $c \in Ops_C$ and $i \leq \text{arity}(c)$, if $b \notin \text{range}((\cdot)_c)$ then $\pi_{c,i}(b) = \bot$.
- All the functions $(\cdot)_c$ are length-regular, and have pairwise disjoint ranges.
- $|(b_1, \ldots, b_n)_c| > \mathcal{L}$ for each $c \in Ops_C$ and $b_1, \ldots, b_n \in BS$.

- For each $k \in \mathbb{N}$ the sets $I(\texttt{nonce})$, $\text{range}(I(E))$, $\text{range}(I(ek))$, $\text{range}(I(dk))$, and $\text{range}((\cdot)_c)$ for $c \in Ops_C$ are disjoint. □

It is easy to see that strong encoder families exist, but we do not go into details of the construction. We now show how to use strong encoder families to extend regular formatting sets to formatting sets that satisfy local encryption soundness conditions.

**Definition 5.12** Given a formatting set $Ops_F$ we say that $Ops_E \cup Ops_F$ *locally satisfies disjointness conditions* if it satisfies condition 2 of definition 5.5 with $I$ in place of $\tilde{I}_k$. □

**Lemma 5.3** *Let $Ops_F$ be a formatting set and $\Gamma$ a typing environment. Assume that*

1. *$Ops_F$ is regular with respect to $\Gamma$.*
2. *$Ops_E \cup Ops_F$ locally satisfies disjointness conditions.*
3. *$Ops_E$ is locally encryption-sound.*

*Then there exists a formatting set $Ops_F^*$ such that $Ops_F \sim_\Gamma Ops_F^*$ and $Ops_E \cup Ops_F^*$ is locally encryption-sound.* □

Proof Let $Ops_F = Ops_C \cup Ops_P$. By regularity assumption there exists $\mathcal{L} \in \mathbb{N}$ such that $I(T) \subseteq I(\texttt{bounded}_\mathcal{L})$ for each return type $T$ of a formatting function $c \in Ops_C$. Let $(\cdot)_c$ for $c \in Ops_C$ be a strong encoder family for $Ops_E \cup Ops_F$ with respect to $\Gamma$ and $\mathcal{L}$. For each $f \in Ops_F$ with $\Gamma(f) = T_1 \times \ldots \times T_n \to T$ define $f^*$ as follows

$$I(c^*)(b_1, \ldots, b_n) = \begin{cases} I(c)(b_1, \ldots, b_n), & \text{if } b_1 \in I(T_1), \ldots, b_1 \in I(T_1), \\ (b_1, \ldots, b_n)_c, & \text{otherwise,} \end{cases}$$

$$I((c_i^{-1})^*)(b) = \begin{cases} I(c_i^{-1})(b), & \text{if } |b| \leq \mathcal{L}, \\ \pi_{c,i}(b), & \text{otherwise.} \end{cases}$$

It is straightforward to check that the set $Ops_F^* = \{f^* \mid f \in Ops_F\}$ satisfies the requirements of the lemma. ■

We would like to replace the formatting functions $Ops_F$ in a process $Q$ by their strengthened versions $Ops_F^*$. We can do so whenever $Q$ is type-safe and never calls a function with arguments of the wrong types. Since $Ops_F$ and $Ops_F^*$ agree for inputs of correct types, the behaviour of a type-safe process will not change. As discussed in section 5.2, type-safety can be guaranteed if we assume that $Q$ is well-typed and uses functions that are $\perp$-propagating and type-safe up to failure. The following lemma makes this precise.

**Lemma 5.4** *Assume that a well-formed pi process $Q_0$ is well-typed with respect to a typing environment $\Gamma$. Assume further that $I|_{Q_0} \models_\perp \Gamma$ and all functions in $Q_0$ are $\perp$-propagating. Let $Ops_F$ be the set of formatting symbols used by $Q_0$ and consider a formatting set $Ops_F^*$ such* *that* $Ops_F \sim_\Gamma Ops^*_F$*. Let* $Q^*_0$ *be obtained from* $Q_0$ *by replacing each formatting symbol* $f$ *by its matching symbol* $f^* \in Ops^*_F$*. Then* $\mathrm{insec}(Q_0, \rho) = \mathrm{insec}(Q^*_0, \rho)$ *for every trace property* $\rho$ *.* □

PROOF We show that $Q_0 \simeq Q^*_0$ (meaning both $Q_0 \lesssim Q^*_0$ and $Q^*_0 \lesssim Q_0$, as in definition 2.11). For an environment $\eta$ and a pi process $P$ let $(\eta, P) \simeq (\eta, P^*)$ whenever $\eta|_V \models \Gamma$, where $V$ is the set of free variables of $P$ and $P^*$ is obtained from $P$ by replacing each formatting symbol $f \in Ops_F$ by its matching symbol $f^* \in Ops^*_F$. We show that $\simeq$ is a simulation relation both ways. First, $(\eta, Q_0) \simeq (\eta, Q^*_0)$ since by assumption $Q_0$ is well-formed and so does not have free variables.

Let us consider the interesting case of a let-binding. Assume that

$$(\eta,\ \mathsf{let}_\bot\ x = e\ \mathsf{in}\ P) \simeq (\eta,\ \mathsf{let}_\bot\ x = e^*\ \mathsf{in}\ P^*)$$

for some environment $\eta$, an expression $e$, and an output process $P$, where $e^*$ and $P^*$ are obtained from $e$ and $P$ by replacing each formatting symbol $f$ by its matching $f^*$. We need to show that if either of the sides reduces then both sides reduce the same way and preserve the invariant. By induction on the structure of $e$ we show that $[\![e]\!]_\eta = [\![e^*]\!]_\eta$ and either $[\![e]\!]_\eta = \bot$ or $[\![e]\!]_\eta \in I(\mathrm{type}_\Gamma(e))$, where $\mathrm{type}_\Gamma$ is the function that deduces the type of $e$ with respect to the typing environment $\Gamma$, as described in section 4.5. Consider the following cases:

- $e = e^* = x$ for some variable $x$. The statement follows since $x$ is a free variable of $P$ and therefore by assumption $\eta(x) \in I(\Gamma(x))$.
- $e = f(e_1, \ldots, e_n)$ and $e^* = f(e^*_1, \ldots, e^*_n)$ with $\Gamma(f) = T_1 \times \ldots \times T_n \to T$. By induction both $e$ and $e^*$ evaluate the same. Assume that $[\![e]\!]_\eta \neq \bot$. Then $[\![e_i]\!]_\eta \neq \bot$ for all $i$ because $f$ is assumed to be $\bot$-propagating. By induction $[\![e_i]\!]_\eta \in I(\mathrm{type}_\Gamma(e_i))$ for each $i$. Since $Q_0$ is well-typed with respect to $\Gamma$, we have $T_i = \mathrm{type}_\Gamma(e_i)$ for each $i$. By assumption that $I|_{Q_0} \models_\bot \Gamma$ we conclude $[\![e]\!]_\eta \in I(\mathrm{type}_\Gamma(e))$.
- The case $e = f(e_1, \ldots, e_n)$ and $e^* = f^*(e^*_1, \ldots, e^*_n)$ where $f \in Ops_F$ and $f^* \in Ops^*_F$ is treated in the same way. We use the fact that the functions in $Ops^*_F$ are $\bot$-propagating by construction. ■

## 5.6 Relating IML and Pi Security

In this section we show that pi security (definition 5.3) implies IML security (definition 2.2). Given a pi process $Q_0$ and an IML process $Q_A$ that represents an attack on $Q_0$ we shall construct an adversary $A$ that carries out the same attack against $\mathrm{Exec}_{\tilde{I}, Q_0}(1^{k_0})$ (in the following simply Exec) under the assumption that $\tilde{I}_{k_0} = I$.

To make sure that the state $Q^*$ maintained by Exec matches the semantic configuration of IML, we shall require that $Q^*$ is *fully reduced* with respect to $Q_0$, as defined below. In a fully reduced process all pending internal computations are completed, all replications are unrolled a sufficient number of times, and all participant subprocess are ready to either receive input or produce output. Example 5.2 will aim to provide more intuition for this definition.

**Definition 5.13 (Fully Reduced Processes)** A pi process $Q^*$ is *fully reduced* with respect to a pi process $Q$ if one of the following holds:

- $Q = \mathsf{in}(c[\ldots], x); P$ and either $Q^* = \sigma(Q)$ for some renaming of variables $\sigma$ or $Q^*$ is fully reduced with respect to $P$.
- $Q = \mathsf{out}(c[\ldots], x); P$ and either $Q^* = \sigma(Q)$ for some renaming of variables $\sigma$ or $Q^*$ is fully reduced with respect to $P$.
- $Q$ is an output process of the form $Q = \lambda P$ and $Q^*$ is fully reduced with respect to $P$.
- $Q = Q^* = 0$.
- $Q = Q_1 | Q_2$ and $Q^* = Q_1^* | Q_2^*$ such that $Q_1^*$ is fully reduced with respect to $Q_1$ and $Q_2^*$ is fully reduced with respect to $Q_2$.
- $Q = !^{i \leq N} Q'$ and $Q^* = (Q_1^* | (Q_2^* | (\ldots (Q_n^* | !^{i \leq N} Q''))))$, where $n \geq I(N)$ and each $Q_i^*$ for $i \leq n$ is fully reduced with respect to $Q'$ (we do not assume anything about $Q''$). □

In the pi execution the adversary $A$ can always maintain the invariant that the process $Q^*$ maintained by Exec is fully reduced with respect to $Q_0$.

Given an *address* $c, n_1, \ldots, n_k$ with a channel $c$ and values of replication indices $n_1, \ldots, n_k \in \mathbb{N}$ that refers to an input process in an IML run, we shall need to locate the same input process within $Q^*$. To enable this, we shall assume that $Q_0$ is *regular*, defined as follows.

**Definition 5.14 (Regular Processes)** An IML process is *regular* if it uses distinct channels for all inputs (but not necessarily for all outputs) and every input and output that occurs under replication indices $i_1, \ldots, i_k$ is of the form $\mathsf{in}(c[i_1, \ldots, i_k], x)$ or $\mathsf{out}(c[i_1, \ldots, i_k], x)$. □

Given an address $c, n_1, \ldots, n_k$ and assuming that $Q_0$ is regular we shall generate a sequence $\mathrm{id}(c, n_1, \ldots, n_k)$ of labels $L$ or $R$ (left or right) that select a subprocess in $Q^*$ as follows. Let $Q_0, \ldots, Q_m$ be a sequence of processes such that $Q_{i+1}$ is an immediate subprocess of $Q_i$ for each $i < m$ and $Q_m = \mathsf{in}(c[\ldots], x); P$ for some $x$ and $P$. Assume that $Q_m$ occurs under replications $i_1 \leq N_1, \ldots, i_k \leq N_k$ in $Q_0$ and $n_j \leq I(N_j)$ for each $j \leq k$. If this condition is not satisfied, set $\mathrm{id}(c, n_1, \ldots, n_k) = \bot$. For each $i < m$ let

$$
\begin{aligned}
&\mathrm{choose}(Q_i) = L && \text{if } Q_i = (Q_{i+1} | Q') \text{ for some } Q', \\
&\mathrm{choose}(Q_i) = R && \text{if } Q_i = (Q' | Q_{i+1}) \text{ for some } Q', \\
&\mathrm{choose}(Q_i) = R^{n_j - 1} L && \text{if } Q_i = !^{i_j \leq N_j} Q_{i+1} \text{ for some } j \leq k, \\
&\mathrm{choose}(Q_i) = \varepsilon && \text{otherwise}
\end{aligned}
$$

Let $\mathrm{id}(c, n_1, \ldots, n_k) = \mathrm{choose}(Q_1) \ldots \mathrm{choose}(Q_{m-1})$.

Given a sequence $l$ of labels $L$ or $R$ we write $Q_l \rightsquigarrow^* Q'$ if $l$ selects $Q'$ within $Q$, that is, $Q_l$ reduces to $Q'$ by the following rules:

$$Q_\varepsilon \rightsquigarrow Q \qquad (Q|Q')_{Ll} \rightsquigarrow Q_l \qquad (Q|Q')_{Rl} \rightsquigarrow Q'_l$$

Assuming $Q_l \rightsquigarrow^* Q'$ we let $Q_{[l]}$ be the context $C$ such that $Q = C[Q']$.

**Example 5.2** Assume that $Q_0$ is of the form

$$Q_0 = 0 \mid (\mathbf{in}(c_1[], x_1);\ \mathbf{out}(c[], x_1);\ \mathbf{new}\ y\colon T;\ !^{i_1 \le N_1}(\mathbf{in}(c_2[i_1], x_2);\ P))$$

with some output process $P$. We shall initialise Exec with $Q_0^* = Q_0$ that is already fully reduced with respect to $Q_0$. Assume that the attacker $Q_A$ sends a message $b$ on channel $c_1$. We translate it into the actions of the adversary $A$ as follows: by definition above $\mathrm{id}(c_1) = R$ and so $A$ sends to Exec the execution context $Q_{[R]} = (0|[])$ followed by $c_1$ and $b$. The process $Q_0^*$ is now updated to

$$Q_1^* = 0 \mid (\mathbf{out}(c[], x_1);\ \mathbf{new}\ y\colon T;\ !^{i_1 \le N_1}(\mathbf{in}(c_2[i_1], x_2);\ P))$$

We send the channel $c$ to Exec, receive back a bitstring $b$ and forward it to $Q_A$. Now the process $Q_1^*$ is updated to

$$Q_2^* = 0 \mid (\mathbf{new}\ y\colon T;\ !^{i_1 \le N_1}(\mathbf{in}(c_2[i_1], x_2);\ P))$$

This process is not fully reduced with respect to $Q_0$. However, assuming that $I(N_1) = 2$, we can send the context $(0|[])$ twice to force $Q_2^*$ to become the fully reduced process

$$\begin{aligned} Q_3^* = 0 \mid (&(\mathbf{in}(c_2[i_1], x_2');\ P') \\ &\mid ((\mathbf{in}(c_2[i_1], x_2'');\ P'') \\ &\quad \mid !^{i_1 \le N_1}(\mathbf{in}(c_2[i_1], x_2);\ P))) \end{aligned}$$

with new variables $x_2', x_2''$ and processes $P'$ and $P''$ obtained from $P$ by renaming of variables. Now if $Q_A$ wants to send a message to the address $c_2, 2$ we need to send to Exec the context

$$\begin{aligned} Q_{[\mathrm{id}(c_2,2)]} = Q_{[RRL]} = 0 \mid (&(\mathbf{in}(c_2[i_1], x_2');\ P') \\ &\mid (\ [] \\ &\quad \mid !^{i_1 \le N_1}(\mathbf{in}(c_2[i_1], x_2);\ P))) \end{aligned}$$

Below we shall generalize the intuition of this example to arbitrary attackers $Q_A$. □

By construction IML processes execute in bounded time: For each $Q \in \mathrm{IML}$ there exists an upper bound $\mathrm{runtime}(Q) \in \mathbb{N}$ such that any IML trace of $\mathrm{initConfig}(Q)$ can be executed in at most $\mathrm{runtime}(Q)$ steps. We refer to [Blanchet, 2008, Appendix B.5] for a proof that also applies in our setting.

**Lemma 5.5 (Pi Security Implies IML Security)** *Let $Q_0$ be a well-formed regular pi process such that all functions in $Q_0$ are $\bot$-propagating. Let $\tilde{I}$ be a uniformly efficient family of* *interpretations such that* $\tilde{I}_{k_0} = I$. *Then for any IML process* $Q_A$ *that does not contain events there exists a polynomial-time adversary* $A$ *such that for any trace property* $\rho$

$$\text{insec}(Q_A|Q_0, \rho) \leq \text{insec}_\pi(Q_0, A, \tilde{I}, \rho, \text{runtime}(Q_A|Q_0), k_0). \qquad \square$$

Proof Let $Q_0$ be as in the statement of the lemma and let $Q_A$ be the IML process representing the attacker. Let Exec $=$ $\text{Exec}_{\tilde{I},Q_0}(1^{k_0})$. We shall construct the adversary $A$ as follows: $A$ will maintain attacker state of the form $\mathbb{C}_E = (\eta, \mathsf{out}(c[e_1, \ldots, e_k], e); Q'_A), \mathcal{Q}_A$. We obtain the initial state by allowing initConfig($Q_A$) to reduce until there are no further reductions. We let $Q^*$ be the protocol process maintained by Exec and known to the adversary.

The attacker $A$ starts by interacting with Exec to make $Q^*$ fully reduced and then iterates as follows: Given $\mathbb{C}_E$ in the form above, let $n_1 = [\![e_1]\!]_\eta, \ldots, n_k = [\![e_l]\!]_\eta$ and $b = [\![e]\!]_\eta$. Let $l = \text{id}(c, n_1, \ldots, n_k)$ and send $Q^*_{[l]}$ followed by $c$ and $b$ to Exec. Interact with Exec to make $Q^*$ fully reduced until asked for a channel name (at this point the subprocess in $Q^*$ that received input from the attacker is ready to produce output). Let $c$' be the channel the output on which in $Q_0$ occurs under the input on $c$. Send $c'$ to Exec and receive a bitstring $b'$. Again interact with Exec to make $Q^*$ fully reduced. Let

$$\mathbb{C}'_E = (\eta, \mathsf{out}(c[e_1, \ldots, e_k], e); Q'_A),\ \mathcal{Q}_A \cup \{(\emptyset, \mathsf{in}(c[n_1, \ldots, n_k], x); \mathsf{out}(c'[n_1, \ldots, n_k], b')); 0\}$$

and allow $\mathbb{C}'_E$ to reduce until there are no further reductions. Repeat.

The construction of $A$ maintains the following invariant: Assume that there is a trace initConfig($Q_A|Q_0$) $\rightarrow^*_p (\eta, P_E), \mathcal{Q}$, where $P_E$ descends from the attacker and there is a matching input process in $\mathcal{Q}$ that descends from $Q_0$. Then there exists a state of Exec and $A$ reachable with probability $p$ and a renaming of variables $\sigma$ such that if $Q^*$ and $\eta_Q$ are the process and the environment maintained by Exec then $\mathcal{Q} = \mathcal{Q}_A \uplus \mathcal{Q}_P$ such that the attacker state $\mathbb{C}_E$ maintained by $A$ is of the form $(\eta, P_E), \mathcal{Q}_A$ and for each $(\eta, Q_P) \in \mathcal{Q}_P$ with $Q_P = \mathsf{in}(c[i_1, \ldots, i_k], x); P_P$ the following holds:

- $Q^*_{\text{id}(c, [\![i_1]\!]_\eta, \ldots, [\![i_k]\!]_\eta)} = \sigma(Q_P)$.
- For each $x \in \text{dom}(\eta)$ we have $\eta_Q(\sigma(x)) = \eta(x)$.

The importance of the assumption that all functions in $Q_0$ are $\bot$-propagating is in the fact that for such functions the pi and the IML evaluation of expressions coincide: $[\![e]\!]^\pi_{\eta,I} = [\![e]\!]_\eta$ for any expression $e$ in $Q_0$ and environment $\eta$. ■

## 5.7 Summary and Soundness

We now prove soundness of our translation from IML processes to ProVerif models. We start by summarizing the translation procedure.

**Definition 5.15 (IML to Pi Translation)** Consider a well-formed input IML process $Q$. We say that $Q$ *yields a pi model* $\tilde{Q}$ with formatting set $Ops_F$ and a typing environment $\Gamma$ if there exist processes $Q_1, \ldots, Q_5$ such that

1. $Q_1$ is the formatting-normal form of $Q$ (section 4.3).
2. $Q_2$ is the formatting abstraction of $Q_1$, that is, $Q_1 \rightsquigarrow Q_2$ by the rules in figure 4.8 (section 4.4).
3. $\Gamma$ is the result of type inference of function types from $Q$ (section 4.5).
4. $Q_3$ is the result of typechecking $Q_2$ with respect to $\Gamma$, that is, $\Gamma, \emptyset \vdash Q_2 \rightsquigarrow Q_3$ by the rules in figure 4.11 (section 4.6).
5. $\Phi$ is the result of fact inference (section 4.7).
6. $Q_4$ is the result of process strengthening of $Q_3$, that is, $\emptyset \vdash Q_3 \rightsquigarrow Q_4$ by the rules in figure 5.4 (section 5.3).
7. $Q_5$ is the result of erasing conditionals from $Q_4$, that is, $Q_4 \rightsquigarrow Q_5$ by the rules in figure 5.5 (section 5.4).
8. $I, f \models \Gamma$ for every encoder $f$ (section 4.9).
9. $\Phi$ implies that the set $Ops_F$ of the formatting operations used by $Q_5$ is a regular formatting set, as described in definition 5.8.
10. $\tilde{Q}$ is obtained from $Q_5$ by removing inline assumptions.
11. $\tilde{Q}$ is a key-safe pi process. □

We are now ready to present the main result of this chapter.

**Theorem 5.2 (IML to Pi Translation is Sound)** *Assume that a well-formed regular IML process $Q$ that satisfies inline assumptions yields a pi model $\tilde{Q}$ with formatting set $Ops_F$ and a typing environment $\Gamma$. Assume that $\tilde{Q}$ symbolically satisfies a trace property $\rho$ with respect to symbolic rewriting rules for $Ops_E \cup Ops_F$ (definition 5.6).*

*Assume that there exists a uniformly efficient family of interpretations $\tilde{I} = (\tilde{I}_k)_{k \in \mathbb{N}}$ that is encryption-sound for $Ops_E \cup Ops_F$ such that $\tilde{I}_{k_0}$ agrees with $I$ for function symbols in $Ops_E$. Assume that $Ops_E \cup Ops_F$ locally satisfies disjointness conditions (definition 5.12). Assume that for all $f \in Ops_E$ the function $I(f)$ is $\perp$-propagating and $I, f \models_\perp \Gamma$.*

*Then there exists an efficient family of interpretations $\tilde{I}^*$ and a pi process $\tilde{Q}^*$ such that*

1. *For any IML process $Q_A$ that does not contain events there exists a polynomial-time adversary $A$ such that*

$$\text{insec}(Q_A|Q, \rho) \leq \text{insec}_\pi(\tilde{Q}^*, A, \tilde{I}^*, \rho, \text{runtime}(Q_A|Q), k_0).$$

---

2. *The function* $\text{insec}_\pi(\tilde{Q}^*, A, \tilde{I}^*, \rho, p, k)$ *is negligible in* $k$ *for all polynomial-time adversaries* $A$ *and all polynomials* $p$. □

Proof Obtain processes $Q_1, \ldots, Q_5$ from definition 5.15 and choose a trace property $\rho$. The translation steps up to $Q_3$ are the same as in definition 4.14. Applying the same argument as in theorem 4.3 we get $Q \lesssim Q_3$. By lemmas 5.1 and 5.2 $Q_3 \lesssim Q_4 \lesssim Q_5$ and finally, since $Q$ satisfies inline assumptions, $Q_5 \lesssim \tilde{Q}$. Thus overall $Q \lesssim \tilde{Q}$. Thus by theorems 2.2 and 2.3 for every attacker $Q_A$

$$\text{insec}(Q_A|Q, \rho) \leq \text{insec}(Q_A|\tilde{Q}, \rho). \tag{5.1}$$

Since $\tilde{I}$ is assumed to be encryption-sound for $\mathit{Ops}_E \cup \mathit{Ops}_F$, it is in particular encryption sound for $\mathit{Ops}_E$. By assumption $\tilde{I}_{k_0}$ agrees with $I$ for function symbols in $\mathit{Ops}_E$, thus $\mathit{Ops}_E$ is locally encryption-sound (definition 5.8). Applying lemma 5.3 we obtain strengthened versions $\mathit{Ops}_F^*$ of the formatting functions such that $\mathit{Ops}_E \cup \mathit{Ops}_F^*$ is locally encryption-sound. We can now construct $\tilde{I}^*$ by combining $\tilde{I}$ (after renaming formatting functions to their strengthened counterparts) with the interpretations of $\mathit{Ops}_E \cup \mathit{Ops}_F^*$ by $I$:

$$\begin{aligned}
\tilde{I}_k{}^*(f^*) &= \tilde{I}_k(f), \text{ for all } f \in \mathit{Ops}_F \text{ and } k \neq k_0,\\
\tilde{I}_k{}^*(f) &= \tilde{I}_k(f), \text{ for all } f \in \mathit{Ops}_E \text{ and } k \neq k_0,\\
\tilde{I}_k{}^*(f) &= I(f), \text{ for all } f \in \mathit{Ops}_E \cup \mathit{Ops}_F^* \text{ and } k = k_0.
\end{aligned}$$

Clearly $\tilde{I}^*$ is encryption-sound.

We now apply lemma 5.4 to replace $\mathit{Ops}_F$ by $\mathit{Ops}_F^*$ in $\tilde{Q}$. The conditions of the lemma are satisfied as follows: By lemma 4.5 $Q_3$ is well-typed with respect to $\Gamma$. This property is clearly not broken by any subsequent transformations, thus $\tilde{Q}$ is well-typed with respect to $\Gamma$. By assumption of the lemma $I, f \models_\perp \Gamma$ for all $f \in \mathit{Ops}_E$. By item 8 $I, f \models \Gamma$ for all encoders $f$ and by construction in section 5.3 $I, f \models_\perp \Gamma$ for all parsers used by $Q_4$ (and thus also by $\tilde{Q}$). Thus overall $I|_{\tilde{Q}} \models_\perp \Gamma$. By assumption every function in $\mathit{Ops}_E$ is $\perp$-propagating, and this is true by construction for all our formatting functions, so that all functions in $\tilde{Q}$ are $\perp$-propagating. We can now apply lemma 5.4 to obtain a process $\tilde{Q}^*$ that uses functions in $\mathit{Ops}_E \cup \mathit{Ops}_F^*$ such that for every attacker $Q_A$

$$\text{insec}(Q_A|\tilde{Q}, \rho) = \text{insec}(Q_A|\tilde{Q}^*, \rho). \tag{5.2}$$

Statement (1) of the theorem now follows from (5.1) and (5.2) together with lemma 5.5 and statement (2) follows from theorem 5.1 together with the observation that if $\tilde{Q}$ symbolically satisfies a trace property with respect to symbolic rewriting rules for $\mathit{Ops}_E \cup \mathit{Ops}_F$ then $\tilde{Q}^*$ satisfies the same property with respect to rewriting rules $\mathit{Ops}_E \cup \mathit{Ops}_F^*$, since both inputs to ProVerif are the same up to renaming of function symbols. ■

Theorem 5.2 can be summarised as follows. Suppose we translate an IML process $Q$ to a pi model $\tilde{Q}$ that uses formatting operations in $Ops_F$. Suppose further that we supply the tuple $(\tilde{Q}, R, \rho)$ to ProVerif where $\rho$ is a correspondence property and $R$ is the set of symbolic rewriting rules for $Ops_E \cup Ops_F$ (definition 5.6). If ProVerif returns a positive verification result then under appropriate assumptions on the cryptographic functions in $Ops_E$ we can conclude that for any attacker $Q_A$ the function $\text{insec}(Q_A|Q, \rho)$ is bounded from above by the negligible function $\text{insec}_\pi(\tilde{Q}^*, A, \tilde{I}^*, \rho, p, k)$ evaluated in $k = k_0$. We can therefore make $Q$ as secure as necessary by simply increasing $k_0$.

As discussed in section 5.2, the last sentence of the preceding paragraph needs to be considered with care since the definition of $\tilde{I}^*$ itself depends on $k_0$—we choose the formatting functions used by the implementation for $k = k_0$ and ideal formatting functions for all other security parameters. However, we can informally argue that $\text{insec}_\pi(\tilde{Q}^*, A, \tilde{I}^*, \rho, p, k)$ should mostly depend on the properties of the cryptographic functions and not on the properties of the formatting functions.

# Chapter 6

# Implementation and Experiments

We implemented our approach in about 12000 lines of OCaml code. Figure 6.1 shows a list of protocol implementations from the Csec Challenge repository [Aizatulin et al., 2011a] that we used to test our method. We list some statistics regarding the sizes of the different stages of the translation: the size of the original C code, the number of generated CVM instructions, the number of lines in the extracted IML model, the sizes of the CryptoVerif and the ProVerif templates written by the user (containing the environment process and the cryptographic assumptions), and the sizes of the automatically generated CryptoVerif and ProVerif processes. The total size of the CryptoVerif input file also includes the automatically generated type declarations and facts about parsing and auxiliary functions ($\Gamma$ and $\Phi$ in the notation of chapter 4) (for instance, the total size of the CryptoVerif input file for RPC-enc is 665 lines), but we feel that the size of the generated process alone is a better measure of the complexity of the model. As we shall see in appendix C the bulk of the generated CryptoVerif process comprises of trivial auxiliary tests that do not play any interesting role in the verification. Figure 6.1 also lists execution times for the CryptoVerif verification (verification with ProVerif is always at least as fast) and analysis outcomes. Most of the times CryptoVerif terminates in under a second. A notable exception is the NSL protocol for which CryptoVerif takes more than a minute to complete. This is not surprising since NSL is the only protocol among the ones we analysed in which the client and the server exchange a total of 3 messages.

**Simple MAC** This is a protocol in which a single payload is concatenated with its MAC. We successfully verify the authenticity of the payload. In CryptoVerif this assumes that MAC is unforgeable against chosen-message attack (UF-CMA).

**Simple XOR** This is the one-time pad example from figure 1.1, but now including both sides of the protocol. CryptoVerif successfully proves secrecy of the payload. We did not analyse the protocol in ProVerif, because properties of XOR are difficult (and in general impossible [Unruh, 2010]) to model within symbolic cryptography.

| | C LOC | CVM LOC | IML LOC | CV LOC | PV LOC | Time | Result |
|---|---|---|---|---|---|---|---|
| Simple MAC | $\sim 250$ | 7K | 78 | $71 + 38$ | $30 + 15$ | 4s | CP |
| Simple XOR | $\sim 100$ | 4K | 56 | $55 + 13$ | — | 3s | C |
| RPC | $\sim 600$ | 56K | 125 | $85 + 60$ | $28 + 25$ | 13s | FCP |
| RPC-enc | $\sim 700$ | 21K | 171 | $159 + 75$ | $60 + 29$ | 9s | CP |
| CSur | $\sim 600$ | 27K | 278 | — | — | 5s | F |
| NSL | $\sim 450$ | 25K | 232 | $160 + 102$ | $118 + 34$ | 10s + 1m16s | FCP* |
| Metering(1) | $\sim 1000$ | 66K | 421 | $171 + 128$ | — | 31s + 2s | FC |
| Metering(3) | — | 95K | 621 | $171 + 220$ | — | 1m21s + 50s | FC |

Figure 6.1: Summary of the analysed implementations. For CryptoVerif and ProVerif we list the sum of the sizes of the user-provided environment (left) and the automatically generated process (right), excluding the sizes of automatically generated function type declarations and facts. When runtime of CryptoVerif is not negligible we split the time column into the execution time of our tools (left) and the execution time of CryptoVerif (right). F—found and fixed flaws, C—verified with CryptoVerif, P—verified with ProVerif, P*—verified with ProVerif and computational soundness applies.

**RPC** This protocol is an implementation of the MAC-based remote procedure call described by Bengtson et al. [2008] and developed and verified by Dupressoir et al. [2014]. We found several flaws in the implementation:

- The functions for sending and receiving network messages take an argument containing the length of the message of type `size_t`. Internally these functions call OpenSSL functions that take an argument of type `int`. The implementation did not check for integer truncation.
- The function for parsing a message did not check that the message contains enough bytes before trying to extract fields from it (at the end of this section we discuss a similar issue that occurred in another implementation).
- The implementation did not impose any bound on the lengths of the request and the response. This was flagged during our verification because we could not prove that the length field does not overflow.

The verification method described by Dupressoir et al. [2014] does not descend into the implementations of functions for constructing and parsing messages, which explains why these flaws remained unnoticed. Upon fixing the flaws we verified the authenticity of the client request and the server response under the assumption that MAC is UF-CMA.

**RPC-enc** This is our running example—a version of the remote procedure call protocol that uses authenticated encryption, described in detail in section 1.1. We verified the authenticity of the request and the response, and the secrecy of the payloads (which is not protected by the MAC-based RPC). In CryptoVerif this assumes authenticated encryption that is in-

```
read(conn_fd, temp, 128);
// BN_hex2bn expects a zero-terminated string
temp[128] = 0;
BN_hex2bn(&cipher_2, temp);
// decrypt and parse cipher_2 to obtain message fields
```

Figure 6.2: A flaw in the CSur example: input may be too short.

distinguishable against chosen plaintext attack (IND-CPA) and provides ciphertext integrity (INT-CTXT). Full CryptoVerif input for this protocol is shown in appendix C.

**CSur** This implementation is developed and analysed in a predecessor paper on C verification [Goubault-Larrecq and Parrennes, 2005]. It is an implementation of a protocol similar to Needham-Schroeder-Lowe. During our verification attempt we discovered a flaw, shown in figure 6.2: The received message in buffer `temp` is being converted to a `BIGNUM` structure `cipher_2` without checking that enough bytes were received. Later a `BIGNUM` structure derived from `cipher_2` is converted to a bitstring without checking that the length of the bitstring is sufficient to fill the message buffer. In both cases the code does not make sure that the information in memory actually comes from the network, which makes it impossible to prove authentication properties. The CSur example has been verified by Goubault-Larrecq and Parrennes [2005], but only for secrecy, and secrecy is not affected by the flaw we discovered. The code reinterprets network messages as C structures (an unsafe practise due to architecture dependence), which is not yet supported by our analysis and so we were not able to verify a fixed version of it.

**NSL** This is our implementation of the Needham-Schroeder-Lowe protocol. We first verified it using ProVerif and applied the computational soundness result of Backes et al. [2009] to obtain a security statement in the computational model, as described in chapter 5. Appendix D shows the extracted ProVerif model.

The computational soundness result assumes that all cryptographic material, including the nonces, is tagged. In general it is difficult to tell whether an assumption of this kind is important or is just an artefact of the soundness proof. When verifying the protocol with CryptoVerif, we discovered that it is most likely the former. Removing the assumption led to the discovery of a potential flaw: If the third message ($B$'s nonce) is sent out without any tagging, it can be decrypted and parsed as the first message of the protocol (an encoding of the concatenation of $A$'s nonce and $A$'s identity). The failure of the parsing may reveal information about the nonce to the attacker (this issue is also discussed in section 4.2). The NSL protocol is often presented without making this point explicit. Once we fixed the problem by explicitly tagging the third message, CryptoVerif successfully verified the protocol.

**Metering** This example is an implementation of a privacy-friendly protocol for smart electricity meters [Rial and Danezis, 2010] developed at Microsoft Research. We were able to

```
unsigned char session_key[256 / 8];
...
// Use the 4 first bytes as a pad to encrypt the reading
encrypted_reading = ((unsigned int) *session_key) ^ *reading;
```

Figure 6.3: A flaw in the metering code: only one byte of the pad is used.

analyse the protocol without modifying its source code. The model that we obtained uncovered a flaw shown in figure 6.3: incorrect use of pointer dereferencing results in three bytes of each four-byte reading being sent unencrypted. This bug is particularly interesting because both the faulty and the fixed dereferencing order have legitimate uses and are not wrong as such. The only way to classify one of them as a bug is by reference to the high-level secrecy property that it violates.

We found two further flaws: one could result in contents of uninitialised memory being sent on the network, the other could lead to 0 being sent (and accepted) in place of the actual number of readings. All flaws have been acknowledged and fixed. An F# implementation of the protocol has been previously verified by Swamy et al. [2010], which highlights the fact that C implementations can be tricky and can easily introduce new bugs, even for correctly specified and proven protocols.

The protocol relies on a number of cryptographic primitives: a signature scheme that is unforgeable against chosen-message attack (UF-CMA), a Diffie-Hellman key-agreement scheme, XOR, and a hash function that needs to be collision-resistant to prove authentication. Secrecy could only be proved under the assumption that the hash function is a random oracle. CryptoVerif shows that whenever the consumer accepts a reading, it does indeed originate from the meter, and that the readings remain secret.

The protocol implementation works with an arbitrary number of messages, that are all batched and signed together. Neither our symbolic execution nor CryptoVerif can deal with loops, so we unroll the main loop of the protocol and only analyse it for a fixed number of messages. The table of results contains numbers for the verification with one and three messages. When trying to verify the protocol for more than one message, we uncovered another potential security flaw that led to a fix in the code. Given two readings $r_1$ and $r_2$, the protocol would generate commitments $C_1$ and $C_2$ by using calls to bignum operations of the OpenSSL crypto library. It would then concatenate the commitments without using their lengths as "`tag`"$|C_1|C_2$. Since bignums in OpenSSL can have arbitrary length, this could lead to a collision where two different bignums $C_1'$ and $C_2'$ concatenate to the same string and lead to the same signature.

Due to the use of XOR and modular exponentiation in the protocol, both of which are difficult to model symbolically, we did not apply ProVerif to the protocol.

**Parsing and Integer Overflow in RPC-enc** We give an account of a particular verification experience with our tools which suggests that getting message parsing right is very error-prone.

```
1 if (msg_len < 5 || msg_len > MAX_MSG1_LENGTH) exit(1));
2 recv(&(ctx->conn_fd), msg, msg_len);
3 unsigned char *p = msg;
4 if (memcmp("p", p, 1)) exit(1);
5 p += 1;
6 ctx->other_len = *((uint32_t *) p);
7 if (msg_len <= 1 + sizeof(uint32_t) + ctx->other_len) exit(1);
8 // extract the name of the client from the message
```

Figure 6.4: Message parsing in RPC-enc.

Our running example RPC-enc uses message formats of the form $c(x, y) = \text{"p"}|\tau_{4u}^{-1}(\mathrm{len}(x))|x|y$. Our first attempt at writing the parsing routine for this format started as shown in figure 6.4. We first check that the expected length of the message is at least 5 bytes (which would correspond to an encoding of two empty bitstrings) and that the message is not too long as prescribed by the protocol. We then receive the message from the network and place it in the buffer pointed to by `msg`. We create a pointer `p` used for parsing, check the tag of the message and advance `p` to point to the length field. We extract the length field and check that it matches the total length of the message before extracting the payloads (not shown here).

This version of the code did not verify because the symbolic execution could not prove that the expression `1 + sizeof(uint32_t) + ctx->other_len` in line 7 does not overflow. In general, failures of the symbolic execution are rather easy to interpret since we can display the problematic symbolic expression together with the line of C code in which every subexpression has been computed.

The overflow issue can be fixed by moving `1 + sizeof(uint32_t)` to the left side of the inequality. Our fix was to check early enough that the first field of the message, which happens to be the identity of the client, does not exceed the maximum allowed length. We added the following line before line 7 in the code:

```
if (ctx->other_len > MAX_PRINCIPAL_LENGTH) exit(1);
```

**Summary** We were able to verify six protocols in the computational model, findings flaws in externally written protocol implementations without modifying their code (we were, of course, developing our own tools as we went along).

The choice of the protocols to study was intended to be representative of the protocols used in practice. NSL is a classic protocol, widely studied in cryptographic literature, as well as being the foundation of the widely deployed Kerberos protocol. RPC and RPC-enc are variants of the protocol that has recently been studied using alternative techniques by Bengtson et al. [2008] and Dupressoir et al. [2014], so it was useful for comparing the methods. We included the CSur protocol since it was the first ever example of a C implementation that got verified for security and it allows us to demonstrate how we improve upon that original attempt.

Our biggest verified protocol, smart metering, supports two claims. First, cryptographic protocols most often have linear structure and are thus supported by our method. Second, our tools are highly automated—we did not need to modify the code of the protocol at all, except for fixing the bugs we found.

In retrospect the fact that the source code of the smart metering protocol is not public made it a poor choice since it makes our experiments not reproducible. We hope to rectify this this by applying our tools to a major open-source protocol, such as PolarSSL.

# Chapter 7

# Future Work

## 7.1 Multipath Symbolic Execution

The main weakness of our method is that it only considers a single path through the program. We would like to make a first step towards full multipath analysis. We follow the approach of CryptoVerif in which verification is a series of transformations that take the original program to an obviously correct one and propose to establish a series of transformations starting from C code. The first step is to generalise our symbolic execution to a program transformation that takes C programs with control flow and removes all pointer manipulation.

The resulting program will still contain destructive updates and will in general have the same control flow structure as the original program. We believe that dealing with these is best left for future work. There are two reasons: First, we would like our method to be usable with a variety of verification backends. Different backends will use different modelling languages, so any further program transformations will depend on the backend. However we do not know of any security verification backend that would use pointers in its language, so a pointer-removing transformation like the one we propose would be generally useful. Second, many of the complexities in transformed programs will be out of reach of the current security verification tools like CryptoVerif. For instance, the TLS protocol uses a counter for each message of the record layer that serves to preserve the order of the messages. This counter is incremented in a loop, and either CryptoVerif would need to be extended in order to treat this accurately, or appropriate approximations need to be made. In each case, it seems appropriate to first simplify several protocol implementations using symbolic execution to see what further theoretical developments would be necessary.

An immediate outcome of our work would be that for the protocols that we analysed so far we would be able to prove absence of non-trivial branching instead of assuming it, thus providing a stronger security guarantee.

The two languages that we translate between are shown in figure 7.1. For simplicity we omit

| C | | IML | |
|---|---|---|---|
| $e \in CExp$, $s_I$ is a symbolic state | | $e \in IExp$ | |
| $P ::=$ | program | $P ::=$ | program |
| $0$ | nil | $0$ | nil |
| $c;\ P$ | C statement | $c;\ P$ | IML statement |
| $\mathsf{while}_{s_I}(e)\{P\};\ P'$ | loop with invariant | $\mathsf{while}\ e\ \mathsf{do}\ P;\ P'$ | loop |
| $\mathsf{if}\,(e)\{P_1\}\{P_2\};\ P'$ | conditional | $\mathsf{if}\ e\ \mathsf{then}\ P_1\ \mathsf{else}\ P_2$ | conditional |

Figure 7.1: The syntax of C and IML.

$$\frac{[\![e]\!]_\eta = \text{true}}{(\eta, \mathsf{while}\ e\ \mathsf{do}\ P; P'), \mathcal{Q} \rightarrow_1 (\eta, P; \mathsf{while}\ e\ \mathsf{do}\ P; P'), \mathcal{Q}} \quad \text{(IWhileTrue)}$$

$$\frac{[\![e]\!]_\eta = \text{false}}{(\eta, \mathsf{while}\ e\ \mathsf{do}\ P; P'), \mathcal{Q} \rightarrow_1 (\eta, P'), \mathcal{Q}} \quad \text{(IWhileFalse)}$$

Figure 7.2: The execution of IML loops.

the translation step from C to CVM (section 3.1) and show symbolic execution as if operating on C programs directly. For now we also ignore the distinction between input and output processes. We start with a C program and translate it into an intermediate model language (IML), similar to the one described in section 2.1, but with an additional loop construct (we do not allow replication inside loops). The execution of IML loops is shown in figure 7.2 (cf. figure 2.4). Both the source and the target language allow the same control flow, the difference lies in the type of expressions and statements. In C the expressions can include reading and writing through pointer variables, whereas in IML all variables hold (arbitrary-length) bitstrings and there is no dereferencing. For the purpose of this section we drop the restriction that input and output processes should alternate.

In chapter 3 we have shown how to define symbolic execution for single C statements, here we would like to show how we intend to generalise it to arbitrary control flow. We shall use a simple solution in which all the paths are unrolled and represented explicitly as a tree of conditionals. This way a C program with 3 conditionals in a row can give rise to an IML program with 7 conditionals. In general this transformation can lead to an exponential increase in the size of the program, but we expect it to work well for protocol implementations because most of the control flow there deals with configuration. Once the configuration parameters are set, most of the conditions in if-statements will become constants, and so branching can be removed. This is done by the rules we describe below.

The symbolic execution that we are about to define starts from a C program $P$ and an initial symbolic state $s$, defined as in section 3.5, and generates an IML program $\tilde{P}$ and a set of symbolic states $S'$ with each $s' \in S'$ representing an outcome of a single path in $P$. Formally we write $[\![P]\!]s = (\tilde{P}, S')$.

$$[\![0]\!]s = (0, \{s\}) \tag{SNil}$$

$$[\![\,\mathsf{exit}\,(); P]\!]s = (\mathsf{out}(\mathit{yield}, ()); 0, \{s\}) \tag{SYield}$$

$$\frac{(s,\ c; P) \xrightarrow{\lambda} (s', P) \text{ as in figure 3.10} \quad [\![P]\!]s' = (\tilde{P}, S')}{[\![c; P]\!]s = (\lambda\tilde{P}, S')} \tag{SStmt}$$

$$\frac{[\![e]\!]s = \text{false}}{[\![\mathsf{while}_{s_I}(e)\{P\}; P']\!]s = [\![P']\!]s} \tag{SWhileFalse}$$

$$\frac{[\![e]\!]s = \text{true}}{[\![\mathsf{while}_{s_I}(e)\{P\}; P']\!]s = [\![P; \mathsf{while}_{s_I}(e)\{P\}; P']\!]s} \tag{SWhileTrue}$$

$$\frac{s_I \sqsubseteq s \quad [\![P]\!](s \sqcap [\![e]\!]s) = (\tilde{P}, S') \quad \forall s' \in S' \colon s_I \sqsubseteq s'}{[\![\mathsf{while}_{s_I}(e)\{P\}]\!]s = (\mathsf{while}\ [\![e]\!]s\ \mathsf{do}\ \tilde{P}, \{s_I \sqcap (\neg[\![e]\!]s)\})} \tag{SWhile}$$

$$[\![\,\mathsf{if}\,(e)\{P_1\}\{P_2\}; P']\!]s = [\![\,\mathsf{if}\,(e)\{P_1; P'\}\{P_2; P'\}]\!]s \tag{SSeqIf}$$

$$\frac{[\![e]\!]s = \text{true} \quad [\![P_1]\!]s = (\tilde{P}_1, S'_1)}{[\![\,\mathsf{if}\,(e)\{P_1\}\{P_2\}]\!]s = (\tilde{P}_1, S'_1)} \tag{SIfTrue}$$

$$\frac{[\![P_1]\!]s = (\tilde{P}_1, S'_1) \quad [\![P_2]\!]s = (\tilde{P}_2, S'_2)}{[\![\,\mathsf{if}\,(e)\{P_1\}\{P_2\}]\!]s = (\mathsf{if}\ [\![e]\!]s\ \mathsf{then}\ \tilde{P}_1\ \mathsf{else}\ \tilde{P}_2, S'_1 \cup S'_2)} \tag{SIf}$$

Figure 7.3: The generalised symbolic execution.

The generalised symbolic execution rules are presented in figure 7.3. The first two rules (SNil) and (SYield) deal with termination. We do not analyse the process that follows an exit statement and replace it by the nil process. The rule (SStmt) lifts the rules of our single-path execution described in section 3.5 into multipath setting. We need to make one change to the algorithm in section 3.5: in order to capture the action of loops in IML we treat C memory locations as IML variables and transform each C operation that updates the memory into an IML assignment. Memory locations are represented by pointer bases (figure 3.7), and so every time a C program writes through the pointer with base $pb$ we add a statement of the form $\mathsf{let}\ pb = e\ \mathsf{in}\ \ldots$ to the IML model. The models will therefore contain destructive updates that change values of already assigned variables. We shall call $pb$ a *location variable* in the following. The location variables only need to be updated once at the end of each branch in each loop—this can be achieved by a separate simplifying transformation.

The next three rules deal with loops. If the condition of a loop is statically known to be false or true, the loop is unrolled by the rules (SWhileTrue) and (SWhileFalse). This will be especially useful when symbolically executing library initialisation routines that typically iterate over statically known arrays of configuration parameters. For a C expression $e$ we write $[\![e]\!]s$ to denote the IML expression that results from symbolically executing $e$ in a state $s$.

The rule (SWhile) executes arbitrary loops and relies on a user-provided invariant, which is just a symbolic state $s_I$. We check that the invariant $s_I$ is an under-approximation of each state $s$ that results from executing the loop, denoted by $s_I \sqsubseteq s$. For two symbolic states $s = (\Phi, \mathcal{A}^s, \mathcal{M}^s, \mathcal{S}^s)$ and $s' = (\Phi', \mathcal{A}^{s'}, \mathcal{M}^{s'}, \mathcal{S}^{s'})$ we let $s \sqsubseteq s'$ whenever 1) $\Phi' \models \Phi$, 2) $\mathcal{A}^s \subseteq \mathcal{A}^{s'}$ 3) $\mathcal{S}^s = \mathcal{S}^{s'} = []$, and 4) $\mathrm{dom}(\mathcal{M}^s) \subseteq \mathrm{dom}(\mathcal{M}^{s'})$ and for all $pb \in \mathrm{dom}(\mathcal{M}^s)$ either $\mathcal{M}^s(pb) = \mathcal{M}^{s'}(pb)$ or $\mathcal{M}^s(pb) = pb$. Condition 1 requires that all facts in $s$ must also be satisfied in $s'$. Condition 2 requires that the allocation tables agree. This condition implies a limitation—a memory location that is allocated in a loop is only accessible in the same iteration. In particular, a buffer allocated by a `malloc` inside a loop is not accessible outside the loop. Condition 3 requires that the loop execution must finish with an empty stack—this is always the case because we only use the stack to store intermediate values when evaluating expressions. Finally, condition 4 requires that each memory location should either remain untouched by the loop, or be replaced by a location variable. The IML model of the loop body would contain an assignment to the location variable, as described above in the discussion of the rule (SStmt). The rule (SWhile) is based on the conventional reasoning rule for loops. Given a boolean IML expression $e$ and a symbolic state $s$ we write $s \sqcap e$ to denote the result of adding $e$ to the facts in $s$.

In practice the user does not need to provide the full description of the invariant state $s_I$. Instead, the user provides a formula $\phi$ that may mention location variables. We then construct $s_I$ by taking the pre-loop state $s$ and replacing memory locations mentioned in $\phi$ by location variables. More precisely, we obtain $s_I$ from $s$ as follows:

- Remove all facts from $s$ that mention location variables contained in $\phi$. These location variables are expected to be updated by the loop, and so the facts about them may no longer hold.
- For each location variable $pb$ mentioned in $\phi$ set $\mathcal{M}^s(pb) := pb$.

The rule (SIf) deals with conditionals and is straightforward. If we know the condition to be always true, we can apply the *pruning* rule (SIfTrue) to remove one of the branches. There is also the corresponding rule for false conditions, which we omit here. The rule (SSeqIf) performs the unrolling of the conditionals that we mentioned above. In general this rule can lead to an exponential increase in the size of the program, but we expect the pruning rules to remove most of the newly introduced branches.

Our treatment of exit statements introduces an over-approximation. For instance, a C program

```
while(TRUE){out("hello"); exit();}
```

would yield the model

**while** true **do** (**out**("hello"); **out**(*yield*, ()); 0)

```
char * buf = malloc(len);
size_t received = 0;
if(buf == NULL) exit();
while_{s_I}(received < len)
  received += recv(buf + received, len - received);
encrypt(ebuf, buf, len); ...
```

---

**let** buf = $\varepsilon$ **in**
**let** received = 0 **in**
**while** (received < len) **do** {
  **in**(tmp);
  **let** buf = buf | tmp **in**
  **let** received = received + len(tmp) **in** 0 }
**let** ebuf = encrypt(buf) **in** ...

Figure 7.4: An example C fragment together with the extracted model. Location variables are given meaningful names that correspond to C variable names. We use the loop invariant $s_I = (\mathit{received} = \mathrm{len}(\mathit{buf})) \wedge (\mathit{received} \leq \mathit{len})$.

that may output "hello" more than once. Such an exit statement introduces an *indirect jump* in the control flow graph (not to be confused with *non-local* jumps in C) that is not dealt with by the symbolic execution presented here. The same reason prevents us from analysing C programs with goto-statements. To account for that we could formulate the generalised symbolic execution for control flow graphs with indirect jumps, or we could transform the C program to only contain while loops without indirect jumps. We leave this for future research.

In order to ensure that IML processes execute in polynomial time, we need to prove that all loops in the C program terminate in polynomial time as well. This is a separate concern and we leave it for future research. The techniques used in the *Terminator* tool by Cook et al. [2006] could possibly be adapted to our setting.

**Example 7.1** An example of a translation is shown in figure 7.4. In the original C program the call to recv updates the memory through the pointer buf + received. In the model the update is performed on values instead. Even though the extracted model is not significantly shorter in this case, it is much simpler to reason about—we do not have to deal with issues of pointer aliasing and memory layout. A further program transformation could show that the loop in the IML model can be replaced by a single input statement that has the same effect. $\square$

The soundness of the multipath symbolic execution can be expressed using the simulation relation developed in section 2.3 as follows:

**Theorem 7.1 (Multipath Symbolic Execution is Sound)** *Assume that* $[\![P]\!]s_0 = (\tilde{P}, S')$ *for some C program* $P$*, IML process* $\tilde{P}$*, the initial symbolic state* $s_0$*, and a set of symbolic states* $S'$*. Then* $P \lesssim \tilde{P}$*.* $\square$

Proof (Sketch) Following section 3.7 we define the state correspondence relation $c \sim_\eta s$ between concrete C states $c$ and symbolic states $s$ with respect to an environment $\eta$. The relation states that $s$ agrees with $c$ when each symbolic expression in $s$ is evaluated using $[\![\cdot]\!]_\eta$.

We then define the simulation relation between an executing process $(\eta, c, P)$ of the C program and an executing process $(\tilde{\eta}, \tilde{P})$ of the IML process as follows: $(\eta, c, P) \lesssim (\tilde{\eta}, \tilde{P})$ when $\eta = \tilde{\eta}$ and there exists a symbolic state $s$ such that $c \sim_\eta s$ and $[\![P]\!]s = (\tilde{P}, S')$ for some set $S'$ of symbolic states.

We sketch the proof of the simulation property (condition 5 in definition 2.11) for loops. Assume that $(\eta, c, \mathsf{while}_{s_I}(e)\{P\}; P') \lesssim (\eta, \mathsf{while}(\tilde{e})\{\tilde{P}\}; \tilde{P}')$. First consider the case $[\![e]\!]_\eta =$ true, such that the C process reduces to $(\eta, c, P; \mathsf{while}_{s_I}(e)\{P\}; P')$. We obtain the symbolic state $s$ from the definition of the simulation relation and show that

$$[\![P; \mathsf{while}_{s_I}(e)\{P\}; P']\!]s = (\tilde{P}; \mathsf{while}(\tilde{e})\{\tilde{P}\}; \tilde{P}', S')$$

for some set $S'$ of symbolic states. To do this we “unroll” the symbolic execution rule (SWhile). This establishes that both executing processes reduce to processes that still simulate each other. The case $[\![e]\!]_\eta =$ false and the other symbolic execution rules are dealt with similarly. ∎

## 7.2 Observational Equivalence

Our work so far only considers *trace* properties that either hold or do not hold for every single trace. These include properties like *authentication* (if a participant accepts a message then the message has been sent by another honest participant) and *weak secrecy* (the attacker never learns the whole secret). In future we would like to add treatment of *observational equivalence* properties that speak about sets of traces. Two processes are considered observationally equivalent if they cannot be distinguished by any context with non-negligible probability. This allows to formulate the property of *strong secrecy*—the attacker should not be able to distinguish between two copies of a process that run with different values of the secret.

Following Blanchet [2008] we can define observational equivalence with respect to the CryptoVerif semantics as follows: let $\Pr[Q \rightsquigarrow 0]_k$ be the probability that the process $Q$ outputs 0 on a special channel *guess* when executed with respect to the security parameter $k$. Two processes $Q$ and $Q'$ are *observationally equivalent*, denoted by $Q \sim \tilde{Q}$, if for every context $C$ the value $\mathrm{diff}_{CV}(Q, \tilde{Q}, C, k) = |\Pr[C\{Q\} \rightsquigarrow 0]_k - \Pr[C\{\tilde{Q}\} \rightsquigarrow 0]_k|$ is negligible in the security parameter. We define $\mathrm{diff}_{\mathrm{IML}}$ in the same way, except that in IML semantics the security parameter is fixed.

In order to define strong secrecy CryptoVerif uses a special construct find that allows one process to directly access the variables of another process. This way, a process $Q$ *preserves the secrecy* of a variable $x$ if $Q_x|Q \sim Q'_x|Q$, where $Q_x$ is the process that upon request reveals the value of $x$ in $Q$ (retrieving it using find) and $Q'_x$ reveals garbage instead.

It is easy to see that our simulation relation developed in section 2.3 is well-suited for proving observational equivalence for IML processes: if $Q \lesssim \tilde{Q}$ and $\tilde{Q} \lesssim Q$ then for every context $C$ each trace of $C\{Q\}$ can be mapped to a trace of $C\{\tilde{Q}\}$ of the same probability and vice versa, therefore $\mathrm{diff}_{\mathrm{IML}}(Q, \tilde{Q}, C) = 0$. The problem comes from the interaction of the find construct with our simulation relation: being able to look inside another process breaks the invariance of simulation against embedding. If a context $C$ can directly access variables inside $Q$ then $Q \lesssim \tilde{Q}$ does not necessarily imply $C\{Q\} \lesssim C\{\tilde{Q}\}$.

CryptoVerif deals with this issue be defining observational equivalence with respect to a set of public variables that may be accessed by the context. A similar approach could work for our simulation relation. Another option would be not to use find in full generality and only allow a limited set of primitives that allow to model strong secrecy and do not break the simulation relation.

Multipath analysis brings additional challenges for observational equivalence. The algorithm that we propose in section 7.1 may yield models with different control flow when an exit statement is executed inside a loop. An example shown in section 7.1 would be the program `while(TRUE){out("hello"); exit();}` such that the extract model can print "hello" multiple times and would therefore not be observationally equivalent. We leave the best way of dealing with these situations for future research.

# Chapter 8

# Conclusion

We presented a method that proves authentication and weak secrecy properties of C protocol implementations. In order to formalise the security properties and prove soundness of our method we develop a theoretical framework in which C programs are embedded in a context written in a process calculus. The context allows to describe both the environment in which the protocol participants execute and the attacker. Our analysis method works by extracting models $\tilde{Q}_1, \dots, \tilde{Q}_n$ from the source code $Q_1, \dots, Q_n$ of each protocol participant, and then showing that for any context $C$ the process $Q = C\{Q_1, \dots, Q_n\}$ is simulated by $\tilde{Q} = C\{\tilde{Q}_1, \dots, \tilde{Q}_n\}$, which implies that $Q$ satisfies all the security properties of $\tilde{Q}$. We then show conditions under which $\tilde{Q}$ can be simplified enough to be analysed with CryptoVerif or ProVerif.

Overall, our work supports the following *thesis*:

> *Process calculus embedding combined with model extraction is an appropriate tool for analysing security of C programs.*

Below we discuss several aspects of the approach that we believe makes it appropriate.

Formalising security properties by embedding C programs in a process calculus context is *precise*. All we need is to describe the semantics of the process calculus and the source language. This is done in a fully formal fashion. The attacker is itself just a process that runs in parallel with the protocol and interacts with the protocol using the communication primitives provided by the process calculus. Thus there is no vagueness about the threat model. We believe that all our definitions and proofs would be rather straightforward to formalise in a theorem prover like Isabelle [Paulson, 1994] or Coq [The Coq development team, 2004].

A popular alternative approach is to define the attacker as an arbitrary machine that interacts with the protocol. The rules of the interaction are then described in a natural language, as in the definition used by Backes et al. [2009] and reviewed in section 5.1.1. Initially we followed the same approach, but found it much harder to use, because for all practical purposes an arbitrary machine cannot be written down explicitly. This introduces a lot of informal arguments about how the machines interact. In our approach we can explicitly write down an attacker, as demonstrated by the examples in section 2.1. The interaction rules are given by the communication rules of the process calculus, which makes them amenable to fully formal treatment.

The simulation relation that we develop for proving security properties (section 2.3) is *convenient*. Simulation is a local property in that it can be checked by considering each participant process in isolation, and is very well suited for proofs by induction on the structure of the participant. At the same time simulation implies strong global properties for arbitrary interactions of the protocol participants with the attacker.

The use of a process calculus for formalisation builds upon abundant *prior work* on verifying security of process calculi. This allows us to reuse CryptoVerif and ProVerif with relative ease, as our security definitions are chosen to be compatible with theirs.

The use of model extraction for verifying security properties achieves high *automation*. There is no need for the user to provide a protocol specification, as for most industrial protocols such a specification does not exist. Symbolic execution as a tool for model extraction does not impose any restrictions on the coding style and can readily be applied to code that was not written with verification in mind, as demonstrated by our analysis of the metering protocol (chapter 6). Currently we restrict our analysis to a single path through the code, but we believe that it can be generalised to cover multiple paths as demonstrated in section 7.1.

**Symbolic versus Computational Verification** When we first extracted models from the C code, we decided to verify them with ProVerif first, as it is simpler to use. We quickly discovered that transferring the result of the ProVerif verification to the computational setting is quite challenging—the computational result that we used requires the tupling and projection operations to fulfil rather strict criteria that are typically not satisfied in practice, such as the requirement that the tupling operations should be defined for bitstrings of arbitrary length. Proving security with ProVerif required quite a bit of additional argumentation, as described in chapter 5. When we later verified our models with CryptoVerif, we found it a much more natural match for the computational semantics of C that we use. Currently all the results that we obtained with ProVerif are subsumed by the results that we get by applying CryptoVerif directly. It is an interesting question whether there is a protocol that supports a computational soundness result, but cannot be verified with CryptoVerif directly.

ProVerif is still very useful, if one is willing to accept an interpretation with respect to a symbolic model of cryptography, as in Dupressoir et al. [2014]. In this interpretation the attacker is weaker—it is restricted to use only the operations that are allowed by the rewriting rules of the symbolic model. Furthermore the properties are guaranteed to hold only for those traces in which there are no collisions, that is, where syntactically distinct symbolic expressions both on the attacker and the protocol side evaluate to different bitstrings.

**Coverage of the Method** We conclude with a short summary of what our method achieves—what kinds of security flaws are currently covered and what kinds are out of reach. The main weakness of the current implementation is that it provides a verification statement only for a single path in the code. It easy to manually check whether the single path assumption is satisfied: our security theorems apply exactly to protocol code that contains no loops, and in which at least one branch of every conditional contains an exit-statement that is not preceded by an attacker-observable output. The types of flaws that are caught by such an analysis include

- Memory safety and integer overflow errors. Even though we do not target denial of service attacks directly, many denial of service attacks exploit memory safety violations, and so would be prevented by our verification.

- Incorrect use of pointers that leads to errors like the one in the metering protocol, where only a single byte of the key is used instead of four bytes.

- Errors in formatting and parsing of messages. For instance, in the metering protocol we discovered a potentially dangerous violation of injectivity in the format of a tupling operation—two different pairs of payloads may be tupled leading to the same result and producing the same signature. When developing our own implementations we found that message parsing code is prone to subtle integer overflow bugs. The reason for this complexity is that we use the information contained in the (potentially malicious) message to parse the message itself.

- Logical flaws in the protocol structure, such as the famous flaw in the NSL protocol that enables the attack by Lowe [1995].

Some classes of bugs occur only in programs that are currently not covered by our method. These include

- The OpenSSL bug in which a certificate is not properly checked, even though subsequent code assumes that it is [CVE, a]. To catch this bug we would need to analyse the certificate checking function, but we cannot do it because the function is recursive.

- Problems related to the state maintained by the protocol, such as the reinitialisation flaw discovered in the TLS protocol in 2009 [CVE, b]. This is a limitation of CryptoVerif and ProVerif: they do not support stateful protocols. Arapinis et al. [2011] show how to apply ProVerif to verify a certain class of protocols with state in their tool StatVerif.

- Information leaks due to error messages, such as the leak exploited by Albrecht et al. [2009]. Protocols that terminate with an error message are currently not analysed, but we expect that the generalisation proposed in section 7.1 will cover those.

- Bugs in protocols that use message formats not covered by our encoding expressions. This mostly affects protocols like HTML that use explicit delimiters to separate fields. We are not aware of a cryptographic protocol that would use this style of formatting.

- Bugs in protocols that use integer overflow intentionally, say, to implement a message counter. Our solving algorithm currently prohibits all integer overflow, as described in section 3.2.

- Timing attacks [Kocher, 1996].

We trust the correctness of CIL, so that the link between C and CVM is made only informally. We also trust the user input that describes the assumptions about the cryptographic primitives used by the implementation and the environment in which the implementation executes. This input includes

- The proxy functions that link the arguments of C functions to the arguments of the cryptographic primitives in the model.

- The CryptoVerif or ProVerif template that describes the types and assumed properties of the cryptographic primitives, and the environment in which the protocol participants execute.

- Special function calls in the C code that specify where an event is raised or an environment variable is read.

Given that the problem we are dealing with is in general undecidable, the quality of the method needs to be evaluated by applying it to a wide range of protocol implementations. Unfortunately, we are not able to release the source code of our largest experimental example—the Microsoft Research metering code—due to licensing restrictions. Next we would like to apply our method to a major open source implementation, such as PolarSSL, to enable a more detailed public scrutiny of our verification.

Who should be using our tools? To an industrial user a “proof of security” is of little interest—it is very difficult for a non-expert to understand what exactly is being proven, and protocols that have been proven correct are sometimes broken in practice later because of the mismatch of the proof model and the reality. There is often lack of incentive to provide security and the most prevalent security flaws are much shallower than the ones we study in this thesis—many businesses are prone to much simpler social attacks. We would therefore expect hackers to be the most likely early adopters of our technology, which may in turn raise awareness and interest among protocol designers. For that to happen, it is still important to demonstrate that our tools are effective in finding security flaws—we hope that analysing PolarSSL allows us to find interesting vulnerabilities.

Adoption of software verification is difficult because it is a niche branch of research and the potential user base is very limited. However, the symbolic execution presented in chapter 3 is of a more general interest: it provides the user with an explanation of how values in memory are computed, which is useful even without respect to security. We would therefore like to separate the symbolic execution part and integrate it into a general-purpose verifier such as Frama-C.

This is made easier by the fact that both our tools and Frama-C are written in OCaml, and Frama-C has an extensible modular architecture.

# Appendix A

# Proxy Function Examples

We show several examples of proxy functions to showcase how we support various important primitives that are likely to be used by most protocol implementation such as network inputs and outputs, randomness generation, or bitstring comparisons. We also include all the functions used by our example in figure 1.1 that lead to the CVM translation shown in figure 3.2.

The functions send_proxy and recv_proxy emit instructions Output and Input. These are abbreviations and result in two CVM instructions being executed. The instruction Output generates the sequence WriteEnv $x$; Out $x$, and the instruction Input generates the sequence In $x$; ReadEnv $x$, both with a fresh variable $x$.

We expect the functions send_proxy and recv_proxy to send and receive their arguments all at once. This does not quite correspond to the standard semantics in which the functions are allowed to process only part of the string (in which case they return the length of the string that has been processed). In order to handle this semantics correctly, these functions would need to be wrapped in a loop that keeps trying to send or receive until the whole message is delivered. We leave accurate treatment of such loops to future work—figure 7.4 in section 7.1 shows how this can be done.

The function *cmp* that is generated by the memcmp_proxy call is treated specially by our solver: there is a rewriting rule that transforms $\tau_{\text{int}}(cmp(e_1, e_2)) = 0$ into the expression $e_1 = e_2$ (the rule (R8) in section 3.2). This is where most of the bitstring comparisons in our models come from.

The function RAND_bytes_proxy encodes an assumption that the function terminates upon failure. We need this assumption since we do not currently capture the dependence between the bitstring placed by the function into the buffer and the return code of the function. This problem will resolve once we transition to the multipath setting.

```
void LoadBuf(void * buf; size_t len) __attribute__((not_instrumented)) {
  cvm("Load");
}
```

```
void StoreBuf(void * buf) __attribute__((not_instrumented)) {
  cvm("Store");
}


void Val(bool is_signed, size_t width){
  char cmd[100];
  sprintf(cmd, "Apply Val(%b, %ld)", is_signed, width);
  cvm(cmd);
}

void assume_len(const unsigned char * len, bool is_signed, size_t width)
{
  cvm("Dup");
  cvm("Len");
  LoadBuf(len, width);
  Val(is_signed, width);
  cvm("Apply EqInt/2");
  cvm("Assume");
}

void *malloc_proxy(size_t size){
  mute_cvm();
  void * ret = malloc(size);
  unmute_cvm();
  LoadBuf(&size, sizeof(size));
  cvm("Malloc");
  StoreBuf(&ret);
  return ret;
}

int memcmp_proxy(void * a, void * b, size_t len){
  mute_cvm();
  int ret = memcpy(a, b, len);
  unmute_cvm();
  LoadBuf(a, len);
  LoadBuf(b, len);
  cvm("Apply cmp/2");
  size_t len = sizeof(ret);
  assume_len(&len, FALSE, sizeof(len));
  StoreBuf(&ret);
  return ret;
}
```

```
void *memcpy_proxy(void * dest, void const * src, size_t len){
  void * ret = dest;
  mute_cvm();
  ret = memcpy(dest, src, len);
  unmute_cvm();
  LoadBuf(src, len);
  StoreBuf(dest);
  return ret;
}

size_t recv_proxy(void *buf, size_t len){
  mute_cvm();
  size_t ret = net_recv(ctx, buf, len);
  unmute_cvm();

  ret = len;

  LoadBuf(&len, sizeof(len));
  cvm("Input");
  StoreBuf(buf);

  return ret;
}

size_t send_proxy(void *buf, size_t len){
  mute_cvm();
  size_t ret = net_send(ctx, buf, len);
  unmute_cvm();

  ret = len;

  LoadBuf(buf, ret);
  cvm("Output");

  return ret;
}

void * otp_proxy(size_t len){
  mute_cvm();
  void * ret = otp(len);
  unmute_cvm();
```

```
  LoadBuf(&len, sizeof(len));
  cvm("Malloc");
  cvm("Dup");
  cvm("Apply null/0");
  cvm("Apply !=/2");
  cvm("Apply truth/1");
  cvm("Assume");
  StoreBuf(&ret);

  cvm("ReadEnv pad");
  assume_len(&len, FALSE, sizeof(len));
  StoreBuf(ret);

  return ret;
}

int RAND_bytes_proxy(void *buf, int num)
{
  mute_cvm();
  int ret = RAND_bytes(buf, num);
  unmute_cvm();

  // Allow the attacker to choose the return value
  size_t ret_len = sizeof(ret);
  LoadBuf(&ret_len, sizeof(ret_len));
  cvm("In");
  StoreBuf(ret);

  // RAND_bytes() returns 1 on success, 0 otherwise.
  if(ret != 1) exit(1);

  LoadBuf(&num, sizeof(num));
  cvm("New");
  StoreBuf(buf);

  return ret;
}
```

# Appendix B

# Extended Solving Example

We demonstrate a typical case of overflow analysis. Imagine that we consider the value $x$ to be a concatenation of the form $x = \tau_{1u}^{-1}(\text{len}(x_1))|\text{'p'}|x_1|x_2$ and would like to prove that the value of $x_2$ is equal to a particular bitstring, say, "secret". We would then try to prove the fact

$$x\{1 + 1 + \tau_{1u}(x\{1,1\}), \text{len}(x) - (1 + 1 + \tau_{1u}(x\{1,1\}))\} = \text{"secret"}$$

The conditions checked by a C implementation would typically involve bitstring operations and might look something like this:

$$\begin{aligned}
&x\{\tau_{1u}(\text{cast}_{\tau_{1s}\to\tau_{1u}}(\tau_{1s}^{-1}(2) \oplus_{\tau_{1s}} \text{cast}_{\tau_{1u}\to\tau_{1s}}(x\{1,1\}))),\\
&\qquad \tau_{1u}(\tau_{1u}^{-1}(\text{len}(x)) \ominus_{\tau_{1u}} (\tau_{1u}^{-1}(2) \oplus_{\tau_{1u}} x\{1,1\}))\} = \text{"secret"},\\
&\mathit{truth}(x\{1,1\} \le_{\tau_{1u}} \tau_{1u}^{-1}(100)),\\
&\mathit{truth}(\tau_{1u}^{-1}(\text{len}(x)) \ge_{\tau_{1u}} \tau_{1u}^{-1}(200)).
\end{aligned}$$

This mix of signed and unsigned representations is quite typical in cases where an implementation uses unsigned types like `size_t` to store lengths, but calls the OpenSSL library which uses the signed type `int` for lengths.

Below we show the application of our rewriting procedure to all 4 involved facts. This has been automatically generated by inserting callbacks into the rewriting procedure to record each step.

$$\begin{aligned}
&\mathbf{1} \quad \mathit{truth}(\tau_{1u}^{-1}(\text{len}(x)) \ge_{\tau_{1u}} \tau_{1u}^{-1}(200))\\
&\rightsquigarrow \tau_{1u}(\tau_{1u}^{-1}(\text{len}(x))) \ge \tau_{1u}(\tau_{1u}^{-1}(200))\\
&\rightsquigarrow \text{defined}(\tau_{1u}(\tau_{1u}^{-1}(\text{len}(x)))) \wedge \text{defined}(\tau_{1u}(\tau_{1u}^{-1}(200)))\\
&\qquad \wedge (\tau_{1u}(\tau_{1u}^{-1}(\text{len}(x))) \ge \tau_{1u}(\tau_{1u}^{-1}(200)))
\end{aligned}$$

$$\rightsquigarrow (1 = \text{len}(\tau_{1u}^{-1}(\text{len}(x)))) \wedge (\tau_{1u}(\tau_{1u}^{-1}(\text{len}(x))) \geq \tau_{1u}(\tau_{1u}^{-1}(200)))$$

$$\rightsquigarrow \text{defined}(1) \wedge \text{defined}(\text{len}(\tau_{1u}^{-1}(\text{len}(x)))) \wedge (1 = \text{len}(\tau_{1u}^{-1}(\text{len}(x)))) \wedge (\tau_{1u}(\tau_{1u}^{-1}(\text{len}(x))) \geq \tau_{1u}(\tau_{1u}^{-1}(200)))$$

$$\rightsquigarrow \text{defined}(\text{len}(\tau_{1u}^{-1}(\text{len}(x)))) \wedge (1 = \text{len}(\tau_{1u}^{-1}(\text{len}(x)))) \wedge (\tau_{1u}(\tau_{1u}^{-1}(\text{len}(x))) \geq \tau_{1u}(\tau_{1u}^{-1}(200)))$$

$$\rightsquigarrow \text{defined}(\tau_{1u}^{-1}(\text{len}(x))) \wedge (1 = \text{len}(\tau_{1u}^{-1}(\text{len}(x)))) \wedge (\tau_{1u}(\tau_{1u}^{-1}(\text{len}(x))) \geq \tau_{1u}(\tau_{1u}^{-1}(200)))$$

$$\rightsquigarrow (\text{len}(x) \geq 0) \wedge (\text{len}(x) \leq (256-1)) \wedge (1 = \text{len}(\tau_{1u}^{-1}(\text{len}(x)))) \wedge (\tau_{1u}(\tau_{1u}^{-1}(\text{len}(x))) \geq \tau_{1u}(\tau_{1u}^{-1}(200)))$$

$$\rightsquigarrow \text{defined}(\text{len}(x)) \wedge \text{defined}(0) \wedge (\text{len}(x) \geq 0) \wedge (\text{len}(x) \leq (256-1)) \wedge (1 = \text{len}(\tau_{1u}^{-1}(\text{len}(x)))) \wedge (\tau_{1u}(\tau_{1u}^{-1}(\text{len}(x))) \geq \tau_{1u}(\tau_{1u}^{-1}(200)))$$

$$\rightsquigarrow \text{defined}(x) \wedge (\text{len}(x) \geq 0) \wedge (\text{len}(x) \leq (256-1)) \wedge (1 = \text{len}(\tau_{1u}^{-1}(\text{len}(x)))) \wedge (\tau_{1u}(\tau_{1u}^{-1}(\text{len}(x))) \geq \tau_{1u}(\tau_{1u}^{-1}(200)))$$

$$(25) \rightsquigarrow \text{defined}(x) \wedge (\text{len}^y(x) \geq 0) \wedge (\text{len}(x) \leq (256-1)) \wedge (1 = \text{len}(\tau_{1u}^{-1}(\text{len}(x)))) \wedge (\tau_{1u}(\tau_{1u}^{-1}(\text{len}(x))) \geq \tau_{1u}(\tau_{1u}^{-1}(200)))$$

$$\rightsquigarrow \text{defined}(x) \wedge (\text{len}^y(x) \geq 0) \wedge \text{defined}(\text{len}(x)) \wedge \text{defined}(256-1) \wedge (\text{len}(x) \leq (256-1)) \wedge (1 = \text{len}(\tau_{1u}^{-1}(\text{len}(x)))) \wedge (\tau_{1u}(\tau_{1u}^{-1}(\text{len}(x))) \geq \tau_{1u}(\tau_{1u}^{-1}(200)))$$

$$\rightsquigarrow (\text{len}^y(x) \geq 0) \wedge \text{defined}(x) \wedge (\text{len}(x) \leq (256-1)) \wedge (1 = \text{len}(\tau_{1u}^{-1}(\text{len}(x)))) \wedge (\tau_{1u}(\tau_{1u}^{-1}(\text{len}(x))) \geq \tau_{1u}(\tau_{1u}^{-1}(200)))$$

$$(25) \rightsquigarrow (\text{len}^y(x) \geq 0) \wedge \text{defined}(x) \wedge (\text{len}^y(x) \leq (256-1)) \wedge (1 = \text{len}(\tau_{1u}^{-1}(\text{len}(x)))) \wedge (\tau_{1u}(\tau_{1u}^{-1}(\text{len}(x))) \geq \tau_{1u}(\tau_{1u}^{-1}(200)))$$

$$(20) \rightsquigarrow (\text{len}^y(x) \geq 0) \wedge \text{defined}(x) \wedge (\text{len}^y(x) \leq (256-1)) \wedge (1 = 1) \wedge (\tau_{1u}(\tau_{1u}^{-1}(\text{len}(x))) \geq \tau_{1u}(\tau_{1u}^{-1}(200)))$$

$$\rightsquigarrow (\text{len}^y(x) \geq 0) \wedge \text{defined}(x) \wedge (\text{len}^y(x) \leq (256-1)) \wedge (1 = \text{len}(\tau_{1u}^{-1}(200))) \wedge (\tau_{1u}(\tau_{1u}^{-1}(\text{len}(x))) \geq \tau_{1u}(\tau_{1u}^{-1}(200)))$$

$$\rightsquigarrow (\text{len}^y(x) \geq 0) \wedge \text{defined}(x) \wedge (\text{len}^y(x) \leq (256-1)) \wedge \text{defined}(1) \wedge \text{defined}(\text{len}(\tau_{1u}^{-1}(200))) \wedge (1 = \text{len}(\tau_{1u}^{-1}(200))) \wedge (\tau_{1u}(\tau_{1u}^{-1}(\text{len}(x))) \geq \tau_{1u}(\tau_{1u}^{-1}(200)))$$

$$\rightsquigarrow (\text{len}^y(x) \geq 0) \wedge \text{defined}(x) \wedge (\text{len}^y(x) \leq (256-1)) \wedge (1 = \text{len}(\tau_{1u}^{-1}(200))) \wedge (\tau_{1u}(\tau_{1u}^{-1}(\text{len}(x))) \geq \tau_{1u}(\tau_{1u}^{-1}(200)))$$

$$\rightsquigarrow (\text{len}^y(x) \geq 0) \wedge \text{defined}(x) \wedge (\text{len}^y(x) \leq (256-1)) \wedge \text{defined}(\tau_{1u}^{-1}(200)) \wedge (1 = \text{len}(\tau_{1u}^{-1}(200))) \wedge (\tau_{1u}(\tau_{1u}^{-1}(\text{len}(x))) \geq \tau_{1u}(\tau_{1u}^{-1}(200)))$$

$$\rightsquigarrow (\text{len}^y(x) \geq 0) \wedge \text{defined}(x) \wedge (\text{len}^y(x) \leq (256-1)) \wedge (200 \geq 0) \wedge (200 \leq (256-1))$$

$$\wedge\ (1 = \text{len}(\tau_{1u}^{-1}(200))) \wedge (\tau_{1u}(\tau_{1u}^{-1}(\text{len}(x))) \geq \tau_{1u}(\tau_{1u}^{-1}(200)))$$

$$\rightsquigarrow (\text{len}^y(x) \geq 0) \wedge \text{defined}(x) \wedge (\text{len}^y(x) \leq (256-1)) \wedge \text{defined}(200) \wedge \text{defined}(0)$$

$$\wedge\ (1 = \text{len}(\tau_{1u}^{-1}(200))) \wedge (\tau_{1u}(\tau_{1u}^{-1}(\text{len}(x))) \geq \tau_{1u}(\tau_{1u}^{-1}(200)))$$

$$\rightsquigarrow (\text{len}^y(x) \geq 0) \wedge \text{defined}(x) \wedge (\text{len}^y(x) \leq (256-1)) \wedge (1 = \text{len}(\tau_{1u}^{-1}(200)))$$

$$\wedge\ (\tau_{1u}(\tau_{1u}^{-1}(\text{len}(x))) \geq \tau_{1u}(\tau_{1u}^{-1}(200)))$$

$$\rightsquigarrow (\text{len}^y(x) \geq 0) \wedge \text{defined}(x) \wedge (\text{len}^y(x) \leq (256-1)) \wedge \text{defined}(200) \wedge \text{defined}(256-1)$$

$$\wedge\ (1 = \text{len}(\tau_{1u}^{-1}(200))) \wedge (\tau_{1u}(\tau_{1u}^{-1}(\text{len}(x))) \geq \tau_{1u}(\tau_{1u}^{-1}(200)))$$

$$\rightsquigarrow (\text{len}^y(x) \geq 0) \wedge \text{defined}(x) \wedge (\text{len}^y(x) \leq (256-1)) \wedge (1 = \text{len}(\tau_{1u}^{-1}(200)))$$

$$\wedge\ (\tau_{1u}(\tau_{1u}^{-1}(\text{len}(x))) \geq \tau_{1u}(\tau_{1u}^{-1}(200)))$$

$$(6) \rightsquigarrow (\text{len}^y(x) \geq 0) \wedge \text{defined}(x) \wedge (\text{len}^y(x) \leq (256-1)) \wedge (1 = 1)$$

$$\wedge\ (\tau_{1u}(\tau_{1u}^{-1}(\text{len}(x))) \geq \tau_{1u}(\tau_{1u}^{-1}(200)))$$

$$(20) \rightsquigarrow (\text{len}^y(x) \geq 0) \wedge \text{defined}(x) \wedge (\text{len}^y(x) \leq (256-1)) \wedge (\text{len}(x) \geq \tau_{1u}(\tau_{1u}^{-1}(200)))$$

$$(25) \rightsquigarrow (\text{len}^y(x) \geq 0) \wedge \text{defined}(x) \wedge (\text{len}^y(x) \leq (256-1))$$

$$\wedge\ (\text{len}^y(x) \geq \tau_{1u}(\tau_{1u}^{-1}(200)))$$

$$(6) \rightsquigarrow (\text{len}^y(x) \geq 0) \wedge \text{defined}(x) \wedge (\text{len}^y(x) \leq (256-1)) \wedge (\text{len}^y(x) \geq 200)$$

**2** $\textit{truth}(x\{1,1\} \leq_{\tau_{1u}} \tau_{1u}^{-1}(100))$

$$\rightsquigarrow \tau_{1u}(x\{1,1\}) \leq \tau_{1u}(\tau_{1u}^{-1}(100))$$

$$\rightsquigarrow \text{defined}(\tau_{1u}(x\{1,1\})) \wedge \text{defined}(\tau_{1u}(\tau_{1u}^{-1}(100)))$$

$$\wedge\ (\tau_{1u}(x\{1,1\}) \leq \tau_{1u}(\tau_{1u}^{-1}(100)))$$

$$\rightsquigarrow (1 = \text{len}(x\{1,1\})) \wedge (\tau_{1u}(x\{1,1\}) \leq \tau_{1u}(\tau_{1u}^{-1}(100)))$$

$$\rightsquigarrow \text{defined}(1) \wedge \text{defined}(\text{len}(x\{1,1\})) \wedge (1 = \text{len}(x\{1,1\}))$$

$$\wedge\ (\tau_{1u}(x\{1,1\}) \leq \tau_{1u}(\tau_{1u}^{-1}(100)))$$

$$\rightsquigarrow \text{defined}(\text{len}(x\{1,1\})) \wedge (1 = \text{len}(x\{1,1\})) \wedge (\tau_{1u}(x\{1,1\}) \leq \tau_{1u}(\tau_{1u}^{-1}(100)))$$

$$\rightsquigarrow \text{defined}(x\{1,1\}) \wedge (1 = \text{len}(x\{1,1\})) \wedge (\tau_{1u}(x\{1,1\}) \leq \tau_{1u}(\tau_{1u}^{-1}(100)))$$

$$\rightsquigarrow (\text{len}(x) \geq (1+1)) \wedge (1 \geq 0) \wedge (1 = \text{len}(x\{1,1\}))$$

$$\wedge\ (\tau_{1u}(x\{1,1\}) \leq \tau_{1u}(\tau_{1u}^{-1}(100)))$$

$$\rightsquigarrow \text{defined}(\text{len}(x)) \wedge \text{defined}(1+1) \wedge (\text{len}(x) \geq (1+1)) \wedge (1 \geq 0) \wedge (1 = \text{len}(x\{1,1\}))$$

$$\wedge\ (\tau_{1u}(x\{1,1\}) \leq \tau_{1u}(\tau_{1u}^{-1}(100)))$$

$$\rightsquigarrow \text{defined}(x) \wedge (\text{len}(x) \geq (1+1)) \wedge (1 = \text{len}(x\{1,1\}))$$

$$\wedge\ (\tau_{1u}(x\{1,1\}) \leq \tau_{1u}(\tau_{1u}^{-1}(100)))$$

$$(25) \rightsquigarrow \text{defined}(x) \wedge (\text{len}^y(x) \geq (1+1)) \wedge (1 = \text{len}(x\{1,1\}))$$

$$\wedge\ (\tau_{1u}(x\{1,1\}) \leq \tau_{1u}(\tau_{1u}^{-1}(100)))$$

$$\rightsquigarrow \text{defined}(x) \wedge (\text{len}^y(x) \geq (1+1)) \wedge \text{defined}(1) \wedge \text{defined}(0) \wedge (1 = \text{len}(x\{1,1\}))$$

$$\land (\tau_{1u}(x\{1,1\}) \leq \tau_{1u}(\tau_{1u}^{-1}(100)))$$
$$\leadsto \text{defined}(x) \land (\text{len}^y(x) \geq (1+1)) \land (1 = \text{len}(x\{1,1\}))$$
$$\land (\tau_{1u}(x\{1,1\}) \leq \tau_{1u}(\tau_{1u}^{-1}(100)))$$
$$\leadsto \text{defined}(x) \land (\text{len}^y(x) \geq (1+1)) \land \text{defined}(1) \land \text{defined}(0) \land (1 = \text{len}(x\{1,1\}))$$
$$\land (\tau_{1u}(x\{1,1\}) \leq \tau_{1u}(\tau_{1u}^{-1}(100)))$$
$$\leadsto \text{defined}(x) \land (\text{len}^y(x) \geq (1+1)) \land (1 = \text{len}(x\{1,1\}))$$
$$\land (\tau_{1u}(x\{1,1\}) \leq \tau_{1u}(\tau_{1u}^{-1}(100)))$$
$$(24) \leadsto \text{defined}(x) \land (\text{len}^y(x) \geq (1+1)) \land (1 = 1) \land (\tau_{1u}(x\{1,1\}) \leq \tau_{1u}(\tau_{1u}^{-1}(100)))$$
$$\leadsto \text{defined}(x) \land (\text{len}^y(x) \geq (1+1)) \land (1 = \text{len}(\tau_{1u}^{-1}(100)))$$
$$\land (\tau_{1u}(x\{1,1\}) \leq \tau_{1u}(\tau_{1u}^{-1}(100)))$$
$$\leadsto \text{defined}(x) \land (\text{len}^y(x) \geq (1+1)) \land \text{defined}(1) \land \text{defined}(\text{len}(\tau_{1u}^{-1}(100)))$$
$$\land (1 = \text{len}(\tau_{1u}^{-1}(100))) \land (\tau_{1u}(x\{1,1\}) \leq \tau_{1u}(\tau_{1u}^{-1}(100)))$$
$$\leadsto \text{defined}(x) \land (\text{len}^y(x) \geq (1+1)) \land (1 = \text{len}(\tau_{1u}^{-1}(100)))$$
$$\land (\tau_{1u}(x\{1,1\}) \leq \tau_{1u}(\tau_{1u}^{-1}(100)))$$
$$\leadsto \text{defined}(x) \land (\text{len}^y(x) \geq (1+1)) \land \text{defined}(\tau_{1u}^{-1}(100)) \land (1 = \text{len}(\tau_{1u}^{-1}(100)))$$
$$\land (\tau_{1u}(x\{1,1\}) \leq \tau_{1u}(\tau_{1u}^{-1}(100)))$$
$$\leadsto \text{defined}(x) \land (\text{len}^y(x) \geq (1+1)) \land (100 \geq 0) \land (100 \leq (256-1))$$
$$\land (1 = \text{len}(\tau_{1u}^{-1}(100))) \land (\tau_{1u}(x\{1,1\}) \leq \tau_{1u}(\tau_{1u}^{-1}(100)))$$
$$\leadsto \text{defined}(x) \land (\text{len}^y(x) \geq (1+1)) \land \text{defined}(100) \land \text{defined}(0) \land (1 = \text{len}(\tau_{1u}^{-1}(100)))$$
$$\land (\tau_{1u}(x\{1,1\}) \leq \tau_{1u}(\tau_{1u}^{-1}(100)))$$
$$\leadsto \text{defined}(x) \land (\text{len}^y(x) \geq (1+1)) \land (1 = \text{len}(\tau_{1u}^{-1}(100)))$$
$$\land (\tau_{1u}(x\{1,1\}) \leq \tau_{1u}(\tau_{1u}^{-1}(100)))$$
$$\leadsto \text{defined}(x) \land (\text{len}^y(x) \geq (1+1)) \land \text{defined}(100) \land \text{defined}(256-1)$$
$$\land (1 = \text{len}(\tau_{1u}^{-1}(100))) \land (\tau_{1u}(x\{1,1\}) \leq \tau_{1u}(\tau_{1u}^{-1}(100)))$$
$$\leadsto \text{defined}(x) \land (\text{len}^y(x) \geq (1+1)) \land (1 = \text{len}(\tau_{1u}^{-1}(100)))$$
$$\land (\tau_{1u}(x\{1,1\}) \leq \tau_{1u}(\tau_{1u}^{-1}(100)))$$
$$(5) \leadsto \text{defined}(x) \land (\text{len}^y(x) \geq (1+1)) \land (1 = 1) \land (\tau_{1u}(x\{1,1\}) \leq \tau_{1u}(\tau_{1u}^{-1}(100)))$$
$$(24) \leadsto \text{defined}(x) \land (\text{len}^y(x) \geq (1+1)) \land (\tau_{1u}^y(x\{1,1\}) \leq \tau_{1u}(\tau_{1u}^{-1}(100)))$$
$$(5) \leadsto \text{defined}(x) \land (\text{len}^y(x) \geq (1+1)) \land (\tau_{1u}^y(x\{1,1\}) \leq 100)$$

**3** $x\{\tau_{1u}(\text{cast}_{\tau_{1s}\to\tau_{1u}}(\tau_{1s}^{-1}(2) \oplus_{\tau_{1s}} \text{cast}_{\tau_{1u}\to\tau_{1s}}(x\{1,1\}))), \tau_{1u}(\tau_{1u}^{-1}(\text{len}(x)) \ominus_{\tau_{1u}} (\tau_{1u}^{-1}(2) \oplus_{\tau_{1u}} x\{1,$
$= \text{"secret"}$

$$(7,8,9) \leadsto x\{\tau_{1s}(\tau_{1s}^{-1}(2) \oplus_{\tau_{1s}} \text{cast}_{\tau_{1u}\to\tau_{1s}}(x\{1,1\})), \tau_{1u}(\tau_{1u}^{-1}(\text{len}(x)) \ominus_{\tau_{1u}} (\tau_{1u}^{-1}(2) \oplus_{\tau_{1u}} x\{1,1\}))\}$$
$$= \text{"secret"}$$

$$
\begin{aligned}
(10,11,12) &\leadsto x\{\tau_{1s}(\tau_{1s}^{-1}(2)) + \tau_{1s}(\text{cast}_{\tau_{1u}\to\tau_{1s}}(x\{1,1\})), \tau_{1u}(\tau_{1u}^{-1}(\text{len}(x)) \ominus_{\tau_{1u}} (\tau_{1u}^{-1}(2) \oplus_{\tau_{1u}} x\{1,1\}))\} \\
&\quad = \text{"secret"} \\
(13) &\leadsto x\{2 + \tau_{1s}(\text{cast}_{\tau_{1u}\to\tau_{1s}}(x\{1,1\})), \tau_{1u}(\tau_{1u}^{-1}(\text{len}(x)) \ominus_{\tau_{1u}} (\tau_{1u}^{-1}(2) \oplus_{\tau_{1u}} x\{1,1\}))\} \\
&\quad = \text{"secret"} \\
(14,15,16) &\leadsto x\{2 + \tau_{1u}(x\{1,1\}), \tau_{1u}(\tau_{1u}^{-1}(\text{len}(x)) \ominus_{\tau_{1u}} (\tau_{1u}^{-1}(2) \oplus_{\tau_{1u}} x\{1,1\}))\} = \text{"secret"} \\
(24) &\leadsto x\{2 + \tau_{1u}^{y}(x\{1,1\}), \tau_{1u}(\tau_{1u}^{-1}(\text{len}(x)) \ominus_{\tau_{1u}} (\tau_{1u}^{-1}(2) \oplus_{\tau_{1u}} x\{1,1\}))\} \\
&\quad = \text{"secret"} \\
(17,18,19) &\leadsto x\{2 + \tau_{1u}^{y}(x\{1,1\}), \tau_{1u}(\tau_{1u}^{-1}(\text{len}(x))) - \tau_{1u}(\tau_{1u}^{-1}(2) \oplus_{\tau_{1u}} x\{1,1\})\} \\
&\quad = \text{"secret"} \\
(20) &\leadsto x\{2 + \tau_{1u}^{y}(x\{1,1\}), \text{len}(x) - \tau_{1u}(\tau_{1u}^{-1}(2) \oplus_{\tau_{1u}} x\{1,1\})\} = \text{"secret"} \\
(25) &\leadsto x\{2 + \tau_{1u}^{y}(x\{1,1\}), \text{len}^{y}(x) - \tau_{1u}(\tau_{1u}^{-1}(2) \oplus_{\tau_{1u}} x\{1,1\})\} = \text{"secret"} \\
(21,22,23) &\leadsto x\{2 + \tau_{1u}^{y}(x\{1,1\}), \text{len}^{y}(x) - (\tau_{1u}(\tau_{1u}^{-1}(2)) + \tau_{1u}(x\{1,1\}))\} \\
&\quad = \text{"secret"} \\
(26) &\leadsto x\{2 + \tau_{1u}^{y}(x\{1,1\}), \text{len}^{y}(x) - (2 + \tau_{1u}(x\{1,1\}))\} = \text{"secret"} \\
(24) &\leadsto x\{2 + \tau_{1u}^{y}(x\{1,1\}), \text{len}^{y}(x) - (2 + \tau_{1u}^{y}(x\{1,1\}))\} = \text{"secret"}
\end{aligned}
$$

$$
\begin{aligned}
\mathbf{4} &\quad x\{1 + 1 + \tau_{1u}(x\{1,1\}), \text{len}(x) - (1 + 1 + \tau_{1u}(x\{1,1\}))\} = \text{"secret"} \\
(24) &\leadsto x\{1 + 1 + \tau_{1u}^{y}(x\{1,1\}), \text{len}(x) - (1 + 1 + \tau_{1u}(x\{1,1\}))\} = \text{"secret"} \\
(25) &\leadsto x\{1 + 1 + \tau_{1u}^{y}(x\{1,1\}), \text{len}^{y}(x) - (1 + 1 + \tau_{1u}(x\{1,1\}))\} = \text{"secret"} \\
(24) &\leadsto x\{1 + 1 + \tau_{1u}^{y}(x\{1,1\}), \text{len}^{y}(x) - (1 + 1 + \tau_{1u}^{y}(x\{1,1\}))\} = \text{"secret"}
\end{aligned}
$$

$$
\begin{aligned}
\mathbf{5} &\quad \text{defined}(\tau_{1u}^{-1}(100)) \\
&\leadsto (100 \geq 0) \wedge (100 \leq (256 - 1))
\end{aligned}
$$

$$
\begin{aligned}
\mathbf{6} &\quad \text{defined}(\tau_{1u}^{-1}(200)) \\
&\leadsto (200 \geq 0) \wedge (200 \leq (256 - 1))
\end{aligned}
$$

$$
\begin{aligned}
\mathbf{7} &\quad \tau_{1s}(\tau_{1s}^{-1}(2) \oplus_{\tau_{1s}} \text{cast}_{\tau_{1u}\to\tau_{1s}}(x\{1,1\})) \leq (256 - 1) \\
(10,11,12) &\leadsto (\tau_{1s}(\tau_{1s}^{-1}(2)) + \tau_{1s}(\text{cast}_{\tau_{1u}\to\tau_{1s}}(x\{1,1\}))) \leq (256 - 1) \\
(13) &\leadsto (2 + \tau_{1s}(\text{cast}_{\tau_{1u}\to\tau_{1s}}(x\{1,1\}))) \leq (256 - 1) \\
(14,15,16) &\leadsto (2 + \tau_{1u}(x\{1,1\})) \leq (256 - 1) \\
(24) &\leadsto (2 + \tau_{1u}^{y}(x\{1,1\})) \leq (256 - 1)
\end{aligned}
$$

$$
\begin{aligned}
\mathbf{8} &\quad \tau_{1s}(\tau_{1s}^{-1}(2) \oplus_{\tau_{1s}} \text{cast}_{\tau_{1u}\to\tau_{1s}}(x\{1,1\})) \geq 0 \\
(10,11,12) &\leadsto (\tau_{1s}(\tau_{1s}^{-1}(2)) + \tau_{1s}(\text{cast}_{\tau_{1u}\to\tau_{1s}}(x\{1,1\}))) \geq 0 \\
(13) &\leadsto (2 + \tau_{1s}(\text{cast}_{\tau_{1u}\to\tau_{1s}}(x\{1,1\}))) \geq 0 \\
(14,15,16) &\leadsto (2 + \tau_{1u}(x\{1,1\})) \geq 0
\end{aligned}
$$

$$(24) \rightsquigarrow (2 + \tau_{1u}^{y}(x\{1,1\})) \geq 0$$

$$\textbf{9} \quad \text{defined}(\text{cast}_{\tau_{1s}\to\tau_{1u}}(\tau_{1s}^{-1}(2) \bigoplus_{\tau_{1s}} \text{cast}_{\tau_{1u}\to\tau_{1s}}(x\{1,1\})))$$

$$\rightsquigarrow (1 = \text{len}(\tau_{1s}^{-1}(2) \bigoplus_{\tau_{1s}} \text{cast}_{\tau_{1u}\to\tau_{1s}}(x\{1,1\})))$$

$$(12) \rightsquigarrow (1 = 1)$$

$$\textbf{10} \quad (\tau_{1s}(\tau_{1s}^{-1}(2)) + \tau_{1s}(\text{cast}_{\tau_{1u}\to\tau_{1s}}(x\{1,1\}))) \leq (128 - 1)$$

$$(13) \rightsquigarrow (2 + \tau_{1s}(\text{cast}_{\tau_{1u}\to\tau_{1s}}(x\{1,1\}))) \leq (128 - 1)$$

$$(14, 15, 16) \rightsquigarrow (2 + \tau_{1u}(x\{1,1\})) \leq (128 - 1)$$

$$(24) \rightsquigarrow (2 + \tau_{1u}^{y}(x\{1,1\})) \leq (128 - 1)$$

$$\textbf{11} \quad (\tau_{1s}(\tau_{1s}^{-1}(2)) + \tau_{1s}(\text{cast}_{\tau_{1u}\to\tau_{1s}}(x\{1,1\}))) \geq -(128)$$

$$(13) \rightsquigarrow (2 + \tau_{1s}(\text{cast}_{\tau_{1u}\to\tau_{1s}}(x\{1,1\}))) \geq -(128)$$

$$(14, 15, 16) \rightsquigarrow (2 + \tau_{1u}(x\{1,1\})) \geq -(128)$$

$$(24) \rightsquigarrow (2 + \tau_{1u}^{y}(x\{1,1\})) \geq -(128)$$

$$\textbf{12} \quad \text{defined}(\tau_{1s}^{-1}(2) \bigoplus_{\tau_{1s}} \text{cast}_{\tau_{1u}\to\tau_{1s}}(x\{1,1\}))$$

$$\rightsquigarrow (1 = \text{len}(\tau_{1s}^{-1}(2))) \wedge (1 = \text{len}(\text{cast}_{\tau_{1u}\to\tau_{1s}}(x\{1,1\})))$$

$$(13) \rightsquigarrow (1 = 1) \wedge (1 = \text{len}(\text{cast}_{\tau_{1u}\to\tau_{1s}}(x\{1,1\})))$$

$$(16) \rightsquigarrow (1 = 1)$$

$$\textbf{13} \quad \text{defined}(\tau_{1s}^{-1}(2))$$

$$\rightsquigarrow (2 \geq -(128)) \wedge (2 \leq (128 - 1))$$

$$\textbf{14} \quad \tau_{1u}(x\{1,1\}) \leq (128 - 1)$$

$$(24) \rightsquigarrow \tau_{1u}^{y}(x\{1,1\}) \leq (128 - 1)$$

$$\textbf{15} \quad \tau_{1u}(x\{1,1\}) \geq -(128)$$

$$(24) \rightsquigarrow \tau_{1u}^{y}(x\{1,1\}) \geq -(128)$$

$$\textbf{16} \quad \text{defined}(\text{cast}_{\tau_{1u}\to\tau_{1s}}(x\{1,1\}))$$

$$\rightsquigarrow (1 = \text{len}(x\{1,1\}))$$

$$(24) \rightsquigarrow (1 = 1)$$

$$\textbf{17} \quad (\tau_{1u}(\tau_{1u}^{-1}(\text{len}(x))) - \tau_{1u}(\tau_{1u}^{-1}(2) \bigoplus_{\tau_{1u}} x\{1,1\})) \leq (256 - 1)$$

$$(20) \rightsquigarrow (\text{len}(x) - \tau_{1u}(\tau_{1u}^{-1}(2) \bigoplus_{\tau_{1u}} x\{1,1\})) \leq (256 - 1)$$

$$(25) \rightsquigarrow (\text{len}^{y}(x) - \tau_{1u}(\tau_{1u}^{-1}(2) \bigoplus_{\tau_{1u}} x\{1,1\})) \leq (256 - 1)$$

$$(21, 22, 23) \rightsquigarrow (\text{len}^{y}(x) - (\tau_{1u}(\tau_{1u}^{-1}(2)) + \tau_{1u}(x\{1,1\}))) \leq (256 - 1)$$

$$(26) \rightsquigarrow (\text{len}^{y}(x) - (2 + \tau_{1u}(x\{1,1\}))) \leq (256 - 1)$$

$$(24) \rightsquigarrow (\text{len}^y(x) - (2 + \tau_{1u}^y(x\{1,1\}))) \leq (256-1)$$

**18** $(\tau_{1u}(\tau_{1u}^{-1}(\text{len}(x))) - \tau_{1u}(\tau_{1u}^{-1}(2) \bigoplus_{\tau_{1u}} x\{1,1\})) \geq 0$

$$(20) \rightsquigarrow (\text{len}(x) - \tau_{1u}(\tau_{1u}^{-1}(2) \bigoplus_{\tau_{1u}} x\{1,1\})) \geq 0$$

$$(25) \rightsquigarrow (\text{len}^y(x) - \tau_{1u}(\tau_{1u}^{-1}(2) \bigoplus_{\tau_{1u}} x\{1,1\})) \geq 0$$

$$(21,22,23) \rightsquigarrow (\text{len}^y(x) - (\tau_{1u}(\tau_{1u}^{-1}(2)) + \tau_{1u}(x\{1,1\}))) \geq 0$$

$$(26) \rightsquigarrow (\text{len}^y(x) - (2 + \tau_{1u}(x\{1,1\}))) \geq 0$$

$$(24) \rightsquigarrow (\text{len}^y(x) - (2 + \tau_{1u}^y(x\{1,1\}))) \geq 0$$

**19** $\text{defined}(\tau_{1u}^{-1}(\text{len}(x)) \bigominus_{\tau_{1u}} (\tau_{1u}^{-1}(2) \bigoplus_{\tau_{1u}} x\{1,1\}))$

$$\rightsquigarrow (1 = \text{len}(\tau_{1u}^{-1}(\text{len}(x)))) \wedge (1 = \text{len}(\tau_{1u}^{-1}(2) \bigoplus_{\tau_{1u}} x\{1,1\}))$$

$$(20) \rightsquigarrow (1 = 1) \wedge (1 = \text{len}(\tau_{1u}^{-1}(2) \bigoplus_{\tau_{1u}} x\{1,1\}))$$

$$(23) \rightsquigarrow (1 = 1)$$

**20** $\text{defined}(\tau_{1u}^{-1}(\text{len}(x)))$

$$\rightsquigarrow (\text{len}(x) \geq 0) \wedge (\text{len}(x) \leq (256-1))$$

$$(25) \rightsquigarrow (\text{len}^y(x) \geq 0) \wedge (\text{len}(x) \leq (256-1))$$

$$(25) \rightsquigarrow (\text{len}^y(x) \geq 0) \wedge (\text{len}^y(x) \leq (256-1))$$

**21** $(\tau_{1u}(\tau_{1u}^{-1}(2)) + \tau_{1u}(x\{1,1\})) \leq (256-1)$

$$(26) \rightsquigarrow (2 + \tau_{1u}(x\{1,1\})) \leq (256-1)$$

$$(24) \rightsquigarrow (2 + \tau_{1u}^y(x\{1,1\})) \leq (256-1)$$

**22** $(\tau_{1u}(\tau_{1u}^{-1}(2)) + \tau_{1u}(x\{1,1\})) \geq 0$

$$(26) \rightsquigarrow (2 + \tau_{1u}(x\{1,1\})) \geq 0$$

$$(24) \rightsquigarrow (2 + \tau_{1u}^y(x\{1,1\})) \geq 0$$

**23** $\text{defined}(\tau_{1u}^{-1}(2) \bigoplus_{\tau_{1u}} x\{1,1\})$

$$\rightsquigarrow (1 = \text{len}(\tau_{1u}^{-1}(2))) \wedge (1 = \text{len}(x\{1,1\}))$$

$$(26) \rightsquigarrow (1 = 1) \wedge (1 = \text{len}(x\{1,1\}))$$

$$(24) \rightsquigarrow (1 = 1)$$

**24** $\text{defined}(x\{1,1\})$

$$\rightsquigarrow (\text{len}(x) \geq (1+1)) \wedge (1 \geq 0)$$

$$(25) \rightsquigarrow (\text{len}^y(x) \geq (1+1)) \wedge (1 \geq 0)$$

**25** $\text{defined}(x)$

**26** $\text{defined}(\tau_{1u}^{-1}(2))$

$$\rightsquigarrow (2 \geq 0) \wedge (2 \leq (256 - 1))$$

Following our procedure for collecting rewritten facts (omitting conditions that are already present in the conjunction that is being rewritten, as permitted by (R3)) and applying theorem 3.1 we see that we need to ask Yices to discharge the following implication:

$$((\phi_1 \Rightarrow \psi_1) \wedge (\phi_2 \Rightarrow \psi_2) \wedge (\phi_3 \Rightarrow \psi_3)) \Longrightarrow \psi,$$

where

$$\begin{aligned}
\phi_1 &= \text{true} \\
\psi_1 &= (\text{len}^y(x) \geq 0) \wedge \text{defined}(x) \wedge (\text{len}^y(x) \leq (256 - 1)) \wedge (200 \geq 0) \wedge (200 \leq (256 - 1)) \\
&\quad \wedge (1 = 1) \wedge (\text{len}^y(x) \geq 200) \\
\phi_2 &= \text{true} \\
\psi_2 &= \text{defined}(x) \wedge (\text{len}^y(x) \geq (1 + 1)) \wedge (1 \geq 0) \wedge (100 \geq 0) \wedge (100 \leq (256 - 1)) \wedge (1 = 1) \\
&\quad \wedge (\tau^y_{1u}(x\{1, 1\}) \leq 100) \\
\phi_3 &= ((2 + \tau^y_{1u}(x\{1, 1\})) \leq (128 - 1)) \wedge ((2 + \tau^y_{1u}(x\{1, 1\})) \geq -(128)) \\
&\quad \wedge (2 \geq -(128)) \wedge (2 \leq (128 - 1)) \wedge (\tau^y_{1u}(x\{1, 1\}) \leq (128 - 1)) \\
&\quad \wedge (\tau^y_{1u}(x\{1, 1\}) \geq -(128)) \\
&\quad \wedge ((\text{len}^y(x) - (2 + \tau^y_{1u}(x\{1, 1\}))) \leq (256 - 1)) \\
&\quad \wedge ((\text{len}^y(x) - (2 + \tau^y_{1u}(x\{1, 1\}))) \geq 0) \wedge (\text{len}^y(x) \geq 0) \\
&\quad \wedge (\text{len}^y(x) \leq (256 - 1)) \wedge ((2 + \tau^y_{1u}(x\{1, 1\})) \leq (256 - 1)) \\
&\quad \wedge ((2 + \tau^y_{1u}(x\{1, 1\})) \geq 0) \wedge (1 = 1) \wedge (2 \geq 0) \wedge (2 \leq (256 - 1)) \\
&\quad \wedge (\text{len}^y(x) \geq (1 + 1)) \wedge (1 \geq 0) \wedge \text{defined}(x) \\
\psi_3 &= (x\{2 + \tau^y_{1u}(x\{1, 1\}), \text{len}^y(x) - (2 + \tau^y_{1u}(x\{1, 1\}))\} = \text{"secret"}) \\
\psi &= (\text{len}^y(x) \geq (1 + 1)) \wedge (1 \geq 0) \wedge \text{defined}(x) \\
&\quad \wedge (x\{1 + 1 + \tau^y_{1u}(x\{1, 1\}), \text{len}^y(x) - (1 + 1 + \tau^y_{1u}(x\{1, 1\}))\} = \text{"secret"})
\end{aligned}$$

We see that overflow safety conditions can be quite subtle. For instance, if we were to check $\mathit{truth}(\tau^{-1}_{1u}(\text{len}(x)) \geq_{\tau_{1u}} \tau^{-1}_{1u}(100))$ (using 100 instead of 200) then the implication would not hold: since we only know $\tau^y_{1u}(x\{1, 1\}) \leq 100$ from $\psi_2$, we would not be able to prove $(\text{len}^y(x) - (2 + \tau^y_{1u}(x\{1, 1\}))) \geq 0$ in $\phi_3$.

# Appendix C

# CryptoVerif Model of RPC-enc

This section shows the full input that we present to CryptoVerif for verification of the RPC-enc example described in section 3.6. The parts marked by (USER) are provided by the user in a template file. The rest of the input is automatically generated and added to the template.

The example highlights two different types of secrecy that we can prove: for the request we ask CryptoVerif to prove strong secrecy. We are then able to claim that the original C code satisfies unconditional weak secrecy of the request. The secrecy of the response is conditional on the client being honest, therefore we cannot ask CryptoVerif to prove strong secrecy, and have to model weak secrecy directly as a correspondence property instead. This is done using a sentinel process that runs in parallel with the server and generates an event if the server response has been leaked.

Names of the form fixed_1024_payload correspond to our $\texttt{fixed}_{1024}$, the name part after the number is allowed for readability, but ignored by our tools. The notation (len(arg0))^[u,4] corresponds to our $\tau_{4u}^{-1}(\text{len}(x_1))$ and (len(arg0))_[u,4] corresponds to our $\tau_{4u}(\text{len}(x_1))$

Conditions like *cond56* that apply to many arguments are a result of NULL checks of malloc results—our models make explicit the dependency of a malloc result on all the preceding calls of malloc, as described in section 3.5.

```
(**********************************
  (USER) RPC-enc protocol.
***********************************)

param N.

channel c_in, c_out.

type nondet [fixed].
```

```
(*******************************
  (USER) IND−CPA INT−CTXT encryption
********************************)

type fixed_16_keyseed [fixed, large].
(*
  Must be bounded because the security definition refers
  to time(kgen).
*)
type fixed_16_key [bounded].
type fixed_16_seed [fixed, large].

(* 1045 = 1024 for payload + 16 for key + 4 for payload length + 1 for tag *)
type bounded_1045_plaintext [bounded].
(*
  Encryption adds 32 bytes to the message.
*)
type bounded_1077_ciphertext  [bounded].

proba Penc.
proba Pencptxt.
proba Pencctxt.

expand IND_CPA_INT_CTXT_sym_enc(fixed_16_keyseed, fixed_16_key, bounded_1045_plaintext,
    bounded_1077_ciphertext, fixed_16_seed,
                              kgen, E, D, injbot, Zbounded_1045_plaintext, Penc, Pencctxt).

fun inverse_injbot(bitstringbot): bounded_1045_plaintext.

(**************************
  (USER) Key lookup
**************************)

type keydb.
type bounded_1024_id.

(* the ids of a designated pair of honest participants *)
const clientID: bounded_1024_id.
const serverID: bounded_1024_id.

(* key database operations *)
fun add_honest(fixed_16_key, keydb): keydb.
(* Returns some fixed default value if key not in the database. *)
```

```
fun lookup(bounded_1024_id, bounded_1024_id, keydb): fixed_16_key.

forall k: fixed_16_key, db: keydb;
  lookup(clientID, serverID, add_honest(k, db)) = k.

(* a host id that carries along the fact that it is compromised *)
fun badHost(bounded_1024_id): bounded_1024_id [compos].

(*
  It is important that in the bad host branch we remove the mention of the honest key,
  so we can show that it isn't leaked anywhere.
*)
forall h:bounded_1024_id, k: fixed_16_key, db: keydb;
  lookup(badHost(h), serverID, add_honest(k, db)) = lookup(badHost(h), serverID, db).

(******************************
  (USER) Misc
********************************)

type fixed_1024_payload [fixed, large].

(* The port that the server is listening to. *)
const port: bitstring.


(*************************
  Formatting Functions
**************************)

(* conc1 := "p"|(len(arg0))^[u,4]|arg0|arg1 *)
fun conc1(fixed_1024_payload, fixed_16_key): bounded_1045_plaintext [compos].

(* conc2 := "p"|(len(arg0))^[u,4]|arg0|arg1 *)
fun conc2(bounded_1024_id, bounded_1077_ciphertext): bitstring [compos].



(*************************
  Parsing Equations
**************************)


(*************************
```

*Arithmetic Functions*
***************************)*

*(∗ arithmetic1 := CastToInt((((1)^[u,8] + (4)^[u,8]) + (len(arg0))^[u,8]) + (len(arg1))^[u,8]) ∗)*
**fun** arithmetic1(bounded_1024_id, bounded_1077_ciphertext): bitstring.

*(∗ arithmetic2 := (len(arg0))^[u,4] ∗)*
**fun** arithmetic2(bounded_1077_ciphertext): bitstring.

*(****************************
*Auxiliary Tests*
***************************)*

*(∗ cond1 := Malloc(len(arg0) + 1, len(arg1) + 1, len(arg2) + 1, len(arg3) + 1, len(arg4) + 1) <> 0 ∗)*
**fun** cond1(bitstring, bitstring, bitstring, bitstring, bitstring): bool.

*(∗ cond10 := Malloc(len(arg0) + 1, len(arg1) + 1, len(arg2) + 1, len(arg3) + 1, len(arg4) + 1, len (arg5), len(arg6), len(arg6) + 1) <> 0 ∗)*
**fun** cond10(bitstring, bitstring, bitstring, bitstring, bitstring, bounded_1024_id, bounded_1024_id): bool.

*(∗ cond11 := ((arg0)_[s,4] <> 0) ∗)*
**fun** cond11(bitstring): bool.

*(∗ cond12 := (Truth_of_bs(arg0)) ∗)*
**fun** cond12(bitstring): bool.

*(∗ cond13 := (len(arg0) <> len(arg1)) ∗)*
**fun** cond13(bounded_1024_id, bounded_1024_id): bool.

*(∗ cond14 := Malloc(len(arg0) + 1, len(arg1) + 1, len(arg2) + 1, len(arg3) + 1, len(arg4) + 1, len (arg5), len(arg6), len(arg6) + 1, len(arg7)) <> 0 ∗)*
**fun** cond14(bitstring, bitstring, bitstring, bitstring, bitstring, bounded_1024_id, bounded_1024_id, bounded_1024_id): bool.

*(∗ cond16 := (Malloc(len(arg0) + 1, len(arg1) + 1, len(arg2) + 1, len(arg3) + 1, len(arg4) + 1, len(arg5), len(arg6), len(arg6) + 1, len(arg7), 1 + len(arg8) + 4 + len(arg9)) = 0) ∗)*
**fun** cond16(bitstring, bitstring, bitstring, bitstring, bitstring, bounded_1024_id, bounded_1024_id, bounded_1024_id, fixed_16_key, fixed_1024_payload): bool.

*(∗ cond17 := (Malloc(len(arg0) + 1, len(arg1) + 1, len(arg2) + 1, len(arg3) + 1, len(arg4) + 1,*

```
    len(arg7), len(arg5), len(arg5) + 1, len(arg6), 1 + len(arg8) + 4 + len(arg9), 1 + 4 + len(arg7)
    + 32 + 1 + len(arg8) + 4 + len(arg9)) = 0) *)
fun cond17(bitstring, bitstring, bitstring, bitstring, bitstring, bounded_1024_id, bounded_1024_id,
    bounded_1024_id, fixed_16_key, fixed_1024_payload): bool.

(* cond19 := (len(arg0) = 0) *)
fun cond19(bounded_1077_ciphertext): bool.

(* cond20 := ((arg0)_[u,4] < 1056) *)
fun cond20(bitstring): bool.

(* cond21 := ((arg0)_[u,4] > 1056) *)
fun cond21(bitstring): bool.

(* cond22 := (Malloc(len(arg0) + 1, len(arg1) + 1, len(arg2) + 1, len(arg3) + 1, len(arg4) + 1,
    len(arg7), len(arg5), len(arg5) + 1, len(arg6), 1 + len(arg8) + 4 + len(arg9), 1 + 4 + len(arg7)
    + 32 + 1 + len(arg8) + 4 + len(arg9), (arg10)_[u,4]) = 0) *)
fun cond22(bitstring, bitstring, bitstring, bitstring, bitstring, bounded_1024_id, bounded_1024_id,
    bounded_1024_id, fixed_16_key, fixed_1024_payload, bitstring): bool.

(* cond23 := (Malloc(len(arg0) + 1, len(arg1) + 1, len(arg2) + 1, len(arg3) + 1, len(arg4) + 1,
    len(arg7), len(arg5), len(arg5) + 1, len(arg6), 1 + len(arg8) + 4 + len(arg9), 1 + 4 + len(arg7)
    + 32 + 1 + len(arg8) + 4 + len(arg9), (arg10)_[u,4], (arg10)_[u,4] − 32) = 0) *)
fun cond23(bitstring, bitstring, bitstring, bitstring, bitstring, bounded_1024_id, bounded_1024_id,
    bounded_1024_id, fixed_16_key, fixed_1024_payload, bitstring): bool.

(* cond24 := (len(arg0) > ((arg1)_[u,4] − 32)) *)
fun cond24(bounded_1045_plaintext, bitstring): bool.

(* cond25 := (len(arg0) = 0) *)
fun cond25(bounded_1045_plaintext): bool.

(* cond26 := (len(arg0) < 1024) *)
fun cond26(bounded_1045_plaintext): bool.

(* cond27 := Malloc(len(arg0) + 1, len(arg1) + 1, len(arg2) + 1, len(arg3) + 1, len(arg4) + 1)
    <> 0 *)
fun cond27(bitstring, bitstring, bitstring, bitstring, bitstring): bool.

(* cond29 := (len(arg0) >= 256) *)
fun cond29(bounded_1024_id): bool.

(* cond30 := (len(arg0) = 0) *)
```

**fun** cond30(bounded_1024_id): bool.

(∗ cond31 := (Malloc(len(arg0) + 1, len(arg1) + 1, len(arg2) + 1, len(arg3) + 1, len(arg4) + 1, len(arg5)) = 0) ∗)
**fun** cond31(bitstring, bitstring, bitstring, bitstring, bitstring, bounded_1024_id): bool.

(∗ cond32 := (Malloc(len(arg0) + 1, len(arg1) + 1, len(arg2) + 1, len(arg3) + 1, len(arg4) + 1, len(arg5), len(arg5) + 1) = 0) ∗)
**fun** cond32(bitstring, bitstring, bitstring, bitstring, bitstring, bounded_1024_id): bool.

(∗ cond33 := ((arg0)_[s,4] <> 0) ∗)
**fun** cond33(bitstring): bool.

(∗ cond34 := ((arg0)_[s,4] <> 0) ∗)
**fun** cond34(bitstring): bool.

(∗ cond35 := ((arg0)_[u,4] < (1040 + 4 + 1 + 32 + 4 + 1)) ∗)
**fun** cond35(bitstring): bool.

(∗ cond36 := ((arg0)_[u,4] > (1040 + 4 + 1 + 32 + 1024 + 4 + 1)) ∗)
**fun** cond36(bitstring): bool.

(∗ cond37 := (Malloc(len(arg0) + 1, len(arg1) + 1, len(arg2) + 1, len(arg3) + 1, len(arg4) + 1, len(arg5), len(arg5) + 1, (arg6)_[u,4]) = 0) ∗)
**fun** cond37(bitstring, bitstring, bitstring, bitstring, bitstring, bounded_1024_id, bitstring): bool.

(∗ cond38 := "p" = arg0{0, 1} ∗)
**fun** cond38(bitstring): bool.

(∗ cond39 := ((arg0{1, 4})_[u,4] > 1024) ∗)
**fun** cond39(bitstring): bool.

(∗ cond4 := (len(arg0) = 0) ∗)
**fun** cond4(bounded_1024_id): bool.

(∗ cond40 := (Malloc(len(arg0) + 1, len(arg1) + 1, len(arg2) + 1, len(arg3) + 1, len(arg4) + 1, len(arg5), len(arg5) + 1, (arg6)_[u,4], (arg7{1, 4})_[u,4]) = 0) ∗)
**fun** cond40(bitstring, bitstring, bitstring, bitstring, bitstring, bounded_1024_id, bitstring, bitstring): bool.

(∗ cond41 := (((arg0)_[u,4] − (1 + 4 + (arg1{1, 4})_[u,4])) > (1040 + 4 + 1 + 32)) ∗)
**fun** cond41(bitstring, bitstring): bool.

(* cond42 := ((arg0{1, 4})_[u,4] <> len(arg1)) *)
**fun** cond42(bitstring, bounded_1024_id): bool.

(* cond43 := Malloc(len(arg0) + 1, len(arg1) + 1, len(arg2) + 1, len(arg3) + 1, len(arg4) + 1, len (arg5), len(arg5) + 1, (arg6)_[u,4], (arg7{1, 4})_[u,4], len(arg8)) <> 0 *)
**fun** cond43(bitstring, bitstring, bitstring, bitstring, bitstring, bounded_1024_id, bitstring, bitstring, bounded_1024_id): bool.

(* cond44 := (Malloc(len(arg0) + 1, len(arg1) + 1, len(arg2) + 1, len(arg3) + 1, len(arg4) + 1, len(arg5), len(arg5) + 1, (arg7)_[u,4], (arg8{1, 4})_[u,4], len(arg6), (arg7)_[u,4] − (1 + 4 + ( arg8{1, 4})_[u,4] + 32)) = 0) *)
**fun** cond44(bitstring, bitstring, bitstring, bitstring, bitstring, bounded_1024_id, bounded_1024_id, bitstring, bitstring): bool.

(* cond45 := (len(arg0) > ((arg1)_[u,4] − (1 + 4 + (arg2{1, 4})_[u,4] + 32))) *)
**fun** cond45(bounded_1045_plaintext, bitstring, bitstring): bool.

(* cond46 := (len(arg0) < (1 + 4)) *)
**fun** cond46(bounded_1045_plaintext): bool.

(* cond47 := ((arg0{1, 4})_[u,4] > 1024) *)
**fun** cond47(bounded_1045_plaintext): bool.

(* cond48 := ((arg0{1, 4})_[u,4] < 1024) *)
**fun** cond48(bounded_1045_plaintext): bool.

(* cond49 := (len(arg0) <= (1 + 4 + (arg0{1, 4})_[u,4])) *)
**fun** cond49(bounded_1045_plaintext): bool.

(* cond5 := (Malloc(len(arg0) + 1, len(arg1) + 1, len(arg2) + 1, len(arg3) + 1, len(arg4) + 1, len( arg5)) = 0) *)
**fun** cond5(bitstring, bitstring, bitstring, bitstring, bitstring, bounded_1024_id): bool.

(* cond50 := "p" = arg0{0, 1} *)
**fun** cond50(bounded_1045_plaintext): bool.

(* cond51 := (Malloc(len(arg0) + 1, len(arg1) + 1, len(arg2) + 1, len(arg3) + 1, len(arg4) + 1, len(arg5), len(arg5) + 1, (arg7)_[u,4], (arg8{1, 4})_[u,4], len(arg6), (arg7)_[u,4] − (1 + 4 + ( arg8{1, 4})_[u,4] + 32), (arg9{1, 4})_[u,4]) = 0) *)
**fun** cond51(bitstring, bitstring, bitstring, bitstring, bitstring, bounded_1024_id, bounded_1024_id, bitstring, bitstring, bounded_1045_plaintext): bool.

(* cond52 := ((len(arg0) − (1 + 4 + (arg0{1, 4})_[u,4])) <> 16) *)

**fun** cond52(bounded_1045_plaintext): bool.

*(∗ cond53 := (Malloc(len(arg0) + 1, len(arg1) + 1, len(arg2) + 1, len(arg3) + 1, len(arg4) + 1, len(arg5), len(arg5) + 1, (arg7)_[u,4], (arg8{1, 4})_[u,4], len(arg6), (arg7)_[u,4] − (1 + 4 + ( arg8{1, 4})_[u,4] + 32), (arg9{1, 4})_[u,4], len(arg9) − (1 + 4 + (arg9{1, 4})_[u,4])) = 0) ∗)*
**fun** cond53(bitstring, bitstring, bitstring, bitstring, bitstring, bounded_1024_id, bounded_1024_id, bitstring, bitstring, bounded_1045_plaintext): bool.

*(∗ cond56 := (Malloc(len(arg0) + 1, len(arg1) + 1, len(arg2) + 1, len(arg3) + 1, len(arg4) + 1, len(arg5), len(arg5) + 1, (arg7)_[u,4], (arg8{1, 4})_[u,4], len(arg6), (arg7)_[u,4] − (1 + 4 + ( arg8{1, 4})_[u,4] + 32), (arg9{1, 4})_[u,4], len(arg9) − (1 + 4 + (arg9{1, 4})_[u,4]), 32 + len( arg10)) = 0) ∗)*
**fun** cond56(bitstring, bitstring, bitstring, bitstring, bitstring, bounded_1024_id, bounded_1024_id, bitstring, bitstring, bounded_1045_plaintext, fixed_1024_payload): bool.

*(∗ cond57 := (len(arg0) > (32 + len(arg1))) ∗)*
**fun** cond57(bounded_1077_ciphertext, fixed_1024_payload): bool.

*(∗ cond58 := (len(arg0) = 0) ∗)*
**fun** cond58(bounded_1077_ciphertext): bool.

*(∗ cond6 := (len(arg0) = 0) ∗)*
**fun** cond6(bounded_1024_id): bool.

*(∗ cond7 := (Malloc(len(arg0) + 1, len(arg1) + 1, len(arg2) + 1, len(arg3) + 1, len(arg4) + 1, len( arg5), len(arg6)) = 0) ∗)*
**fun** cond7(bitstring, bitstring, bitstring, bitstring, bitstring, bounded_1024_id, bounded_1024_id): bool.

*(∗∗∗∗∗∗∗∗∗∗∗∗∗∗∗∗∗∗∗∗∗∗∗∗∗∗*
*Zero Functions*
*∗∗∗∗∗∗∗∗∗∗∗∗∗∗∗∗∗∗∗∗∗∗∗∗∗∗)*

**fun** Zbitstring(bitstring): bitstring.

**fun** Zbitstring_prime(bitstring): bitstring.

**fun** Zbounded_1024_id(bounded_1024_id): bounded_1024_id.

**fun** Zbounded_1024_id_prime(bounded_1024_id): bounded_1024_id.

*(∗ Zbounded_1045_plaintext is already defined in the template ∗)*

**fun** Zbounded_1045_plaintext_prime(bounded_1045_plaintext): bounded_1045_plaintext.

**fun** Zbounded_1077_ciphertext(bounded_1077_ciphertext): bounded_1077_ciphertext.

**fun** Zbounded_1077_ciphertext_prime(bounded_1077_ciphertext): bounded_1077_ciphertext.

**fun** Zfixed_1024_payload(fixed_1024_payload): fixed_1024_payload.

**fun** Zfixed_1024_payload_prime(fixed_1024_payload): fixed_1024_payload.

**fun** Zfixed_16_key(fixed_16_key): fixed_16_key.

**fun** Zfixed_16_key_prime(fixed_16_key): fixed_16_key.

**fun** Zfixed_16_keyseed(fixed_16_keyseed): fixed_16_keyseed.

**fun** Zfixed_16_keyseed_prime(fixed_16_keyseed): fixed_16_keyseed.

**fun** Zfixed_16_seed(fixed_16_seed): fixed_16_seed.

**fun** Zfixed_16_seed_prime(fixed_16_seed): fixed_16_seed.

**const** zero_fixed_1024_payload: fixed_1024_payload.

**const** zero_fixed_16_key: fixed_16_key.

**const** zero_fixed_16_keyseed: fixed_16_keyseed.

**const** zero_fixed_16_seed: fixed_16_seed.

*(****************************
*Primed Functions*
***************************)*

**fun** cond7_prime(bitstring, bitstring, bitstring, bitstring, bitstring, bounded_1024_id, bounded_1024_id): bool.

**fun** cond6_prime(bounded_1024_id): bool.

**fun** cond58_prime(bounded_1077_ciphertext): bool.

**fun** cond57_prime(bounded_1077_ciphertext, fixed_1024_payload): bool.

**fun** cond56_prime(bitstring, bitstring, bitstring, bitstring, bitstring, bounded_1024_id, bounded_1024_id, bitstring, bitstring, bounded_1045_plaintext, fixed_1024_payload): bool.

**fun** cond53_prime(bitstring, bitstring, bitstring, bitstring, bitstring, bounded_1024_id, bounded_1024_id, bitstring, bitstring, bounded_1045_plaintext): bool.

**fun** cond51_prime(bitstring, bitstring, bitstring, bitstring, bitstring, bounded_1024_id, bounded_1024_id, bitstring, bitstring, bounded_1045_plaintext): bool.

**fun** cond5_prime(bitstring, bitstring, bitstring, bitstring, bitstring, bounded_1024_id): bool.

**fun** conc1_prime(fixed_1024_payload, fixed_16_key): bounded_1045_plaintext.

**fun** cond46_prime(bounded_1045_plaintext): bool.

**fun** cond45_prime(bounded_1045_plaintext, bitstring, bitstring): bool.

**fun** cond44_prime(bitstring, bitstring, bitstring, bitstring, bitstring, bounded_1024_id, bounded_1024_id, bitstring, bitstring): bool.

**fun** cond43_prime(bitstring, bitstring, bitstring, bitstring, bitstring, bounded_1024_id, bitstring, bitstring, bounded_1024_id): bool.

**fun** cond42_prime(bitstring, bounded_1024_id): bool.

**fun** cond40_prime(bitstring, bitstring, bitstring, bitstring, bitstring, bounded_1024_id, bitstring, bitstring): bool.

**fun** cond4_prime(bounded_1024_id): bool.

**fun** conc2_prime(bounded_1024_id, bounded_1077_ciphertext): bitstring.

**fun** cond37_prime(bitstring, bitstring, bitstring, bitstring, bitstring, bounded_1024_id, bitstring): bool.

**fun** cond32_prime(bitstring, bitstring, bitstring, bitstring, bitstring, bounded_1024_id): bool.

**fun** cond31_prime(bitstring, bitstring, bitstring, bitstring, bitstring, bounded_1024_id): bool.

**fun** cond30_prime(bounded_1024_id): bool.

**fun** cond29_prime(bounded_1024_id): bool.

**fun** cond27_prime(bitstring, bitstring, bitstring, bitstring, bitstring): bool.

**fun** cond26_prime(bounded_1045_plaintext): bool.

**fun** cond25_prime(bounded_1045_plaintext): bool.

**fun** cond24_prime(bounded_1045_plaintext, bitstring): bool.

**fun** cond23_prime(bitstring, bitstring, bitstring, bitstring, bitstring, bounded_1024_id, bounded_1024_id, bounded_1024_id, fixed_16_key, fixed_1024_payload, bitstring): bool.

**fun** cond22_prime(bitstring, bitstring, bitstring, bitstring, bitstring, bounded_1024_id, bounded_1024_id, bounded_1024_id, fixed_16_key, fixed_1024_payload, bitstring): bool.

**fun** cond19_prime(bounded_1077_ciphertext): bool.

**fun** cond17_prime(bitstring, bitstring, bitstring, bitstring, bitstring, bounded_1024_id, bounded_1024_id, bounded_1024_id, fixed_16_key, fixed_1024_payload): bool.

**fun** cond16_prime(bitstring, bitstring, bitstring, bitstring, bitstring, bounded_1024_id, bounded_1024_id, bounded_1024_id, fixed_16_key, fixed_1024_payload): bool.

**fun** cond14_prime(bitstring, bitstring, bitstring, bitstring, bitstring, bounded_1024_id, bounded_1024_id, bounded_1024_id): bool.

**fun** cond13_prime(bounded_1024_id, bounded_1024_id): bool.

**fun** cond10_prime(bitstring, bitstring, bitstring, bitstring, bitstring, bounded_1024_id, bounded_1024_id): bool.

**fun** cond1_prime(bitstring, bitstring, bitstring, bitstring, bitstring): bool.

*(****************************
*Typecasting*
****************************)*

**fun** cast_bitstring_bounded_1077_ciphertext(bitstring): bounded_1077_ciphertext [compos].
**fun** cast_bounded_1045_plaintext_fixed_1024_payload(bounded_1045_plaintext): fixed_1024_payload [ compos].
**fun** cast_fixed_1024_payload_bounded_1045_plaintext(fixed_1024_payload): bounded_1045_plaintext [

compos].
**forall** x: fixed_1024_payload;
cast_bounded_1045_plaintext_fixed_1024_payload(cast_fixed_1024_payload_bounded_1045_plaintext(x
)) = x.

*(***************************
Auxiliary Facts
***************************)*

**forall** arg6: bounded_1024_id, arg5: bounded_1024_id, arg4: bitstring, arg3: bitstring, arg2: bitstring, arg1: bitstring, arg0: bitstring;
cond7(arg0, arg1, arg2, arg3, arg4, arg5, arg6) = cond7_prime(Zbitstring(arg0), Zbitstring(arg1), Zbitstring(arg2), Zbitstring(arg3), Zbitstring(arg4), Zbounded_1024_id(arg5), Zbounded_1024_id(arg6)).
**forall** arg0: bounded_1024_id;
cond6(arg0) = cond6_prime(Zbounded_1024_id(arg0)).
**forall** arg0: bounded_1077_ciphertext;
cond58(arg0) = cond58_prime(Zbounded_1077_ciphertext(arg0)).
**forall** arg1: fixed_1024_payload, arg0: bounded_1077_ciphertext;
cond57(arg0, arg1) = cond57_prime(Zbounded_1077_ciphertext(arg0), Zfixed_1024_payload(arg1)).
**forall** arg9: bounded_1045_plaintext, arg8: bitstring, arg7: bitstring, arg6: bounded_1024_id, arg5: bounded_1024_id, arg4: bitstring, arg3: bitstring, arg2: bitstring, arg10: fixed_1024_payload, arg1: bitstring, arg0: bitstring;
cond56(arg0, arg1, arg2, arg3, arg4, arg5, arg6, arg7, arg8, arg9, arg10) = cond56_prime(Zbitstring(arg0), Zbitstring(arg1), Zbitstring(arg2), Zbitstring(arg3), Zbitstring(arg4), Zbounded_1024_id(arg5), Zbounded_1024_id(arg6), arg7, arg8, arg9, Zfixed_1024_payload(arg10))
.
**forall** x510: bounded_1077_ciphertext, x509: bounded_1024_id, arg9: bounded_1045_plaintext, arg7: bitstring, arg6: bounded_1024_id, arg5: bounded_1024_id, arg4: bitstring, arg3: bitstring, arg2: bitstring, arg10: fixed_1024_payload, arg1: bitstring, arg0: bitstring;
cond56_prime(arg0, arg1, arg2, arg3, arg4, arg5, arg6, arg7, conc2(x509, x510), arg9, arg10) = cond56_prime(arg0, arg1, arg2, arg3, arg4, arg5, arg6, arg7, conc2_prime(Zbounded_1024_id(x509), Zbounded_1077_ciphertext(x510)), arg9, arg10).
**forall** x513: fixed_16_key, x512: fixed_1024_payload, arg8: bitstring, arg7: bitstring, arg6: bounded_1024_id, arg5: bounded_1024_id, arg4: bitstring, arg3: bitstring, arg2: bitstring, arg10: fixed_1024_payload, arg1: bitstring, arg0: bitstring;
cond56_prime(arg0, arg1, arg2, arg3, arg4, arg5, arg6, arg7, arg8, conc1(x512, x513), arg10) = cond56_prime(arg0, arg1, arg2, arg3, arg4, arg5, arg6, arg7, arg8, conc1_prime(Zfixed_1024_payload(x512), Zfixed_16_key(x513)), arg10).
**forall** arg9: bounded_1045_plaintext, arg8: bitstring, arg7: bitstring, arg6: bounded_1024_id, arg5: bounded_1024_id, arg4: bitstring, arg3: bitstring, arg2: bitstring, arg1: bitstring, arg0: bitstring;
cond53(arg0, arg1, arg2, arg3, arg4, arg5, arg6, arg7, arg8, arg9) = cond53_prime(Zbitstring(

arg0), Zbitstring(arg1), Zbitstring(arg2), Zbitstring(arg3), Zbitstring(arg4), Zbounded_1024_id(arg5), Zbounded_1024_id(arg6), arg7, arg8, arg9).

**forall** x537: bounded_1077_ciphertext, x536: bounded_1024_id, arg9: bounded_1045_plaintext, arg7: bitstring, arg6: bounded_1024_id, arg5: bounded_1024_id, arg4: bitstring, arg3: bitstring, arg2: bitstring, arg1: bitstring, arg0: bitstring;

cond53_prime(arg0, arg1, arg2, arg3, arg4, arg5, arg6, arg7, conc2(x536, x537), arg9) = cond53_prime(arg0, arg1, arg2, arg3, arg4, arg5, arg6, arg7, conc2_prime(Zbounded_1024_id(x536), Zbounded_1077_ciphertext(x537)), arg9).

**forall** x540: fixed_16_key, x539: fixed_1024_payload, arg8: bitstring, arg7: bitstring, arg6: bounded_1024_id, arg5: bounded_1024_id, arg4: bitstring, arg3: bitstring, arg2: bitstring, arg1: bitstring, arg0: bitstring;

cond53_prime(arg0, arg1, arg2, arg3, arg4, arg5, arg6, arg7, arg8, conc1(x539, x540)) = cond53_prime(arg0, arg1, arg2, arg3, arg4, arg5, arg6, arg7, arg8, conc1_prime(Zfixed_1024_payload(x539), Zfixed_16_key(x540))).

**forall** x544: fixed_16_key, x543: fixed_1024_payload;

cond52(conc1(x543, x544)) = cond52(conc1_prime(Zfixed_1024_payload(x543), Zfixed_16_key(x544))).

**forall** arg9: bounded_1045_plaintext, arg8: bitstring, arg7: bitstring, arg6: bounded_1024_id, arg5: bounded_1024_id, arg4: bitstring, arg3: bitstring, arg2: bitstring, arg1: bitstring, arg0: bitstring;

cond51(arg0, arg1, arg2, arg3, arg4, arg5, arg6, arg7, arg8, arg9) = cond51_prime(Zbitstring(arg0), Zbitstring(arg1), Zbitstring(arg2), Zbitstring(arg3), Zbitstring(arg4), Zbounded_1024_id(arg5), Zbounded_1024_id(arg6), arg7, arg8, arg9).

**forall** x567: bounded_1077_ciphertext, x566: bounded_1024_id, arg9: bounded_1045_plaintext, arg7: bitstring, arg6: bounded_1024_id, arg5: bounded_1024_id, arg4: bitstring, arg3: bitstring, arg2: bitstring, arg1: bitstring, arg0: bitstring;

cond51_prime(arg0, arg1, arg2, arg3, arg4, arg5, arg6, arg7, conc2(x566, x567), arg9) = cond51_prime(arg0, arg1, arg2, arg3, arg4, arg5, arg6, arg7, conc2_prime(Zbounded_1024_id(x566), Zbounded_1077_ciphertext(x567)), arg9).

**forall** x570: fixed_16_key, x569: fixed_1024_payload, arg8: bitstring, arg7: bitstring, arg6: bounded_1024_id, arg5: bounded_1024_id, arg4: bitstring, arg3: bitstring, arg2: bitstring, arg1: bitstring, arg0: bitstring;

cond51_prime(arg0, arg1, arg2, arg3, arg4, arg5, arg6, arg7, arg8, conc1(x569, x570)) = cond51_prime(arg0, arg1, arg2, arg3, arg4, arg5, arg6, arg7, arg8, conc1_prime(Zfixed_1024_payload(x569), Zfixed_16_key(x570))).

**forall** x574: fixed_16_key, x573: fixed_1024_payload;

cond50(conc1(x573, x574)) = cond50(conc1_prime(Zfixed_1024_payload(x573), Zfixed_16_key(x574))).

**forall** arg5: bounded_1024_id, arg4: bitstring, arg3: bitstring, arg2: bitstring, arg1: bitstring, arg0: bitstring;

cond5(arg0, arg1, arg2, arg3, arg4, arg5) = cond5_prime(Zbitstring(arg0), Zbitstring(arg1), Zbitstring(arg2), Zbitstring(arg3), Zbitstring(arg4), Zbounded_1024_id(arg5)).

**forall** x590: fixed_16_key, x589: fixed_1024_payload;

cond49(conc1(x589, x590)) = cond49(conc1_prime(Zfixed_1024_payload(x589), Zfixed_16_key(

x590))).
**forall** x594: fixed_16_key, x593: fixed_1024_payload;
cond48(conc1(x593, x594)) = cond48(conc1_prime(Zfixed_1024_payload(x593), Zfixed_16_key(x594))).
**forall** x598: fixed_16_key, x597: fixed_1024_payload;
cond47(conc1(x597, x598)) = cond47(conc1_prime(Zfixed_1024_payload(x597), Zfixed_16_key(x598))).
**forall** arg0: bounded_1045_plaintext;
cond46(arg0) = cond46_prime(Zbounded_1045_plaintext(arg0)).
**forall** arg2: bitstring, arg1: bitstring, arg0: bounded_1045_plaintext;
cond45(arg0, arg1, arg2) = cond45_prime(Zbounded_1045_plaintext(arg0), arg1, arg2).
**forall** x610: bounded_1077_ciphertext, x609: bounded_1024_id, arg1: bitstring, arg0: bounded_1045_plaintext;
cond45_prime(arg0, arg1, conc2(x609, x610)) = cond45_prime(arg0, arg1, conc2_prime(Zbounded_1024_id(x609), Zbounded_1077_ciphertext(x610))).
**forall** arg8: bitstring, arg7: bitstring, arg6: bounded_1024_id, arg5: bounded_1024_id, arg4: bitstring, arg3: bitstring, arg2: bitstring, arg1: bitstring, arg0: bitstring;
cond44(arg0, arg1, arg2, arg3, arg4, arg5, arg6, arg7, arg8) = cond44_prime(Zbitstring(arg0), Zbitstring(arg1), Zbitstring(arg2), Zbitstring(arg3), Zbitstring(arg4), Zbounded_1024_id(arg5), Zbounded_1024_id(arg6), arg7, arg8).
**forall** x632: bounded_1077_ciphertext, x631: bounded_1024_id, arg7: bitstring, arg6: bounded_1024_id, arg5: bounded_1024_id, arg4: bitstring, arg3: bitstring, arg2: bitstring, arg1: bitstring, arg0: bitstring;
cond44_prime(arg0, arg1, arg2, arg3, arg4, arg5, arg6, arg7, conc2(x631, x632)) = cond44_prime(arg0, arg1, arg2, arg3, arg4, arg5, arg6, arg7, conc2_prime(Zbounded_1024_id(x631 ), Zbounded_1077_ciphertext(x632))).
**forall** arg8: bounded_1024_id, arg7: bitstring, arg6: bitstring, arg5: bounded_1024_id, arg4: bitstring, arg3: bitstring, arg2: bitstring, arg1: bitstring, arg0: bitstring;
cond43(arg0, arg1, arg2, arg3, arg4, arg5, arg6, arg7, arg8) = cond43_prime(Zbitstring(arg0), Zbitstring(arg1), Zbitstring(arg2), Zbitstring(arg3), Zbitstring(arg4), Zbounded_1024_id(arg5), arg6, arg7, Zbounded_1024_id(arg8)).
**forall** x653: bounded_1077_ciphertext, x652: bounded_1024_id, arg8: bounded_1024_id, arg6: bitstring, arg5: bounded_1024_id, arg4: bitstring, arg3: bitstring, arg2: bitstring, arg1: bitstring, arg0: bitstring;
cond43_prime(arg0, arg1, arg2, arg3, arg4, arg5, arg6, conc2(x652, x653), arg8) = cond43_prime(arg0, arg1, arg2, arg3, arg4, arg5, arg6, conc2_prime(Zbounded_1024_id(x652), Zbounded_1077_ciphertext(x653)), arg8).
**forall** arg1: bounded_1024_id, arg0: bitstring;
cond42(arg0, arg1) = cond42_prime(arg0, Zbounded_1024_id(arg1)).
**forall** x659: bounded_1077_ciphertext, x658: bounded_1024_id, arg1: bounded_1024_id;
cond42_prime(conc2(x658, x659), arg1) = cond42_prime(conc2_prime(Zbounded_1024_id(x658) , Zbounded_1077_ciphertext(x659)), arg1).
**forall** x668: bounded_1077_ciphertext, x667: bounded_1024_id, arg0: bitstring;

cond41(arg0, conc2(x667, x668)) = cond41(arg0, conc2_prime(Zbounded_1024_id(x667), Zbounded_1077_ciphertext(x668))).

**forall** arg7: bitstring, arg6: bitstring, arg5: bounded_1024_id, arg4: bitstring, arg3: bitstring, arg2: bitstring, arg1: bitstring, arg0: bitstring;

cond40(arg0, arg1, arg2, arg3, arg4, arg5, arg6, arg7) = cond40_prime(Zbitstring(arg0), Zbitstring(arg1), Zbitstring(arg2), Zbitstring(arg3), Zbitstring(arg4), Zbounded_1024_id(arg5), arg6, arg7).

**forall** x688: bounded_1077_ciphertext, x687: bounded_1024_id, arg6: bitstring, arg5: bounded_1024_id, arg4: bitstring, arg3: bitstring, arg2: bitstring, arg1: bitstring, arg0: bitstring;

cond40_prime(arg0, arg1, arg2, arg3, arg4, arg5, arg6, conc2(x687, x688)) = cond40_prime( arg0, arg1, arg2, arg3, arg4, arg5, arg6, conc2_prime(Zbounded_1024_id(x687), Zbounded_1077_ciphertext(x688))).

**forall** arg0: bounded_1024_id;

cond4(arg0) = cond4_prime(Zbounded_1024_id(arg0)).

**forall** x694: bounded_1077_ciphertext, x693: bounded_1024_id;

cond39(conc2(x693, x694)) = cond39(conc2_prime(Zbounded_1024_id(x693), Zbounded_1077_ciphertext(x694))).

**forall** x698: bounded_1077_ciphertext, x697: bounded_1024_id;

cond38(conc2(x697, x698)) = cond38(conc2_prime(Zbounded_1024_id(x697), Zbounded_1077_ciphertext(x698))).

**forall** arg6: bitstring, arg5: bounded_1024_id, arg4: bitstring, arg3: bitstring, arg2: bitstring, arg1: bitstring, arg0: bitstring;

cond37(arg0, arg1, arg2, arg3, arg4, arg5, arg6) = cond37_prime(Zbitstring(arg0), Zbitstring( arg1), Zbitstring(arg2), Zbitstring(arg3), Zbitstring(arg4), Zbounded_1024_id(arg5), arg6).

**forall** arg5: bounded_1024_id, arg4: bitstring, arg3: bitstring, arg2: bitstring, arg1: bitstring, arg0: bitstring;

cond32(arg0, arg1, arg2, arg3, arg4, arg5) = cond32_prime(Zbitstring(arg0), Zbitstring(arg1), Zbitstring(arg2), Zbitstring(arg3), Zbitstring(arg4), Zbounded_1024_id(arg5)).

**forall** arg5: bounded_1024_id, arg4: bitstring, arg3: bitstring, arg2: bitstring, arg1: bitstring, arg0: bitstring;

cond31(arg0, arg1, arg2, arg3, arg4, arg5) = cond31_prime(Zbitstring(arg0), Zbitstring(arg1), Zbitstring(arg2), Zbitstring(arg3), Zbitstring(arg4), Zbounded_1024_id(arg5)).

**forall** arg0: bounded_1024_id;

cond30(arg0) = cond30_prime(Zbounded_1024_id(arg0)).

**forall** arg0: bounded_1024_id;

cond29(arg0) = cond29_prime(Zbounded_1024_id(arg0)).

**forall** arg4: bitstring, arg3: bitstring, arg2: bitstring, arg1: bitstring, arg0: bitstring;

cond27(arg0, arg1, arg2, arg3, arg4) = cond27_prime(Zbitstring(arg0), Zbitstring(arg1), Zbitstring(arg2), Zbitstring(arg3), Zbitstring(arg4)).

**forall** arg0: bounded_1045_plaintext;

cond26(arg0) = cond26_prime(Zbounded_1045_plaintext(arg0)).

**forall** arg0: bounded_1045_plaintext;

cond25(arg0) = cond25_prime(Zbounded_1045_plaintext(arg0)).

**forall** arg1: bitstring, arg0: bounded_1045_plaintext;
cond24(arg0, arg1) = cond24_prime(Zbounded_1045_plaintext(arg0), arg1).
**forall** arg9: fixed_1024_payload, arg8: fixed_16_key, arg7: bounded_1024_id, arg6: bounded_1024_id, arg5: bounded_1024_id, arg4: bitstring, arg3: bitstring, arg2: bitstring, arg10: bitstring, arg1: bitstring, arg0: bitstring;
cond23(arg0, arg1, arg2, arg3, arg4, arg5, arg6, arg7, arg8, arg9, arg10) = cond23_prime( Zbitstring(arg0), Zbitstring(arg1), Zbitstring(arg2), Zbitstring(arg3), Zbitstring(arg4), Zbounded_1024_id(arg5), Zbounded_1024_id(arg6), Zbounded_1024_id(arg7), Zfixed_16_key(arg8), Zfixed_1024_payload(arg9), arg10).
**forall** arg9: fixed_1024_payload, arg8: fixed_16_key, arg7: bounded_1024_id, arg6: bounded_1024_id, arg5: bounded_1024_id, arg4: bitstring, arg3: bitstring, arg2: bitstring, arg10: bitstring, arg1: bitstring, arg0: bitstring;
cond22(arg0, arg1, arg2, arg3, arg4, arg5, arg6, arg7, arg8, arg9, arg10) = cond22_prime( Zbitstring(arg0), Zbitstring(arg1), Zbitstring(arg2), Zbitstring(arg3), Zbitstring(arg4), Zbounded_1024_id(arg5), Zbounded_1024_id(arg6), Zbounded_1024_id(arg7), Zfixed_16_key(arg8), Zfixed_1024_payload(arg9), arg10).
**forall** arg0: bounded_1077_ciphertext;
cond19(arg0) = cond19_prime(Zbounded_1077_ciphertext(arg0)).
**forall** arg9: fixed_1024_payload, arg8: fixed_16_key, arg7: bounded_1024_id, arg6: bounded_1024_id, arg5: bounded_1024_id, arg4: bitstring, arg3: bitstring, arg2: bitstring, arg1: bitstring, arg0: bitstring;
cond17(arg0, arg1, arg2, arg3, arg4, arg5, arg6, arg7, arg8, arg9) = cond17_prime(Zbitstring( arg0), Zbitstring(arg1), Zbitstring(arg2), Zbitstring(arg3), Zbitstring(arg4), Zbounded_1024_id( arg5), Zbounded_1024_id(arg6), Zbounded_1024_id(arg7), Zfixed_16_key(arg8), Zfixed_1024_payload(arg9)).
**forall** arg9: fixed_1024_payload, arg8: fixed_16_key, arg7: bounded_1024_id, arg6: bounded_1024_id, arg5: bounded_1024_id, arg4: bitstring, arg3: bitstring, arg2: bitstring, arg1: bitstring, arg0: bitstring;
cond16(arg0, arg1, arg2, arg3, arg4, arg5, arg6, arg7, arg8, arg9) = cond16_prime(Zbitstring( arg0), Zbitstring(arg1), Zbitstring(arg2), Zbitstring(arg3), Zbitstring(arg4), Zbounded_1024_id( arg5), Zbounded_1024_id(arg6), Zbounded_1024_id(arg7), Zfixed_16_key(arg8), Zfixed_1024_payload(arg9)).
**forall** arg7: bounded_1024_id, arg6: bounded_1024_id, arg5: bounded_1024_id, arg4: bitstring, arg3: bitstring, arg2: bitstring, arg1: bitstring, arg0: bitstring;
cond14(arg0, arg1, arg2, arg3, arg4, arg5, arg6, arg7) = cond14_prime(Zbitstring(arg0), Zbitstring(arg1), Zbitstring(arg2), Zbitstring(arg3), Zbitstring(arg4), Zbounded_1024_id(arg5), Zbounded_1024_id(arg6), Zbounded_1024_id(arg7)).
**forall** arg1: bounded_1024_id, arg0: bounded_1024_id;
cond13(arg0, arg1) = cond13_prime(Zbounded_1024_id(arg0), Zbounded_1024_id(arg1)).
**forall** arg6: bounded_1024_id, arg5: bounded_1024_id, arg4: bitstring, arg3: bitstring, arg2: bitstring, arg1: bitstring, arg0: bitstring;
cond10(arg0, arg1, arg2, arg3, arg4, arg5, arg6) = cond10_prime(Zbitstring(arg0), Zbitstring( arg1), Zbitstring(arg2), Zbitstring(arg3), Zbitstring(arg4), Zbounded_1024_id(arg5),

```
    Zbounded_1024_id(arg6)).
forall arg4: bitstring, arg3: bitstring, arg2: bitstring, arg1: bitstring, arg0: bitstring;
      cond1(arg0, arg1, arg2, arg3, arg4) = cond1_prime(Zbitstring(arg0), Zbitstring(arg1),
    Zbitstring(arg2), Zbitstring(arg3), Zbitstring(arg4)).

(**************************
  Zero Facts
**************************)

forall arg1: bounded_1077_ciphertext, arg0: bounded_1024_id;
      Zbitstring(conc2(arg0, arg1)) = Zbitstring_prime(conc2(Zbounded_1024_id(arg0),
    Zbounded_1077_ciphertext(arg1))).
forall arg1: fixed_16_key, arg0: fixed_1024_payload;
      Zbounded_1045_plaintext(conc1(arg0, arg1)) = Zbounded_1045_plaintext_prime(conc1(
    Zfixed_1024_payload(arg0), Zfixed_16_key(arg1))).
forall x: bitstring;
      Zbounded_1077_ciphertext(cast_bitstring_bounded_1077_ciphertext(x)) =
    Zbounded_1077_ciphertext_prime(cast_bitstring_bounded_1077_ciphertext(Zbitstring(x))).
forall x: bounded_1045_plaintext;
      Zfixed_1024_payload(cast_bounded_1045_plaintext_fixed_1024_payload(x)) =
    Zfixed_1024_payload_prime(cast_bounded_1045_plaintext_fixed_1024_payload(
    Zbounded_1045_plaintext(x))).
forall x: fixed_1024_payload;
      Zbounded_1045_plaintext(cast_fixed_1024_payload_bounded_1045_plaintext(x)) =
    Zbounded_1045_plaintext_prime(cast_fixed_1024_payload_bounded_1045_plaintext(
    Zfixed_1024_payload(x))).
forall x: fixed_1024_payload;
      Zfixed_1024_payload(x) = zero_fixed_1024_payload().
forall x: fixed_16_keyseed;
      Zfixed_16_keyseed(x) = zero_fixed_16_keyseed().
forall x: fixed_16_key;
      Zfixed_16_key(x) = zero_fixed_16_key().
forall x: fixed_16_seed;
      Zfixed_16_seed(x) = zero_fixed_16_seed().

(*******************************
  (USER) <Query>
*******************************)

event client_begin(bounded_1024_id, bounded_1024_id, fixed_1024_payload).
event client_accept(bounded_1024_id, bounded_1024_id, fixed_1024_payload, fixed_1024_payload).
event server_reply(bounded_1024_id, bounded_1024_id, fixed_1024_payload, fixed_1024_payload).
event bad(bounded_1024_id).
```

```
(* Authentication of the server to the client *)
query hClient: bounded_1024_id, hServer: bounded_1024_id, x: fixed_1024_payload, y:
    fixed_1024_payload;
  event client_accept(hClient, hServer, x, y) ==> server_reply(hClient, hServer, x, y).

(* Authentication of the client to the server *)
query hClient: bounded_1024_id, hServer: bounded_1024_id, x: fixed_1024_payload, y:
    fixed_1024_payload;
  event server_reply(hClient, hServer, x, y) ==> client_begin(hClient, hServer, x) || bad(hClient).

(* Strong secrecy of the request *)
query secret request.

(* Weak conditional secrecy of the response *)
event leaked(bounded_1024_id, fixed_1024_payload).
query hClient: bounded_1024_id, resp: fixed_1024_payload;
  event leaked(hClient, resp) ==> bad(hClient).


(**************************
  Model
**************************)

let client =
in(c_in, ());
if cond1(argv0, argv1, argv2, argv3, argv4) then
if cond4(clientID) then
if cond5(argv0, argv1, argv2, argv3, argv4, clientID) then
if cond6(serverID) then
if cond7(argv0, argv1, argv2, argv3, argv4, clientID, serverID) then
if cond10(argv0, argv1, argv2, argv3, argv4, clientID, serverID) then
out(c_out, (port, serverID));
in(c_in, net_connect_result1: bitstring);
if cond11(net_connect_result1) then
if cond12(net_connect_result1) then
if cond13(clientID, xClient) then
if cond14(argv0, argv1, argv2, argv3, argv4, clientID, serverID, xClient) then
if clientID = xClient then
let key1 = lookup(clientID, serverID, db) in
new kS_seed1: fixed_16_keyseed;
let key2 = kgen(kS_seed1) in
event client_begin(clientID, serverID, request);
```

```
if cond16(argv0, argv1, argv2, argv3, argv4, clientID, serverID, xClient, key2, request) then
if cond17(argv0, argv1, argv2, argv3, argv4, serverID, xClient, clientID, key2, request) then
let msg1 = conc1(request, key2) in
new nonce1: fixed_16_seed;
let cipher1 = E(msg1, key1, nonce1) in
if cond19(cipher1) then
let msg2 = arithmetic1(clientID, cipher1) in
let msg3 = conc2(clientID, cipher1) in
out(c_out, (msg3, msg2));
in(c_in, (msg4: bitstring, cipher2: bitstring));
if cond20(msg4) then
if cond21(msg4) then
if cond22(argv0, argv1, argv2, argv3, argv4, serverID, xClient, clientID, key2, request, msg4) then
if cond23(argv0, argv1, argv2, argv3, argv4, serverID, xClient, clientID, key2, request, msg4) then
let injbot(msg6) = D(cast_bitstring_bounded_1077_ciphertext(cipher2), key2) in
if cond24(msg6, msg4) then
if cond25(msg6) then
if cond26(msg6) then
event client_accept(clientID, serverID, request, cast_bounded_1045_plaintext_fixed_1024_payload(
    msg6));
yield .

let server =
in(c_in, ());
if cond27(argv0, argv1, argv2, argv3, argv4) then
if cond29(serverID) then
if cond30(serverID) then
if cond31(argv0, argv1, argv2, argv3, argv4, serverID) then
if cond32(argv0, argv1, argv2, argv3, argv4, serverID) then
new nondet1: nondet;
out(c_out, (port, serverID));
in(c_in, (net_bind_result1: bitstring, var7: bitstring, net_accept_result1: bitstring, msg7: bitstring,
    msg8: bitstring));
if cond33(net_bind_result1) then
new nondet2: nondet;
if cond34(net_accept_result1) then
if cond35(msg7) then
if cond36(msg7) then
if cond37(argv0, argv1, argv2, argv3, argv4, serverID, msg7) then
if cond38(msg8) then
if cond39(msg8) then
if cond40(argv0, argv1, argv2, argv3, argv4, serverID, msg7, msg8) then
if cond41(msg7, msg8) then
```

```
if cond42(msg8, xClient) then
if cond43(argv0, argv1, argv2, argv3, argv4, serverID, msg7, msg8, xClient) then
let conc2(client2, cipher3) = msg8 in
if client2 = xClient then
let key3 = lookup(client2, serverID, db) in
if cond44(argv0, argv1, argv2, argv3, argv4, serverID, xClient, msg7, msg8) then
let injbot(msg9) = D(cipher3, key3) in
if cond45(msg9, msg7, msg8) then
if cond46(msg9) then
if cond47(msg9) then
if cond48(msg9) then
if cond49(msg9) then
if cond50(msg9) then
if cond51(argv0, argv1, argv2, argv3, argv4, serverID, xClient, msg7, msg8, msg9) then
if cond52(msg9) then
if cond53(argv0, argv1, argv2, argv3, argv4, serverID, xClient, msg7, msg8, msg9) then
let conc1(var18, key4) = msg9 in
event server_reply(client2, serverID, var18, response);
if cond56(argv0, argv1, argv2, argv3, argv4, serverID, xClient, msg7, msg8, msg9, response) then
new nonce2: fixed_16_seed;
let msg12 = E(cast_fixed_1024_payload_bounded_1045_plaintext(response), key4, nonce2) in
if cond57(msg12, response) then
if cond58(msg12) then
let msg11 = arithmetic2(msg12) in
out(c_out, (msg12, msg11)); 0 .

(******************************
  (USER) <Environment>
*******************************)

let client' =
  in(c_in, (argv0: bitstring,
          argv1: bitstring,
          argv2: bitstring,
          argv3: bitstring,
          argv4: bitstring));
  out(c_out, ());

  in(c_in, xClient: bounded_1024_id);

  (*
    For proving correspondences it may be more convincing to let the attacker choose
    the payloads, but we need to generate them randomly to check secrecy.
```

```
  *)
  new request: fixed_1024_payload;

  out(c_out, ());

  client .

(* The sentinel used in formulating weak secrecy of the response *)
let sentinel =
  in(c_in, response': fixed_1024_payload);
  if response' = response then
  event leaked(xClient, response);
  yield .

let server' =
  in(c_in, (argv0: bitstring,
            argv1: bitstring,
            argv2: bitstring,
            argv3: bitstring,
            argv4: bitstring));
  out(c_out, ());

  in(c_in, xClient: bounded_1024_id);

  new response: fixed_1024_payload;

  if xClient = clientID then
    out(c_out, ());
    ( server | sentinel )
  else
    let badHost(xClient') = xClient in
    event bad(xClient);
    out(c_out, ());
    ( server | sentinel ) .

process
! N(
  (* get a key database and the payloads from the attacker *)
  in(c_in, adb: keydb);

  (* generate and insert the honest key *)
  new kAB_seed: fixed_16_keyseed;
  let kAB = kgen(kAB_seed) in
```

```
  let db = add_honest(kAB, adb) in

  out(c_out, ());
  ((! N client') | (! N server'))
)
```

# Appendix D

# ProVerif Model of NSL

The ProVerif model for the NSL protocol is shown below. The processes $A$ and $B$ as well as the symbolic rules for the new encoding and parsing expressions $conc_i$ and $parse_i$ are generated automatically from the source IML process. The rules for encryption and decryption, the query, and the environment process (including $A'$ and $B'$) are specified by hand.

We use signatures to model a key server that accepts keys from the attacker, thus allowing key compromise. When a host $h$ is compromised the key server executes the event $bad(h)$. Our correspondence properties are then formulated with respect to the condition that the host is not compromised: a query of the form

```
query
  ev:endB(x, y) ==> ev:beginA(x, y) | ev:bad(x).
```

means that whenever the event $endB(x, y)$ occurs, either the event $beginA(x, y)$ or the event $bad(x)$ occurred before.

The model is key-safe: neither signing nor decryption keys are being sent. Signing and decryption uses honestly generated keys with fresh randomness. The computational soundness result does not allow event parameters, but it applies to the coarser model in which event parameters are removed.

```
(********************************
  Needham-Schroeder-Lowe protocol.
*********************************)

free c.
fun true/0.

(***********************
  Public-key encryption
```

```
*************************)

(*
  NB! The soundness result in CoSP assumes that we can encrypt messages of any length,
  which means that a correct implementation of E must use hybrid encryption.
*)

fun ek/1.
fun dk/1.
fun E/3.
reduc
  D(E(x, ek(a), r), dk(a)) = x.
reduc
  isek(ek(a)) = ek(a).

(* We do not add rewriting for garbage since we are not using ekof and isenc. *)

(************************
  Signatures
*************************)

fun sk/1.
fun vk/1.
fun sig/3.

(* Just a pairing function *)
data cert/2.

(*
  check_key performs signature verification together with
  bitstring comparison, so it is covered by the soundness result.
*)
reduc
  check_key(host, key, sig(sk(r), cert(key, host), r'), vk(r)) = true.

(* We do not add rewriting for garbage since we are not using vkof. *)

(*******************************
  Names of the honest participants
*********************************)

fun hostA/0.
fun hostB/0.
```

```
(*************************
  Concatenation and Parsing
***************************)

data conc1/3.

data conc3/1.

data conc8/2.

data conc15/1.

reduc
  parse3(conc8(x0, x1)) = x1.
reduc
  parse4(conc3(x0)) = x0;
  parse4(conc8(x0, x1)) = x0.
reduc
  parse5(conc15(x0)) = x0.
reduc
  parse7(conc1(x0, x1, x2)) = x0.
reduc
  parse8(conc1(x0, x1, x2)) = x2.
reduc
  parse9(conc1(x0, x1, x2)) = x1.

(******************
  <Query>
******************)

query
  ev:endB(x, y) ==> ev:beginA(x, y) | ev:bad(x).

query
  ev:endA(x, y) ==> ev:beginB(x, y) | ev:bad(y).

(*************************
  Model
*************************)

let A =
in(c, msg1);
```

```
in(c, msg2);
let msg3 = D(msg2, skB) in
if check_key(parse3(msg3), pkX, sigX, pkS) = true then
event beginB(parse3(msg3), hostB);
new nonce1;
new nonce2;
let msg4 = conc3(E(conc1(parse4(msg3), nonce1, hostB), isek(pkX), nonce2)) in
out(c, msg4);
in(c, msg5);
in(c, msg6);
let var1 = parse5(D(msg6, skB)) in
if var1 = nonce1 then
event endB(parse3(msg3), hostB); 0.

let B =
if check_key(hostX, pkX, sigX, pkS) = true then
event beginA(hostA, hostX);
new nonce1;
new nonce2;
let msg1 = conc3(E(conc8(nonce1, hostA), isek(pkX), nonce2)) in
out(c, msg1);
in(c, msg2);
in(c, msg3);
let msg4 = D(msg3, skA) in
let var1 = parse7(msg4) in
if var1 = nonce1 then
let var2 = parse8(msg4) in
if var2 = hostX then
new nonce3;
let msg5 = conc3(E(conc15(parse9(msg4)), isek(pkX), nonce3)) in
out(c, msg5);
event endA(hostA, hostX); 0.

(**************************
  <Environment>
**************************)

let A' =
  in(c, (pkX, hostX, sigX));
  A .

let B' =
  in(c, (pkX, hostX, sigX));
```

```
  B .

let keyServer =
  in(c, (h, k));
  new r3;
  if h = hostA then
    out(c, (pkA, h, sig(skS, cert(pkA, h), r3)))
  else if h = hostB then
    out(c, (pkB, h, sig(skS, cert(pkB, h), r3)))
  else
    event bad(h);
    out(c, (k, h, sig(skS, cert(k, h), r3))) .

process
! (
  new A_seed;
  new B_seed;
  let pkA = ek(A_seed) in
  let skA = dk(A_seed) in
  let pkB = ek(B_seed) in
  let skB = dk(B_seed) in

  new rkS;
  let pkS = vk(rkS) in
  let skS = sk(rkS) in

  out(c, (pkA, pkB, pkS));
  (! A' | ! B' | ! keyServer)
)
```

# References

The verified software initiative: A manifesto. http://qpq.csl.sri.com/vsr/manifesto.pdf/view, 2007. 13

Martín Abadi. Secrecy by typing in security protocols. *Journal of the ACM (JACM)*, 46(5): 749–786, 1999. 14

Martín Abadi and Jan Jürjens. Formal eavesdropping and its computational interpretation. In *TACS '01: Proceedings of the 4th International Symposium on Theoretical Aspects of Computer Software*, pages 82–94, London, UK, 2001. Springer-Verlag. ISBN 3-540-42736-8. 14

Martín Abadi and Phillip Rogaway. Reconciling two views of cryptography (the computational soundness of formal encryption). In *IFIP TCS*, volume 1872 of *Lecture Notes in Computer Science*, pages 3–22. Springer, 2000. ISBN 3-540-67823-9. 14

Mihhail Aizatulin, Henning Schnoor, and Thomas Wilke. Computationally sound analysis of a probabilistic contract signing protocol. In *ESORICS*, volume 5789 of *Lecture Notes in Computer Science*, pages 571–586. Springer, 2009. ISBN 978-3-642-04443-4. 14

Mihhail Aizatulin, François Dupressoir, Andrew D. Gordon, and Jan Jürjens. Verifying cryptographic code in C: Some experience and the Csec challenge. In *Formal Aspects of Security and Trust (FAST 2011)*, Lecture Notes in Computer Science. Springer, 2011a. 5, 141

Mihhail Aizatulin, Andrew D. Gordon, and Jan Jürjens. Extracting and verifying cryptographic models from C protocol code by symbolic execution. In *18th ACM Conference on Computer and Communications Security (CCS 2011)*, 2011b. Full version available at http://arxiv.org/abs/1107.1017. iii, 5

Mihhail Aizatulin, Andrew D. Gordon, and Jan Jürjens. Computational verification of C protocol implementations by symbolic execution. In *ACM Conference on Computer and Communications Security*, pages 712–723, 2012. iii, 5

Martin R. Albrecht, Kenneth G. Paterson, and Gaven J. Watson. Plaintext recovery attacks against SSH. In *IEEE Symposium on Security and Privacy*, pages 16–26, 2009. 3, 89, 157

José Bacelar Almeida, Manuel Barbosa, Jorge Sousa Pinto, and Bárbara Vieira. Deductive verification of cryptographic software. In *NASA Formal Methods Symposium 2009*, 2009. 4

José Bacelar Almeida, Manuel Barbosa, Gilles Barthe, and François Dupressoir. Certified computer-aided cryptography: efficient provably secure machine code from high-level implementations. In *Proceedings of the 2013 ACM SIGSAC conference on Computer & communications security*, pages 1217–1230. ACM, 2013. 14, 16

José Bacelar Almeida, Manuel Barbosa, Jean-Christophe Filliâtre, Jorge Sousa Pinto, and Bárbara Vieira. Caoverif: An open-source deductive verification platform for cryptographic software implementations. *Science of Computer Programming*, 91:216–233, 2014. 17

João Antunes and Nuno Ferreira Neves. Building an automaton towards reverse protocol engineering. In *INFORUM*, 2009. 17

Myrto Arapinis, Eike Ritter, and Mark Dermot Ryan. StatVerif: Verification of stateful processes. In *CSF*, pages 33–47. IEEE Computer Society, 2011. 157

Alessandro Armando, David A. Basin, Yohan Boichut, Yannick Chevalier, Luca Compagna, Jorge Cuéllar, Paul Hankes Drielsma, Pierre-Cyrille Héam, Olga Kouchnarenko, Jacopo Mantovani, Sebastian Mödersheim, David von Oheimb, Michaël Rusinowitch, Judson Santiago, Mathieu Turuani, Luca Viganò, and Laurent Vigneron. The AVISPA tool for the automated validation of internet security protocols and applications. In *CAV*, volume 3576 of *Lecture Notes in Computer Science*, pages 281–285. Springer, 2005. ISBN 3-540-27231-3. 14

Domagoj Babić. *Exploiting Structure for Scalable Software Verification*. PhD thesis, University of British Columbia, Vancouver, Canada, 2008. 12

Domagoj Babić and Alan J. Hu. Calysto: Scalable and Precise Extended Static Checking. In *30th International Conference on Software Engineering, ICSE 2008, Proceedings*, pages 211–220, New York, NY, USA, May 2008. ACM. ISBN 978-1-60558-079-1. doi: http://doi.acm.org/10.1145/1368088.1368118. 12

Michael Backes, Dennis Hofheinz, and Dominique Unruh. CoSP: A general framework for computational soundness proofs. In *ACM Conference on Computer and Communications Security*, pages 66–78, 2009. Preprint on IACR ePrint 2009/080. 6, 14, 117, 118, 119, 120, 121, 122, 123, 124, 125, 127, 143, 155

Thomas Ball, Byron Cook, Vladimir Levin, and Sriram K. Rajamani. SLAM and static driver verifier: Technology transfer of formal methods inside Microsoft. In *IFM*, pages 1–20, 2004. 9, 13

Thomas Ball, Ella Bounimova, Byron Cook, Vladimir Levin, Jakob Lichtenberg, Con McGarvey, Bohus Ondrusek, Sriram K. Rajamani, and Abdullah Ustuner. Thorough static analysis of device drivers. In *EuroSys '06: Proceedings of the 1st ACM SIGOPS/EuroSys European Conference on Computer Systems 2006*, pages 73–85, New York, NY, USA, 2006. ACM. ISBN 1-59593-322-0. doi: http://doi.acm.org/10.1145/1217935.1217943. 13

Manuel Barbosa, David Castro, and Paulo F Silva. Compiling cao: From cryptographic specifications to c implementations. In *Principles of Security and Trust*, pages 240–244. Springer, 2014. 17

Gilles Barthe, Benjamin Grégoire, and Santiago Zanella Béguelin. Formal certification of code-based cryptographic proofs. In *Proceedings of the 36th annual ACM SIGPLAN-SIGACT symposium on Principles of programming languages*, POPL '09, pages 90–101, New York, NY, USA, 2009. ACM. ISBN 978-1-60558-379-2. doi: http://doi.acm.org/10.1145/1480881.1480894. URL http://doi.acm.org/10.1145/1480881.1480894. 14

Gilles Barthe, Benjamin Grégoire, Sylvain Heraud, and Santiago Zanella Béguelin. Computer-aided security proofs for the working cryptographer. In *Advances in Cryptology – CRYPTO 2011*, Lecture Notes in Computer Science. Springer, 2011. 14

Mihir Bellare, Tadayoshi Kohno, and Chanathip Namprempre. Breaking and provably repairing the SSH authenticated encryption scheme: A case study of the encode-then-encrypt-and-mac paradigm. *ACM Transactions on Information and System Security*, 7:206–241, 2004. 3

Jesper Bengtson, Karthikeyan Bhargavan, Cédric Fournet, Andrew D. Gordon, and Sergio Maffeis. Refinement types for secure implementations. In *CSF '08: Proceedings of the 2008 21st IEEE Computer Security Foundations Symposium*, pages 17–32. IEEE Computer Society, 2008. ISBN 978-0-7695-3182-3. doi: http://dx.doi.org/10.1109/CSF.2008.27. 15, 87, 142, 145

Dirk Beyer, Thomas Henzinger, Ranjit Jhala, and Rupak Majumdar. The software model checker Blast. *International Journal on Software Tools for Technology Transfer (STTT)*, 9 (5-6):505–525, October 2007a. doi: 10.1007/s10009-007-0044-z. URL http://dx.doi.org/10.1007/s10009-007-0044-z. 13

Dirk Beyer, Thomas A. Henzinger, and Grgory Thoduloz. Configurable Software Verification: Concretizing the Convergence of Model Checking and Program Analysis. In *Computer Aided Verification*, Lecture Notes in Computer Science, pages 504–518. Springer, 2007b. 13

Karthikeyan Bhargavan, Cédric Fournet, Andrew D. Gordon, and Stephen Tse. Verified interoperable implementations of security protocols. In *CSFW '06: Proceedings of the 19th IEEE workshop on Computer Security Foundations*, pages 139–152. IEEE Computer Society, 2006. ISBN 0-7695-2615-2. doi: http://dx.doi.org/10.1109/CSFW.2006.32. 15

Karthikeyan Bhargavan, Cédric Fournet, Ricardo Corin, and Eugen Zălinescu. Cryptographically verified implementations for TLS. In *CCS '08: Proceedings of the 15th ACM conference on Computer and communications security*, pages 459–468, Alexandria, VA, October 2008. ACM. 15

Karthikeyan Bhargavan, Cédric Fournet, and Andrew D. Gordon. Modular verification of security protocol code by typing. In *ACM Symposium on Principles of Programming Languages (POPL'10)*, pages 445–456, 2010. 10

Karthikeyan Bhargavan, Cédric Fournet, Markulf Kohlweiss, Alfredo Pironti, and Pierre-Yves Strub. Implementing TLS with verified cryptographic security. In *IEEE Symposium on Security & Privacy (Oakland)*, pages 445–462, 2013. 4, 15

Karthikeyan Bhargavan, Cdric Fournet, Markulf Kohlweiss, Alfredo Pironti, Pierre-Yves Strub, and Santiago Zanella-Bguelin. Proving the TLS handshake secure (as it is). Cryptology ePrint Archive, Report 2014/182, 2014. `http://eprint.iacr.org/`. 4, 15

Bruno Blanchet. A computationally sound mechanized prover for security protocols. *IEEE Transactions on Dependable and Secure Computing*, 5(4):193–207, October–December 2008. 1, 5, 14, 19, 24, 74, 75, 78, 79, 81, 90, 113, 136, 152

Bruno Blanchet. Automatic verification of correspondences for security protocols. *Journal of Computer Security*, 17(4):363–434, 2009. ISSN 0926-227X. 1, 5, 14, 117, 119, 122

Bruno Blanchet. Automatic verification of security protocols in the symbolic model: The verifier proverif. In *Foundations of Security Analysis and Design VII*, pages 54–87. Springer, 2014. 14, 117, 122

Bruno Blanchet, Patrick Cousot, Radhia Cousot, Jérôme Feret, Laurent Mauborgne, Antoine Miné, David Monniaux, and Xavier Rival. A static analyzer for large safety-critical software. *CoRR*, abs/cs/0701193, 2007. 12

Bruno Blanchet, Martín Abadi, and Cédric Fournet. Automated verification of selected equivalences for security protocols. *Journal of Logic and Algebraic Programming*, 75(1):3–51, February–March 2008. 119

Manuel Blum and Silvio Micali. How to generate cryptographically strong sequences of pseudorandom bits. *SIAM J. Comput.*, 13(4):850–864, 1984. ISSN 0097-5397. doi: http://dx.doi.org/10.1137/0213053. 5

Chiara Bodei, Mikael Buchholtz, Pierpaolo Degano, Flemming Nielson, and Hanne Riis Nielson. Static validation of security protocols. *J. Comput. Secur.*, 13(3):347–390, 2005. ISSN 0926-227X. 14

J. Burnim and K. Sen. Heuristics for scalable dynamic test generation. In *ASE '08: Proceedings of the 2008 23rd IEEE/ACM International Conference on Automated Software Engineering*, pages 443–446. IEEE Computer Society, 2008. ISBN 978-1-4244-2187-9. doi: http://dx.doi.org/10.1109/ASE.2008.69. 42

Juan Caballero, Heng Yin, Zhenkai Liang, and Dawn Song. Polyglot: automatic extraction of protocol message format using dynamic binary analysis. In *CCS '07: Proceedings of the 14th ACM conference on Computer and communications security*, pages 317–329, New York, NY, USA, 2007. ACM. ISBN 978-1-59593-703-2. doi: http://doi.acm.org/10.1145/1315245.1315286. 17

Cristian Cadar, Daniel Dunbar, and Dawson Engler. KLEE: Unassisted and automatic generation of high-coverage tests for complex systems programs. In *USENIX Symposium on Operating Systems Design and Implementation (OSDI 2008)*, San Diego, CA, December 2008. 2, 10, 13, 16, 41

David Cadé and Bruno Blanchet. From computationally-proved protocol specifications to implementations. In *International Conference on Availability, Reliability and Security (ARES 2012)*, 2012. 16

David Cadé and Bruno Blanchet. From computationally-proved protocol specifications to implementations and application to ssh. *Journal of Wireless Mobile Networks, Ubiquitous Computing, and Dependable Applications (JoWUA)*, 4(1):4–31, 2013a. 16

David Cadé and Bruno Blanchet. Proved generation of implementations from computationally secure protocol specifications. In *Principles of Security and Trust*, pages 63–82. Springer, 2013b. 16

Sagar Chaki and Anupam Datta. ASPIER: An automated framework for verifying security protocol implementations. In *Computer Security Foundations Workshop*, pages 172–185, 2009. doi: 10.1109/CSF.2009.20. 16

Edmund M. Clarke and E. Allen Emerson. Design and synthesis of synchronization skeletons using branching-time temporal logic. In *Logic of Programs, Workshop*, pages 52–71, London, UK, 1982. Springer-Verlag. ISBN 3-540-11212-X. 12

Edmund M. Clarke, Orna Grumberg, Somesh Jha, Yuan Lu, and Helmut Veith. Counterexample-guided abstraction refinement. In *CAV '00: Proceedings of the 12th International Conference on Computer Aided Verification*, pages 154–169, London, UK, 2000. Springer-Verlag. ISBN 3-540-67770-4. 13

Ernie Cohen, Markus Dahlweid, Mark Hillebrand, Dirk Leinenbach, Michał Moskal, Thomas Santen, Wolfram Schulte, and Stephan Tobies. VCC: A practical system for verifying concurrent C. In *Theorem Proving in Higher Order Logics (TPHOLs 2009)*, volume 5674 of *Lecture Notes in Computer Science*, Munich, Germany, 2009. Springer. Invited paper, to appear. 12

Byron Cook, Andreas Podelski, and Andrey Rybalchenko. Terminator: Beyond safety. In *CAV*, pages 415–418, 2006. 151

Ricardo Corin and Felipe Andres Manzano. Efficient symbolic execution for analysing cryptographic protocol implementations. In *International Symposium on Engineering Secure Software and Systems (ESSOS'11)*, LNCS. Springer, 2011. 16

Patrick Cousot and Radhia Cousot. Abstract interpretation: a unified lattice model for static analysis of programs by construction or approximation of fixpoints. In *Conference Record of the Fourth Annual ACM SIGPLAN-SIGACT Symposium on Principles of Programming Languages*, pages 238–252, Los Angeles, California, 1977. ACM Press, New York, NY. 12

Weidong Cui, Marcus Peinado, Karl Chen, Helen J. Wang, and Luis Irun-Briz. Tupni: automatic reverse engineering of input formats. In *CCS '08: Proceedings of the 15th ACM conference on Computer and communications security*, pages 391–402, New York, NY, USA, 2008. ACM. ISBN 978-1-59593-810-7. doi: http://doi.acm.org/10.1145/1455770.1455820. 17

CVE. CVE-2008-5077, 2008a. Available at http://cve.mitre.org/cgi-bin/cvename.cgi?name=CVE-2008-5077. 1, 157

CVE. CVE-2009-3555, 2009b. Available at http://cve.mitre.org/cgi-bin/cvename.cgi?name=CVE-2009-3555. 157

CVE. CVE-2014-0160, 2014c. Available at http://cve.mitre.org/cgi-bin/cvename.cgi?name=CVE-2014-0160. 1

T. Dierks and E. Rescorla. The Transport Layer Security (TLS) Protocol Version 1.2. RFC 5246 (Proposed Standard), August 2008. URL http://www.ietf.org/rfc/rfc5246.txt. 1

Will Dietz, Peng Li, John Regehr, and Vikram Adve. Understanding integer overflow in C/C++. In *Proceedings of the 2012 International Conference on Software Engineering*, ICSE 2012, pages 760–770, Piscataway, NJ, USA, 2012. IEEE Press. ISBN 978-1-4673-1067-3. URL http://dl.acm.org/citation.cfm?id=2337223.2337313. 30

D. Dolev and A.C. Yao. On the Security of Public-Key Protocols. *IEEE Transactions on Information Theory*, 29(2):198–208, 1983. 5, 117

Vijay D'Silva, Daniel Kroening, and Georg Weissenbacher. A survey of automated techniques for formal software verification. *IEEE Trans. on CAD of Integrated Circuits and Systems*, 27 (7):1165–1178, 2008. 13

F. Dupressoir, A.D. Gordon, and J. Jürjens. Verifying authentication properties of C security code using general verifiers. In *Fourth International Workshop on Analysis of Security APIs (ASA-4 FLOC 2010*, 2010. Presentations only. 16

François Dupressoir, Andrew D Gordon, Jan Jürjens, and David A Naumann. Guiding a general-purpose c verifier to prove cryptographic protocols. *Journal of Computer Security*, 22(5):823–866, 2014. 10, 16, 142, 145, 156

Bruno Dutertre and Leonardo De Moura. The Yices SMT Solver. Technical report, Computer Science Laboratory, SRI International, 2006. 8, 42, 47

Limor Fix. Fifteen years of formal property verification in Intel. In *25 Years of Model Checking: History, Achievements, Perspectives*, pages 139–144. Springer-Verlag, Berlin, Heidelberg, 2008. ISBN 978-3-540-69849-4. doi: http://dx.doi.org/10.1007/978-3-540-69850-0_8. 13

Cormac Flanagan, K. Rustan M. Leino, Mark Lillibridge, Greg Nelson, James B. Saxe, and Raymie Stata. Extended static checking for Java. In *PLDI '02: Proceedings of the ACM SIGPLAN 2002 Conference on Programming language design and implementation*, pages 234–245, New York, NY, USA, 2002. ACM. ISBN 1-58113-463-0. doi: http://doi.acm.org/10.1145/512529.512558. 12

Robert W. Floyd. Assigning meanings to programs. In J. T. Schwartz, editor, *Proceedings of a Symposium on Applied Mathematics*, volume 19 of *Mathematical Aspects of Computer Science*, pages 19–31, Providence, 1967. American Mathematical Society. URL http://www.eecs.berkeley.edu/~necula/Papers/FloydMeaning.pdf. 12

Riccardo Focardi. Static analysis of authentication. In *FOSAD*, volume 3655 of *Lecture Notes in Computer Science*, pages 109–132. Springer, 2004. ISBN 3-540-28955-0. 14

Cédric Fournet, Markulf Kohlweiss, and Pierre-Yves Strub. Modular code-based cryptographic verification. In *18th ACM Conference on Computer and Communications Security (CCS 2011)*, 2011. 10

Frama-C. http://frama-c.cea.fr/, 2009. 9, 10, 12, 158

Patrice Godefroid, Michael Y. Levin, and David A. Molnar. Automated whitebox fuzz testing. In *Proceedings of the Network and Distributed System Security Symposium, NDSS 2008*. The Internet Society, 2008. 2, 13, 41

S. Goldwasser and S. Micali. Probabilistic encryption. *Journal of Computer and System Sciences*, 28:270–299, 1984. 5

Andrew D Gordon and Alan Jeffrey. Authenticity by typing for security protocols. *Journal of computer security*, 11(4):451–519, 2003. 14

Jean Goubault-Larrecq and Fabrice Parrennes. Cryptographic protocol analysis on real C code. In *Proceedings of the 6th International Conference on Verification, Model Checking and Abstract Interpretation (VMCAI'05)*, volume 3385 of *Lecture Notes in Computer Science*, pages 363–379. Springer, 2005. 15, 143

C. A. R. Hoare. An axiomatic basis for computer programming. *Commun. ACM*, 12(10): 576–580, 1969. 12

Tony Hoare. The verifying compiler: A grand challenge for computing research. *J. ACM*, 50 (1):63–69, 2003. ISSN 0004-5411. doi: http://doi.acm.org/10.1145/602382.602403. 13

Gerard J. Holzmann. The model checker SPIN. *IEEE Trans. Softw. Eng.*, 23(5):279–295, 1997. ISSN 0098-5589. doi: http://dx.doi.org/10.1109/32.588521. 12

ISO. ISO C Standard 1999. Technical report, 1999. URL `http://www.open-std.org/jtc1/sc22/wg14/www/docs/n1124.pdf`. ISO/IEC 9899:1999 draft. 30, 35, 48, 49, 55, 56

Alan Jeffrey and Ruy Ley-Wild. Dynamic model checking of C cryptographic protocol implementations. In *Proceedings of Workshop on Foundations of Computer Security and Automated Reasoning for Security Protocol Analysis*, 2006. 16

J. Jürjens. Security analysis of crypto-based Java programs using automated theorem provers. In *ASE '06: Proceedings of the 21st IEEE/ACM International Conference on Automated Software Engineering*, pages 167–176. IEEE Computer Society, 2006. ISBN 0-7695-2579-2. doi: http://dx.doi.org/10.1109/ASE.2006.60. 15

Jan Jürjens. Code security analysis of a biometric authentication system using automated theorem provers. In *Computer Security Applications Conference, 21st Annual*, pages 10–pp. IEEE, 2005a. 16

Jan Jürjens. Verification of low-level crypto-protocol implementations using automated theorem proving. In *Proceedings of the 2nd ACM/IEEE International Conference on Formal Methods and Models for Co-Design*, pages 89–98. IEEE Computer Society, 2005b. 16

Jan Jürjens and Mark Yampolskiy. Code security analysis with assertions. In *Proceedings of the 20th IEEE/ACM international Conference on Automated software engineering*, pages 392–395. ACM, 2005. 16

James C. King. Symbolic execution and program testing. *Commun. ACM*, 19(7):385–394, 1976. 2, 13

James Cornelius King. *A program verifier*. PhD thesis, Pittsburgh, PA, USA, 1970. 12

Paul C. Kocher. Timing attacks on implementations of diffie-hellman, RSA, DSS, and other systems. In *Proceedings of the 16th Annual International Cryptology Conference on Advances in Cryptology*, CRYPTO '96, pages 104–113, London, UK, UK, 1996. Springer-Verlag. ISBN 3-540-61512-1. URL `http://dl.acm.org/citation.cfm?id=646761.706156`. 158

R. Küsters and T. Truderung. Reducing protocol analysis with XOR to the XOR-free case in the horn theory based approach. *Journal of Automated Reasoning*, 46(3):325–352, 2011. 6

Ralf Küsters, Tomasz Truderung, and Jürgen Graf. A Framework for the Cryptographic Verification of Java-like Programs. In *IEEE Computer Security Foundations Symposium, CSF 2012*. IEEE Computer Society, 2012. 9, 15, 74

Ralf Küsters, Tomasz Truderung, Bernhard Beckert, Daniel Bruns, Jürgen Graf, and Christoph Scheben. A hybrid approach for proving noninterference and applications to the cryptographic verification of java programs. *Grande Region Security and Reliability Day*, 2013. 15

Ralf Küsters, Enrico Scapin, Tomasz Truderung, and Jürgen Graf. Extending and applying a framework for the cryptographic verification of java programs. In *Principles of Security and Trust*, pages 220–239. Springer, 2014. 15

Chris Lattner. LLVM: An Infrastructure for Multi-Stage Optimization. Master's thesis, Computer Science Dept., University of Illinois at Urbana-Champaign, 2002. 10, 42

Xavier Leroy. A formally verified compiler back-end. *J. Autom. Reason.*, 43:363–446, December 2009. ISSN 0168-7433. doi: 10.1007/s10817-009-9155-4. URL http://portal.acm.org/citation.cfm?id=1666192.1666216. 42

Zhiqiang Lin, Xuxian Jiang, Dongyan Xu, and Xiangyu Zhang. Automatic protocol format reverse engineering through context-aware monitored execution. In *Proceedings of the Network and Distributed System Security Symposium, NDSS 2008, San Diego, California, USA, 10th February - 13th February 2008*. The Internet Society, 2008. 17

List of Tools. Static source code analysis tools for C. http://spinroot.com/static/, 2009. 12

Gavin Lowe. An attack on the Needham-Schroeder public-key authentication protocol. *Inf. Process. Lett.*, 56:131–133, November 1995. ISSN 0020-0190. doi: 10.1016/0020-0190(95)00144-2. URL http://portal.acm.org/citation.cfm?id=219887.219895. 1, 157

The Coq development team. *The Coq proof assistant reference manual*. LogiCal Project, 2004. URL http://coq.inria.fr. Version 8.0. 155

John McCarthy. A basis for a mathematical theory of computation. In *Computer Programming and Formal Systems*, pages 33–70. North-Holland, 1963. 12

George C. Necula, Scott McPeak, Shree Prakash Rahul, and Westley Weimer. CIL: Intermediate Language and Tools for Analysis and Transformation of C Programs. In *Proceedings of the 11th International Conference on Compiler Construction*, CC '02, pages 213–228, London, UK, 2002. Springer-Verlag. ISBN 3-540-43369-4. URL http://portal.acm.org/citation.cfm?id=647478.727796. 8, 10, 42

Roger M. Needham and Michael D. Schroeder. Using encryption for authentication in large networks of computers. *Commun. ACM*, 21(12):993–999, 1978. 13

Nicholas O'Shea. Using Elyjah to analyse Java implementations of cryptographic protocols. http://homepages.inf.ed.ac.uk/s0237477/papers.html, 2008. 15

Kenneth G. Paterson and Gaven J. Watson. Plaintext-dependent decryption: A formal security treatment of ssh-ctr. Cryptology ePrint Archive, Report 2010/095, 2010. http://eprint.iacr.org/. 3

L. C. Paulson. Mechanized proofs for a recursive authentication protocol. In *CSFW '97: Proceedings of the 10th IEEE workshop on Computer Security Foundations*, page 84. IEEE Computer Society, 1997. ISBN 0-8186-7990-5. 17

Lawrence C. Paulson. *Isabelle - A Generic Theorem Prover (with a contribution by T. Nipkow)*, volume 828 of *Lecture Notes in Computer Science*. Springer, 1994. ISBN 3-540-58244-4. 155

Lawrence C. Paulson. Inductive analysis of the internet protocol TLS. *ACM Trans. Inf. Syst. Secur.*, 2(3):332–351, 1999. 17

Birgit Pfitzmann, Matthias Schunter, and Michael Waidner. Cryptographic security of reactive systems (extended abstract). In *Electronic Notes in Theoretical Computer Science*. Electronic, 2000. 14

Project EVA. Security protocols open repository, 2007. http://www.lsv.ens-cachan.fr/spore/. 10

Jean-Pierre Queille and Joseph Sifakis. Specification and verification of concurrent systems in CESAR. In *Proceedings of the 5th Colloquium on International Symposium on Programming*, pages 337–351, London, UK, 1982. Springer-Verlag. ISBN 3-540-11494-7. 12

Alfredo Rial and George Danezis. Privacy-friendly smart metering. Technical Report MSR–TR–2010–150, Microsoft Research, 2010. 5, 143

David A. Schmidt. Data flow analysis is model checking of abstract interpretations. In *POPL '98: Proceedings of the 25th ACM SIGPLAN-SIGACT symposium on Principles of programming languages*, pages 38–48, New York, NY, USA, 1998. ACM. ISBN 0-89791-979-3. doi: http://doi.acm.org/10.1145/268946.268950. 13

Nikhil Swamy, Juan Chen, Cédric Fournet, Karthikeyan Bharagavan, and Jean Yang. Security programming with refinement types and mobile proofs. Technical Report MSR–TR–2010–149, 2010. 144

Alan M. Turing. Checking a large routine. In *Report on a Conference on High Speed Automatic Computation, June 1949*, pages 67–69, Cambridge, UK, 1949. University Mathematical Laboratory, Cambridge University. URL http://www.turingarchive.org/browse.php/B/8. A corrected version is printed by F.L. Morris and C.B. Jones in *Annals of the History of Computing*, (Vol. 6, Apr. 1984). 12

Octavian Udrea, Cristian Lumezanu, and Jeffrey S. Foster. Rule-based static analysis of network protocol implementations. *Inf. Comput.*, 206(2-4):130–157, 2008. ISSN 0890-5401. doi: http://dx.doi.org/10.1016/j.ic.2007.05.007. 17

Dominique Unruh. The impossibility of computationally sound XOR, July 2010. Preprint on IACR ePrint 2010/389. 6, 141

Zhi Wang, Xuxian Jiang, Weidong Cui, Xinyuan Wang, and Mike Grace. Reformat: Automatic reverse engineering of encrypted messages. In *ESORICS*, volume 5789 of *Lecture Notes in Computer Science*, pages 200–215. Springer, 2009. ISBN 978-3-642-04443-4. 17

Christoph Weidenbach, Uwe Brahm, Thomas Hillenbrand, Enno Keen, Christian Theobald, and Dalibor Topić. Spass version 2.0. In *Automated DeductionCADE-18*, pages 275–279. Springer, 2002. 16

Gilbert Wondracek, Paolo Milani Comparetti, Christopher Kruegel, and Engin Kirda. Automatic Network Protocol Analysis. In *15th Symposium on Network and Distributed System Security (NDSS)*, 2008. 17

Thomas Y. C. Woo and Simon S. Lam. A semantic model for authentication protocols. In *SP '93: Proceedings of the 1993 IEEE Symposium on Security and Privacy*, page 178. IEEE Computer Society, 1993. 14

Andrew C. Yao. Theory and application of trapdoor functions. In *SFCS '82: Proceedings of the 23rd Annual Symposium on Foundations of Computer Science*, pages 80–91. IEEE Computer Society, 1982. doi: http://dx.doi.org/10.1109/SFCS.1982.95. 5

T. Ylonen and C. Lonvick. The Secure Shell (SSH) Protocol Architecture. RFC 4251 (Proposed Standard), January 2006. URL http://www.ietf.org/rfc/rfc4251.txt. 1

# Notation

| | | |
|---|---|---|
| $[\![e]\!]_\eta$ | Evaluation of an IML expression $e$ w.r.t an environment $\eta$ | 20 |
| $[\![e]\!]^\pi_{I,\eta}$ | Evaluation of a ProVerif expression $e$ w.r.t an environment $\eta$ and interpretation $I$ | 120 |
| $b\|b'$ | Concatenation of bitstrings $b$ and $b'$ | 20 |
| $b\{n_o, n_l\}$ | Substring of a bitstring $b$ starting at offset $n_o$ of length $n_l$ | 20 |
| $\tau(b)$ | Integer value of a bitstring $b$ w.r.t. an integer type $\tau$ | 20 |
| $\tau^{-1}(n)$ | Bitstring encoding of an integer $n$ w.r.t. an integer type $\tau$ | 20 |
| $\mathrm{insec}(Q, \rho)$ | Insecurity of an IML process $Q$ w.r.t a trace property $\rho$ | 25 |
| $\mathrm{cvinsec}(Q, I, \Gamma, \rho)$ | Insecurity of a CryptoVerif process $Q$ w.r.t. a trace property $\rho$, an interpretation $I$, and a typing environment $\Gamma$ | 78 |
| $Q \lesssim \tilde{Q}$ | Simulation relation between $\mathrm{IML}_C$ processes | 37 |
| $\eta \models \psi$ | A fact $\psi$ is true w.r.t. an environment $\eta$ | 49 |
| $\psi \models \psi'$ | A fact $\psi$ implies a fact $\psi'$ | 49 |
| $\psi \vdash e \rightsquigarrow \tilde{e}$ | Solver rewriting rule | 50 |
| $I, f \models \Gamma$ | A function $f$ is type-safe w.r.t. a typing environment $\Gamma$ | 75 |
| $I, f \models_\perp \Gamma$ | A function $f$ is type-safe up to failure w.r.t. a typing environment $\Gamma$ | 127 |

# Index